\documentclass{ws-ijmpd}
\usepackage[super,compress]{cite}
\usepackage{url}
\usepackage{amsbsy}
\usepackage{amssymb}
\usepackage{amsmath}
\usepackage{graphicx}
\usepackage{float}
\usepackage{color}
\usepackage{colortbl}
\usepackage{hyperref}
\usepackage{braket}

\def\g{\text{g}}
\def\km{\text{km}}
\def\cm{\text{cm}}
\def\G{\text{G}}
\def\K{\text{K}}
\def\sec{\text{s}}

\def\ms{\text{ms}}

\def\rs{\rm s}
\def\nm{\text{nm}}
\def\m{\text{m}}

\def\rp{\text{p}}
\def\rn{\text{n}}
\def\rp{\text{p}}
\def\re{\text{e}}
\def\II{_{\text{\tiny{II}}}}
\def\mf{_{\rm mf}}
\def\NF{_{\text{\tiny{N}}}}
\def\SF{_{\text{\tiny{S}}}}
\def\CB{\mathcal{B}}
\def\cR{\mathcal{R}}
\def\cN{\mathcal{N}}
\def\v{_{\rm v}}
\def\rL{\text{L}}
\def\B{_{\rm B}}

\def\uO{\hat{\boldsymbol{\Omega}}}
\def\uo{\hat{\boldsymbol{\omega}}}
\def\uk{\hat{\boldsymbol{\kappa}}}

\def\n{{\rm n}}
\def\p{{\rm p}}
\def\e{{\rm e}}
\def\s{{\rm s}}

\def\bB{\bar{\mathcal{B}}}

\def\p{_{\rm p}}
\def\n{_{\rm n}}

\def\s{_{\rm s}}
\def\v{_{\rm v}}
\def\D{_{\rm D}}

\def\B{_{\rm B}}
\def\d{{\rm d}}

\def\e{_{\rm e}}

\def\cl{_{\rm c}}

\def\NF{_{\text{\tiny{N}}}}
\def\SF{_{\text{\tiny{S}}}}

\def\NS{_{\text{\tiny{NS}}}}

\def\SC{_{\text{\tiny{S}}}}
\def\NO{_{\text{\tiny{N}}}}
\def\NOO{_{\text{\tiny{N0}}}}
\def\SOO{_{\text{\tiny{S0}}}}

\def\F{_{\rm F}}
\def\ft{_{\rm ft}}

\def\II{_{\text{\tiny{II}}}}

\def\rx{\text{x}}
\def\ry{\text{y}}
\def\rz{\text{z}}

\def\rn{\text{n}}
\def\rp{\text{p}}
\def\re{\text{e}}

\def\rv{\text{v}}

\def\rL{\text{L}}

\def\rd{\text{d}}

\def\mf{_{\rm mf}}

\def\ut{\hat{\boldsymbol{\theta}}}
\def\uz{\hat{\boldsymbol{z}}}
\def\ux{\hat{\boldsymbol{x}}}
\def\uy{\hat{\boldsymbol{y}}}
\def\uO{\hat{\boldsymbol{\Omega}}}
\def\uo{\hat{\boldsymbol{\omega}}}
\def\uk{\hat{\boldsymbol{\kappa}}}

\def\bB{\bar{B}}
\def\Br{\mathbf{r}}

\def\CB{\mathcal{B}}

\def\cN{\mathcal{N}}
\def\cR{\mathcal{R}}

\def\en{\varepsilon_\rn}
\def\ep{\varepsilon_\rp}

\def\cE{\mathcal{E}}

\def\Br{\mathbf{r}}

\def\Bv{\mathbf{v}}

\def\BA{\mathbf{A}}
\def\Bj{\mathbf{j}}

\begin{document}

\markboth{V. Graber et al.}
{Neutron Stars in the Laboratory}

%
\catchline{}{}{}{}{}
%

\title{NEUTRON STARS IN THE LABORATORY}

\author{VANESSA GRABER$^{1,2}$}
\address{vanessa.graber@soton.ac.uk}

\author{NILS ANDERSSON$^{1}$}
\address{N.A.Andersson@soton.ac.uk}

\author{MICHAEL HOGG$^{1}$ \vspace{0.5cm}}

\address{$^{1}$Mathematical Sciences and STAG Research Centre, University of Southampton \\ Southampton SO17 1BJ, United Kingdom \vspace{-0.3cm}}

\address{$^{2}$Department of Physics and McGill Space Institute, McGill University, \\ 3550 rue University, Montreal, QC, H3A 2T8, Canada}

\maketitle

\begin{history}
\received{Day Month Year}
\revised{Day Month Year}
\end{history}

\begin{abstract}
Neutron stars are astrophysical laboratories of many extremes of physics. Their rich phenomenology provides insights into the state and composition
of matter at densities which cannot be reached in terrestrial experiments. Since the core of a mature neutron star is expected to be dominated by
superfluid and superconducting components, observations also probe the dynamics of large-scale quantum condensates. The testing and understanding of
the relevant theory tends to focus on the interface between the astrophysics phenomenology and nuclear physics. The connections with low-temperature
experiments tend to be ignored. However, there has been dramatic progress in understanding laboratory condensates (from the different phases of
superfluid helium to the entire range of superconductors and cold atom condensates). In this review, we provide an overview of these developments,
compare and contrast the mathematical descriptions of laboratory condensates and neutron stars and summarise the current experimental state-of-the-art.
This discussion suggests novel ways that we may make progress in understanding neutron star physics using low-temperature laboratory experiments.
\end{abstract}


\ccode{PACS numbers:}

\tableofcontents


\section{Neutron Stars and Fundamental Physics}

A neutron star is born in the collapsing core of a supernova explosion -- a violent cosmic furnace that reaches a temperature more than
10,000 times that of the Sun's core -- that signals the end of a heavy star's life. The object that emerges as the dust settles challenges
our understanding of many extremes of physics, since matter has been compressed to densities and pressures far beyond our everyday experience,
a super-strong magnetic field has organised itself and the star's core has started cooling towards exotic superfluid and superconducting
states.

\begin{figure}[t]
\begin{center}
\includegraphics[width=13cm]{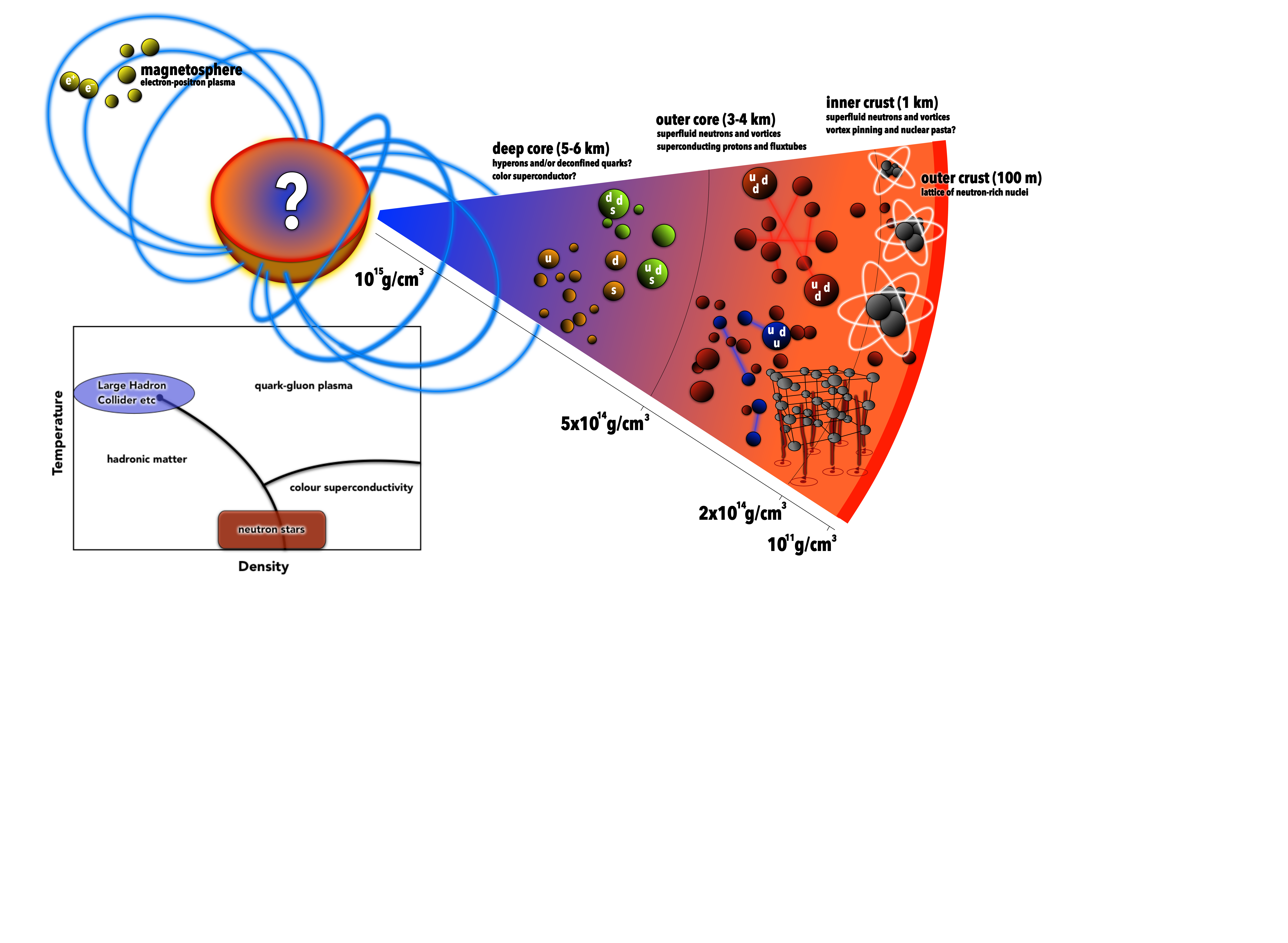}
\caption{Inset: Sketch of the QCD phase diagram. While colliders like the LHC at CERN and RHIC at Brookhaven probe matter at extremely high
temperatures but relatively low densities, neutron star physics relies on the complementary low-temperature, high-density regime for highly
asymmetric matter; a regime that can not be tested by terrestrial experiments. Main image: Schematics of the neutron star interior. The outer
layers consist of an elastic lattice of neutron-rich nuclei. Beyond the `neutron drip' free neutrons form a superfluid that coexists with the
lattice. Above about half the nuclear saturation density, the protons form a liquid and should be in a superconducting state. The composition
and state of matter in the deep core are not well constrained. The stabilising effect of gravitational confinement permits (slow) weak
interactions (such as electron captures) to reach equilibrium, generating matter that is neutron-rich and may have net strangeness. Hyperons
are likely to be present, ultimately giving way to deconfined quarks and a possible colour superconductor. Processes in the pair plasma in the
star's magnetosphere give rise to the lighthouse effect observed by radio telescopes. X-rays associated with explosive nuclear burning on - and
thermal emission from - the star's surface can be used to infer the properties of the outer layers as well as the internal temperature, which
is governed by a range of neutrino processes. Hotspots on the surface of the star star can lead to observable pulsations. Large-scale fluid
motion in the dense interior may generate detectable gravitational waves.}
\label{cross}
\end{center}
\end{figure}

Neutron stars provide a unique exploration space for fundamental physics. The stabilising effect of gravity permits long-timescale weak
interactions (such as electron captures) to reach equilibrium, generating matter that is neutron-rich and which may have net strangeness.
In effect, neutron stars allow us to probe unique states of matter that cannot be created on Earth: nuclear superfluids and strange matter
states such as hyperons, deconfined quarks, and possible colour superconducting phases (see Fig.~\ref{cross} for a schematic illustration).
For these hands-off laboratories progress is made by matching observational data to quantitative theory. Given the variety of
observed phenomena and the fact that neutron stars come in many guises, our current understanding is limited by small-number statistics
and uncertain systematics. However, breakthroughs are anticipated as a new generation of revolutionary telescopes -- e.g.\ Advanced LIGO
for gravitational waves and the Square Kilometer Array (SKA) for radio observations -- comes into operation and reaches design sensitivity.
This is tremendously exciting but, if we want to realise the full  potential of these instruments, we need to make urgent progress on the
corresponding theory. Given the scope of the physics involved, this is  challenging.

One of the main challenges involves the composition and state of matter at densities that can not be reached in terrestrial experiments.
The fundamental interactions that govern matter at extreme densities remain poorly constrained by first principles quantum calculations
(see Drischler et al.~\cite{chiral} for the state-of-the-art).
Each theoretical model leads to a distinct pressure-density-temperature relation for bulk matter (the so-called equation of state), which
in turn generates a unique neutron star mass-radius relation, predicting a characteristic radius for a range of masses and a maximum mass
above which a neutron star collapses to a black hole. The equation of state also uniquely predicts quantities like the maximum spin rate
and moment of inertia. Thus, observational constraints on the equation of state can be used to infer key aspects of microphysics (see
recent discussions concerning the expected capability of LOFT~\cite{loft} and SKA~\cite{ska}), such as the nature of the three-nucleon
interaction or the presence of free quarks at high densities. Determining the equation of state at supranuclear densities is one of the
major challenges for fundamental physics. This information is also key for astrophysics, as the equation of state affects binary merger
dynamics, the timescale for black-hole formation, precise gravitational-wave and neutrino signals, potential gamma-ray burst signatures,
any associated mass loss and r-process nucleosynthesis.

These different aspects ensure that nuclear physicists keep a keen eye on developments in neutron-star astrophysics. Basically, neutron stars
represent a regime that can never be tested by laboratory experiments. Collider experiments like the LHC at CERN and RHIC at Brookhaven probe
matter at extremely high temperatures but relatively low densities, while neutron star physics relies on the complementary low-temperature,
high-density regime for highly asymmetric matter, cf. Fig.~\ref{cross}.

The most accurately measured neutron star parameter (by some margin) is the spin. We have accurate timing solutions for over 2,300 (mainly radio)
pulsars, providing a handle on the strength of the exterior magnetic field, and the star's moment of inertia, through the observed spin-down rate.
However, the fastest observed spin (716 Hz) does not significantly constrain the equation of state other than ruling out unrealistically soft
models. The strongest current constraints on nuclear physics come from binary systems, where orbital parameters may allow an accurate determination
of the neutron star mass~\cite{lat2012}. The currently observed maximum mass (just over two solar masses) constrains the stiffness of the equation
of state, and provides a hint that hyperons (which would have a softening effect) may not dominate the star's core. The neutron star radius is much
more difficult to infer from observations. The most promising results are associated with X-ray emission from the surface of accreting neutron
stars in binary systems and associated burst phenomena, involving explosive burning of accreting material. Although complex systematics need to
be understood (including the composition of the neutron star atmosphere), these mass-radius results are beginning to constrain the theory
(essentially limiting the value of the nuclear symmetry energy and its derivative at the saturation density)~\cite{Lattimer2014}.

While nuclear two-body interactions are well constrained by laboratory experiments, three-body forces represent the frontier of nuclear physics.
At low energies, effective field theory models provide a systematic expansion of the forces involved. Complementary efforts using lattice approaches
to the nuclear forces remain affected by large uncertainties. For example, the appearance of shell closure in neutron-rich isotopes and the position
of the neutron drip-line are sensitive to three-body forces. Exotic neutron-rich nuclei, the focus of present and upcoming experiments, provide
interesting constraints on effective interactions for many-body systems. Nonetheless, the scope of these laboratory systems is limited. While
nuclear masses and their charge radii probe symmetric nuclear matter, the neutron skin thickness of lead tests neutron-rich matter, and giant dipole
resonances and dipole polarisabilities of nuclei also concern largely symmetric matter, all of these laboratory techniques probe only matter at
densities lower than $3\times10^{14} \, \g \, \cm^{-3}$. Neutron stars can reach densities several times higher. For recent discussions of
laboratory constraints on the nuclear symmetry energy, which is known to govern the stiffness of the equation of state at high densities, see Hebeler
et al.~\cite{hebel} and Lattimer~\cite{lat2014}.

At the present time, the state-of-the-art equations of state used in astrophysical models are to a large extent phenomenological. Different approaches
include nuclear potentials (e.g.\ the Urbana/Illinois or Argonne forces) that fit two-body scattering data and light nuclei properties, phenomenological
forces like the Skyrme interaction and microscopic nuclear Hamiltonians that include two- and three-body forces from chiral effective field theories (see
Drischler et al.~\cite{chiral} for recent progress). The challenge for astrophysics modelling in this area is to i) incorporate as much of the predicted
microphysics as possible, and ii) use observations to constrain the unknown aspects.

\begin{figure}[t]
\begin{center}
\includegraphics[width=3.8in]{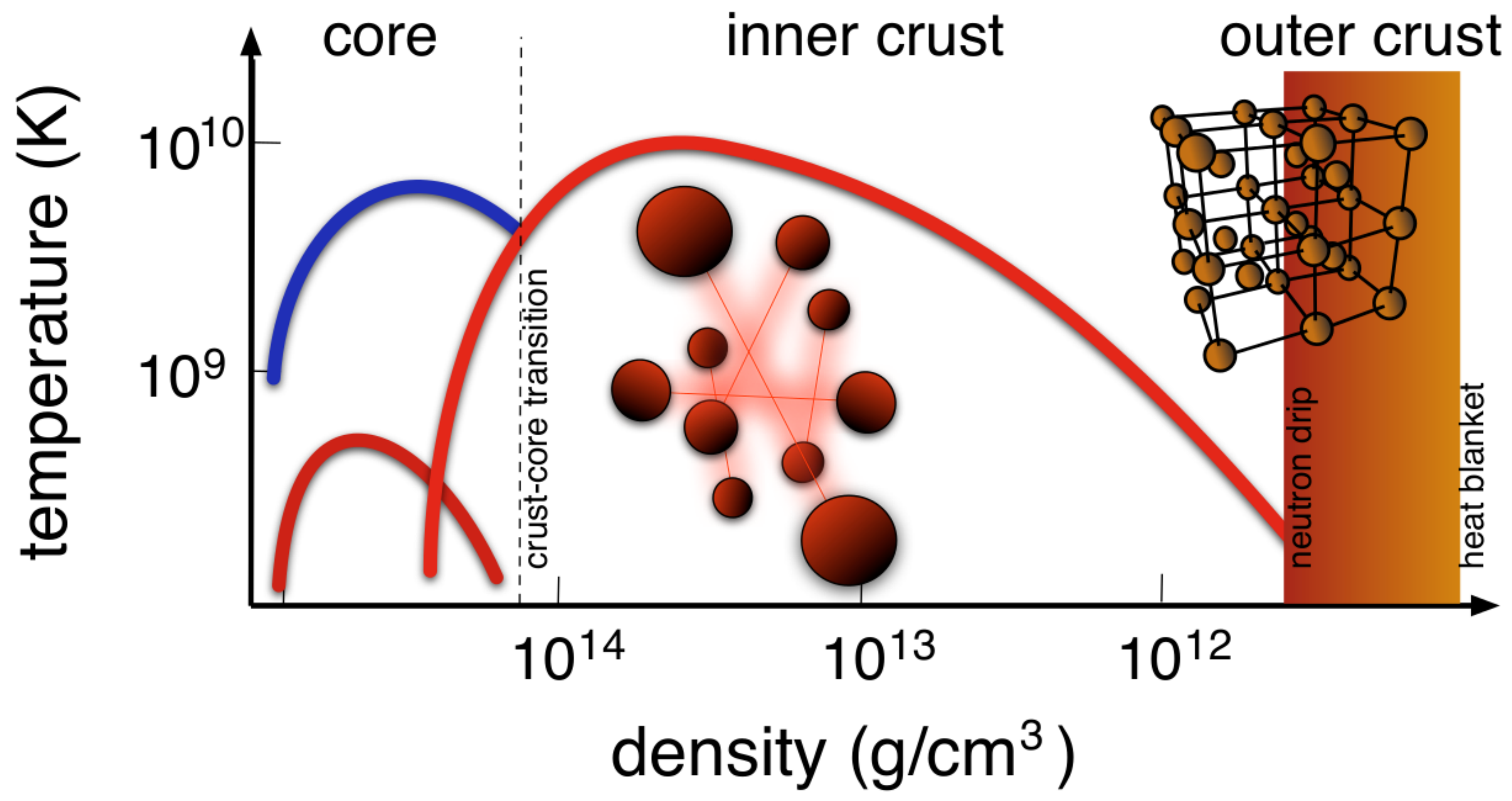}
\caption{Schematic illustration of the critical temperatures for superfluidity of neutrons in both singlet (red, crust) and triplet (red, core) pairing
states and superconducting protons in the singlet state (blue, core). Since neutron stars are well below the Fermi temperatures for the involved
constituents, their cores are expected to be dominated by superfluid and superconducting components. The density-dependent critical temperatures can
be constrained by neutron star cooling data. They also impact on restlessness in the star's spin-down and the enigmatic spin-glitches that are seen in
(predominantly) young pulsars. The glitches provide information on the mobility of superfluid components (the so-called entrainment effect) and the
potential pinning of vortices to nuclei in the star's crust.}
\label{sfgaps}
\end{center}
\end{figure}

The state of matter adds more dimensions to the problem. Mature neutron stars tend to be cold (far below the Fermi temperature of the involved constituents),
making the formation of various superfluid or superconducting phases likely throughout the star's core. The respective parameters (e.g.\ the energy gaps
for Cooper pair formation) have a key influence on the star's long-term dynamics (see Fig.~\ref{sfgaps}), making it much more difficult to track its evolution.
For example, the braking index (essentially the second derivative of the spin-frequency) has only been measured in about a dozen systems and it is not yet
clear whether the diverse results can be explained by magnetic field evolution, the decoupling of an interior superfluid component or other factors. The
problem is complicated by the fact that the spin-evolution is intimately linked to the gradual cooling and the evolution of the magnetic field.

In essence, neutron stars involve a rich palate of exciting physics. In the context of testing and understanding the involved theory, the discussion
usually concerns the interface between the astrophysics phenomenology and nuclear physics. The connections with low-temperature experiments tend to be
ignored. This is somewhat surprising given the dramatic progress in understanding laboratory condensates (from the different phases of superfluid helium
to the entire range of superconductors and cold atom condensates). This apparent gap in the literature provides the motivation for this review. Our aim
is to introduce the key problems, describe the current state-of-the-art and suggest various ways that we may make progress in understanding neutron star
physics using laboratory experiments.


\section{Historical Context: Low-temperature Physics}

As superfluid and superconducting components are expected to be present in a neutron star's interior, we need to understand
these states of matter and how their presence may impact on observations. The former characterises matter behaving like a
fluid with zero viscosity, while the latter describes a state of vanishing electrical resistance accompanied by the expulsion
of magnetic flux. Similarities between these two phases are evident, since both are capable of maintaining particle currents
at constant velocities without any forces being applied. These currents involve large numbers of particles condensed
into the same quantum state. Therefore, superfluidity and superconductivity are characterised as macroscopic quantum phenomena,
closely related to the concept of Bose-Einstein condensation. Understanding their formation and properties is crucial for
developing more realistic neutron star models. Before addressing the mathematical treatment of laboratory condensates in more
detail, we present a brief overview of the research in \textit{low-temperature condensed matter physics} to provide a historical
context for the discussion.

\subsection{Superconductivity}

Superconductivity was discovered by Heike Kamerlingh Onnes in Leiden in 1911~\cite{Onnes1911}, only three years after he had
succeeded in liquefying helium. Cooling several metals such as mercury to low temperatures, Onnes observed that their electrical
resistance disappeared completely. For his ground-breaking work on low-temperature physics and condensed matter he was awarded
the Nobel Prize in Physics in 1913. Two decades later in 1933, Walther Meissner and Robert Ochsenfeld showed that
superconductivity was a unique new thermodynamical state and not just a manifestation of infinite conductivity. Cooling lead
samples in the presence of a magnetic field below their superconducting transition temperature, they expelled the magnetic
field and exhibited perfect diamagnetism~\cite{Meissner1933}. This mechanism is now known as the \textit{Meissner-Ochsenfeld}
effect and the associated phenomenology was first described by the London brothers in 1935~\cite{London1935}. Fritz London
himself developed a semi-classical explanation for the so-called London equations several years later~\cite{London1948} and
was the first one to point out that the quantum nature of particles could play an important role in the superconducting phase transition.

Developed in the 1920s, quantum mechanics has significantly influenced the way scientists interpret the world. This is most
apparent in the definition of an abstract wave function that allows a probabilistic interpretation of physical quantities
such as the momentum and the position of a particle. This wave function is a complex quantity and it was the success of
microscopic theories to relate its properties to the quantum mechanical condensate; the amplitude of the wave function is
directly related to the density of the superconducting particles and the phase is proportional to the superconducting current.
The first microscopic description of superconductivity was developed by John Bardeen, Leon Cooper and John Robert Schrieffer
in 1957~\cite{Bardeen1957}. Their BCS theory is based on the concept of pairing that results from an attractive potential.
Below a critical temperature, the weak attractive interactions between electrons and the lattice in an ordinary metal are
strong enough to overcome the repulsive Coulomb force. This causes the electrons to form Cooper pairs that obey Bose-Einstein
statistics and can condense into a quantum mechanical ground state. BCS theory explained the existence of a temperature-dependent
energy gap (half the energy necessary to break a Cooper pair) and hence the presence of critical quantities above which
superconductivity is destroyed. For their theory, which has also been successfully applied to anisotropic superfluids (see below),
Bardeen, Cooper and Schrieffer obtained the Nobel Prize in 1972.

A second approach to superconductivity that has proven successful to describe the properties of the condensate close to
the transition is the Ginzburg-Landau theory. Formulated in 1950 by Vitaly Ginzburg and Lev Landau~\cite{Ginzburg1950},
it phenomenologically describes a second-order phase transition by means of an order parameter. In the case of
superconductivity, this quantity can be identified with the microscopic electron Cooper pair density. The phase
transition itself is then interpreted as a symmetry breaking, because the density of superconducting pairs changes
drastically at the transition point. Ginzburg and Landau postulated that close to the critical temperature the free
energy of the system could be written as an expansion of the order parameter. Minimising the energy with respect to
the order parameter and the electromagnetic vector potential, one arrives at the Ginzburg-Landau equations. These
introduce two typical lengthscales for superconductivity; the penetration depth, $\lambda$, characterising the scale
for magnetic field suppression and the coherence length, $\xi$, representing the distance over which the order parameter
changes in space (in BCS theory this is equivalent to the dimension of a Cooper pair). The ratio of these two is commonly
referred to as the \textit{Ginzburg-Landau parameter}, $\kappa_{\rm GL} \equiv \lambda / \xi$. Although the approach of
the Russian physicists was purely phenomenological and not based on an analysis of the microscopic features of a
superconductor, Lev Gor'kov showed in 1959 that close to the critical temperature it is possible to derive the Ginzburg-Landau
theory from the microscopic BCS theory~\cite{Gorkov1959}. In 1957, Alexei Abrikosov further investigated the order-parameter
approach and predicted the existence of two classes of superconductors~\cite{Abrikosov1957}. He found that matter characterised
by $\kappa_{\rm GL} > 1/\sqrt{2}$ would be penetrated by magnetic fluxtubes, if the applied magnetic field were to exceed
a critical field strength. These fluxtubes contain normal matter that is screened by circular currents from the surrounding
superconducting material. Up to that point, this state had been considered unphysical, since only $\kappa_{\rm GL} < 1/\sqrt{2}$
superconductors were known. Abrikosov gave this new phase the name type-II superconductor and calculated that the fluxtubes
would arrange themselves in a regular lattice structure (see Sec.~\ref{subsubsec-Abrikosov} for more details). Abrikosov and
Landau were two of the three physicists who obtained the Nobel Prize in Physics in 2003 for their contributions to the modelling
of superconductors.

In the last four decades, superconductors have found increasing commercial success ranging from sensitive
magnetometers based on the Josephson effect~\cite{Josephson1962} (the quantum mechanical tunnelling of Cooper
pairs across a normal barrier between two superconducting wires) to high-field electromagnets. Especially, the
discovery of high-temperature superconductivity in ceramics at around $100 \, \K$ has fuelled new research
efforts~\cite{Bednorz1986}. For their findings in 1986, Georg Bednorz and Alexander M\"uller were given the Nobel Prize
in Physics one year later. This type of superconductivity is still not fully understood, but it is an ongoing research
area that has led to revolutionary ideas. One example is the theory of holographic superconductors based on the
duality between gravity and a quantum field theory (AdS/CFT correspondence) that might give new insight into
the behaviour of experimental condensed matter~\cite{Horowitz2011}.

\subsection{Superfluidity}

\begin{figure}[t]
\begin{center}
 \includegraphics[width=3.5in]{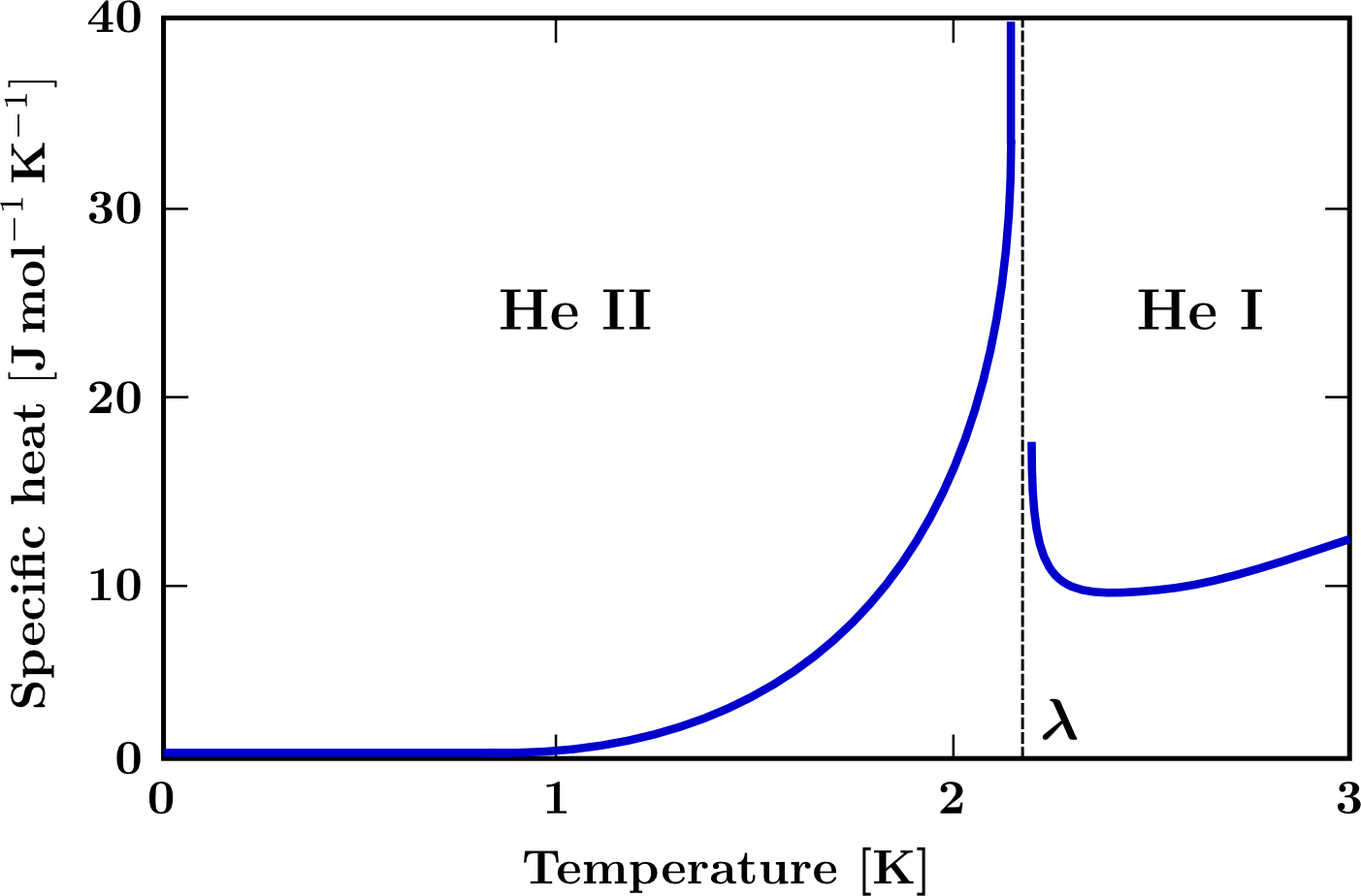}
\caption{Sketch of the specific heat capacity of helium-4 as a function of temperature. At $2.171 \, \K$, the specific
heat changes drastically, marking the superfluid phase transition. Above the Lambda point, helium-4 is usually referred
to as helium I, whereas the superfluid phase is called helium II.}
\label{fig-LambdaSpecificHeat}
\end{center}
\end{figure}

Superfluidity was first observed in liquid helium-4 by Pyotr Kapitsa in Russia~\cite{Kapitza1938} and John Allen and
Don Misener in the United Kingdom in 1937~\cite{Allen1938}. The Soviet physicist was awarded the Nobel Prize in Physics
in 1978 for his experimental findings. Helium forms two stable isotopes, helium-4 and helium-3, that have a relative
abundance of $10^6$:$1$ in the Earth's atmosphere and boiling points at $4.21 \, \K$ and $3.19 \, \K$, respectively.
Right below these temperatures, both isotopes behave like ordinary liquids with small viscosities. However, instead of
solidifying, at $2.171 \, \K$ helium-4 undergoes a transition into a new fluid phase, first detected by Kapitsa, Allen
and Misener as a characteristic change in the specific heat capacity. The observed behaviour resembled the Greek letter
$\lambda$ and the transition temperature was therefore called the \textit{Lambda point} (see Fig.~\ref{fig-LambdaSpecificHeat}).
Above $2.171 \, \K$, helium-4 is named helium I, whereas the superfluid phase is usually referred to as helium II. As
predicted by Lev Pitaevskii~\cite{Pitaevskii1960}, helium-3 also undergoes a superfluid transition at a much lower temperature,
in the mK-regime. To reach such low temperatures, the cooling techniques available in the first half of the twentieth
century were not sufficient and new methods had to be developed. In 1971, more than thirty years after the discovery of
helium II, Douglas Osheroff, Robert Coleman Richardson and David Lee detected two superfluid phases of
helium-3~\cite{Osheroff1972, Osheroff1972a}.

The discovery of superfluid helium-4 stimulated the development of many new experiments and resulted in a lot of theoretical
work analysing the new phase. The first model that was able to explain several observed phenomena was developed by L\'azl\'o
Tisza in 1938~\cite{Tisza1938}. Experiments measuring the viscous drag on a body moving in the superfluid had shown non-viscous
behaviour~\cite{Reppy1964}, while rotation viscometers had revealed viscous characteristics~\cite{Woods1963}. Tisza resolved
this seemingly inconsistent behaviour by introducing a two-fluid interpretation. He assumed that helium II is a mixture of two
physically inseparable fluids, one exhibiting frictionless flow and the other having ordinary viscosity. This phenomenological
approach provided for example an interpretation for the \textit{fountain effect} first observed by Allen Jones in
1938~\cite{Allen1938a} and predicted the existence of \textit{second sound}~\cite{Eselson1974, Khalatnikov2000} (the wave-like
transport of heat).

The two-fluid model was further improved by Lev Landau in the 1940s~\cite{Landau1941}. He put the phenomenological idea
on more solid ground by providing a semi-microphysical explanation that earned him the Nobel Prize in Physics in 1962.
He proposed that a fluid at absolute zero would be in a perfect, frictionless state. Increasing the temperature would then
result in the local excitation of phonons, quantised collisionless sound waves, and additional quasi-particles of higher
momentum and energy that Landau called rotons. Such excitations should behave like an ordinary gas (responsible for the
transport of heat) and form the viscous fluid component, hence providing a basis for the two-fluid model of superfluidity.
His ideas also led Landau to suggest the classic experiment, performed by Elepter Andronikashvili in 1946, that measured the
superfluid fraction of rotating helium II as a function of temperature. It was shown that below $1 \, \K$ almost the entire
sample is in a superfluid state~\cite{Andronikashvili1946}.

Whereas Landau had thought that vorticity entered helium II in sheet-like structures, Lars Onsager and, later independently,
Richard Feynman showed that vorticity enters rotating superfluids in the form of quantised vortices~\cite{Onsager1949,
Feynman1955}. Their ideas are summarised in the Onsager-Feynman quantisation conditions that play a crucial role in deriving
the multi-fluid formalism used to model a neutron star's interior~\cite{Glampedakis2011a}. The problem of rotating superfluid
helium discussed by Onsager and Feynman is equivalent to that of type-II superconductivity in a strong field considered by
Alexei Abrikosov several years later~~\cite{Abrikosov1957}. The first measurement of quantised vortices in rotating helium II
was performed by Henry Hall and William Vinen in 1956~\cite{Hall1956}.

As implied in Landau's interpretation of the two-fluid model, at absolute zero helium II is completely superfluid and
carries no entropy, marking the ground state of the system. Fritz London was the first one to suggest that bosonic helium-4
atoms could become superfluid by Bose-Einstein condensation~\cite{London1938}. This concept had been introduced by
Satyendra Bose and Albert Einstein in 1924 and 1925~\cite{Ketterle2002}. Governed by Bose-Einstein statistics, identical
particles with integer spin such as photons or helium-4 atoms are allowed to share the same quantum state with each other.
At very low temperatures, they tend to occupy the lowest accessible quantum state, resulting in a new phase that is referred
to as a Bose-Einstein condensate (BEC). In the case of superfluid helium II, the Lambda point would then reflect the onset of
this condensation. The original idea of Bose and Einstein was improved by Eugene Gross~\cite{Gross1961} and Lev
Pitaevskii~\cite{Pitaevskii1961} by including interactions of the ground-state bosons. Their work led to the Gross-Pitaevskii
equation that determines the wave function of the condensate and is similar in form to one of the Ginzburg-Landau equations.
London's original proposition gained significant support in 1995, when Carl Wieman and Eric Cornell created the first
atomic Bose-Einstein condensate by cooling a dilute gas of Rubidium-87 atoms to $170 \, \text{nK}$~\cite{Wieman1995}.
Together with Wolfgang Ketterle, whose group created a BEC only a few months later~\cite{Ketterle1995}, Cornell and Wieman
won the Nobel Prize in Physics in 2001. In 1999, the first superfluid transition and the formation of vortices was observed
in a Rubidium-87 boson gas~\cite{Matthews1999, Madison2000}, opening the possibility to study vortex dynamics in such systems.

For helium-3 however, the story is somewhat different, because it is a fermionic particle subject to Pauli's exclusion principle.
Pairing into Cooper pairs is required before any condensation can take place; a mechanism that is similar to the electron pairing
in BCS theory. This explains why fermionic condensates generally appear at lower temperatures than bosonic ones. In contrast
to ordinary superconductivity, the Cooper pairs in helium-3 form in states of non-zero spin and angular momentum so-called
spin-triplet, $p$-wave pairing opposed to spin-singlet, $s$-wave pairing with zero spin and zero angular momentum. This gives
helium-3 an intrinsic anisotropy, resulting in the formation of three different superfluid phases, which are stable under
specific external conditions (see Sec.~\ref{subsec-DifferentIsotopes}). The first Fermi gas analogue of rotating superfluid
helium-3 was observed in 2005 by Zwierlein and collaborators~\cite{Zwierlein2005}.


\section{Modelling Superfluid Flow}
\label{sec-ModellingSF}

Much of the discussion of the dynamics of superfluid systems is  based on the approach taken by Landau to explain the behaviour of
superfluid helium~\cite{Landau1941}. In his original work, Landau assumed that in order to spontaneously excite sound waves such
as phonons or rotons, helium-4 required a flow velocity above a critical value. Landau then showed that these quasi-particles could
move separately from the ground-state particles, which motivated him to combine the excitations to form the \textit{normal}, viscous
fluid. The normal fluid density vanishes at $T=0$ and increases with temperature, ultimately leading to the destruction of superfluidity;
at the Lambda point, the normal fluid density equals the total density and helium is no longer superfluid.

In deriving the two-fluid equations, we start from Landau's model and consider the quantum mechanical condensate at
absolute zero with no viscous counterpart present. Thereafter, the description is extended to account for the second component and
other effects such as vortex formation, mutual friction and turbulence. Finally, the close connection between superfluid helium and
ultra-cold gases is discussed.

\subsection{Wave function and potential flow}
\label{subsec-WaveFunction}

In order to understand the behaviour of the inviscid ground-state component, we draw on the well-known formalism of quantum mechanics. The
condensate at $T=0$ is completely characterised by a single macroscopic wave function. Instead of representing a specific particle,
this wave function is a coherent superposition of all individual superfluid states. The most general case is time- and space-dependent,
\begin{equation}
 	\Psi (\Br , t ) = \Psi_0 (\Br , t ) \,
		\exp \left[ i \varphi (\Br, t ) \right],
		\label{eqn-WaveFunctionPsi}
\end{equation}
where $\Psi_0 (\Br, t )$ and $\varphi(\Br, t )$ are the real amplitude and phase, respectively, and bold symbols denote three
dimensional vectors. The complex wave function, $\Psi(\Br , t )$, is the solution to a Schr\"odinger equation of the form,
\begin{equation}
 	i \hbar \, \frac{\partial \Psi (\Br , t ) }{\partial t} + \frac{\hbar^2}{2 m\cl} \nabla^2 \Psi(\Br , t )
 	- \mu (\Br) \Psi (\Br , t )  =0,
		\label{eqn-Schroedinger}
\end{equation}
with the reduced Planck constant $\hbar$, the fluid's chemical potential $\mu(\Br)$ and the mass $m\cl$ of one bosonic particle that
has condensed into the quantum state. For helium II, $m\cl$ represents the mass of a helium atom, while it equals the mass of a Cooper
pair in the case of a fermionic condensate. The absolute value of the wave function is defined by $|\Psi|^2 \equiv \Psi\Psi^*$, with
$*$ denoting the complex conjugate. Whereas for a single particle wave function, $|\Psi|^2$ denotes the probability of finding this
particle at the point $\Br$ at time $t$, the amplitude of the condensate wave function is related to the number density of bosons
constituting the quantum state, i.e. $ |\Psi (\Br, t )|^2 = |\Psi_0 (\Br, t )|^2 = n\cl (\Br , t )$. Integrating over the volume of
the entire condensate, one thus obtains the total number of indistinguishable particles present in the superfluid ground state at a
specific time, $t$.

A connection between the quantum mechanical description and a hydrodynamical formalism is obtained by substituting the definition of
the wave function~\eqref{eqn-WaveFunctionPsi} into the Schr\"odinger equation~\eqref{eqn-Schroedinger} and separating the resulting
equation into its real and imaginary part. This \textit{Madelung transformation}~\cite{Madelung1927} results in two coupled equations
of motion for the amplitude, $\Psi_0$, and phase, $\varphi$,
\begin{align}
	\hbar \, \frac{\partial \varphi}{\partial t} + \frac{\hbar^2}{2 m\cl} \left(\nabla \varphi\right)^2
		+ \mu - \frac{\hbar^2}{2 m\cl \Psi_0} \, \nabla^2 \Psi_0 &= 0,  \label{eqn-realpart} \\[1.5ex]
	\frac{\partial \Psi_0}{\partial t} + \frac{\hbar}{2 m\cl} \left( 2 \nabla \Psi_0 \cdot \nabla \varphi
		+ \Psi_0 \nabla^2 \varphi\right) &= 0. \label{eqn-imaginarypart}
\end{align}
Multiplying the second equation with $\Psi_0$ and using the chain rule, we arrive at
\begin{equation}
	\frac{\partial |\Psi_0|^2}{\partial t} + \frac{\hbar}{m\cl} \nabla \cdot \left( |\Psi_0|^2  \nabla \varphi \right) = 0.
\end{equation}
This is equivalent to the continuity equation of fluid mechanics, i.e.
\begin{equation}
	\frac{\partial \rho\SF}{\partial t} + \nabla \cdot \mathbf{j}\SF = 0,
		\label{eqn-ContinuitySuper}
\end{equation}
if we substitute the superfluid mass density, $\rho\SF \equiv m\cl n\cl$, and take advantage of the standard definition of the quantum
mechanical momentum density,
\begin{equation}
	 \mathbf{j}\SF = \frac{i \hbar}{2} \left[ \Psi \nabla \Psi^* - \Psi^* \nabla \Psi \right]
		     = \hbar |\Psi_0|^2 \nabla \varphi,
		\label{eqn-Current1}
\end{equation}
where Eqn.~\eqref{eqn-WaveFunctionPsi} has been used to obtain the last equality. We can further identify the momentum density as the
product of the superfluid mass density and a superfluid velocity, i.e. $\mathbf{j}\SF \equiv \rho\SF \mathbf{v}\SF$~\cite{Khalatnikov2000},
which allows us to define the latter as
\begin{equation}
	\mathbf{v}\SF \equiv \frac{\hbar}{m\cl} \, \nabla \varphi.
		\label{eqn-SuperfluidVelocity}
\end{equation}
Note that in addition to this approach of identifying velocities with momenta, it is also possible to treat both variables independently.
While both formalisms are mathematically equivalent and have been employed in the context of neutron star hydrodynamics~\cite{Alpar1984,
Mendell1991, Carter1995a, Prix2004, Andersson2006}, caution is specifically needed when \textit{entrainment}, the non-dissipative
coupling of neutron and proton components in a neutron star's interior, is included (see Sec.~\ref{subsec-chargemultihydro}).

Taking the curl of Eqn.~\eqref{eqn-SuperfluidVelocity}, one finds
\begin{equation}
	\nabla \times \mathbf{v}\SF = 0.
		\label{eqn-PotentialFlow}
\end{equation}
Hence, the condensate is characterised by irrotational, potential flow and the phase of the wave function plays the role of a scalar
velocity potential. As we will see later on, this fundamental property is responsible for the formation of quantised vortex lines in a
rotating superfluid sample.

Moreover, by taking the gradient of Eqn.~\eqref{eqn-realpart}, substituting the superfluid velocity, $\Bv\SF$, and condensate number
density, $n\cl$, and taking the irrotationality into account, we arrive at
\begin{equation}
	\frac{\partial \mathbf{v}\SF}{\partial t} +  \left( \mathbf{v}\SF \cdot \nabla \right) \mathbf{v}\SF
		= - \nabla \tilde{\mu} + \nabla \left( \frac{\hbar^2}{2 m\cl^2 \sqrt{n\cl}} \, \nabla^2 \sqrt{n\cl} \right),
		\label{eqn-EulerFull}
\end{equation}
where $\tilde{\mu} \equiv \mu/m\cl$ is the fluid's specific chemical potential. This equation of motion for the quantum condensate at
$T=0$ resembles the Euler equation of an ideal fluid; the only difference being the second term on the right-hand side. This contribution
reflects the quantum nature of the system and is referred to as the \textit{quantum pressure}. As it captures forces that depend on the
curvature of the amplitude of the wave function, the term is negligible if the spatial variations of $\Psi_0$ occur on large scales,
specifically larger than the coherence length, $\xi$~\cite{Donnelly1991, Pethick2008}. One is then left with the momentum equation for a
perfect fluid;
\begin{equation}
	\frac{\partial \mathbf{v}\SF}{\partial t} +  \left( \mathbf{v}\SF \cdot \nabla \right) \mathbf{v}\SF + \nabla \tilde{\mu} = 0.
		\label{eqn-Euler}
\end{equation}

\subsection{Two-fluid equations}
\label{subsec-TwoFluidModel}

For temperatures $T>0$, the condensate coexists with excitations that constitute the viscous component. Following Landau's model, it is
convenient to continue labelling the hydrodynamical properties of the superfluid component with `S', while the index `N' refers to the
normal part. Any quantities without labels describe parameters of the entire fluid. Assigning a local velocity and density to each of the
constituents of the two-fluid system, the total mass density and mass current density are given by
\begin{align}
	\rho = \rho\NF + \rho\SF, \\[1.5ex]
	\mathbf{j} = \rho\NF \mathbf{v}\NF + \rho\SF \mathbf{v}\SF.
\end{align}
Based on these two relations, the simplest form of the hydrodynamical equations can be derived from conservation laws and the assumption that the
fluid velocities are sufficiently small (for details see for example Roberts and Donnelly~\cite{Roberts1974} or Hills and Roberts~\cite{Hills1977}).
This ensures that dissipation introduced by the viscosity, $\eta$, of the normal fluid and the formation of vortices in the superfluid counterpart
is negligible. Implicitly excluding turbulence makes it possible to treat the fluids individually and neglect any coupling between them.
First of all, the total mass of the sample has to be conserved, leading to the continuity equation
\begin{equation}
	\frac{\partial \rho}{\partial t} + \nabla \cdot \mathbf{j} = 0.
\end{equation}
Additionally assuming that the dissipation mechanisms are weak, every process in the two-fluid system is reversible. This implies that
the entropy per unit mass, $s$, is conserved and results in a second continuity equation. Since entropy and heat are transported by
the normal fluid, we have
\begin{equation}
	\frac{\partial (\rho s)}{\partial t} + \nabla \cdot \left( \rho s \mathbf{v}\NF \right)= 0 ,
\end{equation}
where $\rho s$ is the entropy density and $ \rho s \mathbf{v}\NF $ represents the entropy current density.

In the case of incompressible fluid flow, $\nabla \cdot \mathbf{v} \SF = \nabla \cdot \mathbf{v} \NF =0 $, the conservation of momentum in the entire
system provides a two-fluid Navier-Stokes equation. It can be separated into momentum conservation equations for each individual component by taking
advantage of the Euler equation~\eqref{eqn-Euler}. In the absence of dissipation, the system is in local thermodynamic equilibrium and a small change
in the specific chemical potential is related to changes in the pressure, $p$, and the temperature, $T$, via the Gibbs-Duhem equation, $\d \tilde{\mu}
= \rho^{-1} \d p - s \d T$. Using this relation, one finds~\cite{Roberts1974}:
\begin{align}
	\rho\SF \left[ \frac{\partial \mathbf{v}\SF}{\partial t} +  \left( \mathbf{v}\SF \cdot \nabla \right) \mathbf{v}\SF  \right]
		+ \frac{\rho\SF}{\rho} \nabla p - \rho\SF s \nabla T &=0, \label{eqn-Euler2Fluid} \\[1.5ex]
	\rho\NF \left[ \frac{\partial \mathbf{v}\NF}{\partial t} +  \left( \mathbf{v}\NF \cdot \nabla \right) \mathbf{v}\NF  \right]
		+ \frac{\rho\NF}{\rho} \nabla p + \rho\SF s \nabla T - \eta \nabla^2 \mathbf{v}\NF & = 0.
		\label{eqn-NavierStokes2Fluid}
\end{align}
The former is the Euler equation characterising the fluid fraction condensed into the ground state. At low temperatures, due to existence
of discrete quantum levels, this component cannot exchange energy with the environment and is responsible for the inviscid, frictionless
behaviour of the fluid. For $\rho = \rho\SF$ at absolute zero, Eqn.~\eqref{eqn-Euler2Fluid} is equivalent to Eqn.~\eqref{eqn-Euler}. Finally,
Eqn.~\eqref{eqn-NavierStokes2Fluid} is the equation of motion for the normal constituent. It is composed of all elementary excitations and
has properties similar to that of a classical Navier-Stokes fluid with viscosity $\eta$.

Before turning to the more complicated dynamics of rotating condensates, we clarify that the two-fluid model is a mathematical idealisation.
In reality, the two components are physically inseparable and atoms cannot be designated as belonging to either one of them.

\subsection{Characteristics of a rotating superfluid}
\label{subsec-RotatingSF}

Considering a normal fluid inside a rotating vessel, the motion is characterised by rigid-body behaviour, where the velocity, $\mathbf{v}$, in
the inertial frame is given by
\begin{equation}
	\mathbf{v} =\boldsymbol{\Omega} \times \mathbf{r}.
		\label{eqn-RigidBody}
\end{equation}
Here, $\boldsymbol{\Omega}$ is the container's angular velocity vector and $\mathbf{r}$ the position vector. As a consequence of
shearing, vorticity is created when the fluid is flowing past container walls. The vorticity is defined by
\begin{equation}
	\boldsymbol{\omega} \equiv  \nabla \times \mathbf{v} = 2 \boldsymbol{\Omega}.
		\label{eqn-VorticitySBRotation}
\end{equation}
The second identity is satisfied in the case of rigid-body rotation. Taking the curl of the Navier-Stokes equation and neglecting external
forces, it is possible to show that vorticity transport is described by a diffusion equation.

Although the concept of vorticity had been familiar from viscous hydrodynamics, condensed matter physicists were initially not sure whether
it would be possible to spin up the frictionless component inside a superfluid or not; the main problem being the property of potential
flow as given in Eqn.~\eqref{eqn-PotentialFlow}. For a smooth, irrotational velocity field, $\Bv\SF$, the circulation around an arbitrary
contour $\mathcal{L}$ vanishes, i.e.
\begin{equation}
	 \Gamma = \oint_{\mathcal{L}} \mathbf{v}\SF \cdot \d \mathbf{l}
		 = \int_{\mathcal{A}}  \left( \nabla \times \mathbf{v}\SF \right) \cdot \d \mathbf{S} = 0,
\end{equation}
because Stokes' theorem can be used to rewrite the expression as an integral over the surface $\mathcal{A}$ enclosed by the contour
$\mathcal{L}$. This makes it impossible for an inviscid superfluid to develop circulation in a \textit{classical manner}. The state,
where no superfluid rotation is present, is generally referred to as the \textit{Landau state}~\cite{Kojima1971}.

\begin{figure}[t]
\begin{center}
 \includegraphics[width=3in]{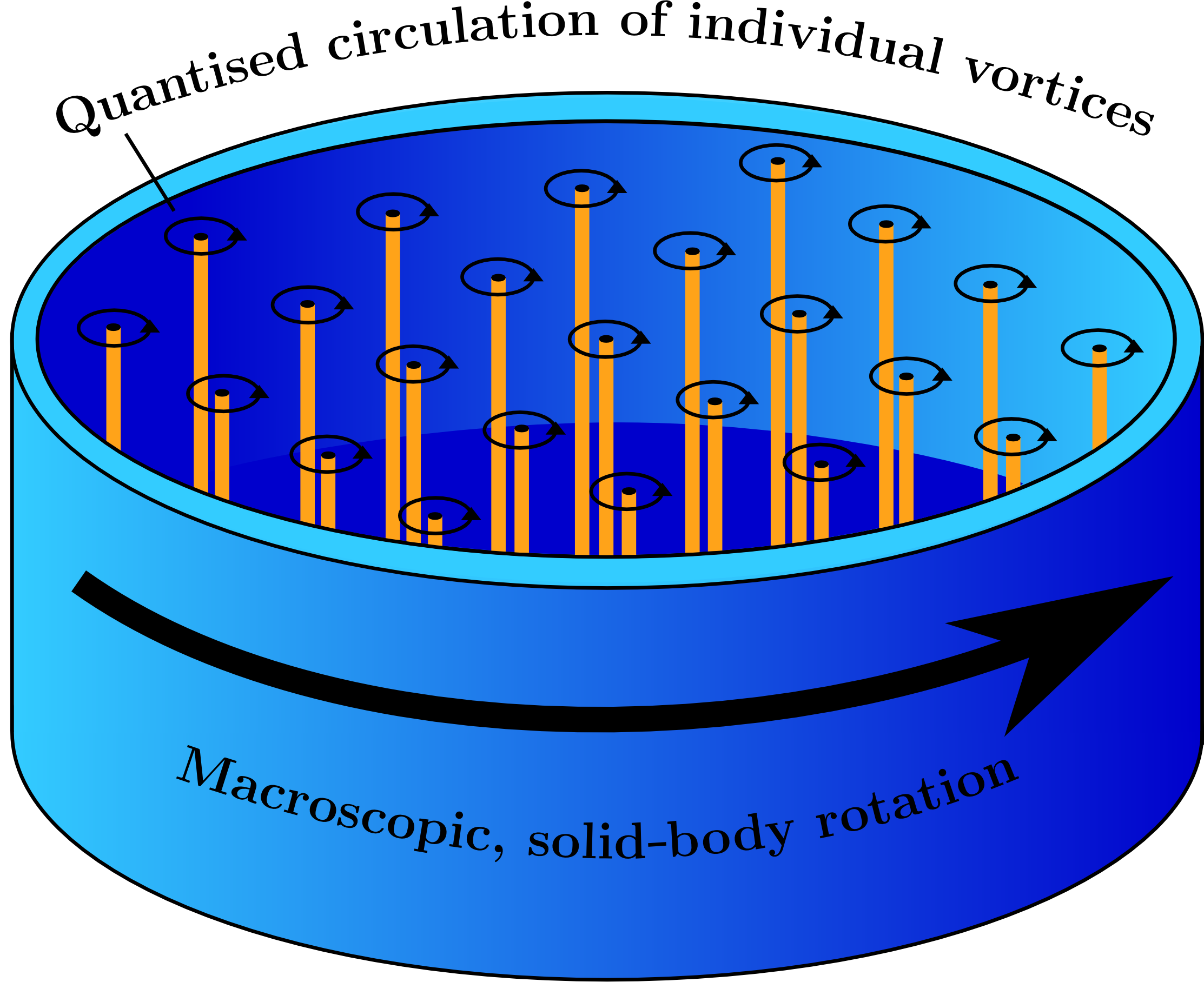}
\caption{Illustration of superfluid rotation. Different to a viscous fluid, a superfluid minimises its energy by forming a
regular vortex array. Each vortex is aligned with the rotation axis and carries a quantum of circulation,
adding up to mimic solid-body rotation on macroscopic scales.}
\label{fig-VorticesMacroscopic}
\end{center}
\end{figure}

However contrary to this discussion, several experiments in the 1960s showed that both components in rotating helium II move
with the same angular velocity, implying that the superfluid component also exhibits rigid-body rotation (see for example
Osborne~\cite{Osborne1950}). The contradiction between theory and observations is resolved by recalling that the quantum mechanical
wave function $\Psi$ is invariant under changes in the phase, $\varphi$, that are multiples of $2 \pi$. Taking this and
Eqn.~\eqref{eqn-SuperfluidVelocity} into account, the circulation is given by
\begin{equation}
	 \Gamma = \oint_{\mathcal{L}} \mathbf{v}\SF \cdot \d \mathbf{l}
		 = \frac{\hbar}{m\cl} \, \oint_{\mathcal{L}} \nabla \varphi \cdot \d \mathbf{l}
		 = \frac{h}{m\cl} \, n
		 \equiv \kappa n,  \hspace{1cm} n \in \mathbb{Z}.
		 \label{eqn-QuantisationCondition}
\end{equation}
The discrete set of phase values introduces a quantisation to the problem and results in the formation of vortices,
singularities at which the circulation is non-zero. $h=2 \pi \hbar$ denotes the Planck constant and the quantity $\kappa$
is defined as the quantum of circulation carried by a single vortex. Each individual vortex has a rotational velocity
profile that is inversely proportional to the distance, $r$, from its centre and additionally a core that is normal
and not superfluid. Using cylindrical coordinates $\{r, \theta,z \}$, one obtains for the superfluid velocity
\begin{equation}
	\mathbf{v}\SF(r) = \frac{\Gamma}{2 \pi r} \, \ut,
\end{equation}
where $\ut$ is the unit vector in $\theta$-direction. The idea of quantisation was pioneered by Onsager~\cite{Onsager1949} and
Feynman~\cite{Feynman1955}. The latter was the first to suggest that vortices could be formed in a regular array, so that the
circulation of all vortices mimics the rotation on macroscopic lengthscales as illustrated in Fig.~\ref{fig-VorticesMacroscopic}.
By forming vortices, the superfluid appears to be moving as a rigid body on macroscopic scales, having a classical moment of
inertia. In this picture, any change in angular momentum is accompanied by the creation (spin-up) or destruction (spin-down)
of vortices. Therefore, the vortex area density, $\cN\v$, is directly proportional to the total circulation within a unit area,
leading to the definition of an averaged vorticity,
\begin{equation}
	\boldsymbol{\omega} \equiv \cN\v \boldsymbol{\kappa},
		\label{eqn-AveragedVorticity}
\end{equation}
where $ \boldsymbol{\kappa} \equiv \kappa \, \uk$ with the unit vector $\uk$ pointing along the direction of the vortices. This
quantisation of the circulation in the form of vortices also serves as the basis for the description of the multi-fluid system
in the interior of the neutron stars (see Sec.~\ref{subsec-chargemultihydro}). For straight vortex lines, this direction coincides
with the rotation axis of the cylindrical container, $\uO= \uk$. With Eqns.~\eqref{eqn-VorticitySBRotation} and
\eqref{eqn-AveragedVorticity}, one finds
\begin{equation}
	\cN\v =  \frac{2 \Omega}{\kappa}.
		 \label{eqn-VortexLineDensity}
\end{equation}

For example, helium II rotating at $1 \, \text{rad} \, \sec^{-1}$ has an average vortex density of $\cN\v \approx 10^4 \, \cm^{-2}$.
The exact shape of the vortex array minimising the energy of the condensate was first calculated by Abrikosov~\cite{Abrikosov1957}.
He considered the case of a strong type-II superconductor (a problem equivalent to that of a rotating superfluid) and found that the
quantised structures form a triangular lattice (see Sec.~\ref{subsubsec-Abrikosov}). Using Eqn.~\eqref{eqn-VortexLineDensity}, one
can determine an average distance, $d\v$, between individual vortices,
\begin{equation}
	d\v \simeq \cN\v^{-1/2} = \left( \frac{\hbar \pi}{\Omega m\cl}\right)^{1/2}.
\end{equation}
For the rotating helium II sample discussed above, one obtains an intervortex spacing of $d\v \approx 0.1 \, \text{mm}$.

\subsection{Mutual friction and HVBK equations}
\label{subsec-MFAndHVBK}

The previous derivation of the two-fluid equations relies on the fact that the two fluid velocities, $\mathbf{v}\NF$ and
$\mathbf{v}\SF$, are small and dissipation can be neglected. If this is however no longer the case, additional terms have
to be included into the equations of motion. This is especially necessary if the individual fluid velocities are large or
the superfluid is rotating and vortices are present, since additional forces couple the two components. Based on an
improved understanding of the underlying physical processes, several extensions to the original two-fluid model have been
suggested. In particular the Hall-Vinen-Bekarevich-Khalatnikov (HVBK) equations~\cite{Hall1956, Bekarevich1961} provide a
more complete model of superfluid hydrodynamics.

Including dissipation leads to the coupling of the momentum equations, because the Gibbs-Duhem relation used in
Sec.~\ref{subsec-TwoFluidModel} is no longer sufficient to capture changes in the specific chemical potential. Considering
non-linear effects, one instead has to substitute $\d \tilde{\mu} = \rho^{-1} \d p - s \d T - \rho\NF (2 \rho)^{-1} \d
\mathbf{w}\NS^2$, where $\mathbf{w}\NS \equiv \mathbf{v}\NF - \mathbf{v}\SF$ is the relative velocity difference between
the two fluids (for details see Roberts and Donnelly~\cite{Roberts1974}). By incorporating these changes into
Eqns.~\eqref{eqn-Euler2Fluid} and~\eqref{eqn-NavierStokes2Fluid}, one arrives at two coupled partial differential equations,
\begin{align}
	\rho\SF \left[ \frac{\partial \mathbf{v}\SF}{\partial t} +  \left( \mathbf{v}\SF \cdot \nabla \right) \mathbf{v}\SF  \right]
		+ \frac{\rho\SF}{\rho} \nabla p - \rho\SF s \nabla T - \frac{\rho\SF \rho\NF}{2\rho} \, \nabla \mathbf{w}\NS^2 & =0,
		\label{eqn-Euler2Fluid2} \\[1.5ex]
	\rho\NF \left[ \frac{\partial \mathbf{v}\NF}{\partial t} +  \left( \mathbf{v}\NF \cdot \nabla \right) \mathbf{v}\NF  \right]
		+ \frac{\rho\NF}{\rho} \nabla p + \rho\SF s \nabla T + \frac{\rho\SF \rho\NF}{2\rho} \, \nabla \mathbf{w}\NS^2
		- \eta \nabla^2 \mathbf{v}\NF & =0.
		\label{eqn-NavierStokes2Fluid2}
\end{align}
These relations can be further improved to capture the dynamics of a rotating superfluid. The presence of vortices has an effect
on the hydrodynamical equations, because they interact with the normal fluid component and cause dissipation. This coupling
mechanism is generally referred to as \textit{mutual friction}. It is for example responsible for spinning up the superfluid as it
communicates the changes in the normal component (coupled viscously to the rotating container) to the frictionless counterpart.
Many advances in understanding the mutual friction force in helium II are based on research performed by Hall and Vinen in the
1960s. They realised that the main mechanism for the dissipative interaction is the collision of excitations with the normal cores
of the vortex lines. For a helium II sample rotating at constant angular velocity, $\mathbf{\Omega} = \Omega \, \uO$, Hall and
Vinen suggested the following form of the mutual friction force~\cite{Hall1956},
\begin{equation}
	\mathbf{F}_{\rm mf} = \CB_{\rm He} \frac{\rho\SF \rho\NF}{\rho} \, \uO \times \left[ \boldsymbol{\Omega}
		\times \left( \mathbf{v}\SF - \mathbf{v}\NF \right) \right]
		+ \CB'_{\rm He} \frac{\rho\SF \rho\NF}{\rho} \, \boldsymbol{\Omega}
		\times \left( \mathbf{v}\SF - \mathbf{v}\NF \right).
		\label{eqn-MutualFrictionHe}
\end{equation}
The two parameters $\CB_{\rm He}$ and $ \CB'_{\rm He}$ reflect the strength of the mutual friction coupling and can be experimentally
determined (see Sec.~\ref{subsec-MutualFrictionHe} for details). Additionally, the fluid velocities $\mathbf{v}\NF$ and $\mathbf{v}\SF$
are no longer mesoscopic quantities but instead obtained by averaging over regions that contain a large number of vortices. Therefore,
this form of the mutual friction between individual vortices and the viscous fluid implicitly relies on an averaging procedure. This way,
the discrete behaviour of the vorticity is smoothed out and the dynamics on small lengthscales are neglected. This process is often called
\textit{coarse-graining}. Hence, accounting for the presence of vortices, other quantities in the hydrodynamical equations also have to
be replaced with their averaged equivalents.

In addition to quasi particle collisions, there is another force acting on the helium vortices that is particularly important for the study
of highly dissipative or turbulent behaviour in superfluids. The original mutual friction force given in Eqn.~\eqref{eqn-MutualFrictionHe}
assumes that vortices are straight and form a regular array. This condition however is not necessarily satisfied, as vortices could be bent
or even form tangled structures~\cite{Eltsov2009}, making it important to include the vortex tension, $\mathbf{T}$. Postulated as an
explanation for experimental results in superfluid helium-4, the mutual friction force in the case of curved vortices, which are sufficiently
far apart so that no reconnections can take place, has the form~\cite{Barenghi1988}
\begin{equation}
	\mathbf{F}_{\rm mf} = - \CB_{\rm He} \frac{\rho\SF \rho\NF}{2\rho} \, \uo \times \left( \boldsymbol{\omega} \times \mathbf{w}\NS
		+ \frac{\mathbf{T}}{\rho\SF}  \right) - \CB'_{\rm He} \frac{\rho\SF \rho\NF}{2 \rho} \, \left( \boldsymbol{\omega}
		\times \mathbf{w}\NS + \frac{\mathbf{T}}{\rho\SF}\right),
		\label{eqn-MutualFrictionHe2}
\end{equation}
where Eqn.~\eqref{eqn-VorticitySBRotation} was used to substitute the averaged vorticity $\boldsymbol{\omega} \equiv \omega \, \uo$
with $\uo=\uO$. Due to its large self-energy (comparable to the tension of a guitar string) a vortex resists bending, which generates
a restoring force trying to bring the vortex back into its equilibrium position. The tension is thus dependent on the vortex curvature:
\begin{equation}
	\mathbf{T}= - \rho\SF \nu\SF \, \boldsymbol{\omega} \times \left( \nabla \times \uo \right)
		=  \rho\SF \nu\SF \left( \boldsymbol{\omega} \cdot \nabla \right) \uo.
		\label{eqn-Tension}
\end{equation}
The parameter $\nu\SF$, which has the dimensions of a kinematic viscosity, is defined by
\begin{equation}
	 \nu\SF \equiv \frac{\kappa}{4 \pi} \, \ln \left( \frac{d\v}{a} \right),
		  \label{eqn-SuperfluidViscosity}
\end{equation}
where $d\v$ denotes the intervortex spacing and $a$ the radius of a vortex core. Note that for the temperature range generally studied in helium II
experiments, $\nu\SF$ is of similar order or larger than the normal component's kinematic viscosity, $\nu\NF = \eta /\rho\NF$~\cite{Barenghi1983}.

Combining the momentum conservation equations~\eqref{eqn-Euler2Fluid2} and~\eqref{eqn-NavierStokes2Fluid2} with the extended version
of the mutual friction~\eqref{eqn-MutualFrictionHe2} and the tension force~\eqref{eqn-Tension}, one finally arrives at the
HVBK equations that describe the hydrodynamics of a rotating two-component superfluid in the presence of averaged
vorticity~\cite{Hall1956, Bekarevich1961, Barenghi1988},
\begin{align}
	\rho\SF \left[ \frac{\partial \mathbf{v}\SF}{\partial t} +  \left( \mathbf{v}\SF \cdot \nabla \right) \mathbf{v}\SF  \right]
		+ \frac{\rho\SF}{\rho} \nabla p - \rho\SF s \nabla T - \frac{\rho\SF \rho\NF}{2\rho} \, \nabla \mathbf{w}\NS^2
		= \mathbf{T} + \mathbf{F}_{\rm mf},
		\label{eqn-Euler2FluidHVBK} \\[1.5ex]
	\rho\NF \left[ \frac{\partial \mathbf{v}\NF}{\partial t} + \left( \mathbf{v}\NF \cdot \nabla \right) \mathbf{v}\NF  \right]
		+ \frac{\rho\NF}{\rho} \nabla p + \rho\SF s \nabla T + \frac{\rho\SF \rho\NF}{2\rho} \, \nabla \mathbf{w}\NS^2 
		- \eta \nabla^2 \mathbf{v}\NF  = - \mathbf{F}_{\rm mf}.
		\label{eqn-NavierStokes2FluidHVBK}
\end{align}
As addressed in Sec.~\ref{subsec-chargemultihydro}, an equivalent set of equations can also be used to study the multi-fluid mixture
in the interior of neutron stars and much of what we know about the manifestation of astrophysical condensates is rooted in the close
analogy between these two mathematical formalisms.

\subsection{Vortex dynamics and turbulence}
\label{subsec-Turbulence}

While the hydrodynamical model provides information about the averaged dynamics of superfluids on macroscopic scales, several phenomena cannot
easily be studied within this framework. In particular,  when vortices are no longer straight and very close to each other, the standard averaging
procedure introduced in Sec.~\ref{subsec-RotatingSF} can no longer be performed, because vortices start to interact and reconnect, which
leads to a turbulent state. The analysis of this new regime of fluid dynamics has significantly advanced in recent decades due to increasing
computational resources.

The modern models of the chaotic flow in superfluids are based on the concept of following individual vortices on mesoscopic scales. In these
\textit{filament approaches}, which were pioneered by Schwarz~\cite{Schwarz1988}, the vortices are reduced to infinitesimally thin three-dimensional
curves $\mathbf{s} (\xi, t)$ (parametrised by the length $\xi$ along the line and the time $t$) of circulation $\kappa$. A description of the
corresponding dynamics is obtained by first assuming that the vortices are only slightly bent and no reconnections take place. Due to this
curvature, the mesoscopic velocity field generated at a specific point on a line also influences the rest of the vortex. Hence, each point on a
vortex moves according to the total superfluid velocity induced at this point plus any additionally forces present. More precisely, the induced
velocity of the superfluid component is given by~\cite{Hanninen2014}
\begin{equation}
	 \Bv\SF ( \Br, t) = \frac{\kappa} {4 \pi} \int_{\mathcal{L}} \frac{(\mathbf{s} - \Br) \times \d \mathbf{s}}{|\mathbf{s} - \Br|^3},
		\label{eqn-BiotSavart}
\end{equation}
with the integral being evaluated over the full vortex length, $\mathcal{L}$. Note that this is of the same form as the \textit{Biot-Savart law}
known from  electromagnetism, which is used to calculate the magnetic field induced by a steady current. Eqn.~\eqref{eqn-BiotSavart}
is singular at any point $\mathbf{s}$ on the vortex line and only well-defined outside the vortex. The singularity can be avoided by introducing
a cut-off to regularise the integral. Ignoring the detailed core structure, a suitable choice would be to cut off the integral at the distance
$a \simeq \xi$. The self-induced contribution to the superfluid flow (caused by the vortex curvature) is responsible for the modification of
the mutual friction as given in Eqn.~\eqref{eqn-MutualFrictionHe2}. By introducing an additional cut-off at large distances from the vortex
(a reasonable estimate is the intervortex spacing $d_\rv$), Eqn.~\eqref{eqn-BiotSavart} can be evaluated and related to the tension, $\mathbf{T}$,
that enters the vortex-averaged force, $\mathbf{F}_{\rm mf}$.

On mesoscopic scales, the motion of a single vortex filament is obtained by balancing the individual forces acting on it. While a more detailed
discussion of the respective forces is postponed to Secs.~\ref{subsec-MesoscopicNSDynamics} and~\ref{subsec-MutualFrictionHe}, a balance of the
\textit{Magnus force} and a \textit{dissipative drag} leads to an equation for the mesoscopic vortex velocity, $\mathbf{u}_{\rm v}$;
\begin{equation}
	\mathbf{u}_{\rm v}
		= \mathbf{v}\SF + \alpha_{\rm He} \,  \hat{\mathbf{s}}' \times \left( \mathbf{v}\NF - \mathbf{v}\SF \right)
		- \alpha'_{\rm He} \, \hat{\mathbf{s}}' \times \left[ \hat{\mathbf{s}}' \times \left( \mathbf{v}\NF - \mathbf{v}\SF \right) \right],
		\label{eqn-MesoscopicVortexVelocity}
\end{equation}
where $\mathbf{v}\SF$ is given by Eqn.~\eqref{eqn-BiotSavart} and a prime denotes a partial derivative with respect to the arc length $\xi$,
implying that $\hat{\mathbf{s}}'$ is the unit tangent of the vortex line. Moreover, $\alpha_{\rm He}$ and $\alpha'_{\rm He}$ form a second
set of mutual friction coefficients, whose connection to the parameters $\CB_{\rm He}$ and $ \CB'_{\rm He}$ is also explained in
Sec.~\ref{subsec-MutualFrictionHe}. While Eqn.~\eqref{eqn-MesoscopicVortexVelocity} allows an analysis of the dynamics of curved
vortices, the filament model does not automatically account for vortex interactions and reconnections, which eventually drive the superfluid
towards a turbulent state. This non-equilibrium behaviour can however be incorporated by introducing an additional algorithmic procedure that
ensures the immediate separation and subsequent reconnection of vortex lines if they get too close to each other or the surface of the sample.
More details of this reconnecting vortex-filament model, which has been successfully applied to capture various features of quantum turbulence
(see H\"anninen and Baggaley~\cite{Hanninen2014} for a review), are provided by Schwarz~\cite{Schwarz1988}.

The mesoscopic approach is providing new insight into the so-called \textit{counterflow behaviour}, which results from the relative motion of the
inviscid and normal fluid component. The first studies of quantum turbulence focused on this chaotic flow regime and were pioneered by Vinen in
the 1950s~\cite{Vinen1957a, Vinen1957b, Vinen1957c, Vinen1958}. By applying a thermal gradient to non-rotating superfluid helium (only affecting the
viscous fluid and hence causing a velocity difference between the normal and the superfluid component), Vinen showed that the energy was dissipated
as a result of the interactions between a turbulent vortex tangle and the excitations. In such a turbulent state the mutual friction
force~\eqref{eqn-MutualFrictionHe2} is no longer suitable to describe the dissipation and an alternative expression needs to be used. The main challenge
remains the calculation of an appropriate average as one cannot simply count the vortices per unit area in the tangle. To circumvent this problem,
Vinen used a phenomenological approach to determine the form of the mutual friction. More precisely, he postulated that for the case of isotropic
turbulence, where vortices do not exhibit a preferred direction, the force per unit volume is~\cite{Vinen1957c}
\begin{equation}
	\mathbf{F}_{\rm mf} = \frac{2}{3} \, L  \rho\SF \kappa \alpha_{\rm He} \left( \Bv\SF - \Bv\NF \right).
		\label{eqn-GorterMellinkMF}
\end{equation}
Here, $L$ is the total length of vortices per unit volume. In order to obtain an estimate for this quantity, one can balance the effects that
increase and suppress turbulence and therefore alter the parameter $L$. Whereas its growth can be attributed to the Magnus effect, the decay
of quantum turbulence on large scales satisfies the same Kolmogorov scaling\cite{Kolmogorov1941} as observed in classical fluids~\cite{Barenghi2014}.
If both mechanisms are in equilibrium, the following steady-state solution for $L$ is found~\cite{Vinen1957c},
\begin{equation}
	L = \left( \frac{2\pi}{ \kappa} \right)^2  \left( \frac{\chi_1}{\chi_2} \right)^2 \alpha_{\rm He}^2 \left( \mathbf{v}\SF - \mathbf{v}\NF \right)^2,
\end{equation}
where $\chi_1$ and $\chi_2$ are dimensionless parameters of order unity. For an isotropic vortex tangle, the mutual friction force is thus
proportional to the cube of the relative velocity, as had previously been suggested by Gorter and Mellink~\cite{Gorter1949}. Note that by
averaging the filament model of Schwarz~\cite{Schwarz1988} over all vortex segments inside a sample, a qualitatively similar result is
obtained for the macroscopic mutual friction force~\cite{Andersson2007a}.

The mesoscopic framework can further help improve our understanding of the stability of superfluid vortices~\cite{Karimaki2012}. Whereas
Vinen's early experiments were performed with non-rotating helium, subsequent studies also examined the counterflow behaviour in rotating
samples which similarly exhibited turbulent characteristics~\cite{Swanson1983}. It has been suggested by Glaberson et al.~\cite{Glaberson1974}
that this could be the result of a hydrodynamical vortex array instability. As soon as the counterflow (applied along the vortex tangent)
exceeds a critical velocity, the vortex lines become unstable to the excitation of \textit{Kelvin waves},  helical displacements
named after Lord Kelvin~\cite{Thomson1880}. Using a simple plane-wave analysis, the dispersion relation associated with the excitation
of a Kelvin mode of wave number $k$ is~\cite{Glaberson1974} (see also Sidery et al.~\cite{Sidery2008})
\begin{equation}
	\omega (k) = 2 \Omega + \nu\SF k^2,
		\label{eqn-DispersionRelation}
\end{equation}
where $\Omega$ denotes the macroscopic angular velocity and the parameter $\nu\SF$ has been defined in Eqn.~\eqref{eqn-SuperfluidViscosity}.
Eqn.~\eqref{eqn-DispersionRelation} displays critical behaviour that can be quantified by minimising $\omega (k)/k$. One obtains the critical
wave number, $k_{\rm c} = \sqrt{2 \Omega / \nu\SF}$, at which the vortex line instability (often referred to as the
\textit{Donnelly-Glaberson instability}) is triggered. The corresponding critical counterflow is
\begin{equation}
	w_{\text{\tiny{NS}},\rm c} = \frac{\omega(k_{\rm c})}{k_{\rm c}} = 2 \sqrt{ 2 \Omega \nu\SF}.
		\label{eqn-DGCounterflowCritical}
\end{equation}
By exceeding this value, an initially regular vortex array can  be destabilised and transformed into a turbulent tangle of vortices, which
drastically changes the rotational dynamics.

We return to the problem of superfluid turbulence in Secs.~\ref{subsec-VortexDynamicsHe} and~\ref{subsec-VortexDynamicsBEC} to compare the
characteristics of laboratory systems and neutron stars.

\subsection{Ultra-cold gases}
\label{subsec-ColdAtoms}

The close analogy between superfluid helium and a bosonic gas at low temperatures is illustrated by the presence of a quantum mechanical
condensate which exhibits macroscopic properties. Since all particles in the BEC occupy the same minimum energy state, a mean-field description
can be employed to obtain the macroscopic wave function as the symmetrised product of the single-particle wave functions. This does however
not account for the interactions between individual bosons. In the limit $T\to 0$, the scattering length in a BEC is typically of the
order of a few $\nm$, whereas the particle separation is about $100 \, \nm$, implying that ultra-cold gases are dilute and two-body scattering
is the dominant interaction mechanism. While such processes are strong, they only come into play if two atoms are very close to each
other. This can be easily captured by including an effective interaction (an additional source term proportional to $|\Psi|^2 \Psi$) into
Eqn.~\eqref{eqn-Schroedinger}. The resulting non-linear Schr\"odinger equation, referred to as the time-dependent Gross-Pitaevskii (GP)
equation~\cite{Gross1961, Pitaevskii1961}, generally applied to model the properties of a Bose-Einstein condensate in the low temperature limit,
then reads
\begin{equation}
 	i \hbar \, \frac{\partial \Psi (\Br , t ) }{\partial t} + \frac{\hbar^2}{2 m\cl} \nabla^2 \Psi (\Br , t )
		- V(\Br) \Psi (\Br , t ) - U_0 |\Psi (\Br , t ) |^2 \Psi (\Br , t )  =0.
		\label{eqn-GPE}
\end{equation}
As before, $\Psi(\Br , t)$ denotes the complex macroscopic wave function and $m\cl$ the boson mass. Furthermore, $V(\Br)$ represents the external
potential confining the BEC and the effective interaction parameter, $U_0$, is related to the scattering length, $a$, via
\begin{equation}
	U_0 = \frac{4\pi \hbar^2 a}{m\cl}.
\end{equation}
The time-independent version of Eqn.~\eqref{eqn-GPE} is similar to the first Ginzburg-Landau equation, which will be given in
Sec.~\ref{subsec-GLNutshell}.

The time-dependent GP equation is particularly useful for studying the dynamics of a BEC, the reason being the close connection between the quantum
mechanical and the hydrodynamical picture. As illustrated for helium II, a Madelung transformation can be applied to express the non-linear Schr\"odinger
equation in terms of two new degrees of freedom, i.e.\ the amplitude, $\Psi_0$, and phase, $\varphi$, of the wave function. By substituting the condensate's
density, $n\cl$, and the gradient of the phase (which is proportional to the condensate's velocity, $\Bv\SF$, as defined in Eqn.~\eqref{eqn-SuperfluidVelocity}),
the behaviour of the wave function can be mapped to the equations of motion for a fluid. Since the non-linear interaction term in Eqn.~\eqref{eqn-GPE}
is real and does not contribute to the imaginary part, the Madelung transformation results in the same continuity equation as the linear Schr\"odinger
equation (see Eqn.~\eqref{eqn-ContinuitySuper}). On the other hand, the second equation of motion can be adjusted by replacing the chemical potential
$\mu \to V(\Br) + U_0 |\Psi (\Br , t ) |^2$. This leads to the following momentum equation:
\begin{equation}
	 \frac{\partial \mathbf{v}\SF}{\partial t} +  \left( \mathbf{v}\SF \cdot \nabla \right) \mathbf{v}\SF
		+ \frac{1}{m\cl} \, \nabla \left(V + U_0 n\cl - \frac{\hbar^2}{2 m\cl \sqrt{n\cl}} \, \nabla^2 \sqrt{n\cl} \right) = 0.
		\label{eqn-EulerGPE}
\end{equation}
The quantum pressure term is again negligible if the typical lengthscale for variations of the macroscopic wave function is much larger than the coherence
length, $\xi$,~\cite{Pethick2008} so we are left with
\begin{equation}
	\frac{\partial \mathbf{v}\SF}{\partial t} +  \left( \mathbf{v}\SF \cdot \nabla \right) \mathbf{v}\SF
		+ \frac{1}{m\cl} \, \nabla \left(V + U_0 n\cl \right) = 0.
		\label{eqn-EulerGPE2}
\end{equation}
This is equivalent to the Euler equation of hydrodynamics in the presence of an external potential and a modified chemical potential. In the context of
neutron star modelling, the former could be identified with the gravitational potential, while the quantity $U_0 n\cl$ has taken the place of the chemical
potential. As this term originated from the addition of an effective interaction in the Schr\"odinger equation, we see that two-body scattering processes
in a BEC produce a pressure-like term in the momentum equation similar to what we would expected from normal fluid dynamics. Note that for a boson gas of
uniform density, $U_0 n\cl$ is indeed equal to the chemical potential~\cite{Pethick2008}. This analogy will  form the basis for the discussion in
Sec.~\ref{sec-ColdGases}.


\section{Modelling Superconductors}

Following Onnes' discovery~\cite{Onnes1911}, the first three decades were dominated by experimental studies aimed at determining the
basic properties of the superconducting phase. In addition to the disappearance of the electrical resistivity below a critical
temperature, the complete expulsion of magnetic flux in the presence of an external field was found to be the main characteristic of
a superconducting sample. We will discuss the Meissner effect, the theoretical work of the London brothers and a quantum mechanical
description that forms the justification for their phenomenological approach. This will be succeeded by a discussion of the differences
between type-I and type-II superconductors and the quantisation of magnetic flux. Finally, we will conclude
with an introduction to the Ginzburg-Landau theory. It provides the possibility to calculate the critical quantities of the phase
transitions in a superconductor and lays the foundation for several analyses of a neutron star's fluid interior. Gaussian units will be
employed throughout.

\subsection{Meissner effect and London equations}
\label{subsec-MeissnerLondon}

In order to determine the behaviour of matter condensing into a superconducting state below a critical temperature, $T_{\rm c}$, one
can study its response to external magnetic fields. The simplest reaction would be the generation of surface currents that flow in a
small sheet of order $\lambda$ (see below) provoking the expulsion of the interior magnetic field. For the purpose of a theoretical
description, a distinction between an external current density, $\mathbf{j}_{\rm ext}$, that generates a macroscopic, averaged field,
$\mathbf{H}$, and so-called \textit{magnetisation currents} affecting only the mesoscopic magnetic induction, $\mathbf{\bB}$, is
beneficial. The electronic supercurrent density, $\mathbf{j}\SC$, present inside a superconductor is of mesoscopic origin and therefore
attributed to the second class. Hence, the exterior field, $\mathbf{H}$, is unaffected by the presence of the superconductor. Moreover,
a macroscopic average of the magnetic induction is defined by $\mathbf{B}$. For a superconducting sample, this quantity could vary smoothly
over macroscopic lengthscales, while in the case of vacuum or a normal metal (where no magnetisation currents are present) one finds
$\mathbf{H}= \mathbf{\bB}=\mathbf{B}$.

Based on experimental observations, Fritz and Heinz London suggested that the mesoscopic electric field, $\mathbf{\bar{E}}$,
and the mesoscopic magnetic induction, $\mathbf{\bB}$, inside a superconductor are governed by the following two equations
\cite{London1935},
\begin{equation}
	\mathbf{\bar{E}} = \frac{m\cl}{n\cl q^2} \, \frac{\partial \mathbf{j}\SC}{\partial t} ,
		\hspace{1.5cm}
	\mathbf{\bB} = - \frac{m\cl c}{n\cl q^2} \, \nabla \times \mathbf{j}\SC,
	\label{eqn-London12}
\end{equation}
where $m\cl$ and $q$ are the mass and charge, respectively, and $n\cl$ is the number density of the charged particles responsible
for the superconducting behaviour. The first London equation captures the perfect conductivity feature, whereas the second one
describes the Meissner effect. This can be seen by combining the second equation from~\eqref{eqn-London12} with Amp\`ere's law,
which locally reads
\begin{equation}
	\nabla \times \mathbf{\bB} = \frac{4 \pi}{c} \, \mathbf{j\SC}. 
		\label{eqn-AmpereLaw01}
\end{equation}
In this relation displacement currents have been neglected. This is possible because in an equilibrium or steady-state
superconductor, the supercurrent density is no longer time-dependent~\cite{Tilley1990}. Hence, the electric field vanishes
in those cases (see first Eqn.~\eqref{eqn-London12}), allowing one to neglect the displacement current that is proportional to
$\partial \mathbf{\bar{E}} / \partial t$. Moreover, no magnetic monopoles are present and the Maxwell equation
\begin{equation}
	\nabla\cdot  \mathbf{\bB} = 0
			\label{eqn-MaxwellNoBMonopole}
\end{equation}
is satisfied. One then arrives at an equation for the mesoscopic magnetic induction,
\begin{equation}
	\lambda^2 \, \nabla^2 \, \mathbf{\bB} = \mathbf{\bB},
		\label{eqn-LondonCombined1}
\end{equation}
where we define the penetration depth as
\begin{equation}
	\lambda \equiv \left(\frac{m\cl c^2}{4 \pi n\cl q^2} \right)^{1/2}.
		\label{eqn-lambda}
\end{equation}
Considering a flat boundary between a superconducting surface and free space that lies in the $z$-direction and a constant
external field, $\mathbf{H}=\mathbf{B} = B_0 \uz$, applied parallel to the boundary, the solution for the magnetic field
inside the superconductor is
\begin{equation}
	\mathbf{\bB} (x) = B_0 \, \exp (-x / \lambda) \, \uz.
		\label{eqn-SolutionLondonCombined1}
\end{equation}
The $x$-direction is perpendicular to the boundary and Eqn.~\eqref{eqn-SolutionLondonCombined1} thus implies that the magnetic
field decays exponentially inside the superconductor. The London penetration depth, $\lambda$, describes how far the field reaches
into the sample and determines the thickness of the surface sheet in which the supercurrents are generated.

The origin of the phenomenological London equation~\eqref{eqn-LondonCombined1} can be enlightened by considering a quantum
mechanical picture, in which the wave function represents the superposition of all superconducting states in the condensate.
As first pointed out by Fritz London~\cite{London1948}, this relies on the usage of a vector potential, $\mathbf{A}$,
defined by
\begin{equation}
	\mathbf{\bB} \equiv \nabla \times \mathbf{A}.
		\label{eqn-Vector}
\end{equation}
Following an approach similar to that used in Sec.~\ref{subsec-WaveFunction} for a superfluid, a quantum mechanical charge current
density can be derived for the charged superconducting condensate. Using the standard formula for minimal coupling, one replaces
\begin{equation}
	\nabla \rightarrow \nabla - \frac{i q}{\hbar c} \mathbf{A},
\end{equation}
in order to obtain
\begin{equation}
	 \mathbf{j}\SC = \frac{i q \hbar}{2m\cl} \left[ \Psi \left( \nabla + \frac{i q}{\hbar c} \mathbf{A} \right) \Psi^*
		- \Psi^* \left( \nabla - \frac{i q}{\hbar c} \mathbf{A} \right)  \Psi \right].
		\label{eqn-Current2}
\end{equation}
Here, $m\cl \equiv 2 m\e$ and $q \equiv - 2 e$, where $m\e$ denotes the mass of an electron and $e\equiv |e|$ is the elementary charge.
Substituting the macroscopic wave function, $\Psi$, given in Eqn.~\eqref{eqn-WaveFunctionPsi} leads to an expression for the charge
current density in a superconductor,
\begin{equation}
	 \mathbf{j}\SC = \frac{q \hbar}{m\cl} \, n\cl \, \nabla \varphi - \frac{q^2 }{m\cl  c} \, n\cl \, \mathbf{A},
		\label{eqn-Current3}
\end{equation}
where $n\cl=|\Psi|^2$ represents the number density of electron Cooper pairs. The first term in
Eqn.~\eqref{eqn-Current3} is equivalent to the result of the superfluid case, while the second term reflects the charge  of
the condensate. Defining the supercurrent density as $\mathbf{j}\SC \equiv q n\cl \mathbf{v}\SC$, one finds a relation for the
velocity of the superconducting particles,
\begin{equation}
	 \mathbf{v}\SC = \frac{\hbar}{m\cl} \, \nabla \varphi - \frac{q}{m\cl c}\, \mathbf{A}.
		\label{eqn-SuperconVelocity}
\end{equation}
Moreover, the quantum mechanical wave function is invariant under specific changes in the phase. Since it can be set to zero in an
appropriate gauge, it is possible to eliminate the term proportional to $\nabla \varphi$ in Eqn.~\eqref{eqn-Current3}. Hence,
the supercurrent density, $\mathbf{j}\SC$, is proportional to the vector potential, $\mathbf{A}$,
\begin{equation}
	 \mathbf{j}\SC = - \frac{q^2 }{m\cl c} \, n\cl \, \mathbf{A}.
		\label{eqn-Current4}
\end{equation}
Taking the curl of this relation and eliminating the current with the help of Amp\`ere's law~\eqref{eqn-AmpereLaw01}, one
finds the following equation valid inside the superconductor,
\begin{equation}
	\lambda^2 \, \nabla^2 \, \mathbf{\bB} = \mathbf{\bB}.
		\label{eqn-LondonCombined2}
\end{equation}
This is exactly the phenomenological London result given in Eqn.~\eqref{eqn-LondonCombined1} that describes the Meissner
effect as the exponential decay of the mesoscopic magnetic induction.

\subsection{London field in rotating superconductors}
\label{subsec-LondonField}

In contrast to a superfluid, a superconducting sample is able to rotate without quantising its circulation, i.e.\ forming vortices.
The quantised fluxtubes themselves are not related to the macroscopic rotation, as these dynamics induce an additional characteristic magnetic
field inside the superconductor, whose axis is parallel to the rotation axis. The so-called \textit{London field}, $\mathbf{b}_{\rm L}$,
is a fundamental property of the superconducting state and can be calculated by combining the definition of the superconducting
velocity~\eqref{eqn-SuperconVelocity} and the condition for rigid-body rotation given in Eqn.~\eqref{eqn-RigidBody}. For vanishing
phase gradients, $\nabla \varphi=0$, the energy of a rigidly rotating superconductor is minimised by a vector potential of the form
\begin{equation}
	   \mathbf{A}_{\rm L} = - \frac{m\cl c}{q}\, \boldsymbol{\Omega} \times \Br,
\end{equation}
which according to Eqn.~\eqref{eqn-Current4} has to be supported by additional surface currents. For a cylindrical geometry and rotation
about the $z$-axis, i.e. $\boldsymbol{\Omega}=\Omega \uz$, this potential corresponds to the following magnetic field,
\begin{equation}
	   \mathbf{b}_{\rm L} = \frac{2m\cl c}{q}\, \boldsymbol{\Omega}.
		\label{eqn-LondonField}
\end{equation}
As discussed in more detail later on, the London field is also present in the neutron star interior. Although small in magnitude (see
Eqn.~\eqref{eqn-EstimateLondonField}), it will have important consequences for the electrodynamical properties of a rotating star (see
Sec.~\ref{subsec-Maxwell}).

\subsection{Two types of superconductors and flux quantisation}
\label{subsection-typeI_II}

\begin{figure}[t]
\begin{center}
\includegraphics[height=4.7cm]{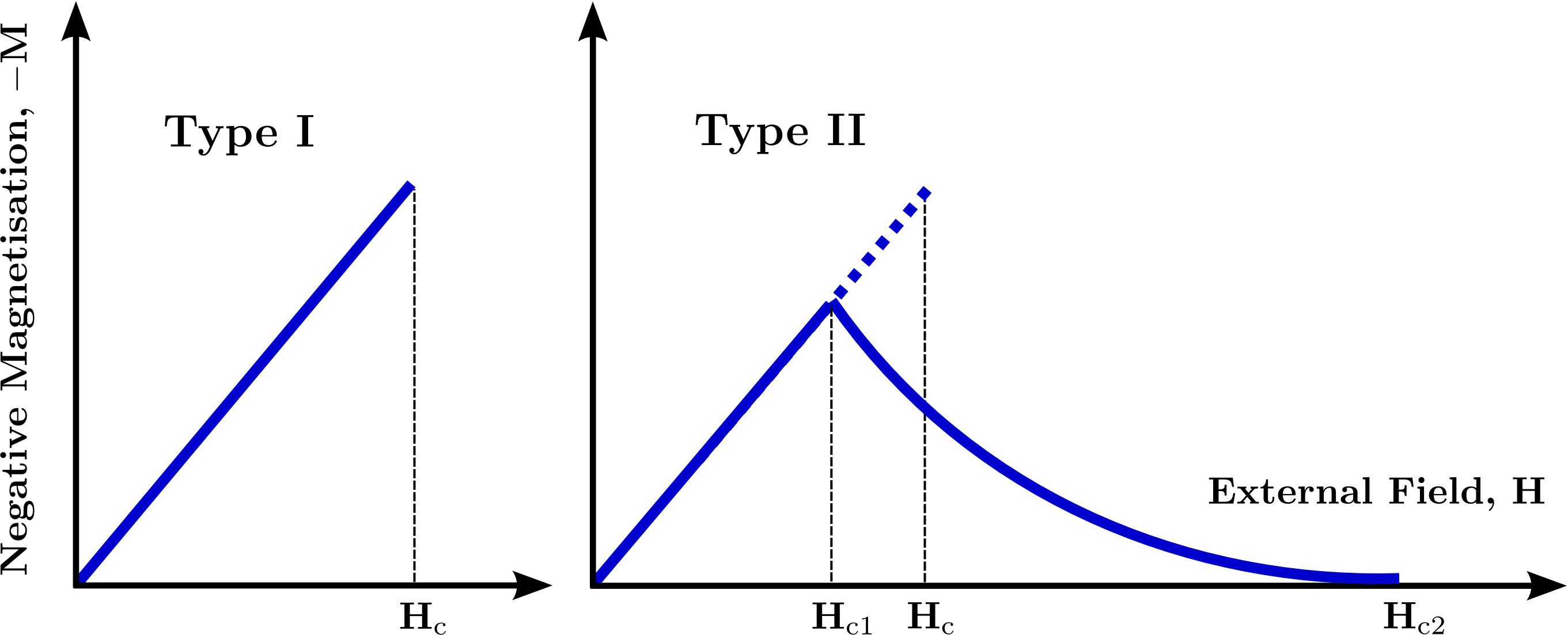}
\caption{Magnetisation curves for a type-I (left) and type-II (right) superconductor with the same thermodynamical critical
field, $H_{\rm c}$. The areas below the curves are equal in both cases and given by the condensation energy, $E_{\rm cond}$.}
\label{fig-magnetisationI}
\end{center}
\end{figure}

The first experiments analysing superconductivity did not provide access to the mesoscopic magnetic induction, $\mathbf{\bB}$,
because they measured the total magnetic flux present in a sample. Hence, one rather obtained information about the spatially averaged
magnetic induction, $\mathbf{B}$. This macroscopic induction is connected to the external field, $\mathbf{H}$, and the average
magnetisation, $\mathbf{M}$, via
\begin{equation}
 	\mathbf{B} = \mathbf{H} + 4 \pi \mathbf{M}.
		\label{eqn-AverageInduction}
\end{equation}
The magnetisation, $\mathbf{M}$, is a function of $\mathbf{H}$ and its behaviour strongly depends on the properties of the medium.
In vacuum or a normal conducting metal, i.e.\ in a superconductor above the transition temperature, $T_{\rm c}$, the average induction
and the external field are equivalent, so the average magnetisation has to vanish. Measurements of $\mathbf{M}$ in superconductors
have revealed features that allow a separation into two distinct classes (\textit{type-I} and \textit{type-II} media) as shown in
Fig.~\ref{fig-magnetisationI}.

For type-I systems, the magnetisation increases linearly with the external field. In this Meissner state, no magnetic flux is present in
the interior of the superconductor, i.e. $\mathbf{B}=0$, and the magnetisation generated by the supercurrents in the surface layer balances
the external field. As soon as $H \equiv |\mathbf{H}|$ reaches the critical value, $H_{\rm c}$, the magnetisation drops to zero. The
superconducting quantum state is destroyed and the material turns normal in a first-order phase transition. The critical magnetic
field, $H_{\rm c}$, is thermodynamically related to the condensation energy, $E_{\rm cond }$. This is the temperature-dependent difference
in the free energy densities of the normal and the superconducting state in the absence of external fields, $f\NOO$ and $f\SOO$,
respectively. It is given by
\begin{equation}
 	E_{\rm cond }(T) = f\NOO(T) - f\SOO(T) = \frac{H_{\rm c}(T)^2}{8 \pi}.
		\label{eqn-CondensationEnergy}
\end{equation}
Depending on the geometry of a type-I superconductor, characterised by the so-called \textit{demagnetisation factor}, it is possible to create
an intermediate state for external fields close to $H_{\rm c}$. Considering for example a superconducting sphere, its averaged surface
field, $B \equiv |\mathbf{B}|$, is not constant. It exceeds the applied field, $H$, in the equatorial plane and is smaller than $H$ close to
the poles~\cite{Tinkham2004}. Therefore, certain regions could have $B > H_{\rm c}$, while others could not, creating a state where
superconducting and normal regions coexist. The size of the corresponding domains depends crucially on the \textit{surface energy},
$\delta$, of the interfaces, which will be considered below.

For a type-II superconductor, the Meissner state does not break down abruptly. Instead, above a critical field, $H_{\rm c1}$, it is
energetically favourable for the medium to let magnetic flux continuously enter in the form of fluxtubes. The quantity responsible for
this behaviour is again the surface energy. It also dictates that the interfluxtube interaction is repulsive and the resulting magnetic
structures are ordered in a triangular array. This was first investigated by Abrikosov~\cite{Abrikosov1957}. Inside the fluxtubes, the
material is in a normal state, which is screened from the superconducting region by additional, circulating supercurrents. As discussed
in Sec.~\ref{subsec-RotatingSF} for the quantised circulation of a rotating superfluid, the quantum mechanical wave function, $\Psi$,
is invariant under changes in the phase, $\varphi$, that are multiples of $2 \pi$. Taking account of this invariance, one can integrate
Eqn.~\eqref{eqn-SuperconVelocity} around a closed contour $\mathcal{L}$ located inside the sample to obtain
\begin{equation}
	 \frac{c \hbar}{q} \, \oint_{\mathcal{L}} \nabla \varphi
		= \frac{c}{q} \, \oint_{\mathcal{L}} \left(m\cl \mathbf{v}\SC + \frac{q}{ c}\, \mathbf{A} \right) \cdot \d \mathbf{l}
		= \frac{c h} {2 e} \, n, \hspace{1cm} n \in \mathbb{Z}.
\end{equation}
In contrast to the superfluid vortex, the velocity profile of a superconducting fluxtube does not have a $1/r$-dependence but decays
exponentially for large distances, $r$, from the core, i.e. $|\mathbf{v}\SC| \sim \exp(-r/\lambda)$. Choosing a contour sufficiently
far away from the centres of individual fluxtubes, the integral over $\mathbf{v}\SC$ vanishes. Moreover, Stokes' theorem can be applied
to rewrite the contour integral into a surface integral over the surface, $\mathcal{A}$, enclosed by $\mathcal{L}$. By using the definition of the
vector potential given in Eqn.~\eqref{eqn-Vector}, a quantisation condition for the total magnetic flux, $\phi$, inside a
superconductor is found,
\begin{equation}
	 \phi = \oint_{\mathcal{L}} \mathbf{A} \cdot \d \mathbf{l}
		= \int_{\mathcal{A}} \left( \nabla \times \mathbf{A} \right) \cdot \d \mathbf{S}
		= \int_{\mathcal{A}} \mathbf{\bB} \cdot \d \mathbf{S}
		= \frac{c h} {2 e} \, n
		\equiv \phi_0 n,
		\label{eqn-FluxQuantisation}
\end{equation}
where $\phi_0$ is the magnetic flux quantum. The flux quanta of all individual fluxtubes have to add up to the total flux
inside the superconductor. Thus, one can define the magnitude of the averaged magnetic induction inside a superconducting sample by
\begin{equation}
	B \equiv \cN\ft \phi_0,
		\label{eqn-AveragedInduction}
\end{equation}
with the fluxtube surface density $\cN\ft$. As in the case of helium II, where the vortex density could be determined from
the angular velocity of the solid-body equivalent, it is possible to determine the number of fluxtubes per unit area in a superconductor
for a given magnetic induction. This again provides the possibility to estimate an average distance, $d\ft$, between individual
fluxtubes,
\begin{equation}
	d\ft \simeq \cN\ft^{-1/2} = \left( \frac{\phi_0}{B}\right)^{1/2}.
		\label{eqn-FluxtubeDistance}
\end{equation}
At the upper critical field, $H_{\rm c2}$, the fluxtubes become so densely packed that their cores start to touch. This destroys
the superconducting properties and the entire sample is turned into a normal conductor in a second-order phase transition. $H_{\rm c2}$
can be much greater than the thermodynamical critical field, $H_{\rm c}$, a fact exploited in high-field superconducting magnets.
If a type-I and type-II superconductor have the same $H_{\rm c}$, then the magnetisation depends on the external field as shown
in Fig.~\ref{fig-magnetisationI}. However, the area under both curves is the same, because it equals the condensation energy, $E_{\rm cond }$.

\subsection{Ginzburg-Landau theory in a nutshell}
\label{subsec-GLNutshell}

One approach to superconductivity that allows the reproduction of many observed phenomena is based on the theory of phase transitions.
It represents a generalisation of the theory developed by the London brothers and was pioneered by Ginzburg and Landau in
1950~\cite{Ginzburg1950}. The basic idea  is that a second-order phase transition can be characterised
by a change in an order parameter. In the case of a superconductor, this role is taken over by the macroscopic wave function, $|\Psi|^2$,
i.e.\ the Cooper pair density. The temperature is the quantity governing the transition. Above the superconducting transition
temperature, $T_{\rm c}$, no Cooper pairs are present, whereas the number of paired states increases drastically below $T_{\rm c}$.
Hence, the phase transition can be interpreted as a symmetry breaking in $|\Psi|^2$. Some of the major successes of the phenomenological
Ginzburg-Landau theory are the derivation of the critical fields of a type-II superconductor and the inclusion of interaction effects
depending non-linearly on the order parameter. Although not obtained from microscopic principles but rather from physical intuition, the
theory is particularly useful for the description of phenomena that are observable on macroscopic scales. Thus, it can also be valuable
in determining the characteristics of the superconducting neutron star interior~\cite{Alford2008,Sinha2015a}.

\subsubsection{Free energy densities and Ginzburg-Landau equations}
\label{subsubsec-FreeEnergy}

Close to $T_{\rm c}$, Ginzburg and Landau assumed that the order parameter would be small and vary only slowly in the spatial coordinate,
$\Br$. This led to the postulate that close to the transition the Helmholtz free energy of a system could be written as an expansion in
the order parameter. For matter turning superconducting in a second-order phase transition, the free energy density, $f\SOO$, in the absence
of an external field is given by
\begin{equation}
	f\SOO(\Br) = f\NOO(\Br) + \alpha \, |\Psi(\Br)|^2 + \frac{\beta}{2} \,  |\Psi(\Br)|^4
	+ \frac{\hbar^2}{2 m\cl} \left| \left( \nabla - \frac{i q}{\hbar c} \, \mathbf{A} (\Br) \right)\Psi(\Br) \right|^2
	+ \frac{\left| \mathbf{\bB}(\Br)\right|^2}{8 \pi},
		\label{eqn-GLBWithout}
\end{equation}
where the phenomenological parameters $\alpha $ and $\beta$ depend on the temperature and $f\NOO$ is the free energy density of
the normal phase in the absence of fields. All other parameters have been defined in the previous subsections. The total free
energy, $F\SOO$, is found by integrating $f\SOO$ over the volume considered. For a vanishing order parameter, i.e.\ the normal
state above $T_{\rm c}$, the free energy density reduces to the expected value $f\NOO+ \bB^2/8 \pi$ with $\bB \equiv
|\mathbf{\bB}|$. Note that the free energy density is not only related to the order parameter, $|\Psi|^2$, but also to the vector
potential, $\mathbf{A}$. This dependence was also found in BCS theory, where the supercurrent is proportional to the potential
(see Eqn.~\eqref{eqn-Current4}), illustrating the similarities between the two models, i.e.\ the fact that the Ginzburg-Landau
theory can be deduced from the microscopic framework for $T\to T_{\rm c}$~\cite{Gorkov1959}.

In the presence of an external field $\mathbf{H}$, $f\SOO$ has to be modified, because the energy density of the normal phase
contains an additional contribution corresponding to $H^2/8 \pi$ ($H \equiv |\mathbf{H}|$), generated by the external currents. Hence,
to calculate the free superconducting energy density, this term has to be subtracted and one obtains
\begin{equation}
	f\SC = f\NO + \alpha \, |\Psi|^2 + \frac{\beta}{2} \,  |\Psi|^4 + \frac{\hbar^2}{2 m\cl}
		\left| \left( \nabla - \frac{i q}{\hbar c} \, \mathbf{A} \right) \Psi \right|^2
		+ \frac{\bB^2}{8 \pi} -\frac{H^2}{8 \pi} ,
			\label{eqn-GLBWith}
\end{equation}
where $f\SC$ and $f\NO$ represent the free energy densities of the superconducting and normal phase in the presence of an external
field $H$, respectively. The spatial dependences have been omitted for clarity. The need for paying attention to the energy density
generated by the external currents becomes superfluous when another thermodynamical potential is considered. In situations where the
external field, $H$, is held constant, it is more convenient to consider Gibbs free energy densities, $g_{\text{\tiny{S,N}}}$,
related to the Helmholtz free energy densities, $f_{\text{\tiny{S,N}}}$, via a Legendre transformation:
\begin{equation}
	  g_{\text{\tiny{S,N}}} = f_{\text{\tiny{S,N}}} - \frac{\bB H}{4 \pi}.
	      	\label{eqn-GibbsEnergy}
\end{equation}
The difference between the two potentials is illustrated by taking into account the definition of the thermodynamical critical field,
$H_{\rm c}$, given in Eqn.~\eqref{eqn-CondensationEnergy}. Applying the external field, $H=H_{\rm c}$, one obtains for the
difference in the Helmholtz energy densities of the normal and the superconducting state,
\begin{equation}
 	f\NO - f\SC = f\NOO + \frac{H_{\rm c}^2}{8 \pi} - f\SOO = \frac{H_{\rm c}^2}{4 \pi} .
	\label{eqn-CalculateGibbsEnergySCI}
\end{equation}
The difference in the Gibbs energy densities on the other hand reduces to
\begin{equation}
 	g\NO - g\SC =f\NOO + \frac{H_{\rm c}^2}{8 \pi} - \frac{H_{\rm c}^2}{4 \pi} - f\SOO = 0.
\end{equation}
In contrast to the Helmholtz energy, the Gibbs energy density remains constant during the phase change of a superconducting medium. The latter
is therefore examined later on in order to calculate the critical fields of superconductivity.

Using the standard Euler-Lagrange equations, it is further possible to minimise $g\SC $ with respect to the complex conjugate wave function,
$\Psi^*$, and the vector potential, $\mathbf{A}$, to arrive at the two Ginzburg-Landau equations~\cite{Tinkham2004},
\begin{equation}
	\frac{\partial g\SC}{\partial \Psi^*} - \sum_{j=1}^3 \frac{\partial}{\partial x_j}
		\frac{\partial g\SC}{\partial \left( \nabla_j \Psi^* \right)}
		= \alpha \, \Psi + \beta \, |\Psi|^2 \, \Psi - \frac{\hbar^2}{2 m\cl} \left(\nabla - \frac{i q}{\hbar c}
		\mathbf{A} \right)^2 \Psi = 0,
		\label{eqn-GL1} \vspace{-0.5cm}
\end{equation}
\begin{equation}
	\frac{\partial g\SC}{\partial A_i} - \sum_{j=1}^3 \frac{\partial}{\partial x_j} \frac{\partial g\SC}{\partial \left( \nabla_j A_i \right)}
		=  \frac{i q \hbar}{2m\cl} \left( \Psi \nabla \Psi^* - \Psi^* \nabla \Psi \right) -
		\frac{q^2}{m\cl c} |\Psi|^2 \mathbf{A} - \frac{c}{4 \pi} \nabla \times \left( \nabla \times \mathbf{A} \right) =0,
		\label{eqn-GL2}
\end{equation}
where $x_j$ and $A_j$ are denoting the spatial components of $\mathbf{r}$ and $\mathbf{A}$, respectively. The first equality is a modified
Schr\"odinger equation for the quantum mechanical wave function, $\Psi$. Employing the vector potential~\eqref{eqn-Vector} and Amp\`ere's
law~\eqref{eqn-AmpereLaw01}, the second relation defines the quantum mechanical current density as given in Eqn.~\eqref{eqn-Current2}.

\subsubsection{Characteristic lengthscales}
\label{subsubsec-GLLengthscales}

The Ginzburg-Landau equations~\eqref{eqn-GL1} and~\eqref{eqn-GL2} introduce two characteristic lengthscales to the problem of
superconductivity; the penetration depth, $\lambda$, and the coherence length, $\xi$. They are defined as
\begin{equation}
 	\lambda \equiv \left( \frac{m\cl c^2}{4 \pi  n\cl q^2} \right)^{1/2}, \label{eqn-PenetrationDepthGL}
\end{equation}
and
\begin{equation}
 	\xi \equiv \left( \frac{\hbar^2}{2 m\cl |\alpha|} \right)^{1/2}.  \label{eqn-CoherenceLengthGL}
\end{equation}
The first quantity is equivalent to the London penetration depth derived within the phenomenological London theory (see Eqn.~\eqref{eqn-lambda}).
It describes the lengthscale on which the Meissner effect suppresses the magnetic induction in the interior of the superconductor. Since the density
of superconducting particles vanishes at the transition temperature, the penetration depth diverges as $T\to T_{\rm c}$. The second quantity represents
the typical distance over which the order parameter, $|\Psi|^2$, varies in space. It is also temperature-dependent and diverges close to the transition
temperature. Comparing to  BCS theory, the coherence length is identified with the dimension of a single Cooper pair. It is defined in terms of the
temperature-dependent energy gap, $\Delta$, and the Fermi velocity, $v\F$, related to the Fermi wave number, $k\F$, via $v\F = \hbar k\F/m\cl^*$. One
therefore has
\begin{equation}
 	\xi_{\rm BCS} \equiv \frac{\hbar v\F}{\pi \Delta} = \frac{\hbar^2 k\F}{m\cl^* \pi \Delta}.
		\label{eqn-CoherenceLengthBCS}
\end{equation}
The effective mass $m\cl^*$ (also referred to as the \textit{Landau effective mass}) characterises a static quantum mechanical ground state.
Note that this effective mass differs from the dynamical effective mass used in the context of neutron stars~\cite{Chamel2006}.

The ratio of the two lengthscales defined in Eqns.~\eqref{eqn-PenetrationDepthGL} and~\eqref{eqn-CoherenceLengthGL} is referred to as the Ginzburg-Landau
parameter, $\kappa_{\rm GL}$. There exists a critical value, $\kappa_{\rm crit} \equiv 1/\sqrt{2}$, that classifies the type of superconductivity. More
precisely
\begin{equation}
 	\text{type-I:} \hspace{0.7cm} \kappa_{\rm GL} \equiv \frac{\lambda}{\xi} < \kappa_{\rm crit},
	      \hspace{1.5cm}
 	\text{type-II:}	\hspace{0.7cm} \kappa_{\rm GL} \equiv \frac{\lambda}{\xi} > \kappa_{\rm crit}.
\end{equation}
The first case is characterised by $\lambda \lesssim\xi $ and a positive surface energy. In the second case ($\lambda \gtrsim \xi $), the surface energy
is negative and the material is in an unstable state. It becomes energetically favourable for the superconductor to divide into regions of order $\xi$
and form a fluxtube array, each fluxtube carrying $\phi_0$. The existence of $\kappa_{\rm crit}$ is thus related to the surface energy and the transition
from an attractive interfluxtube potential in the type-I state to a repulsive interaction for type-II media.

\subsubsection{Critical fields I}
\label{subsubsec-CriticalFieldsI}

The Ginzburg-Landau formalism also provides the means to calculate the critical fields of superconductivity. Firstly, the thermodynamical field,
$H_{\rm c}$, is related to the free parameters, $\alpha$ and $\beta$. An expression is found by equating the Gibbs free energy densities of the
normal and the superconducting state at equilibrium, $H=H_{\rm c}$. For the bulk of a superconducting medium, the free energy density is minimised
by a constant order parameter and a zero vector potential, $\mathbf{A}=0$, which implies that the induction vanishes, $\mathbf{\bB}=0$. The exact
value obtained from Eqn.~\eqref{eqn-GL1} is
\begin{equation}
	|\Psi_{\infty}|^2 \equiv - \frac{\alpha}{\beta} =\frac{|\alpha|}{\beta},
			\label{eqn-InfinityPsi}
\end{equation}
where $\alpha <0$ and $\beta \approx \text{constant}$ for a superconductor~\cite{Tinkham2004}. In the normal phase on the other hand, the minimum
of the energy density is related to a vanishing order parameter, $|\Psi|^2=0$, and $\mathbf{\bB} = \mathbf{B}=\mathbf{H}$. Hence, the Gibbs free
energy densities that have to be equal at the phase transition are
\begin{equation}
	g\SC = f\SC = f\NO - \frac{|\alpha|^2}{2 \beta} - \frac{H^2}{8 \pi}, \hspace{1.5cm}
	g\NO = f\NO - \frac{H^2}{4 \pi}.
		\label{eqn-GibbsNHc}
\end{equation}
Using $H=H_{\rm c}$ leads to the following identity for the critical thermodynamical field,
\begin{equation}
	H_{\rm c} = \left( \frac{4 \pi |\alpha|^2}{\beta} \right)^{1/2}.
		\label{eqn-CriticalFieldHC}
\end{equation}

Additionally, the lower critical field, $H_{\rm c1}$, can be determined. It marks the value at which flux first enters the superconductor. So the Gibbs
energy, $G\SC$, of the sample without fluxtubes must be equal to the case where exactly one fluxtube is present,
\begin{equation}
	G\SC \big|_{\text{no flux}} = G\SC \big|_{\text{one fluxtube}}.
\end{equation}
The total Gibbs energy, $G\SC$, is then obtained by integrating the Gibbs energy density, $g\SC$, over the superconductor's volume, $\mathcal{V}$,
\begin{equation}
	G\SC = \int_{\mathcal{V}} g\SC \, \d V = F\SC - \frac{H}{4 \pi} \int_{\mathcal{V}} \bB \, \d V.
\end{equation}
In the Meissner state, where the fluxtubes are absent and $\bB=0$, the Gibbs energy equals the Helmholtz energy, $G\SC \big|_{\text{no flux}}
=F\SC$. For the single fluxtube, one has instead
\begin{equation}
	 G\SC \big|_{\text{one fluxtube}} = F\SC + \cE\ft L - \frac{H_{\rm c1} \phi_0 L}{4 \pi}.
\end{equation}
Here, $\cE\ft$ denotes the increase in the free energy per unit length due to the presence of a fluxtube  of
length $L$. Hence, the lower critical field is given by \vspace{-0.05cm}
\begin{equation}
	H_{\rm c1} = \frac{4 \pi \cE\ft}{\phi_0},
		\label{eqn-CriticalFieldHC1}  \hspace{1cm} \text{with} \hspace{1cm}
		\cE\ft = \left( \frac{\phi_0}{4 \pi \lambda} \right)^2  \ln \kappa_{\rm GL}. \vspace{-0.05cm}
\end{equation}

\subsubsection{Surface energy}
\label{subsubsec-SurfaceEnergy}

A similar energy statement can be used to calculate the surface energy, $\delta$, which determines how magnetic flux is distributed inside a superconducting
sample to minimise the total energy. More precisely, $\delta$ is obtained by comparing the Gibbs free energies of the pure, flux-free type-I phase and the
coexisting state (in which the magnetic flux is able to penetrate the superconductor) at the thermodynamical critical field, $H=H\cl$. The physical behaviour of
type-I and type-II superconductors is fundamentally different at this point and the surface energy is given by
\begin{equation}
	\delta = G\SC \big|_{H=H\cl, \text{ coexisting}} - G\SC \big|_{H=H\cl, \text{ no flux}}.
\end{equation}
To simplify the problem, a one-dimensional set-up along the $x$-axis is considered. The total Gibbs energies are obtained by integrating the corresponding
densities along this coordinate. At $H=H\cl$, the energy density of the flux-free Meissner state is equal to the Gibbs energy density of the normal
state (see Eqn.~\eqref{eqn-GibbsNHc}) giving
\begin{equation}
	\delta = \int_{-\infty}^{\infty} \, \left( g\SC - g\SC \big|_{\text{no flux}} \right) \d x
		= \int_{-\infty}^{\infty} \, \left( f\SC
		- \frac{\bB(x) H\cl}{4 \pi} - f\NO + \frac{H\cl^2}{4 \pi} \right) \d x,
\end{equation}
where the definition of $g\SC$ given in Eqn.~\eqref{eqn-GibbsEnergy} has been employed. In the coexisting phase, the magnetic induction is no longer
zero but instead a function of $x$. Substituting the Ginzburg-Landau free energy density~\eqref{eqn-GLBWith} for $f\SC$, we obtain
\begin{equation}
	\delta = \int_{-\infty}^{\infty} \, \left( \alpha \, |\Psi|^2 + \frac{\beta}{2} \,  |\Psi|^4
		+ \frac{\hbar^2}{2 m} \left| \left( \nabla - \frac{i q}{\hbar c} \, \mathbf{A}\right)\Psi \right|^2
		+ \frac{\bB^2}{8 \pi} + \frac{H\cl^2}{8 \pi}- \frac{\bB H\cl}{4 \pi} \right) \d x.
			\label{eqn-SurfaceEnergy0}
\end{equation}
This can be further simplified by taking advantage of the first Ginzburg-Landau equation. Multiplying Eqn.~\eqref{eqn-GL1} with $\Psi^*$, integrating over
the $x$-direction and performing an integration by parts gives the following
\begin{equation}
	\int_{-\infty}^{\infty} \, \left( \alpha \, |\Psi|^2 + \beta \,  |\Psi|^4 + \frac{\hbar^2}{2 m}
		\left| \left( \nabla - \frac{i q}{\hbar c} \, \mathbf{A}\right)\Psi \right|^2 \right) \d x = 0.
\end{equation}
Substituting this back into Eqn.~\eqref{eqn-SurfaceEnergy0}, the surface energy reduces to
\begin{equation}
	\delta = \int_{-\infty}^{\infty} \, \left( - \frac{\beta}{2} \,  |\Psi(x)|^4
		+ \frac{\left[\bB(x)-H\cl \right]^2}{8 \pi} \right) \d x.
		\label{eqn-SurfaceEnergy1}
\end{equation}
Generally speaking, this expression has to be integrated numerically while simultaneously solving the Ginzburg-Landau equations to provide expressions
for the order parameter and the vector potential. However, Eqn.~\eqref{eqn-SurfaceEnergy1} can be rewritten in terms of dimensionless quantities, which
allow for a more intuitive interpretation. Using Eqns.~\eqref{eqn-InfinityPsi} and~\eqref{eqn-CriticalFieldHC}, one obtains
\begin{equation}
	\delta = \int_{-\infty}^{\infty} \, \frac{H\cl^2}{4 \pi} \left(- \frac{|\tilde{\Psi}(x)|}{2}
			+ \left[\tilde{B}(x)- \frac{1}{\sqrt{2}} \right]^2 \right) \d x,
	\label{eqn-SurfaceEnergy2}
\end{equation}
where
\begin{equation}
	\tilde{\Psi}(x) \equiv \frac{\Psi(x)}{\Psi_{\infty}}, \hspace{1.5cm} \tilde{B}(x) \equiv \frac{\bB(x)}{\sqrt{2} H\cl}.
	\label{eqn-DimensionlessQuantities}
\end{equation}
Deep inside the normal and superconducting regions, the integrand of Eqn.~\eqref{eqn-SurfaceEnergy2} is constant and of small magnitude. Thus, $\delta$ is
localised around the interface, justifying the name surface energy. The relation also explains the different behaviour of the two types of superconductors.
For $\kappa_{\rm GL} \gg 1$, the field reaches far into the superconducting sample, which results in $\bB \approx H\cl$ and $\tilde{B} \approx 1/\sqrt{2}$.
$\delta$ is thus negative and it becomes energetically favourable for the superconductor to increase the surface of superconducting-normal domain walls.
Hence, it divides into microscopic structures of order $\xi$, which exactly describes the formation of the fluxtube array. For $\kappa_{\rm GL} \ll 1$ however,
the field in the superconducting region vanishes, i.e. $\bB=0$. Since the normalised order parameter is always smaller than one, i.e. $\tilde{\Psi} \leq 1$,
the surface energy, $\delta$, of the interface is positive and regions of macroscopic flux represents the lowest energy state. The type-I sample therefore
forms an intermediate state and does not split into individual fluxtubes.

\subsubsection{Critical fields II}
\label{subsubsec-CriticalFieldsII}

The Ginzburg-Landau theory also allows one to determine the upper critical field, $H_{\rm c2}$. For a decreasing external field, $H_{\rm c2}$ represents the
maximum value at which a sample can still become superconducting in a second-order phase transition. At this point, the order parameter stays small and the
non-linear term in the Ginzburg-Landau equation~\eqref{eqn-GL1} is negligible. Furthermore, screening effects caused by supercurrents remain small, implying
that the averaged and the mesoscopic magnetic induction inside the superconductor are close to the external field, $\mathbf{H}$. This allows the identification
$\mathbf{A} = \mathbf{A}_{\rm ext}$ and results in the decoupling of the two Ginzburg-Landau equations (where $\mathbf{A}_{\rm ext}$ is the vector potential
describing the external field, $\mathbf{H}$). Linearising Eqn.~\eqref{eqn-GL1} leads to
\begin{equation}
	- \left( \nabla - \frac{i q}{\hbar c} \mathbf{A} \right)^2 \Psi = \frac{\Psi}{\xi^2}.
		\label{eqn-LinearisedGL}
\end{equation}
Using Cartesian coordinates $\{ x,y,z \}$ and assuming $\mathbf{H}=H_0 \uz$, a possible choice for the potential would be $\mathbf{A} = x H_0 \uy$. In this
case, the vector potential only depends on the spatial coordinate $x$, suggesting that the solution of Eqn.~\eqref{eqn-LinearisedGL} is of the form $\Psi(x,y,z)
= f(x) \, \text{e}^{i(k_y y+k_z z)}$ (with the wave numbers $k_y$ and $k_z$ in $y$- and $z$-direction, respectively). For an infinite medium, one
then obtains an equation equivalent to a Schr\"odinger equation for a particle in a harmonic oscillator potential,
\begin{equation}
	\frac{\hbar^2}{2m\cl} \left[ - \partial_x^2 + \left( \frac{2 \pi H_0}{\phi_0} \right)^2
		\left( x - \frac{k_y \phi_0}{2 \pi H_0} \right)^2 \right] f(x)
		= \frac{\hbar^2}{2m\cl} \left( \frac{1}{\xi^2} - k_z^2 \right) f(x),
		\label{eqn-ModifiedSE}
\end{equation}
whose solutions correspond to discrete Landau levels (see for example Landau and Lifshitz~\cite{Landau1977}) that are highly degenerate and characterised by the
energy eigenvalues
\begin{equation}
	E_n = \hbar \omega_c \left( \frac{1}{2} + n \right) = \frac{\hbar q H_0}{m\cl c} \left( \frac{1}{2} + n \right).
		\label{eqn-LandauEnergies}
\end{equation}
Here $\omega_c$ is the cyclotron frequency related to the magnitude of the average induction. These quantised energies have to be identical to the right-hand
side of Eqn.~\eqref{eqn-ModifiedSE}, providing a relation for the parameter $H_0$. Its maximum value then corresponds to the upper critical field, $H_{\rm c2}$,
which is obtained for $n=0$ and $k_z=0$:
\begin{equation}
	H_{\rm c2} = \frac{\phi_0}{2 \pi \xi^2}.
	    \label{eqn-CriticalFieldHC2}
\end{equation}
For higher external fields, the medium can no longer condense into a superconducting state but remains normal. For a more detailed derivation see Tinkham
\cite{Tinkham2004}. The eigenfunctions related to the minimum energy state at $H_{\rm c2}$ are
\begin{equation}
	\Psi(x,y) = \exp \left[ - \frac{1}{2 \xi^2} \left( x - \frac{k_y \phi_0}{2 \pi H_0} \right)^2 \right] \, \exp \left[i k_y y \right].
	    \label{eqn-OrderParameterLinearised}
\end{equation}
These linearised solutions for the order parameter play an important role in deriving the structure of the fluxtube array.

\subsubsection{Fluxtube array formation}
\label{subsubsec-Abrikosov}

Studies of the fluxtube arrangement in type-II superconductors were pioneered by Abrikosov~\cite{Abrikosov1957} and his work is briefly summarised in the
following. The original calculation is based on the same concepts as employed in the derivation of the upper critical field. For external magnetic fields
below $H_{\rm c2}$ however, the non-linear term in the Ginzburg-Landau equation~\eqref{eqn-GL1} can no longer be neglected. Using the linearised, decoupled
Ginzburg-Landau equations is thus not sufficient any more. Instead for $H \lesssim H_{\rm c2}$, the effects of the non-linear term can be included by
applying perturbation theory. In this case, the averaged and the mesoscopic magnetic induction in the bulk are no longer equal to the applied field but
rather the sum of the external field and a small correction produced by the circulating supercurrents, $\Bj\s$. Hence, the vector potential, $\BA$,
satisfies
\begin{equation}
	 \mathbf{\bB} =  \mathbf{H} +  \mathbf{\bB}\s = \nabla \times \BA,
		\label{eqn-MagneticInductionLocal}
\end{equation}
where
\begin{equation}
	\nabla \times \mathbf{\bB}\s = \frac{4 \pi}{c} \Bj\s.
		\label{eqn-SupercurrentLocal}
\end{equation}
At this point, it is beneficial to separate the potential into two contributions, i.e. $\BA \equiv \BA_{\rm c2} + \BA_1$. The former part
is the potential generating the upper critical field and the latter perturbative contribution contains all information about the additional
magnetic fields. Combining these relations, one finds
\begin{equation}
	\nabla \times \BA_1 =  \mathbf{H} - \mathbf{H}_{\rm c2} + \mathbf{\bB}\s.
		\label{eqn-LocalPotential}
\end{equation}
Taking as before an external field parallel to the $z$-axis allows the choice $\BA_{\rm c2} = x H_{\rm c2} \uy$. Right below $H_{\rm c2}$, one
would further expect the solution for $\Psi$ to be close to the solution of the linearised equations calculated previously. Hence, $\Psi \equiv
\Psi_0 + \Psi_1$, where the two functions are orthogonal and satisfy the condition
\begin{equation}
	\int \Psi^*_0 \Psi_1 \, \d V = 0.
		\label{eqn-Normalisation}
\end{equation}
This implies that the lowest energy eigenfunctions, $\Psi_0$, and the perturbative contributions, $\Psi_1$, are linear independent. Having calculated
the eigenfunctions of the linearised system (see Eqn.~\eqref{eqn-OrderParameterLinearised}), the $k_y$-dependence of the resulting order parameter
shows that there are infinitely many flux configurations able to generate the energy state $H_{\rm c2}$. Below $H_{\rm c2}$, the non-linearity breaks
this degeneracy and favours a particular fluxtube arrangement. As any regular array should have a lower energy than a random flux distribution, the
wavefunction is expected to be periodic. This behaviour can be ensured by choosing the ansatz $k_y = n k $ with $n \in \mathbb{N}$. The most general
solution of the linearised Ginzburg-Landau equations keeping periodicity in $y$-direction is thus the linear combination
\begin{equation}
	\Psi_0(x,y)  = \sum_n C_n \, \exp \left[ - \frac{1}{2 \xi^2} \,
		\left( x - \frac{n k \phi_0}{2 \pi H_{\rm c2}} \right)^2 \right] \, \exp\left[ i n k y \right].
		\label{eqn-SolutionPsiLin}
\end{equation}
The periodicity in $x$-direction can also be recovered if the coefficients satisfy $C_{n+N} = C_n$ for $N \in \mathbb{N}$. A square lattice is
represented by $N=1$ (implying that all coefficients are equal), whereas $N=2$ together with $C_1 = i C_0$ characterises a triangular lattice.

Substituting the periodic wave function~\eqref{eqn-SolutionPsiLin} and the external potential $\BA_{\rm c2}$ into the left-hand side
of the second Ginzburg-Landau equation~\eqref{eqn-GL2}, one can determine the quantum mechanical current density associated with the linear
solution,
\begin{equation}
	\Bj\s = \frac{q \hbar}{ 2 m\cl} \left( - \partial_y |\Psi_0|^2 \ux + \partial_x |\Psi_0|^2 \uy \right).
		\label{eqn-LinearCurrent}
\end{equation}
In turn, this current generates an additional magnetic induction inside the superconductor. According to Eqn.~\eqref{eqn-SupercurrentLocal}, it reads
\begin{equation}
	\mathbf{\bB\s} = - \frac{q h}{m\cl c} \, |\Psi_0|^2 \uz .
		\label{eqn-InducedField}
\end{equation}
Using these results, it is possible to calculate a more accurate approximate solution to the first Ginzburg Landau equation. Substituting the
decompositions for $\BA$ and $\Psi$ into Eqn.~\eqref{eqn-GL1} and linearising the result gives an equation for $\Psi_1$:
\begin{equation}
	\mathcal{H}_0 \Psi_1 + \beta |\Psi_0|^2 \Psi_0 +
		\frac{1}{c} \BA_1 \cdot \left( \frac{i q \hbar}{m\cl c} \nabla + \frac{q^2}{m\cl c} \BA_{\rm c2} \right) \Psi_0 = 0.
			\label{eqn-LinearisedGL1}
\end{equation}
The zeroth-order operator $\mathcal{H}_0$ is defined by
\begin{equation}
	\mathcal{H}_0 \equiv \alpha - \frac{\hbar^2}{2 m\cl} \left( \nabla - \frac{i q}{\hbar c} \BA_{\rm c2} \right)^2
\end{equation}
and by construction satisfies $\mathcal{H}_0 \Psi_0 = 0$. Taking advantage of the orthogonality of $\Psi_0$ and $\Psi_1$, the
normalisation condition~\eqref{eqn-Normalisation} can then be extended to
\begin{equation}
	\int \Psi^*_0 \, \mathcal{H}_0 \Psi_1 \, \d V = 0,
\end{equation}
which can be rewritten using Eqn.~\eqref{eqn-LinearisedGL1},
\begin{equation}
	\int \left[ \beta |\Psi_0|^4 - \frac{1}{c} \BA_1  \cdot  \left( - \frac{i q \hbar}{2 m\cl c} \, 2 \Psi^*_0  \nabla \Psi_0
		- \frac{q^2}{m\cl c} \BA_{\rm c2} |\Psi_0 |^2 \right) \right]  \d V = 0.
\end{equation}
After integrating the second term by parts and neglecting the surface contribution, the expression in round brackets is equivalent
to the linear quantum mechanical current density, $\Bj\s$ (see Eqn.~\eqref{eqn-GL2}). Using Amp\`ere's law~\eqref{eqn-SupercurrentLocal},
one can further simplify
\begin{equation}
	\int \left( \beta |\Psi_0|^4 - \frac{1}{c} \BA_1 \cdot \Bj\s \right) \, \d V
		= \int \left[ \beta |\Psi_0|^4 - \frac{1}{4 \pi} \BA_1 \cdot \left( \nabla \times \mathbf{\bB}\s \right) \right] \d V = 0.
\end{equation}
Employing a vector identity for the second term and ignoring the total gradient (only contributing at the surface of
the sample and not the bulk), one obtains
\begin{equation}
	 \frac{1}{4 \pi} \BA_1 \cdot \left( \nabla \times \mathbf{\bB}\s \right)  = \frac{1}{4 \pi} \mathbf{\bB}\s \cdot \left( \nabla \times
		\BA_1 \right) = \frac{1}{4 \pi} \mathbf{\bB}\s \cdot \left( \mathbf{H} - \mathbf{H}_{\rm c2} + \mathbf{\bB}\s \right),
\end{equation}
where Eqn.~\eqref{eqn-LocalPotential} has been used to simplify the result. Based on the initial choice for $\mathbf{H}$, all magnetic fields
are aligned in $z$-direction, which allows the right-hand side to be simplified. Using Eqn.~\eqref{eqn-InducedField} then leads to
\begin{equation}
	\int \left[ |\Psi_0|^4 \left( \beta - \frac{ \pi q^2 \hbar^2}{m\cl^2 c^2} \right)
		- \frac{q \hbar}{2 m\cl c} \, |\Psi_0|^2 \left( H_{\rm c2} - H \right)  \right] \d V  = 0.
\end{equation}
As the parameter $\beta$ is related to the Ginzburg-Landau parameter, $\kappa_{\rm GL}$, via~\cite{Tinkham2004}
\begin{equation}
	\kappa_{\rm GL}^2 = \frac{\beta c^2 m\cl^2}{2 \pi \hbar^2 q^2},
\end{equation}
one can rewrite
\begin{equation}
	\int \left[ |\Psi_0|^4 \left( 2 \kappa_{\rm GL}^2 - 1 \right)
		- \frac{m\cl c} {q h}  \, |\Psi_0|^2 \left( H_{\rm c2} - H \right)  \right] \d V  = 0.
		\label{eqn-AbrikosovCalculationHelp}
\end{equation}
This again illustrates the importance of the critical value $\kappa_{\rm crit} = 1/\sqrt{2}$. The linear order  parameter, $\Psi_0$, is
position-dependent and the exact configuration of the fluxtube array has to be known to evaluate the integral. However, it is possible
to deduce more general information about the fluxtube distribution close to the upper critical field by defining a macroscopic average
of a function $f(\mathbf{r})$ by
\begin{equation}
	\langle  f( \mathbf{r}) \rangle \equiv \int f ( \mathbf{r}) \, \d V.
\end{equation}
Eqn.~\eqref{eqn-AbrikosovCalculationHelp} then reduces to
\begin{equation}
	\langle |\Psi_0|^4 \rangle \left( 2 \kappa_{\rm GL}^2 - 1 \right)
		= \frac{m\cl c} {q h} \langle |\Psi_0|^2 \rangle \left( H_{\rm c2} - H \right).
		\label{eqn-AveragedRelation}
\end{equation}
Using Eqns.~\eqref{eqn-MagneticInductionLocal},~\eqref{eqn-InducedField} and~\eqref{eqn-AveragedRelation}, the average magnetic
induction is
\begin{equation}
	\mathbf{B} = \langle \mathbf{\bB} \rangle = \langle \mathbf{H} \rangle + \langle \mathbf{\bB}\s \rangle
		 = \mathbf{H}- \frac{\mathbf{H}_{\rm c2} - \mathbf{H}}{\left( 2 \kappa_{\rm GL}^2 - 1 \right) \beta_A},
\end{equation}
where we followed Abrikosov's work and defined
\begin{equation}
	\beta_A \equiv \frac{\langle |\Psi_0|^4 \rangle}{ \left( \langle |\Psi_0|^2 \rangle \right) ^2} .
\end{equation}
Note that $\beta_A$ is independent of the normalisation condition, reduces to unity for order parameters that are constant in space and becomes larger
for more localised wave functions, i.e. $\beta_A \geq 1$. With the help of Eqn.~\eqref{eqn-AverageInduction}, one can finally obtain an
expression for the magnetisation of the superconductor, i.e.
\begin{equation}
	 \mathbf{M} = - \frac{1}{4 \pi} \frac{\mathbf{H}_{\rm c2} - \mathbf{H}}{\left( 2 \kappa_{\rm GL}^2 - 1 \right) \beta_A},
\end{equation}
which is directly related to the total Gibbs energy via $\partial G / \partial H |_T = - M$. Integrating $M \equiv |\mathbf{M}|$ with respect
to the external field, one therefore arrives at
\begin{equation}
	G(H) = G(H_{\rm c2}) - \frac{1}{4 \pi} \, \frac{\left( H_{\rm c2} - H \right)^2}{\left( 2 \kappa_{\rm GL}^2 - 1 \right) \beta_A}.
\end{equation}
The smaller the parameter $\beta_A$ is, the smaller the Gibbs energy of the system for a given external field $H$ compared to the corresponding
value at $H_{\rm c2}$. As Abrikosov first showed in his seminal paper, $\beta_A$ ultimately determines the structure of the fluxtube array
because the free energy is minimised for the smallest $\beta_A$ value. Comparing a square and a triangular lattice, distinguished by the choice
of coefficients $C_n$ (see Eqn.~\eqref{eqn-SolutionPsiLin}), one finds that in the former case $\beta_A = 1.18$ and in the latter $\beta_A
= 1.16$, making the hexagonal configuration the favourable one.


\section{Astrophysical Condensates}
\label{subsec-Evidence}

Having presented the main aspects of the classical treatment of macroscopic quantum condensates, we now return to neutron star astrophysics and
implications of vortices and fluxtubes on the star's dynamics. Following a brief review of the canonical neutron star structure and the current
understanding of superfluid and superconducting components, astrophysical neutron star observables are discussed in more detail.

\subsection{Basic neutron star structure}

Even though the detailed interior structure of neutron stars remains unknown, it is clear that these stars are a bit like layer cakes (see
Fig.~\ref{cross}), with regions of distinct composition and states of matter separated by phase transitions. Roughly speaking a neutron
star is divided into three regions; the crust, where neutron-rich nuclei form an elastic lattice coexisting with a neutron superfluid,
the outer core, which is dominated by neutrons, protons, electrons and (most likely) muons, and the inner core, where more exotic phases
like hyperons or deconfined quarks may be present.

The neutron star crust is approximately $1 \, \km$ thick\footnote{For a comprehensive summary on the physics of neutron star crusts see Chamel
and Haensel~\cite{Chamel2008b}; our values for the transition densities are taken from their review.}. In thermodynamic equilibrium, the
strong Coulomb forces in the outer crust result in the formation of a $^{56}$Fe lattice~\cite{Shapiro2004}. At about $10^4 \, \g \,
\cm^{-3}$ the atoms are fully ionised, creating a free relativistic electron gas. The lattice energy in this density regime can be
calculated in the Wigner-Seitz approximation~\cite{Wigner1933}, where crustal nuclei are separated into independent spheres centred around
individual lattice sites. Each of these spheres is electrically neutral and it is found that a body-centred cubic (bcc) configuration
minimises the energy of the lattice.

For increasing densities the nuclei become more and more neutron rich and the distance between individual lattice
sites decreases. At $\rho\D \sim 4 \times 10^{11} \, \g \, \cm^{-3}$ (the neutron drip density), it is energetically
favourable for the neutrons to drip out of the nuclei and form a free neutron gas. This state of matter has been
studied intensively using theoretical approaches such as modified liquid-drop models~\cite{Douchin2001}. Quantum
calculations have further shown that the number of protons per nuclei is almost constant with $Z = 32, 40,
50$~\cite{Negele1973}. As the free neutron gas has similar properties to the conducting electrons in metals, nuclear
band theory~\cite{Chamel2005} has also proven very useful in studying the characteristics of the inner crust. Chamel
has shown that the free neutrons are less mobile due to crustal entrainment~\cite{Chamel2013}, an effect that
refers to the coupling of the neutron gas to the nuclear lattice via Bragg scattering. This can be expressed in terms
of an effective mass, which could be considerably higher than the bare mass, having important implications for the
dynamics of the star~\cite{Andersson2012}. Additionally, the crustal dynamics are influenced by macroscopic quantum
effects. Due to an attractive contribution to the nucleon-nucleon interaction, neutrons can form Cooper pairs, giving
rise to superfluidity~\cite{Sauls1989}. It is expected that the crustal superfluid pairs in a spin-singlet, $s$-wave
($^1 S_0$) state, having properties similar to isotropic helium II (see Sec.~\ref{sec-Helium}).

As the density increases further, the number of bound neutrons in the nuclei decreases and the lattice sites move
closer to each other. The transition to the core state, where the lattice structure vanishes completely and pure
nuclear matter dominates, is not sharp but rather smooth. At approximately $10^{14} \, \g \, \cm^{-3}$, the
particles start to form exotic shapes, giving this part of the neutron star crust the name \textit{pasta
phase}~\cite{Ravenhall1983, Horowitz2015}. The spherical nuclei first turn into cylinders (spaghetti) and slabs (lasagne).
Further inside the star, the situation is inverted and the free neutrons form tubes and bubbles enclosed by nuclear
matter. The pasta has the properties of solids and liquids and its behaviour may be described by the theory of
liquid crystals~\cite{Pethick1998}. This layer (although possibly quite thin) comprises most of the crust's mass and
could therefore influence the star's rotational properties. Pons et al.~\cite{Pons2013} suggested that the
observed lack of isolated long-period X-ray pulsars could be explained by a layer of high electrical resistivity
such as the nuclear pasta phase, causing effective dissipation of magnetic energy subsequently resulting in the
saturation of the electromagnetic spin-down behaviour.

The neutron star core contains about $90$\% of the neutron star's mass and has a radius of approximately $9-10 \, \km$.
The crustal structures have completely vanished at densities above $\rho_0/3 \sim 10^{14} \, \g \, \cm^{-3}$ ($\rho_0
\sim 2.8 \times 10^{14} \, \g \, \cm^{-3} $ denotes the nuclear saturation density). For larger densities, information
about the state of nuclear matter has not been experimentally tested on Earth. Usually, existing theories of bulk
matter are extrapolated to the outer core of a neutron star. Up to densities of $2 \rho_0$, matter is thought to consist
of neutrons and a small fraction of protons, relativistic electrons and possibly muons~\cite{Shapiro2004}. The
neutrons in the interior are described by a spin-triplet, $p$-wave ($^3 P_2$) order parameter~\cite{Sauls1989} (a
quantum state comparable to anisotropic helium-3 superfluid created in laboratory experiments), whereas protons are
pairing in a $^1 S_0$ state and most likely exhibit properties of a type-II superconductor~\cite{Baym1969}.

In the inner core (at densities above $2 \rho_0$) the structure of neutron stars is completely unknown. The main problem
for theoretical calculations at these densities is the appearance of new degrees of freedom~\cite{Weber2007}.
Particles such as kaons, pions, hyperons or other exotic species could be generated and change the properties of nuclear
matter. At extremely high densities near the star's centre, $\rho \sim 10^{15} \, \g \,\cm^{-3}$, QCD calculations
even predict transitions to a deconfined quark plasma, creating exotic states such as colour superconductors~\cite{Alford2008a,
Alford2013}. However, it is not known if neutron stars are compact enough to reach such high densities and detailed
observations are required to determine which equation of state adequately captures the properties of the interior.

\subsection{Pairing gaps and superfluidity}

The idea of superfluidity and superconductivity in astrophysical objects was first put forward by Migdal in 1959~\cite{Migdal1959}, several
years before the first pulsar signal was observed~\cite{Hewish1968}. Generalising the theory developed for terrestrial macroscopic quantum
systems to the neutron star case would imply the presence of a neutron superfluid and a proton superconductor, while the relativistic electrons
are in a normal state, as their transition temperature lies well below the typical neutron star temperatures. As nucleons are fermions, they
have to form Cooper pairs before condensing into a superfluid phase. At densities below  $\rho_0/3 \sim 10^{14} \, \g \, \cm^{-3}$, i.e.
in the stars' crust, the particles are most likely to experience pairing in a spin-singlet, $s$-wave state with vanishing angular momentum
($^1S_0$). In the core, where densities above $\rho_0$ are present, the spin-triplet, $p$-wave pairing channel of non-zero angular momentum
($^3P_2$) is the most attractive one, see Fig.~\ref{sfgaps}. The crustal phase is hence similar to helium II, while at high densities the
superfluid neutrons behave like an anisotropic helium-3 superfluid. Moreover, the protons in the core are expected to condense into a $^1S_0$
state and exhibit superconducting properties. The neutron and proton transition temperatures typically range between~\cite{Page2011, Shternin2011,
Ho2012b} (see Fig.~\ref{sfgaps})
\begin{align}
	T_{\rm c n, \, singlet} &\approx 10^{9} - 10^{10} \, \K,  \\[1.7ex]
	T_{\rm c n, \,  triplet} &\approx 10^{8} - 10^{9} \, \K, \\[1.7ex]
	T_{\rm cp, \, singlet} &\approx 10^{9} - 10^{10} \, \K.
\end{align}
For more information on pairing in neutron stars see Sauls~\cite{Sauls1989} and Gezerlis et al.~\cite{Gezerlis2014}.

The presence of quantum condensates has a crucial influence on the rotational and magnetic properties of the compact object. Firstly, a superfluid
interior drastically changes the rotational dynamics in comparison to a normal matter core. As observed in helium II, the bulk spin is controlled
by the dynamics of vortices quantising the circulation. Secondly, in comparison to a normal conductor, the magnetic field is no longer locked to
the charged plasma but instead contained within fluxtubes, so that the evolution of the magnetic field is determined by their behaviour. Numerical
values for the quantum of circulation and magnetic flux are
\begin{align}
	  \kappa = \frac{h}{2 m\n} \approx 2.0 \times 10^{-3} \, \cm^{2} \, \text{s}^{-1}, \\[2.7ex]
	  \phi_0 = \frac{ch}{2 e} \approx 2.1 \times 10^{-7} \, \G \, \cm^{2},
		\label{eqn-ValueKappaPhi}
\end{align}
where $m\n$ denotes the mass of a neutron. Assuming that the two macroscopic quantum states are indeed present in the neutron star's interior,
Eqns.~\eqref{eqn-VortexLineDensity} and~\eqref{eqn-AveragedInduction} can be used to estimate typical values for the neutron vortex and proton
fluxtube surface densities, $\cN\n$ and $\cN\p$, respectively. Normalising the results for a rotation period of $P_{10} \equiv P/(10 \, \ms)$
and a dipole field strength of $B_{12} \equiv B/ (10^{12} \, \G)$, one has
\begin{align}
	 \cN\n = \frac{2 \Omega}{\kappa} = \frac{4 \pi}{\kappa P} \approx  6.3 \times 10^5 P^{-1}_{10} \, \cm^{-2},
			\label{eqn-ArraySurfaceDensityNeutron} \\[2.7ex]
	 \cN\p = \frac{B}{\phi_0} \approx 4.8 \times 10^{18} \, B_{12} \, \cm^{-2},
		\label{eqn-ArraySurfaceDensityProton}
\end{align}
which correspond to the following intervortex and interfluxtube spacings,
\begin{align}
	 d\n \simeq \cN\n^{-1/2} \approx 1.3 \times 10^{-3} \, P^{1/2}_{10} \, \cm, \label{eqn-ArrayDistanceNeutron} \\[3.2ex]
	 d\p \simeq \cN\p^{-1/2} \approx 4.6 \times 10^{-10} \, B^{-1/2}_{12} \, \cm.
		\label{eqn-ArrayDistanceProton}
\end{align}
This implies that there are significantly more fluxtubes per unit area than neutron vortices in the interior of a canonical neutron star.
In addition to new forces that arise from the coupling of vortices or fluxtubes with their respective fluid components and the charged
plasma, the two arrays are also able to interact with each other. In the outer core, fluxtubes and vortices might be strongly interacting,
which could have an influence on the magnetic and rotational evolution as they would be no longer independent~\cite{Ruderman1998,
Jahan-Miri2000}.

\subsection{Superconductivity}
\label{subsec-FurtherSuperCon}

Analogous to experiments with laboratory superconductors, one would expect the magnetism of a superconducting neutron star core to be strongly
influenced by the Meissner effect. However, Baym et al.~\cite{Baym1969} have argued that due to the high electrical conductivity of normal-conducting
nuclear matter the standard diffusion timescale for magnetic flux is extremely large. Flux expulsion would thus act on the order of a million
years, much longer than dynamical timescales of neutron stars. This implies that the macroscopic magnetic induction cannot be expelled from the
interior and the flux is frozen into the matter. Therefore, condensation into the superconducting state has to take place at a constant magnetic
flux. There are two ways to satisfy this constraint, either by creating an intermediate state in a type-I superconductor or a mixed-fluxtube phase
in a type-II superconductor. The physical state realised inside the neutron star depends on the characteristic lengthscales involved. Having defined
the penetration depth in Eqn.~\eqref{eqn-PenetrationDepthGL} and the coherence length in Eqn.~\eqref{eqn-CoherenceLengthBCS}, one can
calculate typical estimates for both parameters.

Compared to laboratory superconductors however, the critical distances have to be modified because of \textit{entrainment}. As first discussed by Andreev
and Bashkin~\cite{Andreev1976}, this effect is a universal characteristic of interacting Fermi liquids. In the neutron star, it results from the
strong nuclear forces present at high densities and causes protons and neutrons to be coupled. As discussed later on, entrainment can be included
into neutron star fluid dynamics by using the concept of effective masses, $m^*$. These effective masses characterise the dynamical response of the
fluid components to a change in momentum. They are therefore different to the static Landau masses entering the coherence length defined in
Eqn.~\eqref{eqn-CoherenceLengthBCS}. However, as addressed in detail by Chamel and Haensel~\cite{Chamel2006} (see also Prix et al.~\cite{Prix2002}),
deviations between the two types of effective masses are very small for neutron star matter, where $\rho_\rp \ll \rho$. Thus, the dynamical effective
masses are employed to determine the coherence lengths.

Substituting for the Cooper pair parameters $q=  2 e$, $m\cl = 2 m\p$ and $n\cl=n\p/2$, where $m\p$ and $n\p$ are the proton mass and number density,
respectively, one can find for the condensate in the zero-temperature limit\cite{Mendell1991a}:
\begin{equation}
	\lambda_* \approx 1.3 \times 10^{-11} \, \left( \frac{m\p^*}{m\p} \right)^{1/2} \, \rho_{14}^{-1/2} \,
		\left( \frac{x\p}{0.05} \right)^{-1/2} \, \cm, \label{eqn-LambdaNS}
\end{equation}
\begin{equation}
	\xi\p \approx 3.9 \times 10^{-12} \, \left( \frac{m\p}{m\p^*} \right) \, \rho_{14}^{1/3} \,
		\left( \frac{x\p}{0.05} \right)^{1/3} \, \left( \frac{10^9 \, \K}{T_{\rm cp}} \right) \, \cm.
		\label{eqn-XiNS}
\end{equation}
Here, $\rho_{14} \equiv \rho /10^{14} \, \g \, \cm^{-3}$ denotes the total normalised mass density, $x_\rp$ the proton fraction and $m\p^*$ the
effective proton mass; in the outer neutron star core typically\cite{Chamel2008}
\begin{equation}
	\frac{m\p^*}{m\p}  \approx 0.6-0.9\ .
		\label{eqn-Estimatempstar}
\end{equation}
Similarly, the neutron condensates in the crust and core can be assigned coherence lengths that correspond to the dimensions of the neutron Cooper
pairs, i.e.\ the sizes of the superfluid vortex cores. For the $p$-wave paired condensate in the outer neutron star core one has
\begin{equation}
	\xi\n \approx 	1.6 \times 10^{-11} \, \left( 1- x\p \right)^{1/3} \, \left( \frac{m\n}{m\n^*} \right)
		\, \rho_{14}^{1/3} \, \left( \frac{10^9 \, \K}{T_{\rm cn}} \right) \, \cm,
		\label{eqn-XiSFNS}
\end{equation}
where the effective mass is $m\n^* \approx m\n$. For comparison, a $s$-wave paired vortex in the crust is about one order of magnitude smaller
\cite{Mendell1991a}.

\begin{figure}[t]
\begin{center}
\includegraphics[width=4in]{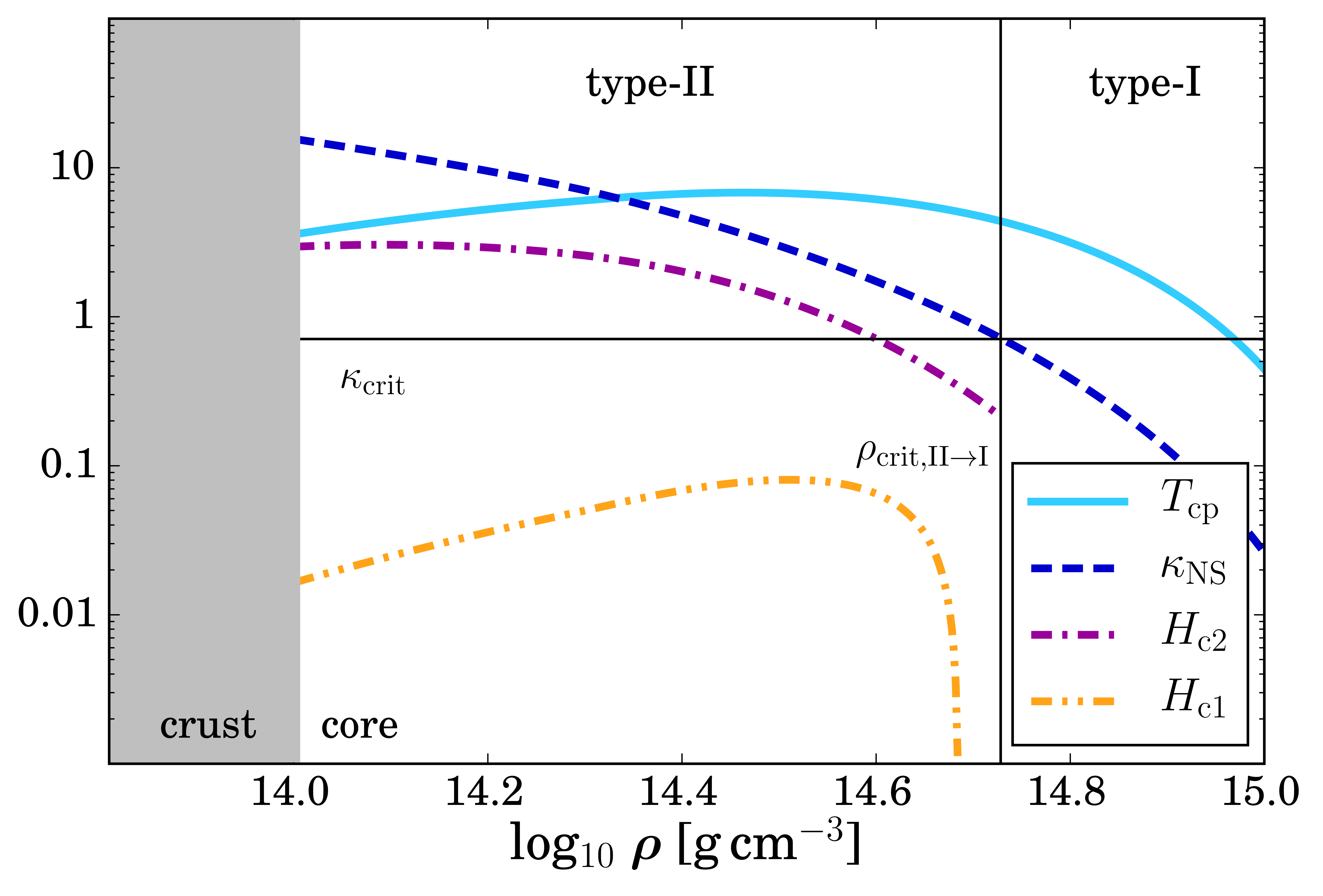}
\caption{Density-dependent parameters of superconductivity. Shown are the transition temperature for proton superconductivity (cyan, solid)
(normalised to $10^{9} \, \K$), the Ginzburg-Landau parameter (blue, dashed) and the two critical fields, $H_{\rm c2}$ (purple, dot-dashed)
and $H_{\rm c1}$ (yellow, dot-dot-dashed) (normalised to $10^{16} \, \G$). The horizontal and vertical line mark $\kappa_{\rm crit}$ and
$\rho_{\rm crit,II \to I}$, respectively. The cross-section is calculated for the NRAPR effective equation of state~\cite{Steiner2005, Chamel2008}.}
\label{fig-CrossSection}
\end{center}
\end{figure}

With Eqns.~\eqref{eqn-LambdaNS} and~\eqref{eqn-XiNS}, the Ginzburg-Landau parameter equates to
\begin{equation}
	\kappa\NS = \frac{\lambda_*}{\xi\p}
		\approx 3.3 \,\left( \frac{m_{\rm p}^*}{m_{\rm p}} \right)^{3/2} \, \rho_{14}^{-5/6} \,
		\left( \frac{x_{\rm p}}{0.05} \right)^{-5/6} \, \left( \frac{T_{\rm cp}} {10^9 \, \K}\right) \gtrsim \frac{1}{\sqrt{2}}.
		\label{eqn-kappaNS}
\end{equation}
For typical neutron star parameters, Eqn.~\eqref{eqn-kappaNS} gives results that are larger than the critical value, $\kappa_{\rm crit}$, obtained
from the phenomenological Ginzburg-Landau approach. The behaviour of $\kappa\NS$ for a typical neutron star equation of state is illustrated in
Fig.~\ref{fig-CrossSection}. In the outer core, the proton fluid is expected to form a type-II superconductor penetrated by a quantised fluxtube
array. Due to the long diffusion timescale of normal matter~\cite{Baym1969b}, this state even forms if the magnetic induction right above the transition
temperature is lower than $H_{\rm c1}$. This is one of the main differences between superconductivity in an astrophysical context and a laboratory
condensate, where the mixed-fluxtube state only prevails for $H_{\rm c1} < B < H_{\rm c2}$ and the flux is expelled from the interior as soon as
$B<H_{\rm c1}$. Using the estimates given in Eqns.~\eqref{eqn-LambdaNS} and~\eqref{eqn-XiNS}, the critical fields defined in
Eqns.~\eqref{eqn-CriticalFieldHC1} and~\eqref{eqn-CriticalFieldHC2} are
\begin{equation}
	H_{\rm c1} \approx 2.0 \times 10^{14} \, \left( \frac{m\p}{m\p^*} \right) \, \rho_{14}
		\, \left( \frac{x_{\rm p}}{0.05} \right) \, \G,
		\label{eqn-HC1NS}
\end{equation}
and \vspace{-0.2cm}
\begin{equation}
	H_{\rm c2} \approx 2.2 \times 10^{15} \, \left( \frac{m_{\rm p}^*}{m_{\rm p}} \right)^2 \, \rho_{14}^{-2/3}
		 \, \left( \frac{x_{\rm p}}{0.05} \right)^{-2/3} \, \left( \frac{T_{\rm cp}} {10^9 \, \K} \right)^2 \, \G.
		\label{eqn-HC2NS}
\end{equation}

\noindent The fields' behaviour as a function of density is also included in Fig.~\ref{fig-CrossSection}. Note that in calculating the
estimate~\eqref{eqn-HC1NS}, $\ln \kappa\NS \approx 2$ has been used. This simplification (generally considered in the context of laboratory
superconductors~\cite{Tinkham2004}) is  an approximation for the outer neutron star core, where $\kappa\NS$ does not change significantly. The
full logarithmic dependence, originating from the fluxtube's energy per unit length, is however accounted for in Fig.~\ref{fig-CrossSection},
since it results in the divergent behaviour of $H_{\rm c1}$ at higher densities. As discussed in more detail in Sec.~\ref{subsec-chargemultihydro},
this is ultimately related to the local distribution of the fluxtubes' magnetic induction; more precisely the divergence of the Bessel function
for small arguments (see Eqn.~\eqref{eqn-SmallrBehaviourBB}). Nonetheless, as just explained, the lower critical field is not of crucial importance
in neutron stars and the outer core of pulsars with typical field strengths between $10^{11}$ to $10^{13} \, \G$ should be in a metastable type-II
state.

Special caution needs to be applied when modelling magnetars. Magnetic fields in these objects could be high enough to create interfluxtube
spacings that are comparable to the fluxtube core radius. This would no longer allow a treatment based on infinitesimally thin, non-interacting
fluxtubes. Moreover, magnetar fields might even reach values above $H_{\rm c2}$. In this case, superconductivity could be completely destroyed,
which would affect  the magnetars' dynamics~\cite{Sedrakian2016}.

The situation might be further complicated by the fact that there is a critical density in the neutron star's inner core at which $\kappa\NS$
should fall below the critical value, $\kappa_{\rm crit}$. In this case, the dominant state of matter would be an intermediate type-I superconductor,
where single proton fluxtubes cannot be present. Using Eqn.~\eqref{eqn-kappaNS}, the transition density can be derived:
\begin{equation}
	\rho_{\rm crit,II\to I} \approx 6.4 \times 10^{14} \,\left( \frac{m\p^*}{m\p} \right)^{9/5} \, \left( \frac{0.05}{x\p} \right)
		\, \left( \frac{T_{\rm cp}} {10^9 \, \K}\right)^{6/5} \, \g \, \cm^{-3}.
		\label{eqn-RhoCrit}
\end{equation}
Above this value, fluxtubes might form bundles and create an intermediate state with macroscopic but irregularly distributed regions of zero and
non-zero magnetic flux~\cite{Sedrakian1995a, Alford2008}. Note however that other phases of matter could be preferred at such high
densities~\cite{Alford2013} and an intermediate type-I state might not form after all.

\subsection{Neutron star phenomenology}

Besides calculations of microscopic parameters, there are observations that support the presence of quantum condensates in neutron stars. Radio pulsars
tend to be extremely stable clocks, but enigmatic spin-glitches have been observed in just over 100 (mainly relatively young) systems. The phenomenon
is thought to be due to the transfer of angular momentum from a superfluid component to the star's crust (to which the magnetic field is anchored).
Superfluidity would help explain the long post-glitch relaxation timescales on the order of months to years. Originally, Baym et al.~\cite{Baym1969}
and Anderson and Itoh~\cite{Anderson1975} proposed that the neutron star's dynamical evolution during and after a glitch could be explained by the
presence of a superfluid component weakly coupled to the crust. Since the original suggestions, a considerable amount of work has gone into modelling
pulsar glitches, but these models are still not (yet) at the level where they can be matched to observations (see Haskell and Melatos~\cite{Haskell2015}
for a recent review). The outcome is sensitive to issues involving the pinning of superfluid vortices to the nuclear lattice in the inner
crust~\cite{Donati2006, barranco, Bulgac2013} and crustal entrainment, i.e.\ how mobile the superfluid component is~\cite{Chamel2012a}. Macroscopic modelling
of glitch dynamics (from trigger to long-term relaxation) is complex~\cite{vortdyn,Sidery2010} but there has been interesting progress on constraining

superfluid parameters for a few regular glitching pulsars~\cite{Andersson2012,Chamel2013,Ho2015a}. Future precision radio timing of the relaxation phase
that follows a pulsar glitch should provide more insight into these issues.

At the present time, the strongest constraint on neutron star superfluidity may be associated with observations of real-time cooling of the compact object
in the Cassiopeia A supernova remnant (the youngest known neutron star in the galaxy). The rapid cooling of this object  can be explained by a relatively
recent transition to superfluidity (triggering neutrino emission due to pair breaking/formation which enhances the slower cooling due to the standard
Urca reactions)~\cite{Page2011, Shternin2011}. This model suggests superfluid pairing gaps broadly in line with theoretical expectations, but systematic
issues regarding the observations remain to be resolved.


\section{Macroscopic Neutron Star Models}
\label{sec-Multifluid}

A few hundred years after birth, neutron stars are in thermal equilibrium, typically having core temperatures in the range of $10^6$ to $10^8
\, \K$~\cite{Ho2012}, several orders of magnitude lager than the Fermi temperature of nuclear matter. Thus, compact objects contain a highly degenerate Fermi
liquid of strongly interacting particles that is often approximated as a ground-state system. For a discussion of this zero-temperature limit and the extension
to a finite-temperature mixture see Andersson et al.~\cite{Andersson2013b} and references therein. With regard to the comparison with terrestrial systems, we
will focus on the neutron star's core where superfluid and superconducting components are expected to be present (for a hydrodynamical description of the
neutron star crust see the work of Pethick et al.~\cite{Pethick2010} and Andersson et al.~\cite{Andersson2011}). The presence of multiple condensates implies
the existence of distinct fluid degrees of freedom and requires a multi-fluid formalism. The simplest representation of the neutron stars' interior is a mixture
of three components, namely relativistic electrons, superconducting protons and superfluid neutrons, denoted by roman indices $\rx = \{\re, \rp, \rn \}$,
respectively.\footnote{Note that it is in principle straightforward to take muons into account and generalise to a four-constituent model since electrons and
muons are strongly coupled on macroscopic length-scales.~\cite{Mendell1991a}} A macroscopic multi-fluid model for this system can be derived by employing the
Lagrangian formalism of Carter, Prix and collaborators~\cite{Carter1995a, Prix2004, Andersson2006}. For a detailed derivation, we refer the reader to recent
work by Glampedakis et al.~\cite{Glampedakis2011a}.

Note that when modelling neutron stars, it is quite common to write equations in a coordinate basis and make use of tensor notation. The main reason for this
is that the equations are then closely related to their general relativistic counterparts. Moreover, it is more straightforward to represent complex systems,
e.g.\ involving elastic matter or entrainment. We will follow this convention in this review, even though it may introduce an element of confusion for the reader
more familiar with the description of laboratory systems. In principle, translating between the two descriptions is relatively simple. In the coordinate basis,
we represent different particle species by the number density $n_\rx$ and the corresponding current density $n^i_\rx = n_\rx v_\rx^i$, where $v_\rx^i$ is the
velocity. It is important to note that repeated constituent indices $\{\rx,\ry,\rz\}$ do not satisfy a summation convention, but vector indices $\{i,j,k\}$ do
satisfy such a convention (as usual). The formulae in this section are written in a general coordinate basis, with $g_{i j} = g_{j i}$ representing the (flat-space)
metric. This metric is used to raise and lower indices, implying that for a vector $\mathbf v$ we have the components $v_i = g_{i j} v^j$ and $v^i = g^{i j} v_j$,
where the inverse metric $g^{i j}$ satisfies $g^{j k} g_{k i} = \delta^j{}_i$ and $\delta^j{}_i$ is the Kronecker-delta tensor. This means that $v^2 = \mathbf{v}
\cdot\mathbf{v} = g_{ij} v^i v^j = v_i v^i$, and so on. Finally, all spatial derivatives are given by the covariant derivative $\nabla_i$ that is compatible with
the metric, i.e. $\nabla_i g_{j k} = 0$.

\subsection{Charged multi-fluid hydrodynamics}
\label{subsec-chargemultihydro}

The large scale dynamics of the outer neutron star core are governed by two momentum equations, one representing the superfluid neutrons and the other representing
the electron-proton conglomerate. The charged fluids can be combined into a single component as long as charge neutrality is satisfied over macroscopic regions so
that $n_\re = n_\rp$ holds. This is expected to be the case, as local charge imbalances can be quickly equilibrated by the electron fluid~\cite{Jackson1999,
Glampedakis2011a}. Moreover, the electron mass is much smaller than the proton mass, $m_\re \ll m_\rp$, which allows one to neglect the electron inertia. In tensor
notation, the corresponding macroscopic Euler equations read
\begin{align} \hspace{-0.21cm}
	\rho\n \bigg[ ( \partial_t + v\n^j \nabla_j ) ( v\n^i + \en w^i_{\rm pn} )
 		+ \nabla^i ( \tilde{\mu}\n + \Phi ) + \en w_{\rm pn}^j \nabla^i \, v^{\rm n}_j \bigg]
 		&= F^i\mf + F^i_{\rm mag,n},
 		\label{eqn-EulerN}
\end{align}
and
\begin{align}
 	\hspace{-0.21cm} \rho\p \bigg[ ( \partial_t + v\p^j \nabla_j ) ( v\p^i + \ep w^i_{\rm np} )
 		+ \nabla^i ( \tilde{\mu} + \Phi ) + \ep  w_{\rm np}^j \nabla^i \, v^{\rm p}_j \bigg]
 		&= - F^i\mf + F^i_{\rm mag, p}.
 		\label{eqn-EulerP}
\end{align}
Here, $v^i_\rx$ denote the average fluid velocities and $\rho_\rx \equiv m  n_\rx$ the mass densities, where $m \equiv m_\rn=m_\rp$ is the baryon mass.
$w_\mathrm{xy}^i \equiv v_\rx^i-v_\ry^i$ is the relative velocity, $\Phi$ the gravitational potential and $\varepsilon_\rx$ the entrainment parameters.
By definition, the latter satisfy the relation $n_\rp \varepsilon_\rp = n_\rn \varepsilon_\rn$. The specific chemical potentials are given by
\begin{equation}
    \tilde{\mu}_\rn \equiv \frac{\mu_\rn}{m}, \hspace{1.5cm} \tilde{\mu} \equiv \frac{\mu_\rp + \mu_\re}{m}.
\end{equation}
The two Euler equations have to be supplemented by three continuity equations for the number densities, reflecting the conservation of mass for each individual
species,
\begin{equation}
    \partial_t n_\rx + \nabla_i \left( n_\rx v_\rx^i \right) = 0,
		\label{eqn-Continuity}
\end{equation}
and the Poisson equation
\begin{equation}
    \nabla^2 \Phi = 4 \pi G \rho,
\end{equation}
where $\rho=\sum_\rx \rho_\rx$ represents the total mass density of the fluid mixture and $G$ the gravitational constant.

The variational formalism employed to derive the hydrodynamical equations explicitly distinguishes the fluid velocities and momenta and provides the possibility
to account for changes caused by the quantum condensates. Compared to the momentum equations of standard plasma physics~\cite{Jackson1999}, superfluid and
superconducting components result in new inertial terms (arising from entrainment) and terms that go beyond the standard electromagnetic Lorentz force. More
precisely, the right-hand sides of Eqns.~\eqref{eqn-EulerN} and~\eqref{eqn-EulerP} contain the total magnetic and mutual friction forces per unit volume,
$F^i_{\rm mag,x}$ and $F^i_{\rm mf}$, respectively. The former one captures the interactions of the vortex/fluxtube magnetic field with the charged fluid (for
details see Glampedakis et al.~\cite{Glampedakis2011a}). Note that the neutron fluid experiences this force due to the entrainment of protons around each vortex,
creating an effective magnetic field. Without entrainment, the magnetic force on the neutrons vanishes. Mutual friction arises due to the dissipative
coupling of the vortex and fluxtube array with the fluid components, analogous to the case for superfluid helium in Sec.~\ref{subsec-MFAndHVBK}.

This system of equations determines how the presence of vortices and fluxtubes influences the macroscopic neutron star dynamics, i.e.\ its rotational and magnetic
evolution. On these large scales, quantum condensates can be treated consistently, as one can take advantage of the large numbers of vortices and fluxtubes and
average over the respective arrays to obtain a smooth-averaged description of the magnetic and mutual friction forces. If the individual quantised structures are
distant enough so that interactions within one array can be neglected, the averaging is achieved by generalising the vorticity definition~\eqref{eqn-AveragedVorticity}
developed for rotating helium II. Assuming that neutron vortices and proton fluxtubes are locally straight and directed along the unit vectors, $\hat{\kappa}^i_\rn$
and $\hat{\kappa}^i_\rp$, the arrays can be associated with surface densities, $\cN_\rn$ and $\cN_\rp$, respectively. Since the vorticities, $\mathcal{W}^i_\rx$, are
related to the circulation of the averaged canonical momenta, the macroscopic quantisation conditions are given by
\begin{align}
	\mathcal{W}_\rn^i &= \epsilon^{ijk} \nabla_j \left( v_k^\rn + \varepsilon_\rn w_k^{\rm pn} \right) = \cN_\rn \kappa^i_\rn ,
		\label{eqn-Quantisation1}  \\[1.8ex]
	\mathcal{W}_\rp^i & = \epsilon^{ijk} \nabla_j \left( v_k^\rp + \varepsilon_\rp w_k^{\rm np} \right) + a_\rp B^i = \cN_\rp \kappa^i_\rp,
		\label{eqn-Quantisation2}
\end{align}
where $\kappa^i_\rx \equiv \kappa \hat{\kappa}^i_\rx$ points along the local vortex/fluxtube direction and we define
\begin{equation}
	a_\rp \equiv \frac{e}{m c} \approx 9.6 \times 10^3 \, \G^{-1} \, \rs^{-1}.
\end{equation}

On macroscopic scales, the total magnetic induction, $B^i$, contains three contributions, i.e.\ the averaged vortex/fluxtube fields and the London field,
\begin{equation}
	B^i = B_\rn^i + B_\rp^i + b_{\rm L}^i .
 		\label{eqn-AveragedField}
\end{equation}
The former terms are given as the product of the surface density, $\cN_\rx$, and the flux carried by a single line of the lattice, $\phi_\rx$, which can be determined
by looking at the mesoscopic dynamics (bars are employed to distinguish the small scale quantities from the macroscopic ones). Considering distances from the vortex
or fluxtube core that are larger than the respective coherence lengths, $\xi_\rx$, the core structure is negligible. Using the corresponding quantisation condition
and a mesoscopic Amp\`ere law (for details see Appendix A1 of Glampedakis et al.~\cite{Glampedakis2011a}), one can derive generalised London equations for the mesoscopic
magnetic inductions, $\bB_\rx^i$,
\begin{equation}
	  \lambda^2_* \nabla^2 \bB_\rx^i - \bB_\rx^i = - \phi_\rx \hat{\kappa}_\rx^i \delta(\mathbf{r} ),
		  \label{eqn-LondonModified}
\end{equation}
where $\delta(\mathbf{r} )$ is the two-dimensional delta function located at each vortex and fluxtube centre, $\phi_\rx$ is defined below and the effective London
penetration depth becomes
\begin{equation}
 	\lambda_* \equiv \left(\frac{1}{4 \pi \rho_\rp a_\rp^2} \, \frac{1 -\en-\ep}{1-\en} \right)^{1/2}.
		\label{eqn-EffectiveLondonDepth}
\end{equation}
Note that this expression reduces to the standard result of superconductivity given in Eqn.~\eqref{eqn-lambda} if entrainment is absent, $\en = \ep=0$. Taking advantage
of the symmetry and using cylindrical coordinates, the inhomogeneous Helmholtz equation~\eqref{eqn-LondonModified} can be solved with a Green's function approach in two
dimensions~\cite{Fetter1969}. This leads to
\begin{equation}
 	\bB_\rx^i = \frac{\phi_\rx}{2 \pi \lambda_*^2} \, K_0 \left( \frac{r}{\lambda_*} \right) \hat{\kappa}_\rx^i.
 		\label{eqn-MesoscopicInduction}
\end{equation}
Here, $K_0(r/\lambda_*)$ is the modified Bessel function of second kind and $r \equiv | \mathbf{r}|$ the radial distance from the vortex or fluxtube core. As a result of
$K_0(r/\lambda_*)$, the mesoscopic inductions exhibit characteristic behaviour for large and small $r$. More precisely, approximating the Bessel function in the respective
limits leads to
\begin{align}
	\bB_\rx (r) &\to \frac{\phi_\rx}{2 \pi \lambda_*^2} \,
	      \left( \frac{\pi \lambda_*}{2 r} \right)^{1/2} \, \text{e}^{-r/\lambda_*} \hspace{1.5cm} \text{for   } r \to \infty,
	      \label{eqn-LargerBehaviourBB} \\[1.8ex]
	\bB_\rx (r) &\approx \frac{\phi_\rx}{2 \pi \lambda_*^2} \,
	      \left[ \text{ln} \left( \frac{\lambda_*}{r} \right) + 0.12 \right] \hspace{1.5cm} \text{for   }  \xi_\rx \ll r \ll \lambda_*.
	      \label{eqn-SmallrBehaviourBB}
\end{align}
In the former case, the magnetic inductions fall off exponentially, while they diverge for small $r$. In reality however, the superfluid and superconducting states break down
in the vortex and fluxtube core, respectively, and normal fluid matter prevails. Thus, $\bB_\rx (r)$ should remain regular at $r=0$, which is usually achieved by introducing
a cut-off at the vortex/fluxtube radius, $r\sim \xi_\rx$, where the Cooper pair density vanishes and $\lambda_*$ becomes infinite.

Integrating the mesoscopic magnetic induction, $\bB_\rx^i$, over a disc of radius $r \gg \lambda_*$ perpendicular to $\hat{\kappa}_\rx^i$ results in the magnetic flux, i.e.
\begin{equation}
	 \int \bar{B}_\rx^i \, \rd S = \phi_\rx \hat{\kappa}_\rx^i.
\end{equation}
As expected, this reproduces the flux quantum, $\phi_\rp \equiv \phi_0$, for the proton fluxtube, whereas for a superfluid vortex we find
\begin{equation}
	\phi_\rn \equiv - \frac{\ep}{1-\en} \, \phi_0.
		\label{eqn-FluxVortex}
\end{equation}
The minus sign originates from the antialignment of $\hat{\kappa}_\rn^i$ and $\bar{B}_\rn^i$. Note that in the absence of entrainment the neutron flux vanishes. With
entrainment, the fluxes are comparable as the entrainment parameters are related to the effective masses via\cite{Andersson2006b}
\begin{equation}
	\ep = 1- \frac{m_\rp^*}{m} \approx 0.1 - 0.4,
		\label{eqn-EstimateEp}
\end{equation}
where the estimate~\eqref{eqn-Estimatempstar} was used, and
\begin{equation}
	\en = 1- \frac{m_\rn^*}{m} = \ep x\p \ll 1,
		\label{eqn-EstimateEn}
\end{equation}
For small proton fractions (a valid approximation in the neutron star interior), the neutron entrainment parameter is hence negligible and we have $\phi_\rn \simeq -
\ep \phi_0$. The two arrays then contribute the following to the total macroscopic induction, $B^i$:
\begin{equation}
	B_\rx^i = \cN_\rx \phi_\rx \hat{\kappa}_\rx^i.
		\label{eqn-BAveraged}
\end{equation}
Despite the fluxes being of similar magnitude, the fluxtube term dominates because $\cN_\rp \gg \cN_\rn$ (see Eqns.~\eqref{eqn-ArraySurfaceDensityNeutron} and
\eqref{eqn-ArraySurfaceDensityProton}).

The final contribution to the induction is the London field, $b_{\rm L}^i$, which is a fundamental property of a superconductor associated with its rotation (see
Sec.~\ref{subsec-LondonField}). In contrast to $B_\rx^i$, the London field is not of microscopic origin but related to the macroscopic electromagnetic current and
equivalent to the magnetic field, $H^i$ (see Sec.~\ref{subsec-Maxwell}). Combining the quantisation conditions~\eqref{eqn-Quantisation1} and~\eqref{eqn-Quantisation2}
with Eqn.~\eqref{eqn-BAveraged}, the London field can be expressed in terms of macroscopic fluid variables. Assuming that the hydrodynamical scales are small enough
to ensure constant entrainment parameters, one obtains
\begin{equation}
	b_\rL^i = - \frac{1}{a_\rp} \, \frac{1 - \en - \ep}{1-\en} \, \epsilon^{ijk} \nabla_j v^{\rm p}_k
		\approx - \frac{1}{a_\rp} \left(1 - \varepsilon_\rp \right) \, \epsilon^{ijk} \nabla_j v^{\rm p}_k,
		\label{eqn-LondonField2}
\end{equation}
which illustrates the close connection between the London field and the rotation of the proton fluid. Taking it to be tightly coupled to the neutron star's crust through
the magnetic field, the protons rotate rigidly with the observed pulsar frequency, i.e. $\epsilon^{ijk} \nabla_j v^{\rm p}_k = 2 \Omega^i$, which allows us to estimate the
magnitude of the London field. For a canonical rotation period, we find
\begin{equation}
	b_{\rm L} \approx 0.1 \left( \frac{m_\rp^*}{m} \right) P^{-1}_{10} \, \G.
		\label{eqn-EstimateLondonField}
\end{equation}
This is many orders of magnitude smaller than the magnetic field strength inferred for neutron stars. Hence, in addition to ignoring the vortex magnetic field contribution,
it is further justified to neglect the London field in Eqn.~\eqref{eqn-AveragedField}. This highlights that the charged condensate dominates the large scale magnetic field
in the neutron star core and mechanisms affecting the fluxtube dynamics could play a dominant role in driving the field evolution~\cite{Graber2015}.

\subsection{Mesoscopic dynamics}
\label{subsec-MesoscopicNSDynamics}

Despite the fact that the variational formalism allows one to calculate the total effective magnetic forces arising from the presence of quantum condensates, it does not
provide sufficient physical insight into how the macroscopic forces are related to the mesoscopic interactions of the arrays. This can however be achieved by using a
different approach based upon the more intuitive concept of individual vortex/fluxtube dynamics; a method that was established by Hall and Vinen~\cite{Hall1956} to describe
the behaviour of superfluid helium and the corresponding mutual friction force (see also Sec.~\ref{subsec-MFAndHVBK}). More precisely by combining the quantisation conditions
\eqref{eqn-Quantisation1} and~\eqref{eqn-Quantisation2} with conservation equations for the vorticities (for details see Glampedakis et al.~\cite{Glampedakis2011a}), one
ultimately obtains a set of momentum equations equivalent to Eqns.~\eqref{eqn-EulerN} and~\eqref{eqn-EulerP}. However, instead of containing the forces $F^i\mf$ and
$F^i_{\rm mag,x}$, the right-hand side reflects the force density acting on the fluids due to the presence of quantised arrays, which is equal to the negative of the averaged
Magnus force density, $F^i_{\rm Mx} = \cN_\rx f^i_{\rm Mx}$. Here, $f^i_{\rm Mx}$ is the Magnus force per unit length, which is proportional to the relative velocity between
the vortices/fluxtubes and the bulk fluids and acts as a lift force on a single quantised structure immersed into the fluid,
\begin{equation}
	f^i_{\rm Mx} \equiv \rho_\rx \kappa \, \epsilon^{ijk} \hat{\kappa}^\rx_j (u_k^\rx - v_k^\rx ).
 		\label{eqn-MagnusForce0}
\end{equation}
This force governs the motion of free vortices and fluxtubes and causes them to be dragged along with the superfluid/superconducting component. However, in a neutron star's
outer core, additional forces are expected to influence the motion of an individual vortex or fluxtube~\cite{Glampedakis2011a}. By balancing all contributions, solving the result
for the Magnus force and substituting this back into the respective Euler equations, it is possible to relate the dynamics on mesoscopic lengthscales to the macroscopic behaviour
of the hydrodynamical fluids.

Firstly, vortices and fluxtubes have a large self-energy, which causes them to resist bending and leads to a tension force per unit length,
\begin{equation}
      f^i_{\rm t x} \equiv \cE_\rx \hat{\kappa}_\rx^j \nabla_j \hat{\kappa}_\rx^i,
		\label{eqn-TensionForce0}
\end{equation}
where $\cE_\rx$ are the vortex and fluxtube energies per unit length which are given by~\cite{Glampedakis2011a}
\begin{equation}
	\cE_\rn \approx \frac{\kappa^2 \rho_\rn}{4 \pi} \, \left( \frac{m}{m_\rn^*} \right) \ln \left( \frac{d_\rn}{\xi_\rn} \right)
		\label{eqn-EnergyVortex}
\end{equation}
and
\begin{equation}
	\cE_\rp \approx \frac{\kappa^2 \rho_\rp}{4 \pi} \, \left( \frac{m}{m_\rp^*} \right) \, \ln \left( \frac{\lambda_*}{\xi_\rp} \right).
		\label{eqn-EnergyFluxtube}
\end{equation}
Apart from the modifications due to entrainment, the tension on a neutron vortex is equivalent to the force defined for superfluid vortices in helium II
(see Eqn.~\eqref{eqn-Tension}). Moreover, the vortices and fluxtubes are magnetised and, thus, experience a conservative Lorentz-type force due to the electromagnetic coupling
with the macroscopic, charged conglomerate. It is given by
\begin{equation}
 	f^i_{\rm em x} \equiv \frac{e n_\rp}{c} \, \phi_\rx \, \epsilon^{ijk} \hat{\kappa}_j^\rx w^{\rm pe}_k .
		\label{eqn-EMForce0}
\end{equation}
As suggested by Sauls et al.~\cite{Sauls1982} and Alpar et al.~\cite{Alpar1984}, the magnetic fields of individual vortices/fluxtubes can further interact
with the electron fluid, resulting in a dissipative force. This coupling depends on the relative velocity between the vortex or fluxtube and the charged particles
and is characterised by a mesoscopic drag
\begin{equation}
	f^i_{\rm d x}  \equiv \gamma_\rx \, ( v^i\e - u^i_\rx ) = \rho_\rx  \kappa \cR_\rx \, ( v^i\e - u^i_\rx ),
		\label{eqn-DragForce0}
\end{equation}
fully determined by the positive coefficient, $\gamma_\rx$, or its dimensionless equivalent, $\cR_\rx$.

Finally, we note that other frictional mechanisms, such as the coupling of vortices and fluxtubes due to magnetic short-range interactions, could be present.
This could give rise to \textit{pinning} between the two arrays and connect the star's rotational and field evolution~\cite{Ruderman1998, Jahan-Miri2000,
Link2003}. However, a detailed microscopic understanding of the pinning interaction is not yet available and thus ignored here. Neglecting the inertia
of the vortex/fluxtube, the force balance per unit length reads
\begin{equation}
	\sum f_\rx ^i = f^i_{\rm M x} + f^i_{\rm t x} + f^i_{\rm em x} + f^i_{\rm d x} =0.
		\label{eqn-ForceBalanceMesoscopic}
\end{equation}
Multiplying this with the surface density, $\cN_\rx$, one obtains a macroscopic, averaged equation for the vortex and fluxtube array,
\begin{equation}
	F^i_{\rm M x} + F^i_{\rm t x} + F^i_{\rm em x} + F^i_{\rm d x} = 0,
		\label{eqn-ForceBalanceMacroscopic}
\end{equation}
where we identify the force per unit volume with $F^i_\rx \equiv \cN_\rx f^i_\rx$ for each mechanisms.

This force balance can now be employed to replace the Magnus force in the Euler equations. Note however that the mesoscopic vortex or fluxtube velocity, $u^i_\rx$,
present in the drag force needs to be expressed in terms of macroscopic fluid variables in order to arrive at a description of the large scale behaviour of the
multi-fluid mixture. The resulting macroscopic drag is referred to as mutual friction, $F^i_{\rm mf x}$, which can be obtained by calculating repeated crossproducts
of Eqn.~\eqref{eqn-ForceBalanceMacroscopic} with $\hat{\kappa}_j^\rx$. Neglecting for simplicity the electromagnetic contribution and the tension, one arrives at the
standard result for the macroscopic mutual friction force per unit volume~\cite{Andersson2006, Glampedakis2011a}:
\begin{equation}
	F^i_{\rm mf x} = \rho_\rx \cN_\rx \kappa \left( \CB_\rx \epsilon^{ijk} \hat{\kappa}_j^\rx \epsilon_{klm} \hat{\kappa}^l_\rx w_{\rm xe}^m
		+ \CB_\rx' \epsilon^{ijk} \hat{\kappa}_j^\rx w^{\rm xe}_k \right).
		\label{eqn-MutualFrictionSimple}
\end{equation}
where the dimensionless mutual friction parameters are defined as
\begin{equation}
	\CB_\rx \equiv \frac{\cR_\rx }{1 + \cR_\rx^2}, \hspace{1.5cm} \CB_\rx' \equiv \cR_\rx \CB_\rx = \frac{\cR_\rx^2}{1 + \cR_\rx^2}.
		\label{eqn-MutualFrictionCoefficientsNS}
\end{equation}
Note that whereas $F^i_{\rm mf x}$ acts on the neutrons and protons, respectively, the electron fluid experiences the opposite forces, $-F^i_{\rm mf x}$. Keeping in mind
that the charged particles move as a single constituent and are described by a combined Euler equation (obtained from adding the individual momentum equations), the two
terms $-F^i_{\rm mf p}$ and $F^i_{\rm mf p}$ cancel each other and only the superfluid contribution, $F^i_{\rm mf} \equiv F^i_{\rm mf n}$, remains in the Euler
equations~\eqref{eqn-EulerN} and~\eqref{eqn-EulerP}.

\subsection{Macroscopic Maxwell equations}
\label{subsec-Maxwell}

In order to describe the electromagnetic response of the fluid mixture in the neutron star core, hydrodynamical equations and quantisation conditions have
to be supplemented by Maxwell's equations. Generalising these to macroscopic scales, one has to be careful to account for the presence of condensates since
the different fields entering Maxwell's equations have to reflect the type-II nature of the superconducting protons. Applying the Lagrangian formalism to derive a
macroscopic Amp\`ere's law  in the presence of quantised arrays, we find that (in contrast to standard MHD where the equality $H^i=B^i$ is satisfied) the averaged
magnetic induction, $B^i$, and the macroscopic magnetic field, $H^i$, are no longer equivalent. Instead the London field replaces the macroscopic field, $H^i=
b_{\rm L}^i$, and the modified Amp\`ere's law (accurately capturing the dynamics of the type-II superconductor) is given by~\cite{Carter2000, Glampedakis2011a}
\begin{equation}
 	\epsilon^{ijk} \nabla_j b^{\rm L}_k = \frac{4 \pi}{c} \, j^i.
 		\label{eqn-AmpereSC}
\end{equation}
This shows that the London field, despite being of small magnitude, is closely connected to the macroscopic electromagnetic current density,
\begin{equation}
	 j^i = e n_\rp w_{\rm pe}^i,
		\label{eqn-Current}
\end{equation}
and plays an important role for the neutron stars' electrodynamics. To understand this difference from standard MHD, we can invoke the classification given for
terrestrial superconductors in Sec.~\ref{subsec-MeissnerLondon}. In laboratory experiments, a distinction between macroscopic electromagnetic currents that generate
a macroscopic field, $H^i$, and magnetisation currents \textit{only} affecting the mesoscopic induction, $\bar{B}^i$, is made. A supercurrent, circulating around
each vortex/fluxtube and generating $\bar{B}^i_\rx$, is thus associated with the second class. It does not contribute to the magnetic field $H^i=b^i_\rL$, which is
created by the current density $j^i$. The macroscopic induction, $B^i$, given in Eqn.~\eqref{eqn-AveragedField} is therefore different from the field, $H^i$. For
comparison, in vacuum or normal conductors, where no magnetisation currents are present and $H^i=B^i=\bar{B}^i$ holds, one obtains the standard result
\begin{equation}
	 \epsilon^{ijk} \nabla_j H_k = \epsilon^{ijk} \nabla_j B_k = \frac{4 \pi}{c} j^i.
		 \label{eqn-AmpereLawStandard}
\end{equation}
In addition to Amp\`ere's law, the averaged magnetic induction has to be divergence free everywhere in the superconducting fluid, i.e.
\begin{equation}
	\nabla_i B^i =0
		\label{eqn-MaxwellB}.
\end{equation}
Finally, the macroscopic Faraday law is given by
\begin{equation}
	\partial_t B^i = - c \, \epsilon^{ijk} \nabla_j E_k.
		\label{eqn-Faraday}
\end{equation}
Instead of defining the macroscopic electric field as the average over the mesoscopic expression, one can also take advantage of the remaining
fluid degree of freedom (the electron Euler equation) to obtain a relation for $E^i$. Neglecting again the electron inertial terms, one finds
\begin{equation}
	E^i = -\frac{1}{c} \epsilon^{ijk} v_j^\re  B_k
		- \frac{m_\re}{e} \nabla^i \left ( \tilde{\mu}_\re + \Phi \right ) - \frac{F^i_\re}{e n_\re},
		\label{eqn-ElectricField}
\end{equation}
where $F^i_\re$ represents the total dissipative force per unit volume experienced by the electrons due to interactions with the surrounding fluid components. By
combining Eqns.~\eqref{eqn-Faraday} and~\eqref{eqn-ElectricField} it is possible to obtain an evolution equation for the magnetic induction that only depends
on macroscopic fluid variables. This procedure formed the basis of recent work by Graber et al.~\cite{Graber2015}, who derived the induction equation for a specific
resistive interaction (the scattering of electrons off the fluxtube magnetic field) and found the coupling to be too weak to explain observed magnetic evolution
timescales.


\section{Laboratory Neutron Star Analogues}
\label{sec-NSAnalogues}

The need to understand implications of macroscopic condensates on neutron stars should be evident. However, despite having been in the focus of research
for nearly half a century, our understanding is far from complete and a lot remains to be learnt. As the star's interior is not directly accessible, the
analysis of future high-precision electromagnetic observations and the possible detection of continuous gravitational waves may play a crucial role in
filling in the missing pieces. In parallel, information from a rather different direction might provide insight into various aspects of neutron star physics.
Although generally given little attention by astrophysicists, well-known terrestrial superfluids and superconductors could serve as versatile analogues.
Having discussed the close connection between the mathematical description of different quantum systems, we now turn our attention to experimental aspects.
We explore the possibility of using laboratory condensates to mimic the behaviour of neutron stars on significantly smaller length scales. Since replicating
the extreme conditions present in a star is out of reach, the aim of laboratory analogues would rather be to create systems that are easily manipulable and
allow the recreation of characteristic neutron star features. The following discussion will address the two-fluid nature, pinning of vortices
and fluxtubes, dynamics of interfaces and instabilities. Briefly introduced previously, superfluid helium, ultra-cold gases and superconductors are prime
candidates to test such behaviour. Their respective advantages and limitations in modelling neutron stars will be reviewed in the following three sections.


\section{Helium}
\label{sec-Helium}

\subsection{Two stable isotopes}
\label{subsec-DifferentIsotopes}

Helium-4 is one of the two naturally occurring, stable isotopes of helium and with a relative abundance of $10^6 : 1$ in the Earth's
atmosphere the more common one. Below its boiling point of $4.21 \, \K$, helium-4 exhibits characteristics of a standard low viscosity
fluid. However, due to the weak interatomic forces in helium, decreasing the temperature further does not lead to a solid state. At the
$2.171 \, \K$ Lambda point, helium-4 instead undergoes a phase transition into a superfluid state, as first observed by Kapitsa,
Allen and Misener in 1937~\cite{Kapitza1938, Allen1938}. This discovery marked the beginning of the era of low temperature experiments.

Many features of superfluid helium-4 (referred to as helium II) can be explained by invoking the presence of two coexisting components.
The two-fluid model, which was developed by Tisza~\cite{Tisza1938} and Landau~\cite{Landau1941} and discussed in Sec.~\ref{sec-ModellingSF},
assumes that helium II is formed of a normal, viscous constituent (representing the fluid excitations) and an inviscid part exhibiting
frictionless flow. This two-fluid interpretation offers the possibility to draw a parallel to neutron star cores by identifying the
superfluid neutrons with the inviscid ground-state component and the combined (charge-neutral) electron-proton conglomerate with the
excitations. Since the concentration of normal to inviscid fluid in helium II is temperature dependent (as first observed by Andronikashvili
in 1946~\cite{Andronikashvili1946}), one would be particularly interested in the temperature range that gives mass ratios similar to the
proton fraction, $x_\rp$. In the neutron star core, this parameter is typically of the order of a few percent. Further note that, while
this laboratory analogue does not allow us to specifically study the influence of superconducting fluxtubes and, thus, address the magnetic
field evolution of a neutron star, it certainly provides a model for the stars' rotational dynamics. On one hand, the normal helium II
component is observed to follow rigid-body rotation (see for example Osborne~\cite{Osborne1950}), a state which also characterises the
angular velocity profile of the charged component in the neutron star's core (coupled to the crust by the strong magnetic
field~\cite{Easson1979a}). On the other hand, the superfluids in both systems rotate by forming quantised vortices. Spin-down experiments
as the ones discussed below could thus help us understand the superfluid's role in generating pulsar glitches.

Recent experiments further provide the possibility of directly imaging the flow behaviour of helium II~\cite{Guo2014}. By inserting small
particles such as liquid neon atoms~\cite{Celik2002} or hydrogen molecules~\cite{Bewley2006} into the superfluid and tracing their motion,
vortices have been visualised. This approach has proven particularly useful in studying superfluid turbulence~\cite{Kivotides2008, Vinen2014}
and could improve our understanding of the stars' fluid motion.

The second stable, naturally occurring helium isotope is helium-3~\cite{Vollhardt1990}. This isotope only constitutes a very small fraction
of the noble gas in the atmosphere and irradiation of lithium-6 with neutrons is necessary to synthesise larger quantities. Because of these
obstacles, experiments with helium-3 did not commence until the 1960s, several decades after the first helium-4 experiments had been performed.

\begin{figure}[t]
\begin{center}
\includegraphics[width=3.2in]{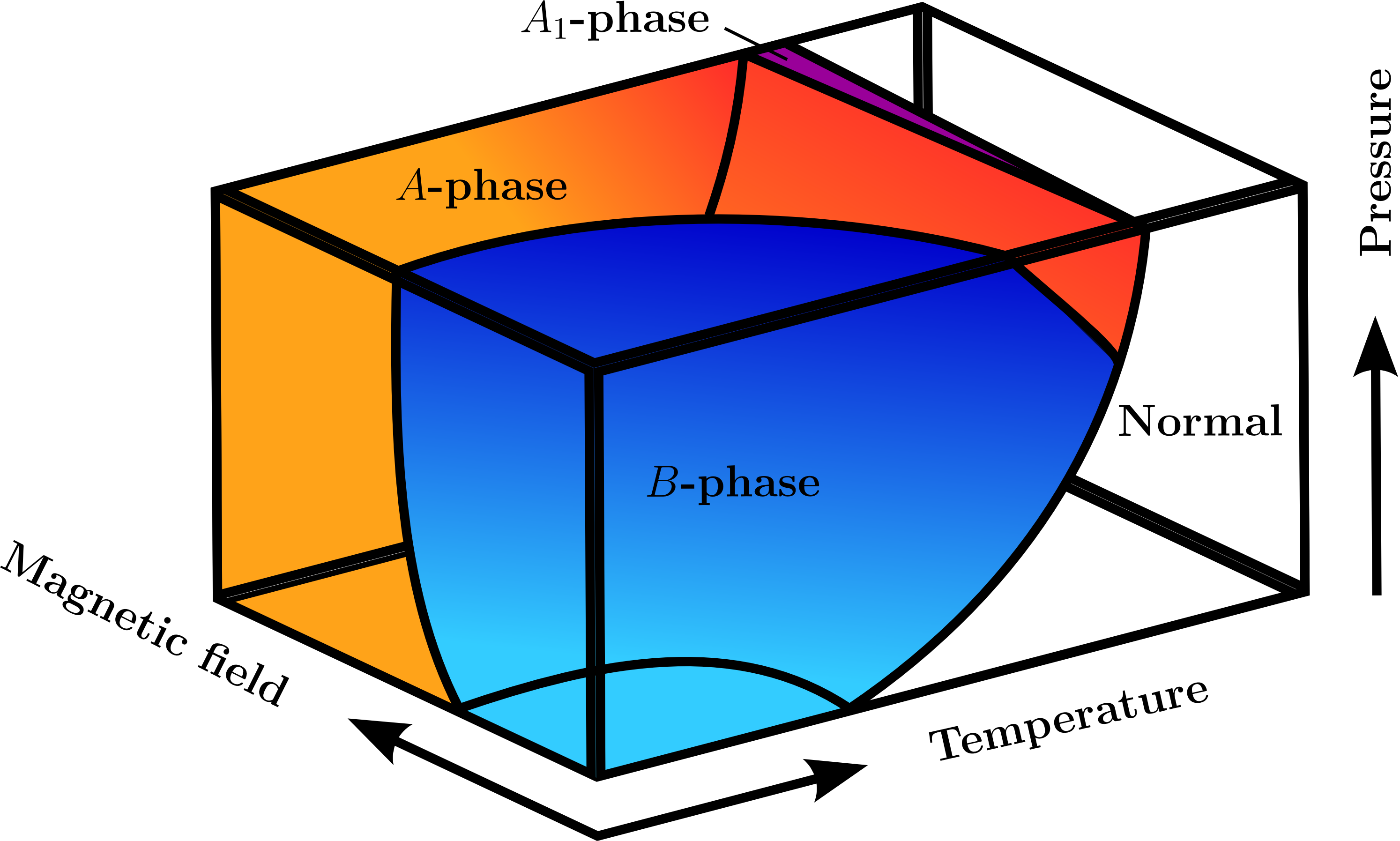}
\caption{Phase diagram of helium-3 illustrating the presence of three different phases depending on the temperature, pressure and magnetic
field. In the absence of a magnetic field, only the helium-$A$ (red) and helium-$B$ (blue) phases are stable. While the former occupies a
small temperature range above a critical pressure, the latter dominates the phase diagram and is stable down to the lowest temperatures
observed. In the presence of an external field, the $A$-phase is stable even for zero pressure and replaces the $B$-phase for sufficiently
high field strengths. Additionally, the $A_1$-phase (purple) develops in a very narrow region between the normal and the superfluid zones.}
\label{fig-PhasediagramHe3}
\end{center}
\end{figure}

Below the $3.19 \, \K$ boiling point helium-3 behaves like a fluid with low viscosity. Similar to helium-4 it does not solidify at normal
pressures and lower temperatures. It instead shows superfluid behaviour below $3 \, \text{mK}$, as first reported by Osheroff et al.\ in
1972~\cite{Osheroff1972, Osheroff1972a}. Though the origin of these phase transitions is very different to that of helium II. Whereas
spin-$0$ helium-4 atoms are described by Bose-Einstein statistics and turn superfluid by condensing into the quantum mechanical ground state
\cite{London1938} (governed by a single complex wave function), helium-3 atoms are fermions of spin $1/2$ due to the presence of an
unpaired nucleon. Thus, they are subject to Fermi-Dirac statistics and obey Pauli's principle. As dictated by BCS theory~\cite{Bardeen1957},
fermions have to form Cooper pairs before any condensation can take place, which results in the much lower transition temperatures. However,
in contrast to standard superconductivity, where the attractive interaction between two electrons is mediated by the underlying lattice, no
such crystal network is present in the case of helium-3. Hence, superfluidity has to be an intrinsic property of the atoms. More precisely,
Cooper pairs in a conventional superconductor form in a spin-singlet, $s$-wave state, governed by a total spin $|\mathbf{S}|=0$ and an
orbital angular momentum $|\mathbf{L}|=0$. They are characterised by a pair wave function that is anti-symmetric under the exchange of the
two electron spins and determined by a single complex amplitude, i.e.
\begin{equation}
	\Psi = \Psi_{\uparrow \downarrow} ( \ket{\uparrow \downarrow} - \ket{\downarrow \uparrow} ).
\end{equation}
Note that the superfluid neutrons in the inner neutron star crust and the protons in the stars' interior are expected to be represented by
a similar order parameter. On the other hand, helium-3 atoms pair in the spin-triplet, $p$-wave state with quantum numbers $|\mathbf{S}|=1$
and $|\mathbf{L}|=1$, where the spin part of the Cooper pair wave function is symmetric. As there are three corresponding spin substates for
$|\mathbf{S}|=1$, the most general helium-3 wave function is thus a linear combination of the form
\begin{equation}
	\Psi = \Psi_{\uparrow \uparrow} \ket{\uparrow \uparrow} + \Psi_{\uparrow \downarrow}( \ket{\uparrow \downarrow} + \ket{\downarrow \uparrow} )
	+ \Psi_{\downarrow \downarrow} \ket{\downarrow \downarrow}.
		\label{eqn-Helium-3PairAmplitude}
\end{equation}
In principle, such an order parameter is characterised by a total of nine substates, i.e.\ nine complex-valued amplitudes, as a result of the
three spin times three orbital substates. Due to these additional degrees of freedom, helium-3 Cooper pairs have an internal structure that
results in the formation of different superfluid phases. Three distinct phases have been observed, commonly referred to as $B$, $A$ and
$A_1$~\cite{Vollhardt1997}. They populate different regions of the phase diagram as illustrated in Fig.~\ref{fig-PhasediagramHe3} depending on
which state has the lowest energy for a given temperature, pressure and magnetic field. If no magnetic field is applied, only the helium-$B$ and
the helium-$A$ phase are stable. While the former dominates the phase diagram and is stable down to the lowest temperatures observed, the latter
occupies only a small range of temperatures above a critical pressure of $\sim 22 \, \text{bar}$. In the presence of an external field however,
the $A$-phase is stable even for zero pressure and replaces the $B$-phase for sufficiently high magnetic field strengths. Additionally, the
$A_1$-phase develops in a very narrow region between the normal and the superfluid zones.

The emergent phase diagram is a direct result of the respective order parameters. For the $B$-phase all three spin substates contribute equally
to Eqn.~\eqref{eqn-Helium-3PairAmplitude} resulting in a single complex amplitude, while the Cooper pairs' total angular momentum vanishes,
$|\mathbf{J}|=|\mathbf{S}+\mathbf{L}|=0$, as a consequence of the orbital amplitude. Due to this, the wave function becomes invariant under
simultaneous rotations of the spin and orbital axes, further leading to an isotropic energy gap. Hence, despite the intrinsic anisotropy of the
helium-3 superfluids, the $B$-phase is stable for all temperatures and resembles conventional superfluids and superconductors. This becomes for example
evident in the formation of vortices. As explained in Sec.~\ref{sec-ModellingSF}, the velocity field of a superfluid (governed by a single complex
wave function) is proportional to the gradient of the phase and hence curl free. In order to rotate, the superfluid quantises its circulation
and forms vortex lines around which the phase changes by a factor of $2 \pi$ or multiples of it. The circulation of all regularly distributed
vortices subsequently adds up to mimic solid body rotation on large scales. This was originally invoked to explain rotation in helium II but has
also been observed in the $B$-phase of rotating helium-3~\cite{Manninen1992}. Note however that the vortex core structure in the latter case is
very different from that in helium-II as a result of the order parameter's underlying complexity. For a detailed discussion of vortices in helium-3
see for example Lounasmaa and Thuneberg~\cite{Lounasmaa1999} and references therein. The behaviour of helium-$A$ is further complicated because the
pair amplitude~\eqref{eqn-Helium-3PairAmplitude} only contains two spin-states, namely $\ket{\uparrow \uparrow}$ and $\ket{\downarrow \downarrow}$,
preserving the inherent anisotropy. This gives the Cooper pairs' spin wave function an intrinsic direction, $\hat{\mathbf{d}}$, that is displayed on
macroscopic scales; a similar preferred axis, $\hat{\mathbf{l}}$, is also present for the orbital angular momentum. The interplay of effects modifying
the orientation of both axes causes their continuous variation in space referred to as \textit{texture}, which essentially maps out the topology
of the complex order parameter~\cite{Vollhardt1990}. One important consequence of this texture is that the superfluid velocity field does no longer
have to be curl-free. Instead changes in the orientation of $\hat{\mathbf{l}}$ can be related to the superfluid's circulation, generating the rich
vortex architecture observed in anisotropic helium-$A$. These include planar defects such as domain walls~\cite{Lounasmaa1999} or coreless vortex
structures of double integer quantisation~\cite{Blaauwgeers2000}, which can be detected by employing modern nuclear magnetic resonance (NMR) spectra.
This technique is non-invasive and allows accurate mapping of topological defects in the order parameter. For completeness, we mention that the pair
wave function of the third experimentally detected phase, $A_1$, is only composed of a single spin substate, $\ket{\uparrow \uparrow}$, implying
that the superfluid itself is magnetic. In the remainder of this section however, we only focus on the importance of helium-$B$ and helium-$A$ in
studying possible laboratory neutron star analogues.

Having briefly alluded to the complex vortex structures in helium-3, one would similarly expect the $p$-wave paired superfluid in a neutron star's
outer core to exhibit diverse features~\cite{Masuda2016a}. One example is the persistent core magnetisation of $^3 P_2$ vortices~\cite{Sauls1982}
(a direct result of the complicated order parameter structure), which leads to the possibility of coupling the electrons and the superfluid in the
stars' interior. However, it is not well understood if there are other ways for the vortex anisotropy to manifest itself on hydrodynamical scales.
As helium-$A$ is one of the few terrestrial anisotropic superfluids, it provides the unique opportunity to study the superfluid in the neutron star's
core. Despite this advantage, we note that the rich spectrum of observed phenomena also significantly complicates a comparison between the two systems
and raises the question of how far the analogy can be extended. Drawing direct conclusions from experiments with laboratory condensates for neutron
star dynamics should thus always be done with caution.

Before analysing possible helium neutron star analogues, we point out that despite the fundamental differences in the
formation of the superfluid phases and the underlying microscopic theories of helium-4 and helium-3, the two-fluid model is also applicable to
describe the macroscopic characteristics of the latter. This is related to the close connection between the symmetries of the Fermi system and
its hydrodynamical variables. It suggests the presence of an inviscid component responsible for the frictionless behaviour and a normal component
representing the quasi-particle excitations, allowing the modelling of the neutron stars' two-fluid behaviour. Experiments in the 1980s confirmed
the existence of persistent currents and the onset of dissipation above a critical velocity in both helium-$B$ and helium-$A$ (see for example
Gammel et al.~\cite{Gammel1984, Gammel1985}). As in helium II, the ratio of superfluid to normal fluid is a function of temperature in the $B$-phase,
whereas in the $A$-phase the ratio also depends on the applied field. This provides an additional possibility to tune the latter condensate into
a state of interest for neutron star experiments.

\subsection{Spin-up and spin-down experiments}
\label{subsec-SpinupExperiments}

Early efforts of studying the spin-up and spin-down behaviour of superfluid helium-4 contained in closed vessels were undertaken in the late 1950s
(see for example Hall~\cite{Hall1957} and Walmsley and Lane~\cite{Walmsley1958}). These experiments measured the torques necessary to accelerate and
decelerate containers filled with helium II and additionally monitored the fluid's response by immersing closely spaced discs into the medium. The
studies discovered that acceleration and retardation are asymmetric processes. While the fluid responded with a delay to setting the vessel (initially
at rest) into motion, it instantly reacted when the container rotation was stopped. These effects were interpreted as the manifestation of the quantised
vortex array. Because the asymmetry of spin-up and spin-down was also influenced by the surface roughness of the container walls and discs, it was
further suggested that pinning and nucleation of vortices could play an important role for the dynamics.

Following these initial endeavours, a systematic analysis of rotating helium II was carried out by the Georgian physicists Tsakadze and Tsakadze in the
1970s~\cite{Tsakadze1972, Tsakadze1973, Tsakadze1974, Tsakadze1980}. These spin-up experiments were performed shortly after the initial observations of
glitches in the Vela and the Crab pulsar~\cite{Radhakrishnan1969, Reichly1969, Boynton1969, Richards1970} and represent the first (and only) attempts
to model neutron star physics with laboratory analogues. However, rather than exploring the underlying mechanism for glitches, the experiments were
aimed at validating the assumption of a superfluid component inside the star by studying the vessel relaxation after an initial increase in the angular
rotation (see below).

\begin{figure}[t]
\begin{center}
 \includegraphics[height=7.5cm]{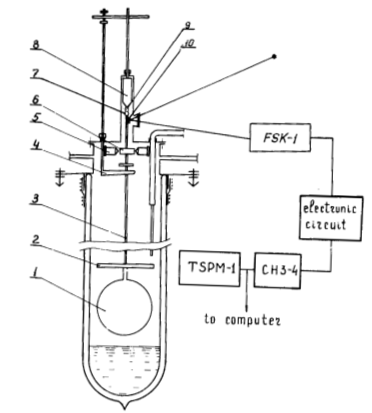}
\caption{Schematic setup of the helium II experiments performed by Tsakadze and Tsakadze. The test neutron star is represented by a
hollow glass sphere (1), which together with a brass disk (2) is rigidly connected to a thin steel rod (3) and magnetically suspended (7-9).
The support device (4) is used to lower the sphere into a bath of helium. The freely suspended components (1-3) are then suddenly accelerated
using the electric motor (5-6), while the rotation period is measured by a focused light beam reflected off a mirror (10). The figure is
reproduced from Tsakadze and Tsakadze~\cite{Tsakadze1980}. Copyright $\copyright$ 1980, Plenum Publishing Corporation.}
\label{fig-TsakadzeSetup}
\end{center}
\end{figure}

Despite the fact that these experiments were performed more than 30 years ago, they still mark the highlight of research on laboratory neutron
star analogues and have remained the only ones specifically focusing on the idea of reproducing the neutron stars' rotational evolution. We will
therefore discuss a few more aspects in detail. The schematic setup of the experiments is shown in Fig.~\ref{fig-TsakadzeSetup}. The test
neutron star is represented by a hollow glass sphere (1) that has a radius of $3.4 \pm 0.05 \, \cm$.\footnote{The information about the size of
the container is somewhat contradicting. In Tsakadze and Tsakadze~\cite{Tsakadze1980}, $3.4 \pm 0.05 \, \cm$ is given as both the radius and the
diameter of the sphere. According to Reisenegger~\cite{Reisenegger1993}, the smaller size should be adopted to match the measured relaxation
timescales.} Together with a brass disk (2), which is attached to increase the moment of inertia of the system, the sphere is rigidly connected
to a thin steel rod (3) and magnetically suspended (7-9) in order to reduce friction. The support device (4) is used to lower the sphere into a
bath of helium, filling it with the fluid before each experiment. The freely suspended components (1-3) were then suddenly accelerated to a
desired angular frequency using an electric motor (5-6). Once the motor is switched off, the system is allowed to evolve freely under the
frictional forces present. Its rotation period is measured by a focused light beam reflected off a mirror (10) fixed to the rod (3). Since the
helium fluid follows the initial acceleration of the container as a result of the coupling to the walls, the vessel is first observed to spin-down
abruptly due to angular momentum conservation. Subsequently its angular velocity exhibits exponential decay with a small damping parameter and
eventually reaches a linear deceleration regime. According to Tsakadze and Tsakadze this transition marks the characteristic relaxation timescale,
$t_0$, of the container spin-down.

As the experiment was performed for various temperatures, vessel configurations, initial angular velocities and velocity jumps, a formula for
the relaxation timescale as a function of the external parameters could be extracted. For a sphere of radius $R$ rotating at a frequency $\Omega_0$,
exposed to a jump $\Delta \Omega$, Tsakadze and Tsakadze give~\cite{Tsakadze1980}
\begin{equation}
	t_0 \approx \frac{A}{\Omega_0} 	\left( \frac{m_{\rm He} R^2 \Omega_0}{\hbar} \right)^{\beta} \left(\frac{\rho\NF}{\rho} \right)^{-\alpha} \,
		\text{ln} \left( 1 + C \Delta \Omega \right).
			\label{eqn-spindownHeIIexperiment}
\end{equation}
Here, $m_{\rm He}$ denotes the mass of a helium atom, whereas $\rho$ and $\rho\NF$ represent the total and normal component's mass density.
The constants are best fitted as $A=1.0\pm 0.1$, $\beta = 0.40 \pm 0.05$, $\alpha = 0.25 \pm 0.01$ and $C = (5.0 \pm 0.2) \, \sec$.
Eqn.~\eqref{eqn-spindownHeIIexperiment} specifically shows that the relaxation timescale increases as the temperature (and therefore the number
of excitations constituting the viscous fluid component) decreases, providing direct evidence for the validity of the two-fluid model. Tsakadze
and Tsakadze~\cite{Tsakadze1973} further suggested that the relaxation following a pulsar glitch could be governed by a similar expression
if corresponding neutron star parameters are adapted. Comparing timescales from glitch observations, they found general agreement, which supported
the idea of astrophysical quantum condensates.

The two physicists also modified the experimental setup to investigate different coupling strengths and pinning. By introducing impurities in the
form of crushed Plexiglas crystals into the fluid, the friction between
the normal component and the container walls was increased, resulting in shorter relaxation timescales. Glueing the crystals to the inside of the
sphere (effectively providing more nucleation and pinning sites), a larger number of vortices were generated and, thus, the dissipation increased.
This also caused shorter relaxation times compared to the experiments with smooth walls. Such a degree of control over the pinning strength and the
possibility to study its effect on the macroscopic rotational evolution of the fluid components could be particularly useful for improving the
understanding of superfluid pinning in the inner neutron star crust. Large uncertainties prevail for the strength of the vortex-lattice interaction,
which is crucial for determining the crustal mutual friction mechanisms, and helium II experiments could provide more insight into the interplay
between the mesoscopic coupling and the large-scale dissipation. The Georgian physicists further repeated their experiment with a mixture of
helium-4 and helium-3~\cite{Tsakadze1977}. As the temperature was set below the Lambda point but above the transition temperature for helium-3
superfluidity, the normal helium-3 atoms act as an additional viscous fluid, increasing the dissipation. Due to stronger friction, the
relaxation timescales were shorter the more helium-3 was dissolved into helium II.

\begin{figure}[t]
\begin{center}
\includegraphics[width=4.5in]{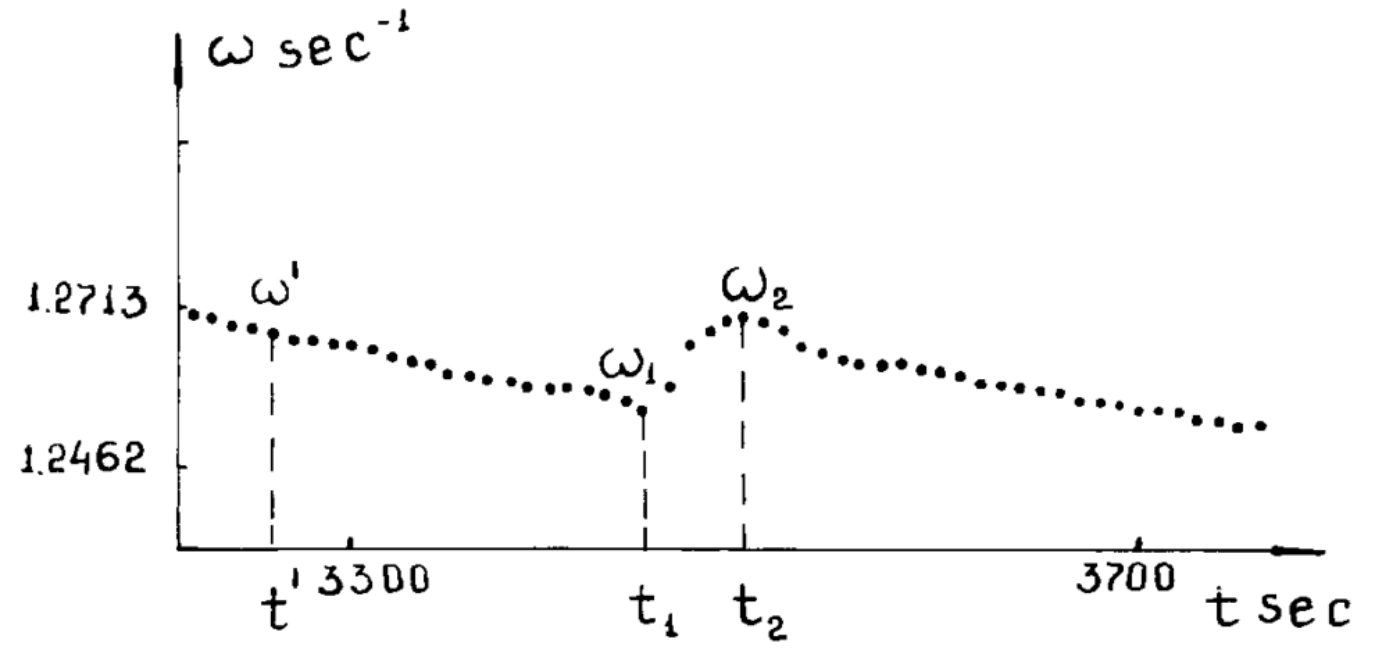}
\caption{Original measurements of the rotational velocity of a rotating cylinder filled with helium II. After an initial acceleration at
$t=0$, the vessel is spinning down and observed to accelerate between the times $t_1$ and $t_2$. The figure is reproduced from Tsakadze and
Tsakadze~\cite{Tsakadze1980}. Copyright $\copyright$ 1980, Plenum Publishing Corporation.}
\label{fig-TsakadzeSpinup}
\end{center}
\end{figure}

Furthermore, Tsakadze and Tsakadze detected spontaneous acceleration during periods of observation, which lasted for over an hour. Using a
cylindrical Plexiglas container with a diameter of $1.5 \, \cm$, an external pulse was applied to spin up the vessel, which was then let to
evolve freely. While the spin-down was initially observed to be linear, it was suddenly impeded by a jump in the rotation frequency as shown in
Fig.~\ref{fig-TsakadzeSpinup}. This phenomenon was explained in terms of the dynamics of the vortex array. As for the pulsar glitch mechanism,
the superfluid does not follow the spin-down of the container and forms a metastable state with a non-equilibrium number of vortices. When the
superfluid component and the container are recoupled, a large number of vortices decays,  leading to the acceleration of the container due to the
conservation of angular momentum. While the star-quake model~\cite{Ruderman1969, Baym1971} had been in the focus of the astrophysics
community up to this point, the new experimental results were pointing towards a superfluid-related glitch mechanism, since quake-like disruptions
had not been generated in the glitching helium II samples.

Despite numerous improvements of laboratory techniques in the last 40 years, the research performed by Tsakadze and Tsakadze has not been repeated
or improved. While there appears to have been little interest for  helium II spin-up experiments  in the low-temperature-physics community, the
benefits of studying this terrestrial neutron star analogue have been pointed out by several astrophysicists. Reisenegger~\cite{Reisenegger1993}
for example examined the laminar spin-up of helium II by analytically solving the fluid equations of motion in the presence of vortices for a
simplified geometry. While the analysis agreed quantitatively with the smooth container experiments, detailed comparison was not possible. More
recently, van Eysden et al.~\cite{VanEysden2011, VanEysden2014b} modelled the dynamics of the superfluid by including the back-reaction torque
exerted by the container. They derived a self-consistent, analytical solution of the HVBK equations and found good agreement with the Tsakadze
and Tsakadze data. However, more detailed experiments would be needed in order to infer reliable information about the physics of neutron stars.

\begin{figure}[t]
\begin{center}
\includegraphics[width=4.2in]{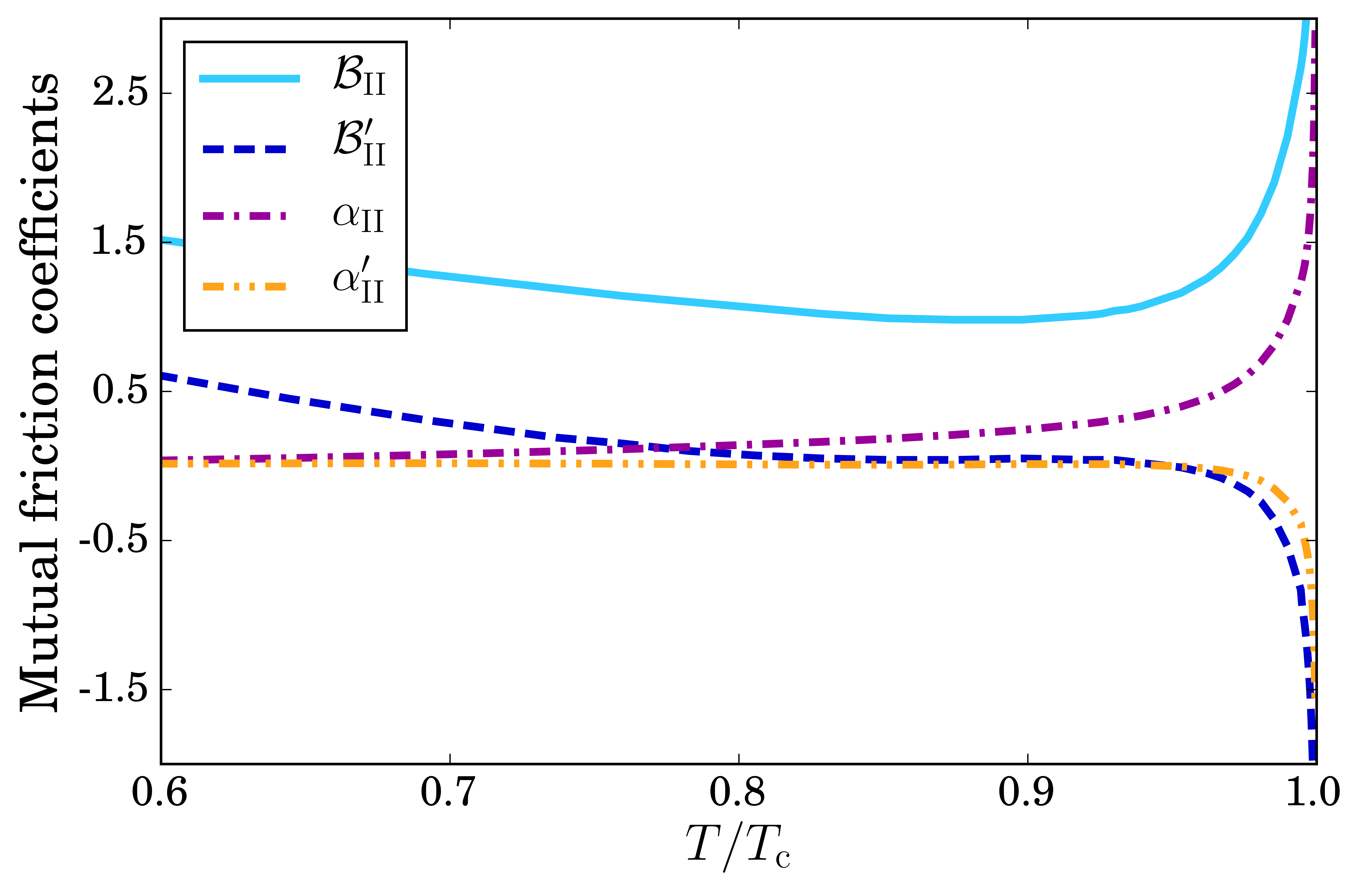}
\caption{Experimental ($\CB\II, \CB'\II$) and modified ($\alpha\II, \alpha'\II$) dimensionless mutual friction coefficients of helium II as a function
of the reduced temperature, $T/T_{\rm c}$, where $T_{\rm c} = 2.171 \, \K$ is the superfluid transition temperature. The data is taken from Barenghi et
al.~\cite{Barenghi1983}.}
\label{fig-HeII_MutualFriction}
\end{center}
\end{figure}

\subsection{Mutual friction}
\label{subsec-MutualFrictionHe}

In a rotating superfluid, the normal and the inviscid components are coupled by forces that result from interactions of vortices with the viscous
fluid. This type of dissipative coupling, called mutual friction, was first investigated by Hall and Vinen in the 1960s~\cite{Hall1956} in the
context of uniformly rotating helium II, which is permeated by an ordered array of straight vortices. They developed a mathematical formalism
allowing one to include the dissipation by introducing a macroscopic, averaged mutual friction force that results from coarse-graining over regions
containing large numbers of vortices. As given in Eqn.~\eqref{eqn-MutualFrictionHe}, the force for a straight vortex array is
\begin{equation}
 	\mathbf{F}_{\rm mf} = \CB\II \frac{\rho\SF \rho\NF}{2\rho} \, \omega \, \uo \times \left[ \uo \times \left( \mathbf{v}\SF -
		\mathbf{v}\NF \right)\right] + \CB'\II \frac{\rho\SF \rho\NF}{2 \rho} \, \omega\, \uo
 		\times \left( \mathbf{v}\SF - \mathbf{v}\NF \right).
 		\label{eqn-MutualFrictionHeExperiment}
\end{equation}
Here $\rho\SF$, $\rho\NF$ and $\rho$ are the superfluid, normal fluid and total mass density, respectively, and $\boldsymbol{\omega} \equiv \omega \uo$
denotes the averaged vorticity, while $\mathbf{v}\SF$ and $\mathbf{v}\NF$ represent the averaged velocities of the inviscid and viscous component. As
mutual friction modifies the propagation of second sound in helium II, the dimensionless coefficients $\CB\II$ and $\CB'\II$ can be directly determined
in rotating container experiments. More precisely, $\CB\II$ is related to the excess attenuation of second sound caused by the presence of quantised
vortices, while $\CB'\II$ is responsible for the coupling of modes that would be degenerate in the absence of rotation. This type of experiment can be
performed in the range of $1.3$ to $2.171 \, \K$, which is equivalent to a reduced temperature of $0.6 <T/T_{\rm c} < 1$. Using results of various studies
(discussed in detail by Barenghi et al.~\cite{Barenghi1983}), the behaviour of the phenomenological coefficients $\CB\II$ and $\CB'\II$ and the modified
mutual friction coefficients, usually denoted by $\alpha\II$ and $\alpha\II'$ in the literature~\cite{Donnelly1991, Vinen2002, Finne2003}, is illustrated
in Fig.~\ref{fig-HeII_MutualFriction}. The latter are defined as
\begin{equation}
	\alpha\II \equiv \CB\II \frac{\rho\NF}{2\rho}, \hspace{2cm}
	\alpha'\II \equiv \CB'\II \frac{\rho\NF}{2\rho},
		\label{eqn-ModifiedMFCoefficients}
\end{equation}
which corresponds to the following mutual friction force,
\begin{equation}
 	\mathbf{F}_{\rm mf} = \alpha\II \rho\SF \, \omega \, \uo \times \left[ \uo \times \left( \mathbf{v}\SF -
 		\mathbf{v}\NF \right)\right] + \alpha'\II \rho\SF \, \omega\, \uo \times \left( \mathbf{v}\SF - \mathbf{v}\NF \right).
 		\label{eqn-MutualFrictionHeExperimentII}
\end{equation}
In the absence of dissipation, vortices are free and their motion is simply governed by the Magnus force, which causes them to move with the superfluid.
In the presence of a viscous drag however, the vortex motion is modified and balancing the two forces allows one to express the averaged vortex velocity
in terms of the macroscopic fluid variables. Generalising Eqn.~\eqref{eqn-MesoscopicVortexVelocity} to the case of straight vortices, one has
\begin{equation}
	\mathbf{u}_{\rm v} = \mathbf{v}\SF - \frac{1}{\rho\SF \, \omega} \, \uo \times \mathbf{F}_{\rm mf}
		= \mathbf{v}\SF + \alpha\II \uo \times \left( \mathbf{v}\NF - \mathbf{v}\SF \right) + \alpha'\II \left( \mathbf{v}\NF - \mathbf{v}\SF \right).
		\label{eqn-MacroscopicVortexVelocity}
\end{equation}
As seen in Fig.~\ref{fig-HeII_MutualFriction}, the values for $\alpha\II$ and $\alpha'\II$ are of the same order for a large temperature range,
implying that mutual friction induces changes of similar degree to the vortex velocity components parallel and perpendicular to the superfluid velocity.

\begin{figure}[t]
\begin{center}
\includegraphics[width=4in]{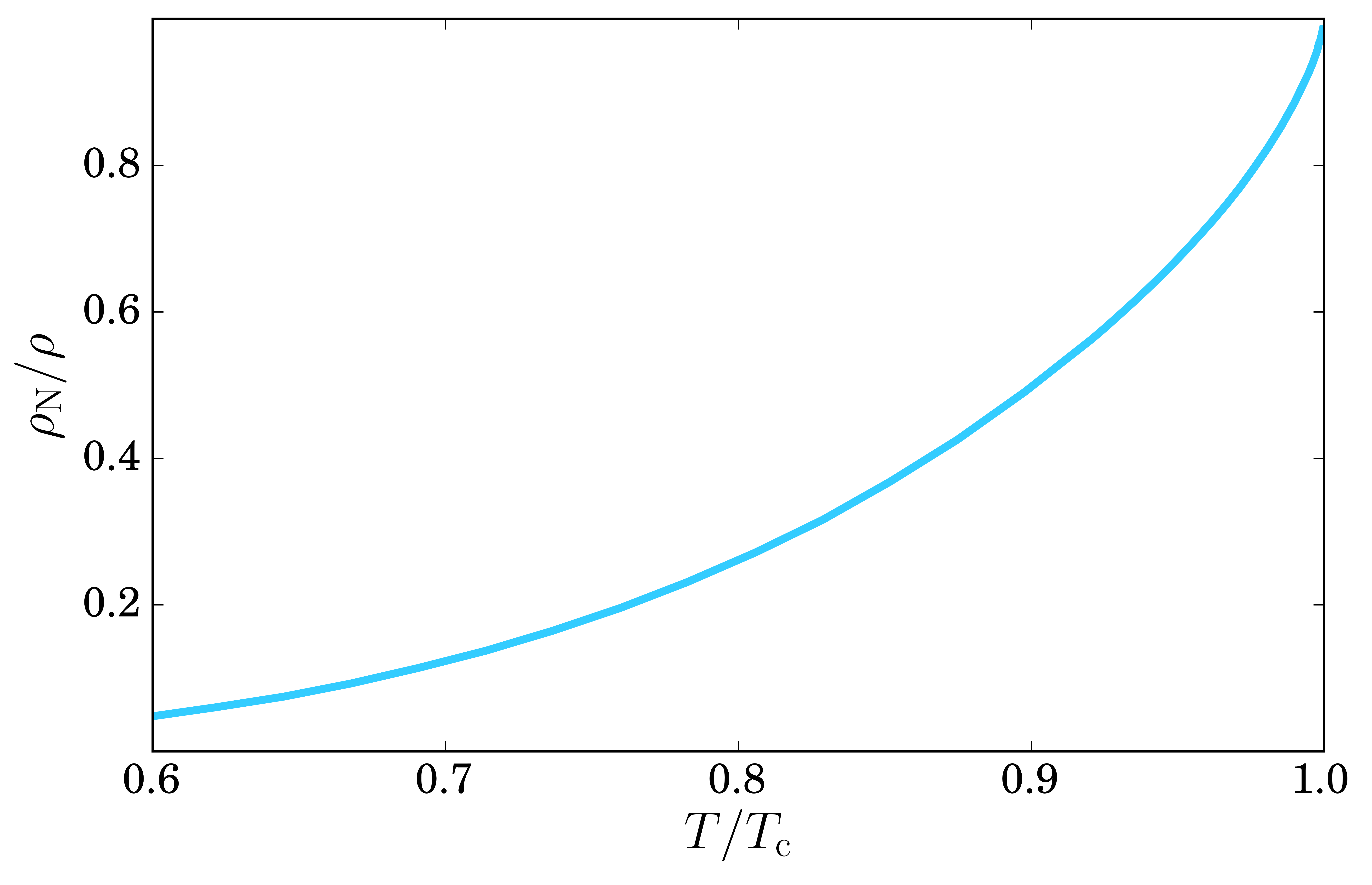}
\caption{Viscous fluid fraction, $x\NF$, of helium II as a function of the reduced temperature, $T/T_{\rm c}$. $T_{\rm c} = 2.171 \, \K$ is the superfluid
transition temperature. The data is taken from Barenghi et al.~\cite{Barenghi1983}.}
\label{fig-HeII_xn}
\end{center}
\end{figure}

To allow for a comparison of coupling strengths in helium II and neutron stars, Eqn.~\eqref{eqn-MutualFrictionHeExperimentII} has to be compared to the
neutron star mutual friction (see Eqn.~\eqref{eqn-MutualFrictionSimple});
 \begin{equation}
	\mathbf{F}_{\rm mf}  = \CB \rho_\rn \, \cN_\rn \kappa \uk  \times \left[ \uk \times \left( \mathbf{v}_\rn - \mathbf{v}_\re \right) \right]
		+ \CB' \rho_\rn \, \cN_\rn \kappa \uk \times \left( \mathbf{v}_\rn - \mathbf{v}_\re \right),
		\label{eqn-MutualFrictionNSExperiment}
\end{equation}
where $\cN_\rn \kappa \uk$ corresponds to the averaged vorticity. Identifying the neutrons with the inviscid component in helium II and the charged-particle
conglomerate with its excitations, $\rho\SF$ and $x\NF$ are equal to $\rho_\rn$ and $x_\rp$, respectively. This gives the following relations between the
friction coefficients in neutron stars and laboratory systems:
\begin{equation}
	\CB  = \CB\II \frac{x\NF}{2} = \alpha\II, \hspace{2cm}
	\CB' = \CB'\II \frac{x\NF}{2} = \alpha'\II,
		\label{eqn-RelationMFCoefficients}
\end{equation}
with the viscous fluid fraction defined as $x\NF \equiv \rho\NF / \rho$ (for experimental results see Fig.~\ref{fig-HeII_xn}). Hence, the numerical estimates
for the dissipation strengths in the neutron star superfluids have to be compared to the modified helium II mutual friction parameters $\alpha\II$ and $\alpha'\II$,
if the analogy between the two systems is to be exploited.

One instantly notices that the experimental values for helium II show little agreement with the ones invoked for neutron stars. While the standard coupling
mechanisms in neutron stars are generally attributed to the weak mutual friction limit (with positive parameters $\CB$ and $\CB'$), the coupling in superfluid
helium is much stronger. This implies that the dynamics in neutron star cores are well approximated by the free vortex limit and vortices are dragged
along with the superfluid component, whereas in helium II the two components do not move together. Close to the Lambda point, $\alpha\II$ and $\alpha'\II$
are of order unity with both coefficients expected to diverge as $(T_{\rm c} - T)^{-1/3}$ for $T \to T_{\rm c}$~\cite{Mathieu1976}. Moreover, $\alpha'\II$
turns negative for temperatures above $2.07 \, \K$, which would suggest rather different physical behaviour. Also note that the fraction of normal fluid in
helium II is significantly larger than the proton fraction in neutron star cores. Only for very low temperatures does $x\NF$ take
values between $2$ and $10\%$, the range one would want to examine. Experimental data below $1.3 \, \K$ would therefore be important in order to develop
laboratory neutron star models with helium II. However, the standard rotating helium experiments measuring the speed of second sound are no longer applicable
at such low temperatures, because the viscous fluid concentration is too low to provide reliable results. Instead, experiments that measure the drag on vortex
rings, which are attached to individual ions, have been designed. While these studies would in principle allow access to the parameters $\CB\II$ and $\CB'\II$,
no conclusive data is available and the values in Fig.~\ref{fig-HeII_MutualFriction} have been restricted to the data from second sound experiments.

When analysing vortex-averaged dissipation, the phenomenological parameters $\CB\II$ and $\CB'\II$ provide little information about the underlying mesoscopic
or microscopic mechanisms in helium II. However, as first suggested by Landau~\cite{Landau1941}, dissipation in helium II could result from the interactions of
thermally excited quasi-particles with individual vortices. It is thus possible to relate the large-scale dynamics to the small-scale physics by considering
a mesoscopic coupling force of the form~\cite{Barenghi1983}
\begin{equation}
	\mathbf{f}_{\rm d}  = \rho\SF \kappa \cR\II \, ( \mathbf{v}_{\rm q} - \mathbf{u}\v)
		+ \rho\SF \kappa \cR'\II \uo \times ( \mathbf{v}_{\rm q} - \mathbf{u}\v).
		\label{eqn-DragForceHe}
\end{equation}
Here, $\mathbf{v}_{\rm q}$ denotes the corresponding quasi-particle velocity and $\mathbf{u}\v$ the vortex velocity. The force is proportional to the relative
velocity and characterised by two mesoscopic friction coefficients, $\cR\II$ and $\cR'\II$. Different theories are available to calculate these parameters,
since the coupling mechanism depends on two crucial lengthscales, i.e.\ the size of the region responsible for mutual friction and the quasi-particles' mean free
path~\cite{Sonin1987}. For low temperatures, the former lengthscale is smaller and the scattering of rotons and phonons off rectilinear vortices causes dissipation.
Using the scattering theory for non-interacting quasi-particles, the mesoscopic coefficients can be obtained. They typically satisfy $|\cR'\II|\gg
\cR\II$~\cite{Barenghi1983}. For temperatures close to the Lambda point, the vortex core size increases considerably and becomes larger than the mean free path, which
raises the need for phenomenological approaches like the time-dependent Ginzburg-Landau theory~\cite{Sonin1987}. Utilising a force balance equation to eliminate
the vortex velocity in Eqn.~\eqref{eqn-DragForceHe}, the macroscopic and mesoscopic coefficients can be related to each other. Therefore, $\cR\II$ and $\cR'\II$
are controlled by $\CB\II$ and $\CB'\II$ and vice versa, providing a possibility to constrain mesoscopic theoretical models with experimental data. Studying
mutual friction in helium II could thus provide useful information about the coupling mechanisms in neutron star, where direct observations are not feasible and
microscopic interactions can only be studied from a theoretical point of view. In particular, one could learn how to determine a suitable average over mesoscopic
lengthscales in order to match the measured macroscopic dissipation strengths. Such studies might even help to deduce information about the vortex arrangement and
deviations from a straight vortex array.

One difference between the two formalisms is immediately evident. Direct comparison between the helium II drag in Eqn.~\eqref{eqn-DragForceHe} and the ansatz for
the force in neutron stars shows that an additional term proportional to $\cR'\II$ is included in the former system. It acts in the direction orthogonal to the
relative velocity and the local orientation of the vortex. This transverse drag component, which is generally not included in the neutron star case, is needed in
order to explain the experimental data, in particular the negative values of $\alpha'\II$ close to the Lambda point. As can be seen from Fig.~\ref{fig-HeII_MutualFriction},
the simple relationship between the neutron star mutual friction coefficients, i.e. $\CB' \approx \cR \CB$, does not hold in superfluid helium-4. By introducing a
second parameter $\cR'\II$, this behaviour can be captured. This suggests that the neutron star problem tends to be oversimplified, and that a second mesoscopic
drag term $\cR'$ should perhaps be accounted for.

\begin{figure}[t]
\begin{center}
\includegraphics[width=3.5in]{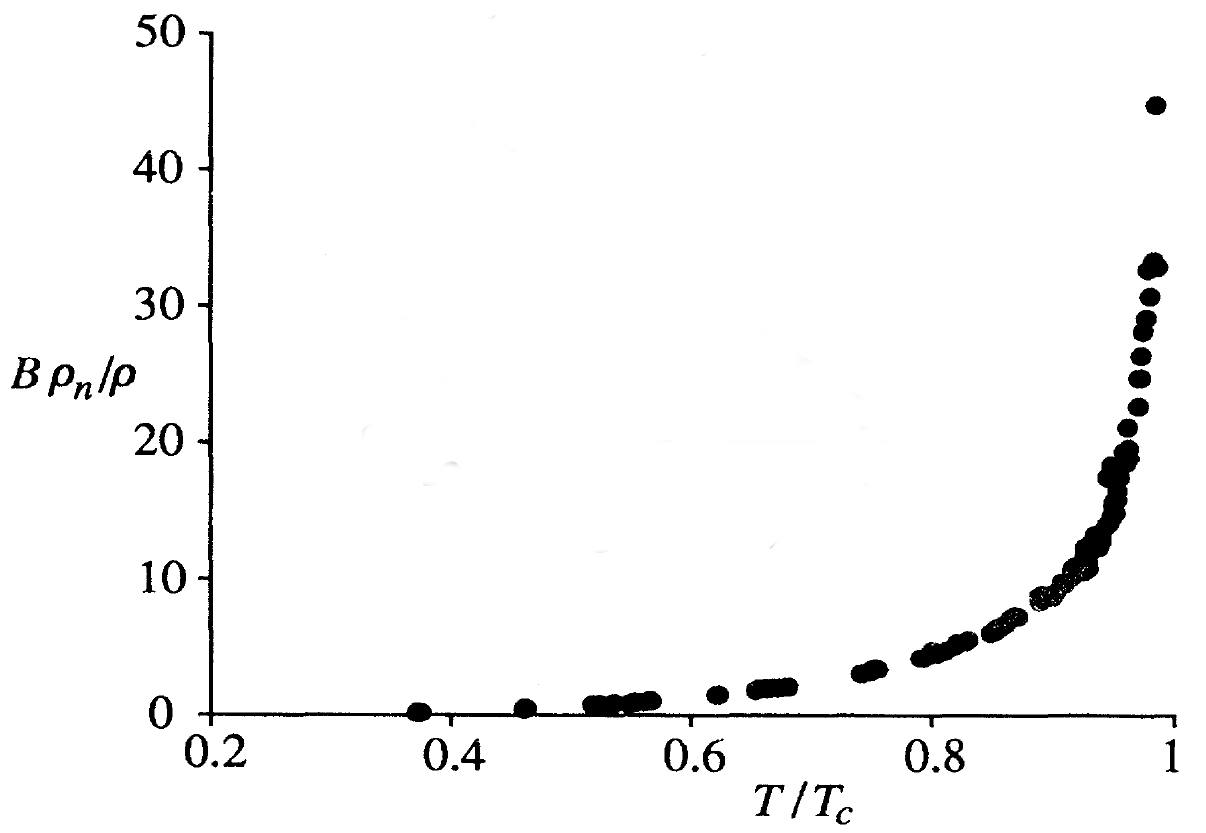}
\caption{Behaviour of the mutual friction parameter ($B_{\rm 3B} x\NF = 2 \alpha_{\rm 3B}$) in the $B$-phase as a function of the
reduced temperature, $T/T_{\rm c}$. The measurements are taken at a pressure of $1.6 \, \text{bar}$. In this case, the critical temperature
for the superfluid transition is $T_{\rm c} \sim 1.1 \, \text{mK}$. The figure is reproduced with permission from Bevan et al.~\cite{Bevan1995}.
Copyright $\copyright$ 1995, American Physical Society.}
\label{fig-He3B_1bar_alpha}
\end{center}
\end{figure}

Although helium II is the best studied system regarding mutual friction, several attempts have been undertaken at measuring drag parameters in helium-3.
These were not only complicated by the presence of the different phases but also by the fact that the viscosity of the normal component is four orders
of magnitude larger than that of helium II~\cite{Finne2003}. The latter problem causes second sound to become highly damped, eliminating such studies as
a tool to investigate mutual friction in helium-3 and raising the need for new techniques. Whereas earlier efforts~\cite{Hall1984, Krusius1993} had only
provided limited information, Bevan et al.~\cite{Bevan1995, Bevan1997} designed an experiment allowing one to determine the coefficients of the $B$-phase
and the $A$-phase. Taking advantage of the normal fluid's large viscosity, the following idea was exploited: Separating two regions of helium-3, a vibrating
Kapton film was used to set the superfluid component into motion, while the normal fraction remained stationary due to its viscous properties. As the
film's vibrational modes are influenced by the vortex array, the mutual friction strength  could be deduced by analysing the mode frequencies. This way,
experimental estimates of the modified coefficients $\alpha_{\rm 3}$ and $\alpha'_{\rm 3}$ were obtained for various temperatures and pressures.

\begin{figure}[t]
\begin{center}
\includegraphics[width=3.5in]{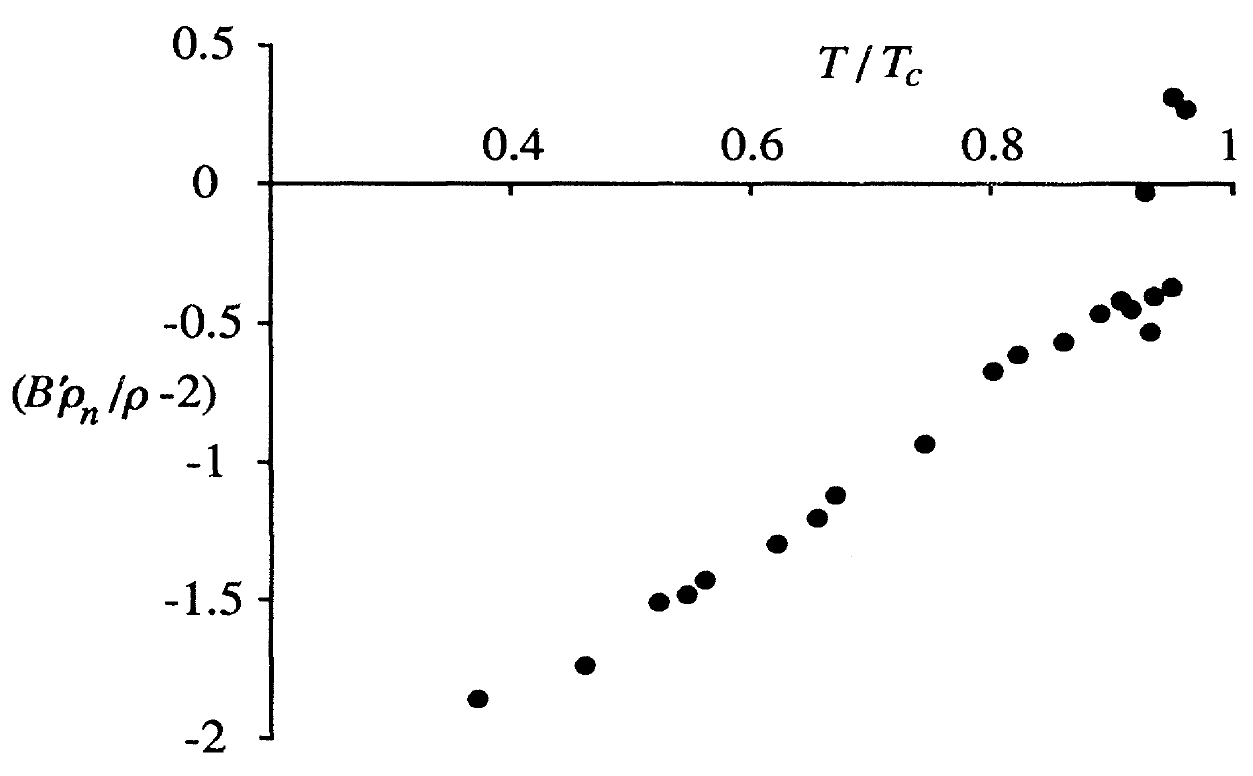}
\caption{Behaviour of the mutual friction parameter ($B'_{\rm 3B} x\NF - 2 = 2 (\alpha'_{\rm 3B} - 1)$) in the $B$-phase as a function of the
reduced temperature, $T/T_{\rm c}$. The measurements are taken at a pressure of $1.6 \, \text{bar}$. In this case, the critical temperature
for the superfluid transition is $T_{\rm c} \sim 1.1 \, \text{mK}$. The figure is reproduced with permission from Bevan et al.~\cite{Bevan1995}.
Copyright $\copyright$ 1995, American Physical Society.}
\label{fig-He3B_1bar_alphaprime}
\end{center}
\end{figure}

\begin{figure}[t]
\begin{center}
\includegraphics[width=3.5in]{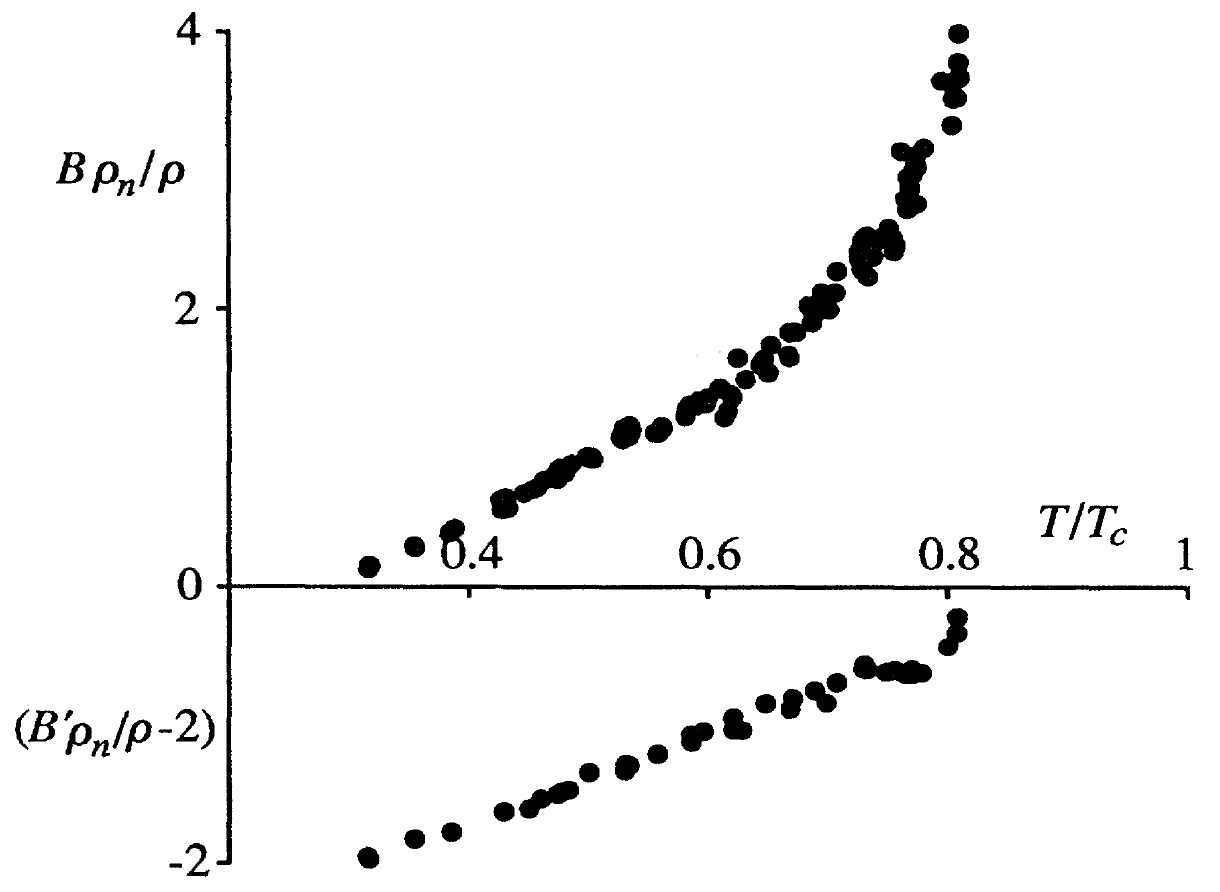}
\caption{Behaviour of both $B$-phase mutual friction parameters as a function of the reduced temperature, $T/T_{\rm c}$. The measurements are taken at a pressure
of $29.3 \, \text{bar}$, implying that the critical temperature for the superfluid transition is $T_{\rm c} \sim 2.4 \, \text{mK}$. Above $T \sim 0.8 T_{\rm c}$
the $A$-phase dominates. The figure is reproduced with permission from Bevan et al.~\cite{Bevan1995}. Copyright $\copyright$ 1995, American Physical Society.}
\label{fig-He3B_29bar}
\end{center}
\end{figure}

\begin{figure}[t]
 \begin{center}
\includegraphics[width=3.3in]{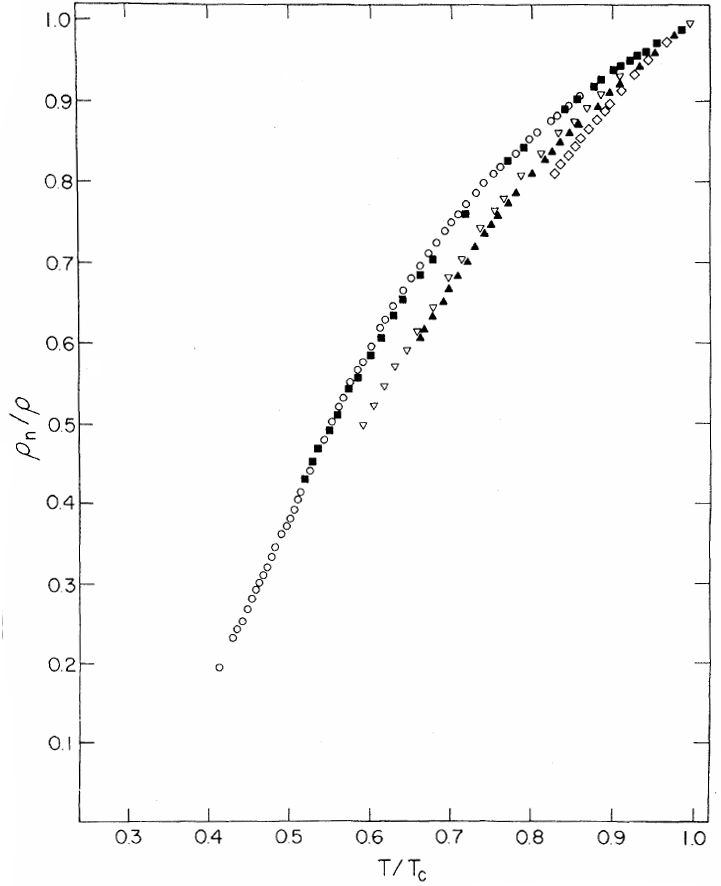}
\caption{Temperature-dependence of the normal-fluid fraction in the helium-3 $B$-phase at different pressures: circles ($29 \, \text{bars}$), closed
squares ($20 \, \text{bars}$), inverted open triangles ($10 \, \text{bars}$), closed triangles ($5 \, \text{bars}$), and diamonds ($2 \, \text{bars}$).
The figure is reproduced from Archie et al.~\cite{Archie1979}. Copyright $\copyright$ 1979, American Physical Society.}
\label{fig-He3B_xN}
\end{center}
\end{figure}

The original $B$-phase measurements at $1.6 \, \text{bars}$ and $29.3 \, \text{bars}$ are illustrated in Figs.~\ref{fig-He3B_1bar_alpha}-\ref{fig-He3B_29bar}
as a function of the reduced temperature. The critical temperature for the superfluid transition is pressure-dependent and given by $T_{\rm c}
\sim 1.1 \, \text{mK}$ in the former case and $T_{\rm c} \sim 2.4 \, \text{mK}$ in the latter case~\cite{Greywall1986}. Note further that for pressures
above $22 \, \text{bar}$, both helium-3 phases are present. For $29.3 \, \text{bars}$, the transition temperature $T_{AB}$ is located at around $2 \,
\text{mK}$~\cite{Greywall1986}, which implies that the $B$-phase only exists up to $T/T_{\rm c} \sim 0.8$ as can be seen in Fig.~\ref{fig-He3B_29bar}. The data
shows that both mutual friction coefficients vanish in the limit $T \to 0$ as expected, while $\alpha_{\rm 3B}$ diverges close to the transition temperature and
$\alpha'_{\rm 3B}$ approaches $1$. Similar to helium II, the coupling between the normal and the superfluid component in helium-3 is much stronger than predicted
for neutron stars. Measurements for the $A$-phase lead to even stronger dissipation~\cite{Bevan1997}.As before, the macroscopic dissipation parameters can be
related to mesoscopic drag coefficients. In contrast to the previous discussion however, it is less clear what kind of interactions between quasi-particles and
vortices generate the coupling on small scales. One cannot simply transfer the theoretical predictions for helium II to helium-3 due to the fundamental differences
in vortex formation~\cite{Sonin1987}. Instead, dynamical features seem to be well explained using the theories available for superconductors~\cite{Kopnin2002},
which are also characterised by Fermi-Dirac statistics and exhibit quantum properties by forming Cooper pairs. Analysing the dissipative coupling in helium-3
could again help to understand mutual friction in neutron stars, where the dominating coupling mechanisms are not well known. To exploit this analogy further,
the ratio of superfluid to normal fluid component should ideally be $2$ to $10\%$. Experimental results for the viscous fluid fraction in helium-3 at various
pressure are for example discussed by Alvesalo et al.\ and Archie et al.~\cite{Alvesalo1975, Archie1979}. As seen from Fig.~\ref{fig-He3B_xN}, the normal fluid
fraction in the $B$-phase and the proton fraction in neutron stars are comparable for temperatures below $\sim 0.4 T_{\rm c}$. This region is accessible with
experiments~\cite{Walmsley2011}, presenting an advantage of helium-3 over helium-4.

The experimental drag parameters are strongly dependent on the properties of the superfluid. While the underlying data for
Figs.~\ref{fig-HeII_MutualFriction}-\ref{fig-He3B_29bar} is determined for straight vortices able to move freely through the container, the mutual
friction coefficients could be very different when pinning~\cite{Sonin1981} or turbulence~\cite{Vinen2002} are present. External forces that keep
vortices at rest could be particularly important at higher temperatures close to the Lambda point. Dissipation in the superfluid is also expected to
significantly increase when a turbulent state is formed~\cite{Vinen1957c}. There, mutual friction is no longer described by
Eqn.~\eqref{eqn-MutualFrictionNSExperiment} but depends on the cube of the relative velocities as postulated by Gorter and Mellink~\cite{Gorter1949}
(see Sec.~\ref{subsec-Turbulence} for details). While measurements of transport coefficients under these conditions are very difficult and detailed
experimental data is not available, superfluid turbulence in helium might provide more insight into how such a chaotic state could influence the
neutron star dynamics and is therefore discussed in more detail in the next section.

Before addressing the behaviour of vortices, we raise one critical issue with helium as a laboratory neutron star analogue. In addition to mutual friction,
neutron star dynamics are strongly influenced by entrainment. Number densities in the outer core  are expected to reach $10^{38} \, \cm^{-3}$, which
corresponds to interparticle spacings of $10^{-13} \, \cm$. At such short distances, the strong nuclear force couples the neutron and proton fluids.
However, this process is non-dissipative and cannot simply be reproduced in weakly-coupled single-component condensates such as helium II, where the
particle density is typically of order $10^{22} \, \cm^{-3}$ resulting in a distance of $10^{-8} \, \cm$. Interpenetrating liquids are necessary to
recreate any phenomena resulting from entrainment, which could for example be achieved by studying mixtures of superfluid helium-3 and superfluid
helium-4. Note that this situation was originally considered when entrainment was discovered by Andreev and Bashkin~\cite{Andreev1976}. However, due
to the strong interactions between the two isotopes, helium mixtures only contain a small fraction of helium-3, which has so far prohibited the
experimental realisation of simultaneous superfluidity in both species~\cite{Tuoriniemi2002, Rysti2012}.

\subsection{Vortex dynamics}
\label{subsec-VortexDynamicsHe}

\begin{figure}[t]
\begin{center}
\includegraphics[width=4.2in]{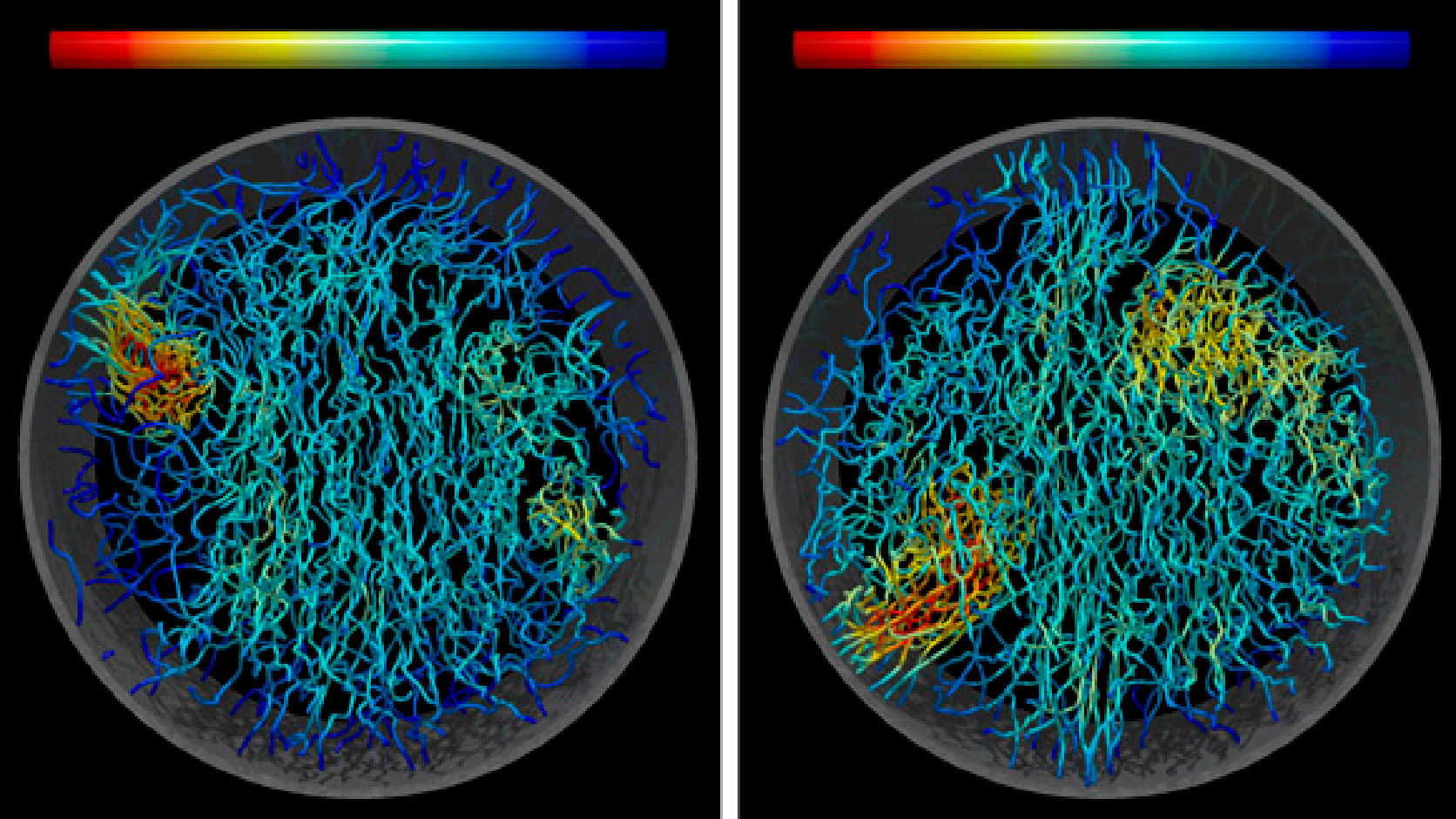}
\caption{Top view of a vortex-line simulation for the spin-up of the $B$-phase superfluid in a tilted, rotating cylinder. Initially, only one
vortex is present. The snapshots show vortex configurations at $t = 1,100 \, \sec$ (left) and $t = 1,140 \, \sec$ (right). The
colour code mirrors the relative amplitude of the averaged vorticity. In both configurations coherent structures appear in the form of vortex
bundles (orange and red regions). The figure is reproduced with permission from H\"anninen and Baggaley~\cite{Hanninen2014}.}
\label{fig-VortexLineSim}
\end{center}
\end{figure}

Whereas the macroscopic formalism (providing vortex-averaged information about the dynamics of vortices located within comparatively large fluid elements)
allows one to discuss the superfluids' influence in a more classical manner and correctly predicts several observed phenomena, some aspects are difficult
to study. The subjects of interface physics, turbulence and instabilities are especially challenging~\cite{Finne2006, Barenghi2014}. In these areas however,
the experimental and theoretical methods for analysing individual vortices have been greatly improved in the last decade. For a recent review on vortex
studies in superfluid helium and BECs see Tsubota et al.~\cite{Tsubota2013}. One theoretical tool that has been very valuable in modelling the behaviour
of superfluids are \textit{vortex-line simulations}~\cite{Schwarz1988, Hanninen2014}. Within this filamentary model, vortices are regarded as line
defects. This implies that the entire vortex configuration determines the velocity of the superfluid component, which is governed by a Biot-Savart-type law
as introduced in Eqn.~\eqref{eqn-BiotSavart},
\begin{equation}
	 \Bv\SF ( \Br, t) = \frac{\kappa} {4 \pi} \int_{\mathcal{L}} \frac{(\mathbf{s} - \Br) \times \d \mathbf{s}}{|\mathbf{s} - \Br|^3},
		\label{eqn-BiotSavart_LabSection}
\end{equation}
where the line integral is taken along all vortices. The motion of a single vortex is then obtained by balancing the various forces acting on it.
Snapshots of such a simulation, modelling the spin-up of the helium-3 $B$-phase confined to a tilted container, are shown in Fig.~\ref{fig-VortexLineSim}.
While vortex-line simulations are useful to self-consistently model laboratory condensates (containing only several thousand vortices), large computational
costs make it difficult to apply this technique to neutron stars, where significantly more vortices and fluxtubes are present. However, the study of vortex
dynamics in laboratory condensates could provide crucial information for the development of better neutron star models, a fact which has generally been
ignored. Terrestrial experiments could be particularly valuable for understanding how non-classical phenomena such as instabilities and turbulence influence
neutron stars.

Several aspects of vortex dynamics are similar in superfluid helium-4 and helium-3. Differences arise however due to the mutual friction properties
and the variations in vortex size, since vortices are generally fatter in the helium-3 phases. While for helium II, the core dimension is of the order
of the coherence length and given by $\sim 0.1 \, \nm$,  the coherence length in the isotropic $B$-phase is $\sim 10 \, \nm$~\cite{Finne2006}. In the
anisotropic $A$-phase, the vortex cores are three orders of magnitude larger than in the $B$-phase~\cite{Mineev1986}, because the characteristic
lengthscale is no longer the coherence length but the so-called healing length of the spin-orbital coupling. With a radius of about $10 \, \mu \m$,
$A$-phase vortices are not localised but instead stretch over macroscopic regions. The increase in helium-3 core sizes creates several experimental
advantages over helium-4. Not only are the critical velocities for the onset of vortex formation lower in the former superfluid but also the interactions
of vortices with the container walls are very different. The latter makes pinning generally negligible~\cite{Finne2006} and allows better control over
the vortices' motion, which is of great importance for laboratory neutron stars. Note at this point that the vortex dimensions in all three helium
condensates are several orders of magnitude larger than those in neutron stars (see Sec.~\ref{subsec-FurtherSuperCon}) and one has to be careful with
directly inferring information about the star's physics from terrestrial experiments. Instead, the observed features should be interpreted as indications
of similar phenomena in the astrophysical context that subsequently need to be studied in more detail.

\subsubsection{Interface behaviour}
\label{subsubsec-InterfacesHe}

The canonical picture of neutron star structure invokes the presence of distinct interfaces. In particular, the crust-core boundary connecting
the $^1 S_0$ and the $^3 P_2$ neutron superfluid phases and the possible type-II to type-I transition of the superconducting protons at high densities are
expected to have a crucial influence on the stars' dynamics. However, the physics of these interfaces are only poorly understood and, hence, laboratory
experiments may provide valuable insight. Superfluid helium-3 plays a unique role in this endeavour, as two-phase samples provide the possibility of studying
vortex behaviour at a stable first-order interface. The advantage is that the order parameter's phase remains continuous across the interface, allowing
vortices to cross the boundary. This is different to the case of two phase-separated superfluid layers, where vortices terminate at the boundary and exhibit
little interaction. As addressed previously, the vortices in the $A$-phase have very different properties as they are much bigger than those in the $B$-phase
and could carry double the quantisation. This raises the question of how the vortices stretch across the interface and how they influence each other
during the rotational evolution.

Walmsley et al.~\cite{Walmsley2011} discuss experimental NMR measurements and vortex-line simulations of a rotating two-phase sample that shows very unusual
vortex behaviour. These features could have important implications for neutron star dynamics and will therefore be reviewed in more detail. The two helium-3
superfluids are contained in a cylindrical, smooth-walled quartz container that can be set into motion by rotating the surrounding cryostat. The cylinder has
a length of $110 \, \text{mm}$, a diameter of $6 \, \text{mm}$ and a small superconducting solenoid is attached around its middle, which generates an axial
magnetic field that stabilises the $A$-phase. The presence of the anisotropic superfluid thus splits the sample into two identical $B$-phase regions creating
two $AB$-interfaces. The time evolution and distribution of vortices in the $B$-phases are monitored with two NMR detectors secured to the bottom and the top
of the cylinder. The experiments are performed at $T = 0.2 \, T_{\rm c}$, where unexpected vortex characteristics are most dominant. At this temperature, the
mutual friction coefficients are given by $\alpha_{\rm 3B} \approx 4.3 \times 10^{-3}$ and $\alpha_{\rm 3B}' \approx 0$ for the $B$-phase and $\alpha_{\rm 3A}
\approx 2$ and $\alpha_{\rm 3A}' \approx 0.8$ for the $A$-phase, respectively~\cite{Walmsley2011}. Using this set-up, the vortex response to a change in the
container's angular rotation period is studied. The authors note that the jump in the rotational velocity is sufficiently small to ensure that the interface
remains stable -- as will be explained below, the superfluid Kelvin-Helmholtz instability would become active if a critical velocity is exceeded. The distinct
vortex features, discussed in the following, have been simultaneously confirmed by vortex-line simulations and non-invasive NMR measurements.

\begin{figure}[t]
\begin{center}
\includegraphics[width=4in]{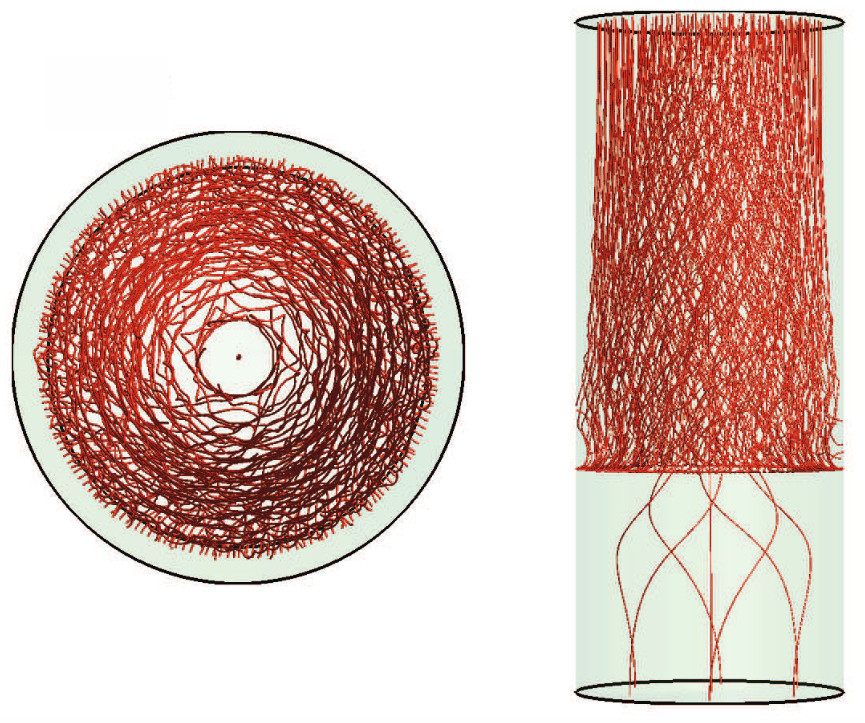}
\caption{Vortex-line simulation for the spin-down behaviour of a two-phase helium-3 sample. Starting from an equilibrium configuration with straight
vortices stretching across the interface, both phases evolve freely. Due to differences in mutual friction the $A$-phase (bottom) responds quickly
to the external change, while the $B$-phase (top) responds slower. The left figure shows a radial cross-section of the $B$-phase layer with an almost
vortex-free centre. The right figure illustrates the formation of a turbulent vortex tangle increasing the dissipation. Reproduced with permission from
Walmsley et al.~\cite{Walmsley2011} Copyright $\copyright$ 2011, American Physical Society.}
\label{fig-WalmsleyFigure}
\end{center}
\end{figure}

The spin-down behaviour is investigated by bringing the container from an equilibrium configuration abruptly to rest. Initially the two phases are corotating
at an angular velocity $\Omega_0$ with straight vortices stretching across the interfaces; despite carrying different units of quantisation the vortices
of both helium-3 phases interconnect across the boundaries~\cite{Finne2006}. After applying the external change, both layers evolve freely. However, due to the
different strengths in mutual friction, the superfluids do not react to the container's spin-down on the same timescale. Since strong coupling prevails in the
$A$-phase, it responds very quickly to the external change, whereas the $B$-phase reacts much slower. Thus, the $A$-phase contains significantly fewer vortices
than the $B$-phase, as can be seen from Fig.~\ref{fig-WalmsleyFigure}. Moreover, the interface region between the two states crucially influences the dynamics
as it introduces new boundary conditions at the surface. Three main observations can be made.

Firstly, the $A$-phase vortices spiral outwards in a laminar manner to annihilate at the container walls, creating an additional pull on the ends of the $B$-phase
vortices. At the interface, this causes the formation of a vortex sheet because the vortex ends bend parallel to the boundary and terminate at the container
walls. Away from the interface, the additional force causes the $B$-phase vortices to develop a helically twisted tangle. This turbulent state promotes more
reconnections in the superfluid bulk, effectively increasing the dissipation. Consequently, it is found that the simple law capturing the averaged azimuthal
$B$-phase flow in absence of the $A$-phase,\cite{Walmsley2011}
\begin{equation}
	\Omega(t) = \frac{\Omega_0}{1+t/\tau},
\end{equation}
where a single timescale $\tau \approx 740 \,\sec$ fits all data of the laminar decay, does no longer hold in the two-phase experiments. Instead, the spin-down
of the $B$-phase is faster in the presence of the $A$-phase.

Secondly, the extra force on the $B$-phase vortices depletes the region closest to the rotation axis faster than the rest of the container, implying that the
superfluid is at rest in the middle of the container. While the centre is almost vortex free, a large number of vortices forming the vortex tangle can be found
in a cylindrical shell closer to the container walls (see the cross-section in Fig.~\ref{fig-WalmsleyFigure}). The averaged circulation of this shell exceeds
the initial solid-body rotation value, which suggests that the superfluid fraction in the $B$-phase no longer rotates as a solid body. Instead, it exhibits
a differential rotation profile along the radial direction of the cylinder.

Finally, comparison with spin-up experiments from rest shows that spin-down and spin-up are not symmetric phenomena. As the critical velocity for vortex
formation is about one order of magnitude lower in the $A$-phase~\cite{Finne2006}, vortices are first generated in the anisotropic superfluid and a vortex sheet
develops on this side of the two-phase sample. The $B$-phase response to the external perturbation crucially depends on the number of remnant vortices. If several
lines are present, the spin-up of the isotropic phase is laminar. If however the $B$-phase is initially vortex-free, the superfluid is spun-up via a sudden
burst of vortex formation leading to significantly faster spin-up. This kind of behaviour is not observed in the spin-down experiments.

It is difficult to asses to which extent this behaviour can be mapped to the rotational dynamics of neutron stars, but one would expect the strength of the
mutual friction to play an important role in this analogy. As explained before, the interactions between the normal and the superfluid components in helium-3
are much stronger than in neutron stars and, hence, the characteristic timescales are shorter in the laboratory system. However, the main reason the two-phase
helium-3 sample exhibits remarkable vortex dynamics during the spin-down and spin-up is the relative difference of the coupling strengths across the interface.
Mutual friction is about three orders of magnitude stronger in the $A$-phase than in the $B$-phase. While detailed information about the neutron star mutual
friction mechanisms are not available, coupling strengths are also likely to jump several orders of magnitude across the neutron star crust-core boundary as
a result of the changes in composition~\cite{Alpar1984, Andersson2006b, Haskell2012}. Based upon this simple criterion, complicated vortex characteristics
should also be present in neutron stars. Thus, the general assumption of a straight, regular vortex array, which enables an averaging procedure using a constant
vortex surface density, no longer holds. As observed in the helium-3 experiments, the presence of an interface breaks the cylindrical symmetry and spin-up or
spin-down of the superfluids can no longer be treated as a two-dimensional problem. The break-down of solid-body rotation further suggests that the superfluid
neutrons in the interior could be differentially rotating. This would change the neutron stars' macroscopic rotational properties and could, for example,
provide a possibility to store supplementary angular momentum. Such an additional reservoir would impact on observational features and could for example
be related to the possible observation of an anti-glitch in a magnetar~\cite{Archibald2013} or the evolution of pulsar braking indices~\cite{Ho2012}. Finally,
a twisted vortex tangle located in a spherical shell would also increase the dissipation and result in a faster spin-down of the superfluid component. As
discussed further in the following, the presence of a turbulent state would thus have significant influence on the observable parameters of neutron star.

\subsubsection{Turbulence and instabilities}
\label{subsubsec-InstabilitiesHe}

Turbulence, representing the chaotic regime of fluid flow, has long been studied in classical fluids and is one of the most complex problems of classical
physics. A comprehensive discussion of this field of research is given by Lesieur~\cite{Lesieur2008}. Since the 1950s, turbulence and instabilities have also
been analysed in superfluid helium-4 and, in the last two decades, additional experiments studying the non-classical dynamics of helium-3 and quantum gases
have been developed. The ingredient that classical and superfluid systems have in common is that a single hydrodynamical equation is no longer sufficient to
accurately capture the non-linear dynamics. Instead, multiple models characterising the behaviour on different lengthscales are needed. However, turbulence
in superfluids is strongly influenced by the quantum nature creating features unobservable in classical fluids. Hence, the chaotic flow in superfluids is
generally referred to as \textit{quantum turbulence}. For a recent introduction to the subject see Barenghi et al.~\cite{Barenghi2014}. Compared to classical
turbulence, the main differences in quantum turbulence arise due to the two-fluid nature. While the viscous component experiences standard turbulence reflected
as quasi-classical behaviour on large scales, the vortices of the inviscid component generate new features on small scales.

Early studies of quantum turbulence performed with non-rotating helium II samples focused on the \textit{thermal counterflow} behaviour. As first suggested by
Feynman~\cite{Feynman1955}, the injection of a heat current, acting on the normal component of helium II but not the superfluid fraction and thus generating
a velocity difference between the two, leads to turbulence without a classical analogue. In 1957, it was observed that above a critical, temperature-dependent
velocity superflow indeed became dissipative~\cite{Vinen1957a}. Vinen~\cite{Vinen1957b, Vinen1957c, Vinen1958} suggested this to be the result of interactions
between an isotropic turbulent vortex tangle and the viscous fluid component and derived a mutual friction force similar to the one previously postulated by
Gorter and Mellink~\cite{Gorter1949}, i.e.
\begin{equation}
	\mathbf{F}_{\rm mf} \propto \rho\SF \alpha_{\rm He}^3 \left( \Bv\SF - \Bv\NF \right)^2 \left( \Bv\SF - \Bv\NF \right).
\end{equation}
For a more detailed mathematical discussion we refer the reader to Sec.~\ref{subsec-Turbulence}. Subsequently, counterflow turbulence was also studied in
rotating superfluids. Motivated by experiments showing that a small axial counterflow changes the vortex array properties~\cite{Cheng1973}, Glaberson et
al.~\cite{Glaberson1974} suggested that dissipation could be related to a hydrodynamical instability of the vortex array, referred to as the \textit{Donnelly-Glaberson
instability}. It is triggered once the thermal counterflow along the vortex axes exceeds the critical value (see also Eqn.~\eqref{eqn-DGCounterflowCritical})
\begin{equation}
	w_{\text{\tiny{NS}},\rm c} = 2 \sqrt{ 2 \Omega \nu\SF},
		\label{eqn-DGCounterflowCriticalExperiment}
\end{equation}
at which the vortices become unstable to the excitation of kelvin waves ($\nu\SF$ is defined in Eqn.~\eqref{eqn-SuperfluidViscosity}). Experiments by Swanson
et al.~\cite{Swanson1983} have confirmed the existence of this critical velocity and further shown that rotation stabilises the superfluid since the onset of
turbulence in absence of rotation was governed by lower critical velocities than in the rotating case. For a numerical analysis of these vortex instability
features and the growth of superfluid turbulence see for example Tsubota et al.~\cite{Tsubota2004}.

In addition to this new type of turbulence, an analogue to classical turbulence has been observed~\cite{Vinen2002}. Helium II experiments in the late
1990s~\cite{Maurer1998} found no counterflow turbulence but local pressure fluctuations, following the statistical Kolmogorov law. In this case,
the energy spectrum function $E(k)$ depends on the wave number $k$ as\cite{Kolmogorov1941}
\begin{equation}
	E(k) = C \epsilon_k^{2/3} k^{-5/3},
		\label{eqn-KolmogorovLaw}
\end{equation}
where $C$ is a universal constant of order $1$ and $\epsilon_k$ the energy dissipation rate per unit mass. In this regime, an extensive range of coupling
strengths has been investigated by dragging a grid through the superfluid (see for example Stalp et al.~\cite{Stalp1999}), a method usually employed for
studying turbulence in classical fluids. These experiments showed that the two-component fluid behaves like a single fluid on macroscopic scales exhibiting
quasi-classical flow properties~\cite{Stalp2002}. It has been suggested that this is caused by vortices forming bundles mimicking the behaviour of classical
\textit{eddies}~\cite{Barenghi2014b}, which results in a homogeneous and isotropic turbulent state that exhibits classical Kolmogorov decay. However, as soon
as the characteristic lengthscale of these eddies decreases to distances comparable to the intervortex spacing, quantum effects become important again.
Analyses of hydrogen tracer particles have revealed that the velocity field on these microscopic scales shows a power-law behaviour, which differs from the
Gaussian velocity distribution of classical turbulence~\cite{Paoletti2008}. Kelvin waves excited by the reconnections of vortices play a crucial role for this,
as they are expected to distribute the energy in a cascading manner to lengthscales smaller than the intervortex spacing~\cite{Walmsley2014}. These helical
displacements propagating along the vortex lines have only recently been observed on reconnecting helium II vortices~\cite{Fonda2014}.

Another type of instability that has been discovered in the context of helium II is the \textit{Rayleigh-Taylor instability}. This phenomenon, which generally
acts at an interface between two fluids of different densities and initiates the mixing of the two, is known to play a crucial role for many astrophysical
scenarios such as supernova explosions or accretion processes~\cite{Shapiro2004}. In laboratory helium-4 experiments, the instability has been observed in
the form of crystallisation waves at the superfluid-solid interface during the pressure-controlled growth of a helium crystal immersed in the superfluid
phase~\cite{Burmistrov2009}. The Rayleigh-Taylor instability could thus be of importance at the neutron stars' crust-core interface, where crustal lattice
nuclei are in contact with the neutron superfluids.

Besides the standard turbulence experiments studying helium-4, non-linear dynamics have also been investigated in the $B$-phase of helium-3~\cite{Eltsov2007,
Bradley2008, Eltsov2009}. Note that for the anisotropic $A$-phase the dissipation is so large that one would not expect superfluid turbulence to play any role
in the temperature ranges currently accessible~\cite{Finne2003}. Most noticeably, superfluid helium-3 contained in a cylindrical container exhibits two regimes
of vortex behaviour. While helium II had only been observed to display turbulent characteristics below the Lambda point, the $B$-phase showed laminar spin-down
behaviour above a temperature of about $0.6 \, T_{\rm c}$ and turbulent behaviour below~\cite{Finne2004}. Three main properties~\cite{Finne2003} help stabilise
the dynamics of the fermionic superfluid down to low temperatures: the viscosity of its normal fluid exceeds that of helium II by about four orders of magnitude;
due to the large vortex core size pinning is negligible and its mutual friction coupling is stronger than in helium II. The last difference is of particular
importance as it has been proven experimentally and theoretically (see for example Finne et al.~\cite{Finne2003} and Eltsov et al.~\cite{Eltsov2010}) that the
spin-down behaviour is governed by a single dimensionless parameter, $\text{Re}$, which only depends on the modified mutual friction coefficients,
\begin{equation}
	\text{Re} \equiv \frac{1-\alpha_{\rm He}'}{\alpha_{\rm He}}.
	\label{eqn-SuperfluidReynoldsNo}
\end{equation}
The two regimes are separated by $\text{Re}_{\rm crit} \sim 1$. For $\text{Re} \gg 1$ turbulence dominates as the inertial terms drive the dynamics, whereas
mutual friction stabilises the superfluid for $\text{Re} \lesssim 1$ and laminar behaviour is observed. In analogy with classical fluid dynamics, the parameter
$\text{Re}$ is sometimes referred to as the \textit{superfluid Reynolds number}. For the helium $B$-phase, $\text{Re}_{\rm crit}$ appears right in the experimental
temperature regime (corresponding to $T \sim 0.6 \, T_{\rm c}$), while for helium II the transition lies very close to the Lambda point making the laminar regime
almost inaccessible. This highlights one of the main advantages of helium-3 over helium-4 for the studies of non-linear fluid dynamics, as it allows one to
perform detailed studies of the onset of turbulence and vortex instabilities~\cite{Finne2006b}.

One instability that has been analysed in great detail is the \textit{Kelvin-Helmholtz instability}. Using a magnetically stabilised two-phase sample of helium-3
similar to the set-up discussed for the study of interfaces, Blaauwgeers et al.~\cite{Blaauwgeers2002} examined the shear flow between two superfluids. Spinning
up the sample from rest, vortices are first formed in the anisotropic $A$-phase as a result of the stronger mutual friction and organised as a vortex sheet at
the interface, while the $B$-phase remains vortex-free. The average circulation in both layers is different, creating a discontinuity in the tangential superfluid
velocities. This state of two superfluids moving relative to each other is stable and non-dissipative up to high relative velocities and, thus, provides the
perfect environment for investigating the Kelvin-Helmholtz instability~\cite{Volovik2002}. In the classical case, an instability between two immiscible,
inviscid fluid layers of density $\rho_1$ and $\rho_2$ occurs when the difference between the velocity components parallel to the interface, i.e. $|v_1 - v_2|$,
satisfies the condition~\cite{Kelvin1910}
\begin{equation}
	\frac{\rho_1 \rho_2}{\rho_1 + \rho_2} \left(v_1 - v_2 \right)^2 = 2 \sqrt{ \sigma_{\rm s} F}.
		\label{eqn-KelvinHelmholtzclassical}
\end{equation}
The interface becomes unstable once the inertial effects can no longer be balanced by the interface's surface tension $\sigma_{\rm s}$ and an external field $F$, which is
generally taken to be the gravitational force $F = g | \rho_1 - \rho_2|$ ($g$ is the gravitational acceleration). Waves of wave vector $k= \sqrt{F / \sigma_{\rm s}}$ are
then excited on the interface. Compared to studies using classical fluids, the superfluid set-up has the advantage that viscosity does not obscure the instability.
However, Eqn.~\eqref{eqn-KelvinHelmholtzclassical} no longer applies in the superfluid case, but has to be modified due to the two-fluid nature of the quantum
condensates. The corresponding instability criterion reads~\cite{Volovik2002}
\begin{equation}
		\rho_{\text{\tiny{S}} 1} \left(v_{\text{\tiny{S}} 1} - v\NF \right)^2
			+ \rho_{\text{\tiny{S}} 2} \left(v_{\text{\tiny{S}} 2} - v\NF \right)^2= 2 \sqrt{ \sigma_{\rm s} F},
		\label{eqn-KelvinHelmholtzSuperfluid}
\end{equation}
where $\rho_{\text{\tiny{S}} 1}$, $\rho_{\text{\tiny{S}} 2}$, $v_{\text{\tiny{S}} 1}$ and $v_{\text{\tiny{S}} 2}$ would correspond to the superfluid mass densities
and the superfluid velocities of the two helium-$3$ phases, respectively, while $v\NF$ represents the normal fluids', i.e.\ the container's, velocity.
The instability threshold is thus not related to the relative velocity of the two superfluid components but instead depends on the velocity difference of
the normal and inviscid constituents on both sides of the interface~\cite{Abanin2003}. In the case of neutron star cores, where multiple quantum states are present,
a combination of the different superfluid and superconducting velocities should enter the instability criterion~\cite{Mastrano2005}. Every time the threshold is
reached, the Kelvin-Helmholtz instability results in a wave-like distortion of the $AB$-interface, suggested to cause the \textit{injection} of vortex tangles
into the $B$-phase (see for example Finne et al.~\cite{Finne2006}). For temperatures above $\sim 0.6 \,T_{\rm c}$, each vortex loop quickly turns into a straight
line connecting across the interface to the $A$-phase defects, with a similar number of vortices being created each time the instability is triggered~\cite{Blaauwgeers2002}.
Below $\sim 0.6 T_{\rm c}$ however, the instability acts in a very non-linear way, explosively injecting a large number of vortices into the isotropic phase. A
detailed, temperature-dependent analysis of the Kelvin-Helmholtz instability thus confirms the laminar and turbulent spin-down regimes separated by the different
strengths of mutual friction.

Insight from helium experiments could also provide important information for neutron stars, where vortex dynamics are generally assumed to be laminar and little is known
about how turbulence and superfluid instabilities manifest themselves. As a first step, one could try to classify the neutron star spin-down behaviour by simply
calculating the superfluid Reynolds number associated with the mutual friction mechanisms. Since the coupling is expected to be rather weak with $\CB' \approx B^2
\ll 1$, Eqn.~\eqref{eqn-SuperfluidReynoldsNo} would result in $\text{Re} \gg 1$, which implies that the neutron star interior should be strongly influenced by turbulence.
Despite the fact that this criterion neglects effects such as the stars' rapid rotation that could suppress the development of non-linear dynamics~\cite{Ruderman1974},
turbulence could significantly alter the vortex motion. Peralta et al.~\cite{Peralta2005,Peralta2006} for example study the onset of the Donnelly-Glaberson
instability in the neutron star core, which excites unstable Kelvin waves that result in a distortion of the initially straight neutron vortices. As briefly mentioned
in Sec.~\ref{subsec-Turbulence}, the frictional coupling in a vortex tangle is very different from the standard force considered above in
Eqn.~\eqref{eqn-MutualFrictionNSExperiment}. Instead of being proportional to the velocities, the mutual friction force depends on the cube of the relative
velocities~\cite{Gorter1949, Vinen1957b, Vinen1957c, Vinen1958}. Andersson et al.~\cite{Andersson2007a} have however recently argued that such a turbulent state might only
exist locally and not globally, because a fully developed vortex tangle is isotropic and the averaged vorticity of a macroscopic fluid element would vanish. Hence, for
neutron stars, where the superfluid has to form vortices in order to support the observed bulk rotation, a disordered vortex tangle would have to retain some rigid-body
characteristics to some extent. This could for example be achieved in form of a \textit{polarised turbulent state}~\cite{Andersson2007a}, where each fluid element
contains tangled and straight vortices; the latter being responsible for the macroscopic rotation. The corresponding mutual friction would then be a superposition
of both structures, which is in agreement with observations made in counterflow studies of rotating helium-II~\cite{Swanson1983, Tsubota2004}.

Despite the fact that detailed knowledge of the turbulent state in neutron star interiors is not available, implications for macroscopic observables could be
significant. A modified frictional coupling due to the presence of turbulent vortex tangles would generally lead to dissipation timescales that differ from the ones
usually considered. This would affect various hydrodynamical phenomena such as the post-glitch relaxation or the damping of neutron star
free precession~\cite{Jones2001} and oscillations modes~\cite{Lindblom1995, Andersson2003a}. Moreover, it has been suggested that timing noise could result from an
instability of the vortex array being imperfectly pinned to the proton fluxtubes due to thermal activation~\cite{Link2012a} or a variation of the crustal rotation phase
caused by the turbulent core superfluid exerting a fluctuating torque on the crust~\cite{Melatos2013}. Additionally, the presence of superfluid instabilities are relevant,
since they could explain the origin of pulsar glitches. Several mechanisms have been studied in the literature. As discussed by Mastrano and Melatos~\cite{Mastrano2005},
the Kelvin-Helmholtz instability could act at the neutron star crust-core interface and rapidly transfer circulation between the $^1 S_0$ and $^3 P_2$ superfluids,
similar to what is observed in helium-3 experiments~\cite{Blaauwgeers2002}. This would result in the spin-up of the rigidly rotating neutron star component, which would
manifest itself as a discrete jump in the angular rotation frequency of the crust, observed as the glitch by a distant observer.

Similar to the Kelvin-Helmholtz instability, \textit{two-stream instabilities} could further serve as a possible trigger mechanism in differentially rotating neutron
stars~\cite{Andersson2003}. The main difference for this type is that the interacting fluids are interpenetrating and not separated by an interface. As before however, the
non-linear dynamics set in as soon as a critical velocity lag is reached. The instability mechanism itself may be mediated through various coupling mechanisms depending
on the level of complexity included. Andersson et al.~\cite{Andersson2004} discuss the simplified case of a neutron and proton fluid rotating at different angular velocities
(as present in the neutron star core) enclosed by an infinitesimally thin spherical shell. Here, entrainment is responsible for the coupling of the two components and a
perturbative analysis of the hydrodynamical equations shows that the instability is mediated through inertial r-modes~\cite{Andersson2000, Andersson2001}. This process
is particularly interesting because it has been shown that these modes are not only dynamically unstable on small scales but also suffer a global instability, which might
not be completely damped by shear viscosity~\cite{Glampedakis2009, Andersson2013}. Thus, the r-mode instability could trigger the unpinning of a large number of vortices
and lead to observable glitches. Finally note that this instability has not yet been observed in helium experiments and it is unclear whether one would expect to find
unstable r-modes at all, as the coupling mechanisms in laboratory systems differ significantly from those in neutron stars~\cite{Hogg2014}.

To conclude our discussion of terrestrial neutron star models using helium, we highlight another exciting way to exploit the analogy between both systems. By
combining helium-3 with \textit{aerogel} (a mixture which has also attracted a lot of attention in the low-temperature physics community~\cite{Chan1996, Halperin2004}),
one can study the influence of disorder on a three-dimensional quantum liquid. These aerogels are very porous media formed of strand-like structures that are generated
from silica clusters in a gelation process. The strands' diameters are typically of the order of a few nanometres, smaller than the coherence length of pure helium. The
advantage of this system is that (due to the different lengthscales) helium-3 superfluidity can be controlled through the aerogel's porosity. It was for example shown
that for a $98$\% porous solid, helium-3 exhibits the typical characteristics of a superfluid phase transition~\cite{Porto1995, Sprague1995}. While the $B$-phase is
suppressed and an $A$-phase like state seems to dominate, the exact nature of the various superfluid phases in aerogel is not known and recent analyses~\cite{Bennett2011,
Pollanen2012} revealed a rather complex phase diagram. However, these experiments present the unique possibility to investigate the normal-superfluid interface as
it should be possible to grow aerogel with a continuously changing porosity. As this would provide information about the pairing behaviour of quantum systems undergoing
such transitions, it could allow insight into the properties of the neutron star protons turning superconducting at the crust-core boundary. It should be further
possible to study the behaviour of superfluid vortices in disordered aerogel. This has obvious analogies to the inner neutron star crust, where a superfluid neutron
liquid is thought to coexists with the nuclear pasta, and experiments could thus give information about the interactions between the two components.


\section{Ultra-cold Gases}
\label{sec-ColdGases}

It was first suggested by Fritz London~\cite{London1938} that the transition of helium-4 atoms into a superfluid state could be the result of
condensation into the quantum mechanical ground state. Since bosons obey Bose-Einstein statistics, when confined in an external potential, they enter
the lowest energy state available. At low temperatures, this results in the formation of a new macroscopic quantum phase referred to as a Bose-Einstein
condensate (BEC). For a general introduction see Chevy and Dalibard~\cite{Chevy2013}. The state of weakly-interacting bosons was first observed in 1995 by
Wieman and Cornell~\cite{Wieman1995} and Ketterle~\cite{Ketterle1995}, who created BECs by cooling dilute gases of Rubidium atoms to nanokelvin temperatures.
Despite the fact that this branch of low temperature physics is relatively new, it became evident soon after the initial detection that these macroscopic
quantum condensates could serve as a perfect testing tool for various areas of physics, ranging from solid state physics to many-body physics all the
way to astrophysics~\cite{Pethick2008}.

The main purpose of this section is to examine if laboratory BECs could be used to probe neutron star dynamics. The following discussion will hence
focus on characteristics that are expected to play a crucial role in compact stars, such as the multi-fluid nature, vortex motion, superfluid turbulence
and interfaces. We conclude with a brief analysis of ultra-cold Fermi gases.

\subsection{General properties of BECs}
\label{subsec-GeneralProperties}

The most important feature when designing laboratory neutron stars is the presence of distinct components. Whereas the condensate itself occupies the ground
state of the external potential, it spatially coexists with a second component, formed by the BEC's collective excitations~\cite{Pethick2008}. This is
equivalent to the two-fluid model of helium, where two interpenetrating fluids are invoked to explain superfluid behaviour, and hence the characteristics of
superfluidity are also expected to exist in BECs. The corresponding transition in a bosonic gas was first observed in 1999~\cite{Matthews1999, Madison2000}
and associated hydrodynamical features such as macroscopic superfluid flow~\cite{Raman1999} or the propagation of first and second sound have also been
experimentally confirmed~\cite{Jin1996, Andrews1997, Shenoy1997}. Additionally, ultra-cold gases offer another possibility to mimic the neutron stars'
multi-component nature, as multiple condensates can be positioned on top of each other. This was realised shortly after the first detection of BECs. In
1997, Myatt et al.~\cite{Myatt1997} generated two overlapping condensates using two Rubidium-87 BECs in different spin states, which could be analysed using
absorption imaging. Examination of the interaction properties revealed that the two clouds exhibited mutual repulsion. Following this initial realisation
of a binary BEC, which further serves as an analogue of interpenetrating helium-3 and helium-4 superfluids~\cite{Ho1996}, a lot of effort has been put into
theoretically and experimentally investigating multi-component BECs formed of different Alkali atoms. It was found that these systems display a rich variety
of phenomena~\cite{Kasamatsu2005, Woo2007, Navarro2009}. For instance, it has been demonstrated that depending on the atomic interaction, some binary mixtures
are miscible~\cite{Papp2008}, whereas others show immiscible behaviour~\cite{Hall1998, Ao1998}. In order to model neutron stars (where the matter is
homogeneously distributed and coupling between different components is expected to take place), BECs with strong mutual repulsion are less relevant since
they display an inhomogeneous matter distribution in their ground state. However, the main advantage of these systems is the great amount of control one has
over the experiment, as binary condensates can be prepared in almost every desired density profile by varying the mixing of the states.

Furthermore, observational features become very diverse when the condensates are no longer stationary but dynamical. One example is the study of shock wave
propagation in BECs, where pulsed and tightly focused laser beams are used to generate blast waves. Compared to classical, dissipative shock waves, the equivalent
in BECs is highly non-linear and dispersive causing very different wave structures and shock speeds~\cite{Hoefer2006}. The dynamics further change significantly
when rotation is added to the picture. This becomes especially important if the condensate is in a superfluid state, as vortex formation takes place (see
Sec.~\ref{subsec-VortexDynamicsBEC}).

Another tool for studying the divers properties of BECs is a close examination of their excitation modes, which can be understood as coherent fluctuations in
the condensate's density. These excitations were observed shortly after the generation of the first BECs and exhibit several parallels with helium II
phonons~\cite{Jin1996, Mewes1996}. Hence, understanding the frequencies of excitations and their evolution~\cite{Fedichev1997, Liu1997, Cornell2002} would give insight
into the interactions between individual atoms. It is also possible to directly image the matter distribution and thus study the excitation modes experimentally.
Therefore, BECs are important testing systems, because they could provide clues on how asteroseismology (the analysis of neutron star oscillations) could be used
to obtain information about compact objects. This would be particularly valuable when analysing how continuous gravitational waves, generated by the fluid modes
of rotating neutron stars, could help to constrain the interior physics~\cite{Andersson2003a, Gaertig2011, Lasky2015b}.

At this point, it seems inevitable to return to the problem of entrainment (the non-dissipative coupling of neutrons and protons), which greatly influences
the neutron stars' dynamics. While it is certainly possible to derive a mathematical formalism for a binary BEC with entrainment and determine how observables
(such as the excitation modes) would be modified~\cite{Hogg2014}, it is not obvious if such a strong, non-dissipative coupling could be present in real systems.
As the particle densities in laboratory BECs typically reach $10^{12}$ to $10^{15} \, \cm^{-3}$~\cite{Pethick2008}, which corresponds to an interparticle
distance of $10^{-4}$ to $10^{-5} \, \cm$, the prospect of recreating entrainment in a weakly-coupled BEC is rather poor. However, recent theoretical studies
have examined the superfluid drag behaviour between two BECs confined to optical lattices~\cite{Kaurov2005, Hofer2012}. An optical lattice consists of a
spatially periodic potential created by the interference of laser beams that is superimposed on the condensates to trap individual atoms~\cite{Gemelke2014}.
These lattices are easy to tune and provide the unique possibility to study transport properties of binary condensates, such as the non-dissipative entrainment
coupling between two BECs, which originates from the interspecies interaction on short ranges. Regardless of these developments, studies of entrainment in
BECs are still in their infancy and actual experiments are needed before any parallels with neutron star physics can be drawn.

Despite the problem of probing entrainment in bosonic condensates, BECs are brilliant laboratory systems allowing a lot of flexibility, which is crucial when
designing neutron star experiments. The main reason fine tuning in BECs is very straightforward is due to so-called \textit{Feshbach resonances}. For a recent
review see Chin et al.~\cite{Chin2010}. These resonances (effectively generating bound states between atoms in the condensate) are named after Herman Feshbach,
who studied similar many-body resonances in nuclear physics collisions~\cite{Feshbach1958}. The existence of these resonances was theoretically predicted in
1993~\cite{Tiesinga1993} and experimentally confirmed by various groups in 1998~\cite{Inouye1998, Courteille1998, Roberts1998}. In the case of atomic gases,
the resonances between the particles allow one to change the scattering length, i.e.\ the interaction strength of the condensate, by simply changing the external
magnetic fields. This creates an extraordinary degree of control, which could be very valuable when analogies between neutron stars and BECs are exploited.

Before moving on to a more detailed discussion of vortex dynamics, we note that a series of experiments has unveiled another phenomenon, which can be related
to neutron stars~\cite{Roberts2001, Donley2001}. While it does not concern the properties of an old, equilibrium star, it is somewhat similar to the formation
of the star itself and, thus, illustrates the close analogy between BECs and compact stars. In these experiments, the external magnetic field was tuned to obtain
negative scattering lengths (corresponding to an attractive self-interaction), which caused the condensate to become unstable. This was most remarkable as the
BEC showed characteristics of a collapse: after shrinking slightly, the condensate underwent an \textit{explosion}, expelling a large number of particles and
leaving a small, cold and stable remnant behind. Although the energy scales for this process are obviously smaller than the ones associated with neutron star
formation, the collapse of a BEC is called a \textit{bosenova}~\cite{Cornell2002} due to the apparent similarities with a core-collapse supernova.

\subsection{Vortex dynamics}
\label{subsec-VortexDynamicsBEC}

If bosonic condensates are to serve as laboratory neutron star analogues, the presence of quantised vortices is crucial. In the case of helium experiments,
the fluid is confined to a container and rotation of the two components is obtained by setting the vessel into motion. However, superfluid BECs are trapped
in an external potential and not enclosed in a container and can thus no longer simply be accelerated. Instead, different methods have been employed in order
to set condensed clouds into motion. Two approaches are in operation for the generation of rotating BECs. The first one uses optical beams with a suitable
inhomogeneous topology to \textit{imprint} a phase onto an existing condensate. At points where the local density vanishes, the condensate is then forced to
forms vortices. Alternatively, an optical stirring beam or a distortion in the magnetic field can be applied to rotate the cloud, which subsequently
leads to the generation of quantised vortices. The imprinting method was adopted by Matthews et al.~\cite{Matthews1999}  in 1999 to create vortices in one of
the constituents of a two-component BEC. Using an interference technique~\cite{Hall1998}, it was shown that the phase of a single vortex changed by a factor
of $2 \pi$ as given by the standard quantisation condition. These experiments also allowed an analysis of the vortex stability and the decay behaviour. Shortly
after the successful phase-imprinting experiments, Madison et al.~\cite{Madison2000} observed quantised rotation in a single-component Rubidium-87 BEC after
stirring the condensate with a focused laser beam. Similar to rotating helium II experiments, it was demonstrated that vortices are formed as soon as the
stirring frequency exceeds a critical value. However, the advantage of BECs is that vortices can be easily visualised using absorption imaging, because they
correspond to holes in the resulting density distribution.

Further studies of rapidly rotating, ultra-cold BECs have revealed many interesting vortex features (for recent summaries see Cooper~\cite{Cooper2008} and
Tsubota et al.~\cite{Tsubota2013}). Two general aspects observed are that the process of vortex formation is highly non-linear and the regular vortex lattice
only becomes apparent once the equilibrium steady state has been reached. The resulting defects are also not necessarily singly quantised but other structures
such as vortex sheets~\cite{Woo2007} or alternative, highly distorted but stable patterns~\cite{Butts1999} have been predicted. In this respect, BECs show
several similarities with the anisotropic $A$-phase in helium-3. In order to draw information about the dynamics of neutron star vortices from BEC experiments,
one would have to ensure that vortices are point-like defects, regularly distributed in a hexagonal array as generally invoked for the stars' interior.

In order to assess whether attributes of BEC vortices could be relevant for neutron stars, the first step would be to compare the size of a BEC and the
dimensions of its vortices. Typical BEC clouds extend from $10$ to $100 \, \mu \m$~\cite{Barenghi2014}, whereas the vortex core size (determined by the
healing length similar to the anisotropic helium-3 phase) is given by $\sim 0.5 \, \mu \m$~\cite{Cooper2008} and the number of vortices in bosonic
condensates reaches up to several hundred. This is very different to the case of neutron stars, where significantly more vortices and fluxtubes are
expected to be present. Moreover, the dimension of a BEC vortex is of similar order as the intervortex spacing, implying that ultra-cold atoms probe a
regime that is in great contrast to the conditions of the neutron star interior. In the latter system, it is usually assumed that individual vortices and
fluxtubes are distant enough to not influence each other. Hence, deducing information about the macroscopic dynamics of non-interacting vortices might be
difficult with BECs. However, these laboratory condensates are excellent testing grounds for probing the mesoscopic dynamics of vortices and how they interact
with each other and possible pinning sites. The latter is particularly important for neutron stars, because vortex-fluxtube pinning is thought to play an
important role for their rotational and magnetic evolution~\cite{Epstein1988}. Regardless of this fact, a detailed mesoscopic theory of pinning is not
available yet and exploring this phenomenon in more detail would be beneficial. While BECs are not in contact with a surface, which the vortices could
pin to, one can take again advantage of optical lattices~\cite{Reijnders2004}. Superimposing regular energy barriers could mimic the pinning potential present
in a star's crust or core and would allow one to study the interaction of vortices with these potentials. This type of pinning has been recently
observed for a single-component condensate~\cite{Tung2006} and theoretically extended to a binary BEC~\cite{Mink2009}.

\begin{figure}[t]
\begin{center}
\includegraphics[width=5in]{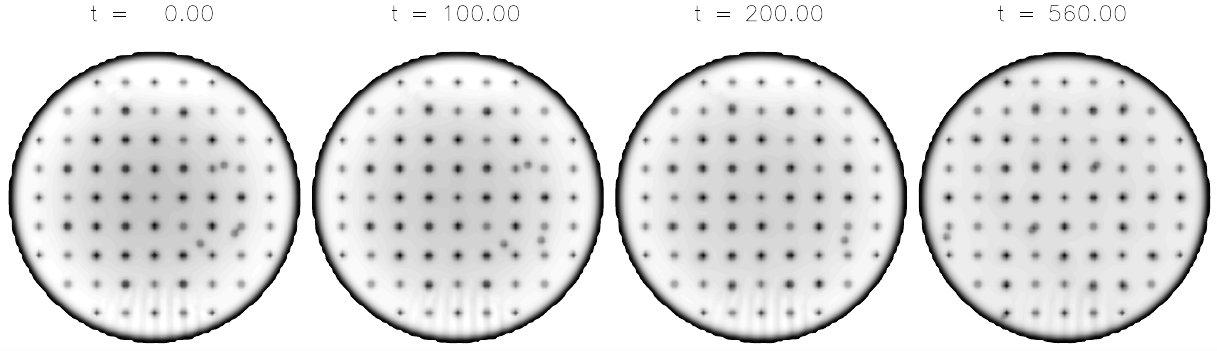}
\caption{Snapshots of the superfluid density during the modelled spin-down of a BEC at different times, $t=0, 100, 200, 560$ and $810$ in arbitrary
units. A light grey colour corresponds to a low and dark grey to a high density. The rectangular structures indicate the presence of pinning centres,
where dark points mark occupied and light points mark unoccupied sites. Dots that are not part of the array are moving vortices. Note that the
vortices initially populate the centre of the cylinder and move outwards as the container is spinning down. The figure is reproduced with permission
from Warszawski and Melatos~\cite{Warszawski2012}.}
\label{fig-BECSimulation}
\end{center}
\end{figure}

One example, where the analogy between vortex pinning in neutron stars and BECs has already been exploited, is the modelling of pulsar glitches. As explained
in Sec.~\ref{subsec-ColdAtoms}, in the low temperature limit, a weakly-interacting BEC is well described by the time-dependent GP equation (see also
Eqn.~\eqref{eqn-GPE}),
\begin{equation}
 	i \hbar \, \frac{\partial \Psi}{\partial t} + \frac{\hbar^2}{2 m\cl} \nabla^2 \Psi 	- V \Psi - \frac{4\pi \hbar^2 a}{m\cl} |\Psi |^2 \Psi =0,
		\label{eqn-GPE_Experiment}
\end{equation}
where $\Psi$ denotes the wave function, $m\cl$ the boson mass, $V$ the external potential and the interaction parameter depends on the scattering length $a$.
The time-evolution of Eqn.~\eqref{eqn-GPE_Experiment} provides information about the motion of vortices. Despite the fact that many-body forces in neutron
stars crusts are not weak as in a bosonic condensate, Warszawski and Melatos~\cite{Warszawski2011, Warszawski2013} suggest that the GP equation could also
be used to model the pinned, decelerating crustal superfluid. The authors have shown that collective motion of BEC vortices in the presence of a regular
pinning potential can trigger glitch-like events, which have exponentially distributed waiting times and sizes that follow a power-law distribution. These
characteristics have also been seen in pulsar glitches and indicate the presence of self-organised critical processes such as observed in earthquakes or
BEC vortex avalanches~\cite{Bak1988}. Typical simulation snapshots of the superfluid density during the condensate's spin-down are shown in
Fig.~\ref{fig-BECSimulation}. However, BEC simulations only deal with up to several thousand vortices and not the $\sim 10^{16}$ vortices expected in neutron
star crusts. So one has to be careful when generalising the dynamics of small BECs to the much larger system.

\subsubsection{Turbulence and instabilities}
\label{subsubsec-InstabilitiesBE}

Being less than a decade old, the analysis of instabilities and superfluid turbulence in BECs is a young field. Accordingly, experimental results
are not as extensive as for helium. However, several observations have been made, which show relevance for laboratory neutron stars and will be discussed
below. One benefit of studying non-linear dynamics with BECs is that the turbulent behaviour can be imaged. Because instabilities cause a dephasing
of the condensate's wave function, time of flight experiments can be used to directly probe the velocity distribution of the condensate and, thus, provide
information about the structure of superfluid turbulence~\cite{Dunningham2005}.

In order to model non-linear evolution (like the neutron star two-stream instability, expected to arise from the differential rotation between the neutrons
and the charged conglomerate) with ultra-cold condensates, a similar instability mechanism is needed. The presence of any superfluid hydrodynamical
instability would first of all require the existence of a macroscopic, inviscid flow. This was detected in 1999~\cite{Raman1999, Onofrio2000}, when
experiments showed that a laser beam could be moved through a BEC without generating dissipation. Heat production and the subsequent breakdown of the superfluid
state was only observed once a critical stirring velocity was exceeded. This is equivalent to the frictionless flow observed in early helium II counterflow
experiments. The advantage of BECs is that ultra-cold atomic clouds are less complex (as surface effects and strong coupling are absent) and, therefore, easier
to analyse. The initial observation of BEC superflow stimulated further experiments studying the macroscopic drag and onset of dissipation by creating
persistent flow patterns~\cite{Ryu2007} and directly probing the flow fields around potential barriers immersed into the fluid~\cite{Engels2007}. In the latter
case, it was shown that the condensate becomes unstable for intermediate flow speeds, developing instabilities in the form of solitons~\cite{Hakim1997}.

\begin{figure}[t]
\begin{center}
\includegraphics[width=4.8in]{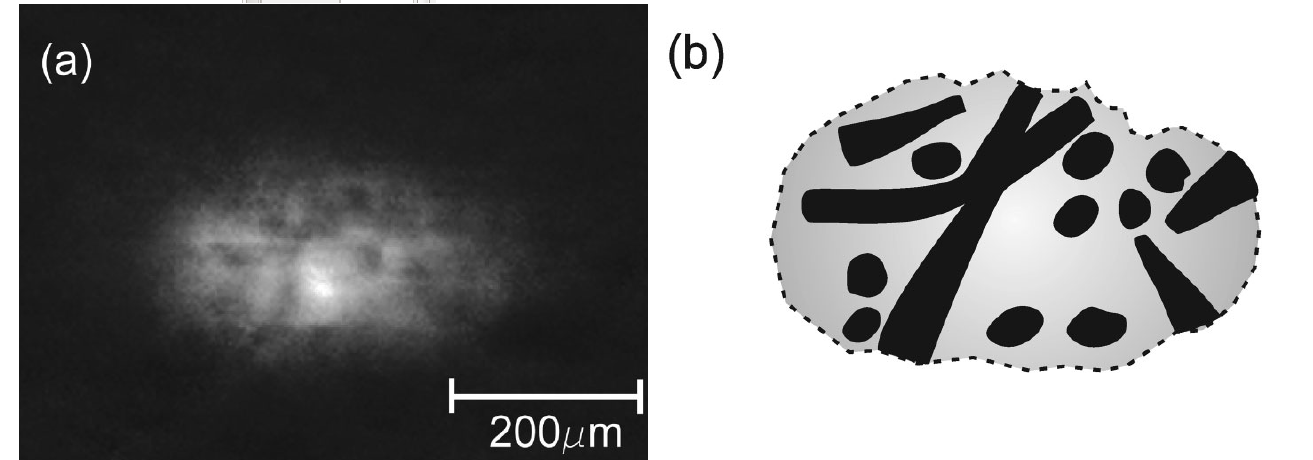}
\caption{(a) Snapshot of the atomic density in a BEC after a $15 \, \ms$ phase of expansion, showing an unordered distribution of vortex structures.
This illustrates the first observation of a turbulent tangle. (b) Schematic diagram of the vortex tangle as inferred from the snapshot in (a). The
figure is reproduced with permission from Henn et al.~\cite{Henn2009}. Copyright $\copyright$ 2009, American Physical Society.}
\label{fig-VortexTangleBEC}
\end{center}
\end{figure}

Fully developed turbulence has recently been observed in a single-component BEC by Henn et al.~\cite{Henn2009}, who reported the direct observation of an
entangled vortex state (see Fig.~\ref{fig-VortexTangleBEC}), which had previously been identified in numerical simulations~\cite{Kobayashi2007}, and the
corresponding change in the hydrodynamic properties of the condensate. In the experiment, the turbulent vortex state was created by applying an oscillatory
perturbation to the magnetic trapping potential, generating excitations which in turn are thought to lead to vortex formation. Although it is not understood
how this proceeds in detail, Henn et al.\ suggest that the Kelvin-Helmholtz instability could be responsible, as the excited BEC is surrounded by a thermal
cloud of atoms~\cite{Parker2005}, creating an interface. Coupling of the superfluid BEC to this thermal cloud is also predicted to result in the decay of
the turbulent state\cite{Marago2000, Parker2005}, which follows the quasi-classical Kolmogorov statistics on large scales as defined in
Eqn.~\eqref{eqn-KolmogorovLaw}. While this power-law behaviour has also emerged in numerical BEC simulations~\cite{Kobayashi2007, Zamora-Zamora2015}, it has
not yet been confirmed experimentally. Another feature of superfluid turbulence that has been recreated numerically is the non-classical distribution of the
small-scale velocity field, which has already been detected in helium experiments~\cite{Paoletti2008}. White et al.~\cite{White2010} study the decay of a
vortex tangle for realistic experimental parameters by evolving the GP equation and find that the velocity distribution of the BEC follows a power-law and
does not exhibit the classical Gaussian statistics.

Returning to the question if two-stream instabilities could be triggered in laboratory BECs, one again has to consider two interacting condensates represented
as $i,j \in \{ 1,2 \}$. By studying the corresponding GP equations coupled via an additional interaction term, i.e.
\begin{equation}
 	i \hbar \, \frac{\partial \Psi_i }{\partial t} + \frac{\hbar^2}{2 m_{\text{c} i}} \nabla^2 \Psi_i - V_i \Psi_i
		- \left( \frac{4\pi \hbar^2 a_i}{m_{\text{c}i}} |\Psi_i |^2 + \frac{2\pi \hbar^2 a_{ij}} {m_{\text{c}ij}} |\Psi_j |^2 \right) \Psi_i =0,
		\label{eqn-GPE_ExperimentBinary}
\end{equation}
where $a_{ij}$ represents the interspecies scattering length and $m_{\text{c}ij}^{-1} \equiv m_{\text{c}i}^{-1} + m_{\text{c}j}^{-1}$ the reduced mass, it has been
shown theoretically that for the simple case of linear relative flow between two condensates, the binary mixture becomes dynamically unstable once a critical
velocity is reached~\cite{Law2001, Abad2015}. This is exactly what one would expect for linear relative motion between the neutrons and protons in the neutron star
core~\cite{Andersson2004}. As mentioned in Sec.~\ref{subsec-ColdAtoms}, this similarity has its foundation in the hydrodynamical description of both systems.
Instabilities between two counter-moving BECs have also been modelled numerically (see for example Takeuchi et al.~\cite{Takeuchi2010}) and observed in experiments,
where the superfluids are confined inside a narrow channel~\cite{Hamner2011, Hoefer2011}. In order to develop an idea whether an analogue of the superfluid r-mode
instability could be observable in BECs, the concept of relative flow has to be extended to rotating BECs. In this case, the dynamics get much more complicated and
no experimental data is available. However, numerical investigations have predicted the presence of different types of instabilities~\cite{Corro2009}. Some of these
could be relevant for neutron stars as they hint at turbulence and vortex nucleation in the form of ripples or catastrophic events, which also show some resemblance
to the instability phenomena previously discussed for helium. For example, it has been determined that similar to vortex formation in a single-component BEC, the
Kelvin-Helmholtz and Rayleigh-Taylor instabilities could generate vortices in immiscible binary BECs~\cite{Kobyakov2014}. It has further been suggested that binary
BECs consisting of Rubidium-85 and Rubidium-87 could be used to study the onset and the dynamics of these interface instabilities by tuning the interaction strength
of the two condensates through Feshbach resonances~\cite{Gautam2010}. This would allow more flexibility than currently available in helium experiments. Overall, it
seems promising that future studies of non-linear evolution in bosonic condensates will provide more insight into instability formation and how this affects superfluid
turbulence, which could contribute to the development of more accurate neutron star models.

\subsubsection{Interface behaviour}
\label{subsubsec-InterfacesBEC}

One advantage of using helium for neutron star modelling is the presence of stable, phase-coherent interfaces. By creating two-phase samples of helium-3,
it is possible to study vortex behaviour across such interfaces. If BECs are to be used as laboratory neutron star analogues, it would be beneficial to realise
similar conditions in ultra-cold condensates. The interfaces between two BECs mentioned above are generally phase-separated and not suitable for this purpose,
since vortices are simply terminated and not connected across the interfaces. However, it may be possible to mimic phase-coherent neutron star interfaces by employing
topological defects of spinor BECs. Here, different to standard bosonic condensates, the spin is a manipulable degree of freedom and not fixed by the external
magnetic field. This can be achieved by confining the atoms in optical instead of magnetic traps. Spinor BECs, first observed in 1998~\cite{Stenger1998}, exhibit
the standard turbulent characteristics of superfluids~\cite{Fujimoto2012, Villasenor2014} but can additionally be deformed to more complex structures. Particularly
promising seems the behaviour of vortices at the interface of two spin-1 condensates distinguished by their magnetic phases. It has been suggested that coherent
interfaces could be constructed by phase-imprinting vortices on each side that would cross in a continuous manner~\cite{Borgh2012, Borgh2013}.

\begin{figure}[t]
\begin{center}
\includegraphics[width=4in]{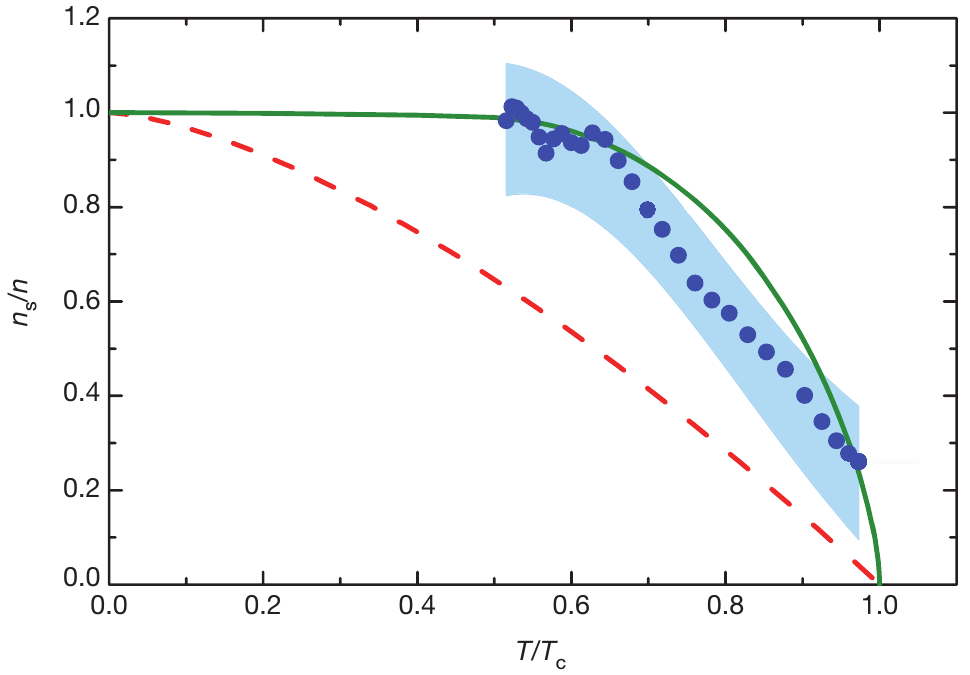}
\caption{Superfluid density fractions for various macroscopic quantum systems. The blue data points and the shaded uncertainty region represent the superfluid
fraction of a uniform, resonantly interacting Fermi gas as a function of the reduced temperature, $T/T_{\rm c}$. For comparison the superfluid fraction of
helium II (green solid line) and the theoretical expression, $1-(T/T_{\rm c})^{3/2}$, for the condensed fraction of an ideal Bose gas (dashed red line) are
given. The figure is reproduced with permission from Sidorenkov et al.~\cite{Sidorenkov2013}. Copyright $\copyright$ 2013, Nature Publishing Group.}
\label{fig-SFDensityFractions}
\end{center}
\end{figure}

\subsection{Fermi gases}
\label{subsec-FermiGases}

Before concluding, we briefly mention that due to advances in experimental methods not only bosonic but also fermionic condensates are available to study the properties
of superfluids. An introduction to Fermi gases (first observed in 2003~\cite{Greiner2003, Jochim2003, Zwierlein2003}) is given by Giorgini et al.~\cite{Giorgini2008}.
Two years after the initial detection, the superfluid state was discovered by observing the presence of quantised vortices~\cite{Zwierlein2005}. The corresponding
superfluid transition at a critical temperature (accompanied by the characteristic Lambda-shaped change in the thermodynamical quantities such as the specific heat)
has only recently been recorded by Ku et al.~\cite{Ku2012}. Hence, fermionic condensates exhibit similar features to bosonic gases~\cite{Zwierlein2014} and the benefit
of studying these systems is analogous to the aspects discussed before. One similarity is the observations of first~\cite{Joseph2007} and second sound~\cite{Sidorenkov2013}
velocities. As for superfluid helium, the latter phenomenon characterises the wave-like transport of heat and is closely related to the two-fluid nature of the system.
Measurements of the second sound velocity thus allow the extraction of the superfluid density fraction in an ultra-cold Fermi gas. Results for the uniform case are
illustrated in Fig.~\ref{fig-SFDensityFractions}.

The main difference to bosonic gases is that fermions obey the Pauli exclusion principle and cannot condense into the minimum energy state. Instead, fermions have
to form Cooper pairs as governed by BCS theory. This causes the superfluid behaviour of Fermi gases to become visible at much lower temperatures than BEC superfluidity.
The same characteristic was encountered when discussing the formation of the anisotropic helium-3 phase. Due to the presence of pairing, interactions play an important
role in superfluid Fermi gases. In contrast to bosonic condensates, fermionic ones are strongly coupled and, therefore, inherently closer to the nuclear matter present
in neutron stars. Hence, fermionic gases could be the perfect candidate for designing laboratory neutron stars. A recent comparison between cold Fermi atoms and low-density
neutron matter in the stars' crust~\cite{Gezerlis2008, Gezerlis2014} has pointed out that both systems have superfluid transition temperatures comparable to the Fermi
temperatures and should thus display similar physics. One instance clearly illustrating the benefit of this analogy is the numerical simulation of vortex pinning.
Solving the GP equation for a unitary Fermi gas, where the scattering length is much larger than the interatomic distance, provides a way to determine the detailed
mesoscopic vortex motion initiated by interactions with a pinning site (see for example Bulgac et al.~\cite{Bulgac2013}). This could help to considerably improve
the understanding of pinning in neutron stars, where detailed models are not available. Moreover, recent work has shown that the unitary Fermi gas also offers unique
possibilities to study quantum turbulence~\cite{Wlazowski2015}. In this system, microscopic vortex dynamics such as reconnections and crossings can not only be modelled
theoretically but new techniques could allow the direct imaging of these phenomena~\cite{Ku2014}.

Finally, we point out that mixtures of Bose and Fermi superfluids have recently been realised by cooling a bosonic and a fermionic lithium isotope below both transition
temperatures~\cite{Ferrier-Barbut2014}. It was possible to measure the energy exchange between the two fluids and determine their coupling strength, which was observed to
be rather weak. In essence similar to a mixture of helium-3 and helium-4 but significantly easier to control, a lot of attention is currently given to investigate such
double superfluid systems~\cite{Kinnunen2015}, which could also prove beneficial for modelling neutron star physics. One example is the recent observation of a superfluid
two-stream instability by Delehaye et al.~\cite{Delehaye2015}, which as mentioned previously is also thought to affect the dynamics of neutron stars~\cite{Andersson2013}.
At very low temperatures, two interpenetrating lithium clouds of different spin were kept in a magneto-optical trap and set into motion by displacing their centres of mass.
This excited dipole modes of different frequency in the Bose and Fermi component, causing relative motion between the two condensates. Delehaye et al.\ observed undamped
mode behaviour for slow relative motion, whereas for higher relative velocities the oscillations were damped. The existence of a critical velocity for the onset of
dissipative dynamics is typical for the presence of an instability. For the case that dissipation is caused by the creation of quasi-particles, the critical velocity has
been calculated to be equal to the sum of the sound speeds of both components~\cite{Castin2015}. While not contradicting this result, the experimental data could
not provide conclusive evidence and more work is needed to understand the small-scale physics of this instability in a Bose-Fermi mixture.

Despite these promising results, the study of ultra-cold Fermi gases is still a very new field of research that mainly focuses on investigating the fundamental properties.
Building neutron star analogues with these condensates is therefore more likely to be a task for the coming decades.


\section{Superconductors}
\label{sec-SCAnalogue}

Superconductors were the first systems to be observed to exhibit macroscopic quantum behaviour and the discovery of superconductivity by Onnes in 1911~\cite{Onnes1911}
sparked a new era of theoretical and experimental research that is still ongoing today. The theoretical advances most notably involve BCS theory~\cite{Bardeen1957} and
the phenomenological Ginzburg-Landau theory~\cite{Ginzburg1950}, which are still widely applied. The former has proven particularly useful in describing the microphysics
of superconductors, whereas the latter is used to study the macroscopic physics close to the transition temperature. Since the first discovery, many different materials
have been found to undergo a phase transition into the superconducting state and large numbers of experiments have been conducted to study their behaviour. Superconductors
are generally classified as \textit{conventional} if BCS theory can explain their properties and \textit{unconventional} if this is not the case. So-called heavy-fermion
superconductors~\cite{Steglich1979} and cuprate superconductors (compounds of copper and oxygen)~\cite{Bednorz1986} belong to the second category. The latter exhibit very
high transition temperatures not described within BCS theory and are attributed to the class of high-$T_{\rm c}$ superconductors, which have attracted a lot of interest
due to their potential use in industrial applications. For a review of the wide range of superconducting substances see Hirsch~\cite{Hirsch2015}.

On one hand, the vast variety of superconductors and experimental data available opens up many possibilities for designing neutron star analogues. On the other hand, it
is difficult to filter out which features could provide helpful information in the first place. The following discussion can thus only be viewed as a small tasting
sample of possible analogies between laboratory superconductors and neutron stars. In particular, the analysis will focus on fluxtube dynamics in type-II superconductors,
the state expected to dominate the outer neutron star core. We address pinning, resistive phenomena, instabilities, interfaces and how such studies could be
transferred to neutron stars to improve our understanding of their dynamics.

\subsection{General properties}
\label{subsec-SCProperties}

Theoretical models of neutron stars heavily rely on the use of fluid dynamics and in particular the presence of distinct, interacting fluid components. Hence, superconducting
analogues of neutron stars should be able to reflect such behaviour. The first phenomenological theory of superconductivity by Fritz and Heinz London~\cite{London1935} indeed
relied on the two-fluid description. As discussed in Sec.~\ref{subsec-MeissnerLondon}, based upon experimental observations, the London brothers postulated relations
between the mesoscopic electromagnetic fields $\mathbf{\bar{E}}$ and $\mathbf{\bB}$, the number density $n\cl$ and the current density $\mathbf{j}\SC$ of the component
responsible for the superconducting properties, i.e.
\begin{equation}
	\mathbf{\bar{E}} = \frac{m\cl}{n\cl q^2} \, \frac{\partial \mathbf{j}\SC}{\partial t} ,
		\hspace{1.5cm}
	\mathbf{\bB} = - \frac{m\cl c}{n\cl q^2} \, \nabla \times \mathbf{j}\SC.
	\label{eqn-London12_Experiment}
\end{equation}
This bears some obvious resemblance with Landau's two-fluid model~\cite{Landau1941}, usually invoked to explain observations of superfluid helium. Combining the latter relation
with Amp\`ere's law, a differential equation for the mesoscopic magnetic induction can be derived. It's solution describes the exponential magnetic field decay inside the
superconductor, i.e. $\bB(x) = B_0 \, \text{exp} (-x/\lambda)$ with the characteristic London penetration depth $\lambda$. Eqns.~\eqref{eqn-London12_Experiment} further
illustrate how (analogous to the superflow in helium II) a small electric current flows through a superconducting sample without creating a voltage. However, while Landau's
two-fluid interpretation was based on semi-microphysical considerations, the London equations did not provide any insight into the microphysics of superconductors. The
development of quantum mechanics played a crucial role in improving the understanding of small-scale processes, ultimately leading to a full microscopic theory of conventional
superconductivity. An equivalent formalism for the microphysics of superfluids has not yet been developed.

The first person to suggest that the quantum nature of particles could play an important role in the superconducting phase transition was Fritz London himself~\cite{London1948}.
Using a standard vector potential defined through $\mathbf{\bB} \equiv \nabla \times \mathbf{A}$, he showed that the supercurrent density $\mathbf{j}\SC$ is directly
proportional to $\mathbf{A}$ (see Sec.~\ref{subsec-MeissnerLondon} for more details),
\begin{equation}
	\mathbf{j}\SC = - \frac{c}{4 \pi \lambda^2} \, \mathbf{A}.
		\label{eqn-supercurrent_experiment}
\end{equation}
In 1953 Pippard~\cite{Pippard1953} built upon this result after he had demonstrated (by adding impurities to a superconductor) that $\lambda$ is not constant but dependent
on the material parameters. He accounted for this change by introducing a non-local modification to Eqn.~\eqref{eqn-supercurrent_experiment} on the order of a second scale
$\xi$ (which he called coherence length),
\begin{equation}
	\mathbf{j}\SC = - \frac{3}{16 \pi^2} \, \frac{c}{\xi_0 \lambda^2} \, \int \frac{\Br \left( \Br \cdot \mathbf{A} \right)}{r^4} \, \text{e}^{-r/\xi} \, \d V.
		\label{eqn-supercurrent_modified}
\end{equation}
Here, $\xi_0$ is a constant with the dimensions of length and the volume integral is evaluated over the entire superconductor. The supercurrent is thus no longer directly
proportional to the vector potential, but instead related to an average of $\mathbf{A}$ over a region of order $\xi$, which depends on the superconductor's degree of
impurity. Hence, Pippard was the first to illustrate the significance of the wave function's non-local, macroscopic properties.
Encouraged by these results, Bardeen, Cooper and Schrieffer published their microscopic description of superconductivity in 1957~\cite{Bardeen1957}. The central
idea of BCS theory is the presence of an attractive force, which causes two fermions to form a Cooper pair. In standard metals these fermions are electrons, which
are coupled to the lattice and can thus bind by exchanging virtual phonons. These electron Cooper pairs can then condense into the ground state and subsequently
form a superconducting state, if the attractive interaction is stronger than the repulsive Coulomb force. The corresponding phase transition is observed at a critical
temperature, $T_{\rm c}$. It had been previously pointed out that the existence of a critical temperature could also be explained by invoking an energy gap, $\Delta$,
at the Fermi level~\cite{Bardeen1955}. This implies that a finite energy (more precisely $2 \Delta$) is required to break a Cooper pair and excite electrons out
of the ground state. The energy gap is temperature-dependent as it takes its maximum at $T=0$ and vanishes at $T_{\rm c}$, where the proton coherence length diverges
and the superconductivity breaks down. In these limits, BCS theory predicts the following behaviour for $s$-wave pairing~\cite{Tinkham2004}
\begin{equation}
      \Delta(T = 0) \approx 1.764 \, k\B T_{\rm c}, \hspace{1.5cm}
      \Delta(T \to T_{\rm c}) \approx 3.06 \, k\B T_{\rm c} \left( 1- \frac{T}{T_{\rm c}}\right)^{1/2}.
\end{equation}
For $T \to T_{\rm c}$, the energy gap is proportional to the order parameter of the Ginzburg-Landau theory, illustrating the close connection between the two
formalisms near the transition temperature, $T_{\rm c}$.

With the microscopic picture in mind, one can re-evaluate the idea of a two-fluid model for superconductors. Similar to the case of helium II, at zero temperature
the system occupies the ground state, whereas a gas of quasi-particle excitations is present above $T=0$. This in principle provides the microphysical
justification for a macroscopic two-component model. Bardeen himself examined in 1958~\cite{Bardeen1958} if Landau's model of superfluidity could be applied to
fermionic superconductors and pointed out that the analogy is limited due to the presence of the metallic lattice. Whereas superfluid dynamics simply depend on the
two components' relative velocity, this can only be an approximation for the superconducting case, where the flow of normal electrons is constantly disrupted by
scattering off the lattice and other defects. Hence, the two-fluid description of superconductors is limited to small relative velocities (where the free-electron
approximation holds) and phenomena such as second sound that are related to the system's two-fluid nature are most likely not observable~\cite{Tilley1990}. Despite
these constraints, the two-fluid model has for example proven useful in calculating the microwave surface impedance (shown to depend on the ratio of normal to
superconducting electrons~\cite{Linden1994}) and the transport properties of granular superconductors\footnote{Granular superconductors are composed of microscopic
superconducting grains, separated by normal regions. Quantum mechanical Josephson tunnelling between these weakly-coupled grains generates the macroscopic
superconducting state. Many high-$T_{\rm c}$ superconductors are of this granular structure.}~\cite{DosSantos2006}. In both cases, good agreement with experimental
data has been obtained. The two-fluid model has also been able to account for the excitations of collective modes~\cite{Bray1975, Pethick1979}, which have been
observed as propagating phase fluctuations in thin superconducting aluminium films~\cite{Carlson1975}. As the normal electrons are effectively immobilised by
scattering with the lattice and impurities, these oscillations are very similar to the fourth sound in helium II, where the normal fluid motion is impeded by its
large viscosity and the behaviour of the two components is characterised by a phonon-like mode.

The preceding discussion clearly illustrates that the two-fluid model is seldomly applied to laboratory superconductors. This makes it difficult to address the
characteristics of the multi-component neutron star interior. While there could be some parallels to the crustal superfluid (which coexists with the nuclear lattice),
terrestrial superconductors are less suitable to perform detailed investigations of a neutron star's multi-fluid dynamics. In particular, creating mixtures of
two macroscopic condensates seems out of reach as a result of the solid character of the superconducting samples. Due to these disadvantages, experiments with helium
and ultra-cold gases have considerably more potential to improve our understanding of interacting, interpenetrating superfluids and superconductors.

\subsection{Fluxtube dynamics}
\label{subsec-FluxtubeDynamicsSC}

While laboratory superconductors appear less relevant to study phenomena related to the neutron stars' multi-fluid nature, another aspect can be analysed in great
detail: the physics of fluxtubes. Most conventional, heavy-fermion and high-$T_{\rm c}$ superconductors are of type-II~\cite{Kopnin2002} and permeated by quantised
fluxtubes if an external magnetic field $H >H_{\rm c1}$ is applied. Their magnetic properties are especially important for the industrial design of superconducting
wires and magnets, which has resulted in extensive research on the properties of type-II materials. All these studies are based on Abrikosov's seminal work from
1957~\cite{Abrikosov1957}, which demonstrated for the first time the existence of a class of superconductors, not completely expelling magnetic flux from their
interior but instead forming a regular mixed state. The experience, gained from studying the fluxtube lattice in laboratory system over decades (for a review see
Brandt~\cite{Brandt1995}), could greatly benefit the development of neutron star analogues, where the fluxtubes' motion is expected to govern the magnetic field
evolution but only little is known about the mechanisms affecting them.

\begin{figure}[t]
\begin{center}
\includegraphics[width=4in]{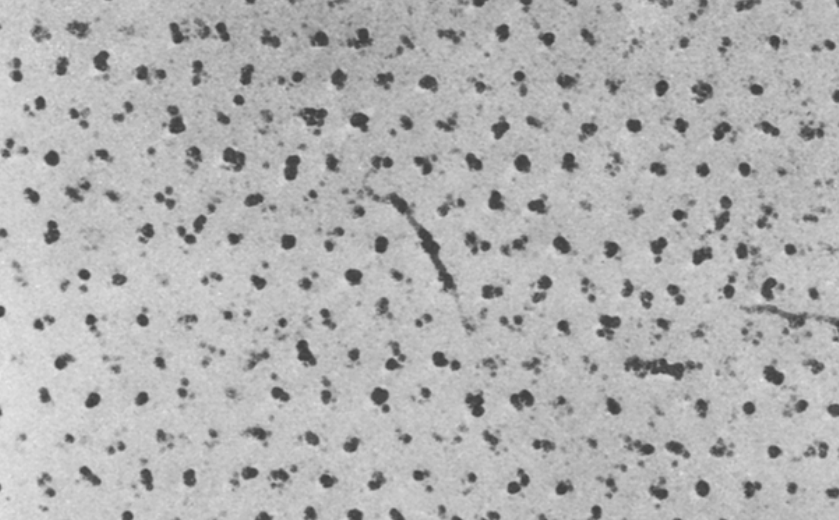}
\caption{Electron micrograph showing the fluxtube lattice of a type-II superconductor at $1.1 \, \K$. The dark points are small cobalt particles distributed
with the decoration method. The figure is reproduced with permission from Essmann und Tr\"auble~\cite{Essmann1967}. Copyright $\copyright$ 1967, Elsevier B.V.}
\label{fig-FLL}
\end{center}
\end{figure}

One advantage of terrestrial superconductors is the possibility to image the magnetic flux structures. It has been possible since the mid 1960s to obtain direct
evidence of the fluxtube lattice by using the \textit{decoration method} (see for example Essmann und Tr\"auble~\cite{Essmann1967}), which is based on evaporating
a metal wire. The resulting magnetised smoke falls onto the surface of a superconductor and settles on the points of highest magnetic field, which coincide with
the ends of fluxtubes. The resulting pattern is then observed with an electron microscope as shown in Fig.~\ref{fig-FLL}. Still applied today, this visualisation
technique not only provides information about the flux distribution on the surface but also allows the extraction of bulk properties~\cite{Marchetti1995}. It has
also proven useful in studying high-$T_{\rm c}$ superconductors~\cite{Grigorieva1994}. Today, many more methods are available to analyse the interior of type-II
superconductors ranging from neutron scattering and magnetic force microscopy to scanning tunnelling spectroscopy~\cite{Hess1989, Brandt1995, Peets2007}. Using
these approaches, it has for example been possible to detect the motion and pinning of a single fluxtube, visualise pinning defects~\cite{Hyun1989} and obtain
time-resolved images of fluxtube dynamics~\cite{Leiderer1993} and high-resolution pictures of individual fluxtube cores~\cite{Suderow2014}.

The typical size of superconducting fluxtubes is given by the coherence length, which is of the order of $10 \, \nm$~\cite{Tilley1990}. Comparison with quantised
structures in superfluid helium shows that the fluxtubes are several orders of magnitude larger than helium II vortices but of similar dimension as the $B$-phase
ones in helium-3, where BCS theory is also invoked to explain the macroscopic quantum behaviour. As previously discussed, these dimensions are many orders of
magnitude larger than the typical proton and neutron coherence lengths in neutron star cores. This could cause problems for superconducting laboratory neutron star
analogues. However, because little conclusive evidence has been acquired in the five decades since the presence of quantised magnetic structures was first
suggested~\cite{Baym1969}, understanding the dynamics of vortices and fluxtubes is one of the most important problems of modern neutron star astrophysics. Hence,
any system that could provide more information about these mechanisms is worth a more detailed investigation. In the following, various aspects of fluxtube physics
in terrestrial media will be explored.

Before proceeding with an analysis of the individual features, we raise two notes of caution. Firstly, almost all experiments are performed in two dimensions. Using
thin films (which are easy to manufacture) simplifies the studies because it ensures that fluxtube bending can be neglected. Moreover, laboratory systems are mainly
dominated by two-dimensional phenomena as a result of the underlying crystal structure~\cite{Machida1995}. In particular, the physics of high-temperature superconductors
can be described by invoking weakly-coupled layers~\cite{Lake2005}. While this two-dimensional geometry is suitable to describe the local dynamics of the proton
fluxtube array in neutron stars, it is certainly insufficient to account for the macroscopic magnetic field characteristics. Thus, experiments with three-dimensional
samples would be desirable. However, these have only recently started to attract attention and only limited studies are available (addressing the magnetic properties
of superconducting spheres~\cite{Doria2007, Broun2009, Velez2009} or fluxtube motion in layered systems~\cite{Kolton2000, Auslaender2008, Pleimling2011}). Additionally,
the proton superconductor in neutron stars is rapidly rotating at an angular speed $\Omega$, which is associated with a magnetic field (see
Sec.~\ref{subsec-chargemultihydro} for details)
\begin{equation}
	   b_{\rm L} \approx - \frac{2m c}{e}\, \Omega.
		\label{eqn-LondonField_Experiments}
\end{equation}
Although this characteristic London field is of small magnitude in neutron stars, it has important implications for the stars' electromagnetism as discussed
in Sec.~\ref{subsec-Maxwell}. Measurements of the London field in terrestrial superconductors have been performed in metallic~\cite{Hildebrandt1964},
high-temperature~\cite{Verheijen1990} and heavy-fermion~\cite{Sanzari1996} systems. However (apart from a few exceptions~\cite{VlaskoVlasov1998, Avila2001, Cortes2011})
most experiments investigating the dynamics of fluxtubes are performed with static films and non-rotating external fields, thus not providing information about the
influence of rotation on the superconductor's properties. For these two reasons, the following discussion primarily concerns aspects which could help to improve our
knowledge of mesoscopic neutron star physics.

\subsubsection{Pinning}
\label{subsubsec-Pinning}

The first evidence that pinning affects the motion of fluxtubes in superconductors was obtained by analysis of their electromagnetic properties. It was observed
that impurities modify the current-carrying characteristics of type-II media while not significantly altering their transition temperatures~\cite{Tilley1990}. The effect
of impurities or defects is particularly apparent when magnetisation curves of \textit{pure} and \textit{dirty} type-II superconductors are compared. Whereas the curves
are reversible in the former case, their behaviour is irreversible in the latter case and strong hysteresis is exhibited for a cycle in the applied magnetic field.
The first description, successful in modelling the hysteresis behaviour of superconductors with large Ginzburg-Landau parameter exposed to high fields, was developed
by Bean in the early 1960s~\cite{Bean1962, Bean1964}. Ignoring the mesoscopic fluxtube physics, Bean postulated that a maximum, critical current density is flowing in
the surface layer of a superconductor (its penetration depth depending on the strength of the applied field). The field then decays linearly towards the centre of the
sample, which is field free. Depending on the magnetisation history, the superconductor thus responds differently and hysteretic behaviour is observed. This simple
so-called \textit{critical-state model} (which gives excellent agreement with experiments~\cite{Kim1962, Bean1964} and is still widely used in the engineering
community~\cite{Chapman2010}) can also be interpreted as an average over the mesoscopic quantisation in the limit of large numbers of fluxtubes~\cite{Altshuler2004}.
Applying a magnetic field, fluxtubes are pushed into the superconductor. In pure systems, they move freely and are able to uniformly distribute in the sample, which
explains the reversible magnetisation curves. In dirty systems however, structures are present impeding the fluxtubes' motion and preventing them from moving to the
centre of the superconductor. Hence, the fluxtube distribution is much denser at the surface generating a metastable state, which results in irreversible magnetisation
curves. This characteristic magnetic field penetration of a dirty type-II superconductor following Bean's critical state model has recently been confirmed with
diamond-magnetometric measurements~\cite{Alfasi2016}.

Since the 1960s, pinning has been studied in much more detail and many reviews have been published on the subject (see for example Campbell and Evetts~\cite{Campbell1972},
Dew-Hughes~\cite{Dew-Hughes1974} and Blatter et al.~\cite{Blatter1994}). The reason for this lies mainly in the importance of pinning for industrial applications, which
can be explained in the following way: the interaction between fluxtubes of parallel magnetic field orientation, $\hat{\mathbf{B}}$, is repulsive and a single fluxtube
in a lattice experiences a Lorentz force per unit length,
\begin{equation}
      \mathbf{f}_\rL \equiv \mathbf{j}\SF \times \frac{\Phi_0}{c} \hat{\mathbf{B}}.
	      \label{eqn-LorentzForceFLL}
\end{equation}
Here, $\mathbf{j}\SF$ denotes the total supercurrent density generated by all other fluxtubes, so that an individual fluxtube can only be in equilibrium if the Lorentz
force vanishes (implying $\mathbf{j}\SF = 0$ at its centre). This is realised in a regular array where the triangular arrangement has the lowest free energy (see
Sec.~\ref{subsubsec-Abrikosov}). Eqn.~\eqref{eqn-LorentzForceFLL} also shows that any additional currents, $\mathbf{j}_{\rm ext}$, flowing through a medium will
disturb the equilibrium configuration, create a net force per unit volume, $\mathbf{f}_{\rm ext}$, acting on the fluxtubes and cause them to move,
\begin{equation}
      \mathbf{f}_{\rm ext} \equiv \mathbf{j}_{\rm ext} \times \frac{\mathbf{B}}{c},
	      \label{eqn-EffectiveForce}
\end{equation}
where $\mathbf{B}$ denotes the averaged magnetic induction. As explained in the next section, moving fluxtubes dissipate energy and generate heat. This can be very
problematic for superconducting wires and magnets, where high fields and currents are desired. By using impure materials however, the fluxtubes can pin to inhomogeneities
which obstruct their motion. Due to the benefits of decreasing dissipation and preventing heat generation, pinning is an important field of modern superconductivity
research.

The great benefit of studying pinning physics with laboratory superconductors is the amount of experimental control these systems offer. The different ways to modify the
pinning properties seem endless, ranging from standard techniques such as irradiation, doping~\cite{Tilley1990}, varying the sample thickness~\cite{Martinoli1978} or the
introduction of holes~\cite{Moshchalkov1994, VanLook2002} to more advanced methods like nanofabrication. These days, the interior of superconductors can be doped with
small nanoparticles~\cite{Miura2011, MacManus-Driscoll2004} and the pinning surfaces customised by using electron beams to deposit small particles~\cite{Dobrovolskiy2011}.
The possibility of creating any desired structure with high resolution could prove very important for studying the unknown pinning characteristics of neutron star
fluxtubes and vortices.

\begin{figure}[t]
    \begin{center}
	\includegraphics[width=5in]{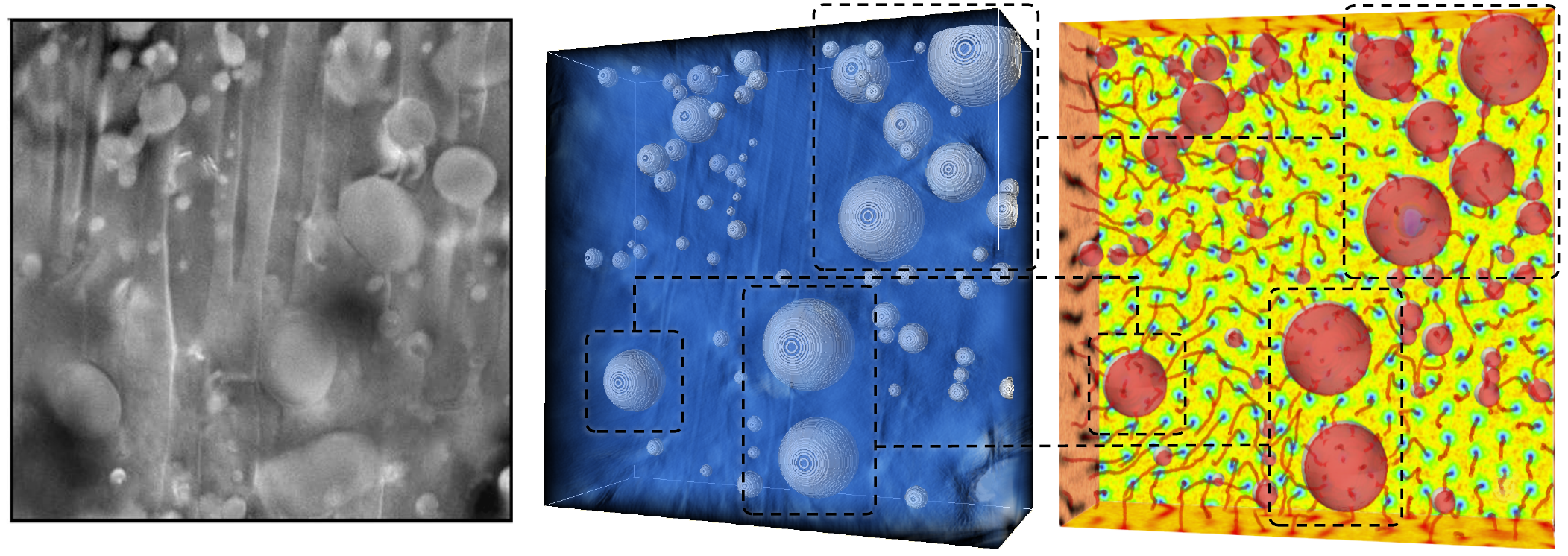}
\caption{Modelled fluxtube motion based upon measurements of the pinning landscape in a high-$T_{\rm c}$ superconductor. The left figure is the three-dimensional
scanning transmission electron microscope (STEM) tomogram of a superconducting sample of dimensions $534 \times 524 \times 129 \, \nm^3$. The box contains
$\sim 71$ almost spherical particles with sizes ranging from $12.2$ to $100 \, \nm$. The middle figure combines the first one with a numerical reconstruction of the model
volume. The final figure on the right represents a snapshot of the time-dependent Ginzburg-Landau simulation showing the behaviour of the order parameter. Isosurfaces
of the order parameter close to the normal state are marked in red and illustrate the motion of fluxtubes and the positions of pinning sites. The colour in the background
represents the amplitude of the order parameter with yellow showing superfluid and blue identifying normal regions. The left figure is reproduced with permission from
Ortalan et al.~\cite{Ortalan2009}. Copyright $\copyright$ 2009, Elsevier B.V. The middle and the right figure are reproduced with permission from Sadovskyy et
al.~\cite{Sadovskyy2015}. Copyright $\copyright$ 2016, American Physical Society.}
\label{fig-RealPinningLandscape}
\end{center}
\end{figure}

Above all, the studies of laboratory superconductors demonstrate that pinning is a very common phenomenon since fluxtubes cannot only be locked to defects and impurities
but also to dislocations, vacancies, grain boundaries, interstitials, rough surfaces and layered structures. Generally speaking, the pinning centres are most effective
if they are of similar dimension as the fluxtubes' normal cores. Among other aspects, understanding the pinning phenomena theoretically has involved efforts to analyse
the depinning of fluxtubes from random pinning potentials~\cite{Blatter1994} and calculate detailed pinning forces from microscopic BCS principles~\cite{Thuneberg1984,
Thuneberg1989} or the phenomenological Ginzburg-Landau theory (see Brandt~\cite{Brandt1995} and references therein). In recent years, it has further become possible to
directly image the defects in superconductors and measure the corresponding pinning forces~\cite{VanderBeek2002, Shapoval2010}. Besides such small-scale approaches (studying
individual fluxtubes) pinning in type-II superconductors has also been examined from a macroscopic perspective. The properties of a fluxtube lattice affected by pinning
have for instance been modelled with a mean-field formalism,~\cite{Brandt1995, Chapman1995, Chapman2010} where flux transport is governed by a non-linear diffusion equation.
Looking for example at a simplified two-dimensional geometry with $\mathbf{B} = B(x,y,t) \uz$, the evolution of the magnetic induction inside a sample is obtained by
solving~\cite{Chapman2010}
\begin{equation}
	\frac{\partial B}{\partial t} = \nabla \cdot \left( D \nabla B \right)
	\label{eqn-nonlinear_diffusion}
\end{equation}
with suitable boundary conditions. Note that the diffusion coefficient $D$ can depend on the spatial coordinates, the time and the magnetic induction itself. Alternatively,
fluxtube dynamics have been modelled by means of a collective theory~\cite{Larkin1974, Larkin1979, Kerchner1983}, which is based on the statistical averaging of pinning
forces exerted by a configuration of random pinning centres within a characteristic volume, $V$. This volume results from the assumption that weak, randomly distributed
pinning sites are able to destroy the long-range order of an elastic lattice but do not affect its short-range structure. Hence, the microscopic pinning forces ($F_{\text{pin},
i}$ is the force exerted by the $i$th pinning centre) are uncorrelated inside the region $V$ and their squares can be simply added up. This provides a way to
calculate the average of the pinning force density squared, i.e. $W \equiv n_{\rm pin}  \langle F_{\text{pin}, i}^2 \rangle $ with $n_{\rm pin}$ denoting the density of
pinning centres. The collective pinning force density affecting an elastic lattice without dislocations is then given by $\langle f_{\rm pin, collective} \rangle \approx
(W /V)^{1/2}$. In three dimensions, one for example obtains~\cite{Larkin1979, Brandt1995}
\begin{equation}
	\langle f_{\rm pin, collective} \rangle \propto \frac{W^2}{r_{\rm pin}^3 c_{\rm el}},
\end{equation}
where $r_{\rm pin}$ is the range of the pinning forces (typically of the order of the coherence length $\xi$) and the parameter $c_{\rm el}$ depends on the elastic
properties of the fluxtube lattice (for more details see e.g.\ Brandt~\cite{Brandt1995}). Within this collective picture, (de)pinning events are thus related to the motion
of large numbers of fluxtubes, located within \textit{bundles} associated with the characteristic volume, instead of individual fluxtubes. Corresponding theoretical
predictions agree well with experiments~\cite{White1993, Yaron1994, Chandra2015}. Similar macroscopic models of the quantum condensates and the associated characteristics
could also be applicable to the neutron star's interior. The collective motion of vortices is for example expected to play an important role in the generation of glitches.
Hence, laboratory studies could provide valuable input to model these jumps more accurately and allow one to test macroscopic formalisms against observational data.

Before continuing with the dynamical fluxtube processes, we point out the recent work by Sadovskyy et al.~\cite{Sadovskyy2015}, illustrating how experiments and
modern theoretical calculations can complement each other in the study of type-II media. Sadovskyy et al.\ numerically reconstruct the scanning transmission
electron microscope (STEM) measurements of pinning defects in a high-$T_{\rm c}$ superconductor~\cite{Ortalan2009} to calculate the motion of fluxtubes in a realistic
pinning landscape (see Fig.~\ref{fig-RealPinningLandscape}). By numerically solving the time-dependent Ginzburg-Landau equations~\cite{Sonin1987, Crabtree2000},
they account for features such as pinning defects, fluxtube flexibility, long-range fluxtube repulsion, fluxtube cutting and reconnections. The simulated critical
properties are in very good agreement with the experimental results, demonstrating the impact such combined approaches could have. Keeping in mind that laboratory
systems further offer the possibility to manufacture arbitrary pinning landscapes, the method provides the unique opportunity to perform detailed studies of the
pinning interaction and deduce the fluxtube-averaged properties of the superconducting sample. Such analyses could provide specifically useful to improve our
understanding of the vortex-fluxtube coupling in neutron star interiors and its large-scale implications.

\subsubsection{Flux creep, flux flow and Hall effect}
\label{subsubsec-ForceSC}

In the 1960s, Abrikosov's fluxtube interpretation became increasingly popular and several phenomena that had previously been unexplained were interpreted in terms of the
fluxtube picture. Two important examples are the \textit{flux-creep} and \textit{flux-flow} behaviour, which represent the non-linear and linear regimes of fluxtube motion
in conventional superconductors. The former was first examined by Kim et al.~\cite{Kim1962}, who found that close to the transition temperature flux could \textit{leak}
through a superconductor and create a measurable resistance if the external current, $\mathbf{j}_{\rm ext}$, exceeded a critical value. More precisely, the experiments
revealed that persistent currents decayed proportional to the logarithm of time and the creep behaviour was faster when the transport current was increased~\cite{Kim1963}.
Based on Bean's critical state model, a theoretical description of this phenomenon was developed by Anderson and Kim~\cite{Anderson1962, Anderson1964}. Whereas in the
critical state at $T=0$, fluxtubes only move when the Lorentz force exceeds the pinning force, Anderson and Kim suggested that creep dynamics are related to the thermal
energy of the lattice. At finite temperatures fluxtubes start to vibrate, which can result in the unpinning of several fluxtubes within a distance of the order $\lambda$.
Such a \textit{bundle} will move as a unit and jump to an adjacent pinning site. This thermal activation of fluxtubes over a pinning barrier occurs at a
rate~\cite{Anderson1962, Anderson1964}
\begin{equation}
  \nu \approx \nu_0 \, \text{e}^{-U/ k_{\rm B} T},
  \label{eqn-fluxjumprate}
\end{equation}
where $\nu_0$ is a vibration frequency of the bundle and $U$ the effective activation energy. This process decreases exponentially with the temperature, is discrete and stochastic
and has the advantage to not require a detailed specification of the nature of the pinning centres. In the absence of an external current, the net jump rate of all flux bundles
vanishes. If however a small transport current is applied, thermal flux-creep is significantly enhanced, because the Lorentz force given in Eqn.~\eqref{eqn-LorentzForceFLL} acts
as an additional driving force effectively lowering the barrier for thermal activation, i.e.\ the energy $U$ in Eqn.~\eqref{eqn-fluxjumprate}. The flux distribution thus
changes slowly over time, which dissipates energy and leads to the observed resistance and characteristic logarithmic behaviour of persistent currents. Based upon this creep theory
for superconductors, Anderson and Itoh~\cite{Anderson1975} proposed that a similar mechanism could be operating on the superfluid vortices in neutron stars, explaining the
noisiness of pulsar rotation frequencies. The idea has also been taken up by Alpar et al.~\cite{Alpar1984b, Alpar1984c}, who considered vortex creep in the neutron star crust
to develop a more realistic model of pulsar glitches. By comparing observational data to the theoretical framework, the creep theory provided estimates of interior characteristics
such as the temperature or the pinning energy between the vortices and the crustal nuclei.

Increasing the transport current, $\mathbf{j}_{\rm ext}$, further, the Lorentz force will eventually exceed the pinning force, which enables the fluxtubes to flow through
the sample. This is analogous to the scenario invoked for large pulsar glitches, where neutron vortices are thought to be pinned to the crustal lattice until the pinning
force is overcome by the Magnus force and a large number of vortices are released simultaneously. However, in superconductors the Lorentz force is observed to replace the
Magnus force as the driving source of fluxtube motion. The two forces can be experimentally distinguished in the following way: As the Lorentz force acts perpendicular to
the applied current and the magnetic induction $\mathbf{B}$ (see Eqn.~\eqref{eqn-EffectiveForce}), the resulting mean fluxtube velocity $\mathbf{u}_{\rm ft}$ is
transverse to $\mathbf{j}_{\rm ext}$ and induces an average electric field,
\begin{equation}
  \mathbf{E}_{\rm ind} \equiv - \mathbf{u}_{\rm ft} \times \frac{\mathbf{B}}{c},
\end{equation}
which is parallel to the transport current and generates a longitudinal voltage~\cite{Tilley1990}. If the Magnus force would dominate the dynamics,
the fluxtubes would feel no net force and be dragged with the current, similar to the motion of free superfluid vortices. This would induce a transverse voltage, which is
characteristic for the conservative \textit{Hall effect}. Experiments have shown that for most superconductors the longitudinal voltage is several orders of magnitude larger
than the transverse \textit{Hall voltage}, implying that the Lorentz force determines the fluxtube motion~\cite{Strnad1964, Reed1965} and not the Magnus force as would be
the case in a superfluid. This difference in superconductors arises because the metallic lattice and impurities complicate the dynamics, acting as scattering centres for the
normal electrons and impeding the fluxtube flow. As noted by multiple authors, only in very pure superconductors in the limit of $T \to 0$ (where electron-electron scattering
is dominant), one would expect the fluxtube motion to be almost free and governed by the Magnus force~\cite{DeGennes1964, Vinen1967, Kopnin1976, Sonin1987}.

Experiments have further demonstrated that at low temperatures the voltage in the dissipative flux-flow regime changes linearly with the applied magnetic field and the
\textit{flux-flow resistance}, $R\SF$, thus obeys the empirical law~\cite{Kim1965}
\begin{equation}
	R\SF \equiv R\NF \, \frac{B} {H_{\rm c2}}.
		\label{eqn-FFResistance}
\end{equation}
Here, $R\NF$ is the normal-state resistance, $B$ the magnitude of the magnetic induction and $H_{\rm c2}$ the upper critical field. Applying different treatments to the
superconducting sample shows that the flux flow is not affected by the surface conditions~\cite{Vinen1967a}. Therefore, the fluxtubes' motion is independent of the pinning
characteristics and the resistive physics are solely determined by the superconductor's bulk properties. More precisely, the phenomenological relation~\eqref{eqn-FFResistance}
indicates that strong dissipation is directly connected to the fluxtubes, as the factor $B / H_{\rm c2}$ represents the fractional volume their normal cores occupy in the
superconductor~\cite{Kim1965}. This illustrates that similar to superfluid helium or the multi-component mixture in neutron stars, the dissipative effects are determined by
drag forces acting on the individual fluxtubes.

The first theoretical calculation of the corresponding microscopic drag coefficient was provided by Tinkham~\cite{Tinkham1964}. He pointed out that the order parameter at
any specific point would oscillate between zero (normal fluxtube cores) and a constant value (superconducting regions) as the fluxtubes are moving through the sample. Since
the equilibrium is not instantaneously restored, Tinkham showed that the ensuing relaxation process would cause dissipation. A process of similar order, Bardeen and
Stephen~\cite{Bardeen1965} considered the induction of electric fields by the moving fluxtubes, which generate dissipative eddy currents in their normal cores. Based upon
electron-lattice scattering, they derived the corresponding drag coefficient. Whereas for Tinkham's mechanism, the dissipation is proportional to the amplitude of the order
parameter, the dissipation is related to the supercurrent density and, thus, the phase of the order parameter for the second mechanism. Both processes and the flux-flow
resistance~\eqref{eqn-FFResistance} can be recovered by performing a more detailed study of the non-equilibrium physics in type-II superconductors using a time-dependent
extension of the Ginzburg-Landau theory~\cite{Cyrot1973, Gorkov1975}. However, Kopnin~\cite{Kopnin2002} has pointed out that such models neglect additional dissipation
mechanisms due to the relaxation of quasi-particle excitations created by the moving fluxtubes. Other formalisms such as a semi-classical Boltzmann theory~\cite{Kopnin2002}
or a time-dependent microscopic theory~\cite{Blatter1999} are necessary to correctly capture all microscopic processes, which is particularly important when the strength of
the Hall effect is calculated. The same models have been invoked to explain the strong mutual friction in superfluid helium-3 (see also Sec.~\ref{subsec-MutualFrictionHe}),
which is conceptually similar to a superconductor despite having a more complicated structure of the order parameter and could also give insight into the dissipation
mechanisms affecting a neutron star's interior magnetic field  evolution.

Discussing the flux-creep, flux-flow and Hall regimes shows that laboratory superconductors provide the means to directly observe the dissipative fluxtube behaviour. Using
the experimental data allows one to investigate the validity of theoretical models, which helps to improve our understanding of the underlying microphysics. The methods
applied to terrestrial systems thus proceed analogous to the study of the neutron stars' properties, where macroscopic characteristics are used to constrain the interior
physics. However, many questions concerning the magnetic fields of neutron stars are still unanswered. Laboratory superconductors could therefore help to examine the
mesoscopic fluxtube dynamics in more detail and, for example, assist in determining the mechanism dominating the stars' field evolution.

\subsubsection{Lattice melting}
\label{subsubsec-LatticeMelting}

While conventional superconductors are well described by the flux-creep and flux-flow models, high-temperature systems exhibit more complicated fluxtube physics,
predominantly caused by two factors. First of all, high-$T_{\rm c}$ superconductors have a strongly layered crystal structure, which results in the formation of weakly-coupled,
two-dimensional fluxtubes such as layered pancakes~\cite{Clem1991, Guikema2008}. Amongst other aspects, this highly anisotropic behaviour decreases the strength of the pinning
potentials, favouring thermal depinning~\cite{Brandt1995}. Secondly, due to the high transition temperatures, thermal fluctuations play a more important role than in
conventional media. In high-temperature superconductors, these differences can for example lead to the activation of creep over a large temperature range, referred to as
\textit{giant flux creep}~\cite{Tinkham2004}. As thermal fluctuations destroy the long-range order of an elastic lattice, this is generally described within the collective
pinning theory~\cite{Larkin1974, Larkin1979, Kerchner1983}. This theory, as discussed previously, is based on the statistical summation over random pinning sites.

The same approach can also account for the fact that the resistivity of high-$T_{\rm c}$ superconductors does not decrease exponentially for $T \to 0$ as predicted by the
conventional Anderson-Kim model~\cite{Anderson1964}, always leading to a non-zero resistance. Instead, it drops rapidly at low temperatures indicating an abrupt change
in the fluxtube dynamics that is usually interpreted as evidence for a second phase transition well below the superconducting one~\cite{Fisher1989}. In analogy with other
crystalline structures, this is expected to correspond to a first-order \textit{melting transition}, which results from the destruction of the long-range order and
characterises the change from a solid to a liquid fluxtube phase. The discontinuity in the resistance of high-$T_{\rm c}$ superconductors would thus mark the \textit{freezing}
of the fluxtube lattice as $T \to 0$. Clear experimental evidence for such a melting transition was first obtained by measuring the macroscopic induction of a pure
superconductor~\cite{Zeldov1995}. Similar to the expansion that accompanies the freezing of water into ice, the fluxtube density in a superconductor changes discontinuously
at a first-order transition for any given applied magnetic field. As the fluxtube liquid is denser than the solid, the former has a higher average magnetic induction
than the latter and $B$ increases upon melting. The corresponding change was observed experimentally~\cite{Zeldov1995} and the melting behaviour found to agree well with
theoretical predictions obtained by invoking the \textit{Lindemann criterion}. It predicts melting once the amplitude of thermal fluxtube vibrations reaches a specific
fraction $c_{\rm L}$ (typically of order $0.1$~\cite{Tinkham2004}) of the interfluxtube spacing $d\ft$ and gives for example a power-law behaviour for the melting
field as a function of temperature close to $T_{\rm c}$,
\begin{equation}
  B_{\rm m}(T) \propto \left( 1 - \frac{T}{T_{\rm c}} \right)^{\alpha},
\end{equation}
where $\alpha \approx 1.55$ was obtained in the experiment~\cite{Zeldov1995}. The melting of a two-dimensional lattice has further been directly recorded with scanning
tunnelling microscopy~\cite{Guillamon2009}, which unambiguously showed the transition into an ordered, isotropic liquid. The first-order nature however
is only observed in very pure samples as inhomogeneities and strong pinning introduce additional disorder into the system. As the temperature decreases, the melting transition
is thus complicated and thought to be of second order, causing a change from the fluxtube liquid to a \textit{glass-like} phase~\cite{Nattermann2000}.

Fluxtube lattice melting may also be present in conventional superconductors~\cite{Bossen2012}. Despite predicted to appear very close to the superconducting transition
temperature and being difficult to observe~\cite{Brandt1989}, several experiments have found indications for a melting transition in conventional type-II
systems\cite{Berghuis1993, Lortz2006}. This would suggest that a similar mechanism could also be manifested in the proton superconductor of neutron stars if the interior
is sufficiently hot. Whereas this scenario is rather unlikely for old, rotation-powered pulsars, the decay of magnetic energy in young high-field magnetars might act
as a potential heat source with the ability to locally exceed the critical temperature and initiate the melting transition into a fluxtube liquid. As observed in laboratory
systems, this would be accompanied by a local increase in the fluxtube density, i.e.\ the magnetic induction, that in turn could result in additional dissipation. While
highly speculative, this would significantly change the flux-carrying properties of the region compared to the rest of the star and could potentially shorten evolution
timescales of the large-scale magnetic field.

\subsubsection{Instabilities}
\label{subsubsec-InstabilitiesSC}

While laboratory superconductors are observed to exhibit non-linear features, this regime is different to the non-linearity of helium and ultra-cold quantum
gases discussed previously. In particular, the chaotic behaviour of superfluids is mainly influenced by their two-fluid nature, leading to turbulent dynamics due to
counterflow or two-stream instabilities. Laboratory type-II superconductors, on the other hand, do not exhibit strong turbulence~\cite{Chapman1995, Barenghi2014} as a
result of several factors. Firstly, due to the presence of the lattice, Landau's two-fluid model is less satisfactory in describing the physics of superconductors and
one would hence not expect the mechanisms for superfluid turbulence to apply to the charged condensates. Moreover, pinning and the strong two-dimensional character of
many superconductors stabilise these systems and suppress the development of turbulent behaviour on large scales. Hence, laboratory superconductors are not suitable
analogues for the study of the neutron stars' fluid instabilities. However, there are several elements of laboratory type-II systems (in particular related to
instabilities of individual fluxtubes), which could provide new information on the local, non-linear behaviour of proton fluxtubes and the transfer to a macroscopic
model of a star's dynamics.

One phenomenon thought to influence type-II superconductors is equivalent to the Donnelly-Glaberson instability of superfluid vortices in helium II~\cite{Glaberson1974,
Tsubota2004}. Applying an axial heat current, vortices become unstable to helical displacements and Kelvin waves are excited along the vortex lines. Similarly,
Clem~\cite{Clem1977} has shown that a single, unpinned fluxtube is unstable against helical perturbations if a current (parallel to the fluxtube's axis) exceeds a
critical value. The existence of this critical current has been confirmed experimentally~\cite{Irie1989}. Generalising the mesoscopic to the macroscopic dynamics
has for example been achieved by modifying the averaged vortex-density model to include small-scale helical instabilities~\cite{Chapman1995}. Additionally,
Brandt~\cite{Brandt1981} has shown directly that the full fluxtube lattice experiences an instability to helical deformations by balancing the driving force due to
an external current with the restoring force resulting from the lattice elasticity. The instability growth is further influenced by pinning, which stabilises the
fluxtube array and can suppress the instability on large scales. Nonetheless, even for strong pinning the instability excites helical deformations at a critical
wave number of $k_{\rm c} \approx 2 \pi /2.2 d\ft $ along the fluxtube segments in between pinning sites once a longitudinal current density of~\cite{Brandt1981}
\begin{equation}
  j_{\rm c} \approx 0.47 \, \frac {B d\ft} {\lambda^2}
\end{equation}
is exceeded. Here, $B$ denotes the magnitude of the average magnetic induction, $d\ft$ the interfluxtube spacing and $\lambda$ the London penetration depth. Note that
Charbonneau et al.~\cite{Charbonneau2007} propose that the same helical instability could be acting on the proton fluxtubes in the neutron star core, destroying the
regularity of the lattice (i.e.\ the type-II state). While not providing specific details about the phase that would be formed instead, Charbonneau et al.\ suggest
that the intermediate state of a type-I superconductor is one possibility. This would subsequently solve the problem discussed by Link~\cite{Link2003}, who pointed
out the incompatibility of type-II superconductivity with observational indications of long-period precession in pulsars~\cite{Stairs2000, Shabanova2001, Haberl2006}.

While the helical instability in laboratory systems has not been directly observed yet, experimental evidence for this mechanism has been recovered. Imagine an initially
straight fluxtube lattice located inside a cylindrical container that becomes unstable once the critical axial current is exceeded. The distorted fluxtubes would start
to spiral outwards in a helical manner~\cite{Brandt1995}. The helices would grow until the fluxtubes hit the sample surface (leading to a measurable change in the magnetic
flux density on the container walls~\cite{Leblanc1993, LeBlanc2003}) or start cutting through each other (resulting in additional dissipation~\cite{Clem1982}). Experiments
have also made it possible to determine the cutting force between two inclined fluxtubes~\cite{Blamire1986, Palau2008}. This force is independent of the applied magnetic
field, i.e.\ independent of the interfluxtube spacing, only weakly dependent on the temperature and generally found to be rather low ($10^{-14} \, \text{N}$ per intersection
in agreement with microscopic calculations~\cite{Pardo2007}). This illustrates that fluxtube cutting could be very important for the dynamics of a superconductor, as it
for example suggests that fluxtubes could cut through each other in order to avoid pinning centres. For more information on flux cutting in laboratory superconductors
see the recent review by Campbell~\cite{Campbell2011}. Analysing the cutting of fluxtubes could also help to improve our understanding of the short-range magnetic coupling
between the two coexisting lattices in the outer neutron star core. While it is not clear how much energy it costs to cut neutron vortices through proton fluxtubes and vice
versa, the interaction could lead to strong dissipation in the interior. If the energy costs for vortex-fluxtube cutting are too large, the two quantised arrays could even
be locked together. Better models of laboratory fluxtube cutting could thus help to understand the neutron stars' magneto-rotational evolution~\cite{Ding1993}.

\begin{figure}[t]
\begin{center}
\includegraphics[width=3.5in]{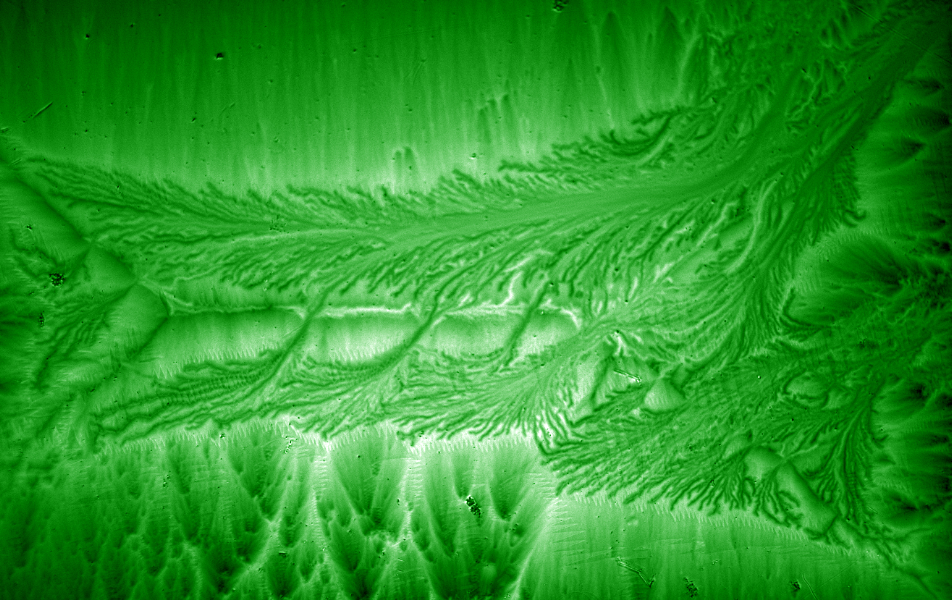}
\caption{Magneto-optical image of the collapsed meta-stable state in a thin superconducting film. Following a thermomagnetic instability, fluxtubes
are suddenly redistributed forming avalanche-type patterns. The figure is reproduced with permission from Eliasson~\cite{Eliassen2008}.}
\label{fig-Avalanche}
\end{center}
\end{figure}

In addition to the helical instability, other dynamical instabilities of the fluxtube lattice in type-II superconductors have been studied. Theoretical
models~\cite{Wertheimer1967, Mints1977, Olson1997} generally neglect the detailed microphysics and instead consider Bean's macroscopic, critical state model (originally
introduced to explain the irreversible behaviour of dirty superconductors). The models are based on the assumption that by changing external parameters such as the
temperature or the applied magnetic field, the temperature inside the superconductor locally increases, which causes the pinning force and, thus, the critical current
density to decrease. Subsequently, the fluxtubes start to move and the metastable state becomes unstable, leading to the uncorrelated propagation of flux through the
sample. These sudden bursts of collective motion, also known as \textit{flux jumps}, lead to dissipation and create measurable voltages~\cite{Field1995}.

As fluxtube motion in the presence of pinning centres is characterised by a non-linear diffusion equation~\cite{Brandt1995, Chapman2010}, the catastrophic flux propagation
is closely related to the concept of self-organised criticality originally discussed by Bak et al.~\cite{Bak1988}. Again ignoring the microphysics and simply considering
the statistical properties of the non-linear behaviour, flux jumps can be interpreted as avalanches similar to those observed in sand-pile experiments~\cite{Wiesenfeld1989}
or those discussed previously in the context of BECs and pulsar glitches. These systems have in common that after being driven to the threshold of instability, they organise
themselves and exhibit dynamics showing a power law. In the case of a type-II superconductor, the overall activity and sizes of avalanches follow this power
law~\cite{Richardson1994, Field1995}, characteristic for scale-invariant processes.

Laboratory superconductors have the advantage that self-critical behaviour can be easily visualised. In early experiments, heating was applied to a small fraction of the sample
surface (triggering a thermomagnetic instability and creating dendritic fluxtube structures), which were recorded with magneto-optical imaging~\cite{Leiderer1993}. The distinct
lightning-strike pattern has also been found in numerical simulations of several hundred fluxtubes exposed to a strong pinning landscape~\cite{Olson1997}. Today, the imaging
methods have significantly improved~\cite{Johansen2002, Altshuler2004, Eliassen2008, Mikheenko2016}, providing high-resolution pictures of the fluxtube avalanches as illustrated
in Fig.~\ref{fig-Avalanche}. However, while it has been possible to image different stages of the collective behaviour, time-resolved observations of entire avalanches
on nanosecond scales have not been achieved yet~\cite{Vestgarden2012}.

\subsubsection{Interfaces and the Meissner effect}
\label{subsubsec-InterfacesSC}

One key unknown of neutron stars is the influence of interfaces on their dynamics. Typical equations of state predict a layered structure with the crust-core
transition located at densities of about  $\sim 10^{14} \, \g \, \cm^{-3}$. While this change is expected to be smooth and not present as a sharp interface, the charged
protons have to undergo a transition from a normal resistive to a type-II superconducting state. The detailed microphysics of this are not understood but could include a
highly resistive \textit{pasta layer}~\cite{Ravenhall1983, Horowitz2015} and the formation of current sheets~\cite{Vainshtein2000, Braithwaite2015}. Moreover at higher
densities, protons prefer to be in a type-I state implying the presence of another transition region in the star's inner core, where an intermediate phase of macroscopic
normal and flux-free Meissner regions would exist~\cite{Sedrakian1995a, Alford2008}. By using superconductors to design laboratory analogues, one could take advantage of
experiments in order to shed light onto how such interfaces manifest themselves.

\begin{figure}[t]
\begin{center}
\includegraphics[width=5in]{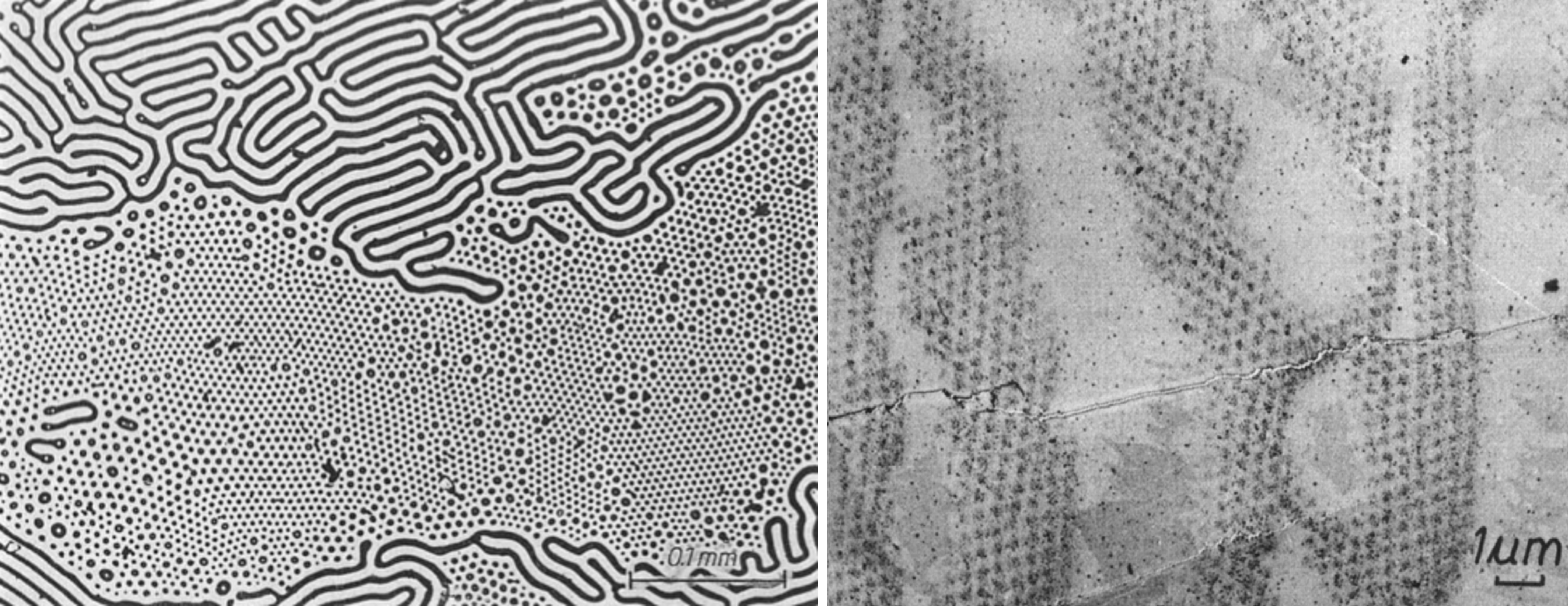}
\caption{Images of the intermediate state of a conventional type-I (left) and type-II (right) superconductor obtained with the decoration method. The type-I system (Ta)
shows regular and irregular multi-quantum flux structures, where the dark domains indicate normal-conducting behaviour. The type-II medium (Pb-Tl alloy) has $\kappa_{\rm GL}
\approx 0.73$ and similarly exhibits flux-free regions and normal ones consisting of a regular lattice structure. The left figure is reproduced with permission from Brandt
and Essmann~\cite{Brandt1987}. Copyright $\copyright$ 1987, WILEY-VCH Verlag GmbH \& Co. KGaA. The right figure is reproduced with permission from Essmann~\cite{Essmann1971}.
Copyright $\copyright$ 1971, Elsevier B.V.}
\label{fig-TypeIandII}
\end{center}
\end{figure}

The transition between the two superconducting states in the inner neutron star core is considered first. Laboratory systems exhibiting different types of superconducting
behaviour within one sample have been observed for decades. Images of the intermediate states in a conventional type-I and type-II superconductor are given in Fig.~\ref{fig-TypeIandII}.
For the former, multi-quantum fluxtubes~\cite{Huebener1974} and lamella structures surrounded by flux-free regions are observed. Ge et al.~\cite{Ge2014} have recently used
scanning Hall microscopy to determine the number of flux quanta inside such lamellae and found them to be of integer value. On the other hand, type-II systems characterised
by a low Ginzburg-Landau parameter can show features very similar to the intermediate type-I state, the main difference being the presence of a regular fluxtube lattice inside
the flux-carrying regions. Whereas in type-I media the diverse flux patterns arise due to demagnetisation effects in combination with the sample geometry (see also
Sec.~\ref{subsection-typeI_II}), the behaviour in type-II systems with $\kappa_{\rm GL} \approx 1/\sqrt{2}$ is thought to be a result of the underlying microphysics and not the
lengthscales of the Ginzburg-Landau model itself. More precisely, at $\kappa_{\rm crit}$ the attractive and repulsive contributions to the interfluxtube coupling forces cancel each
other~\cite{Kramer1971, Bogomolny1976}. This causes microscopic corrections to dominate the fluxtube interaction potential and has been suggested to lead to a small attractive
force at long scales. Combined with a repulsive force on short scales, this would cause the alternation of macroscopic flux-free and flux-containing domains seen in type-II
systems with $\kappa_{\rm GL} \approx 1/\sqrt{2}$. More recently, alternating domain characteristics have also been discovered in \textit{multi-component superconductors}. Within
the Ginzburg-Landau model, a system with $i$ components (described by complex fields $\Psi_i$) is governed by an energy functional~\cite{Babaev2013}
\begin{equation}
	f = \sum_i \left[ \alpha_i \, |\Psi_i|^2 + \frac{\beta_i}{2} \,  |\Psi_i|^4
	+ \frac{\hbar^2}{2 m_{\text{c} i}} \left| \left( \nabla - \frac{i q_i}{\hbar c} \, \mathbf{A} \right)\Psi_i \right|^2 \right]
	+ \frac{\left| \nabla \times \mathbf{A} \right|^2}{8 \pi} + f_{\rm mix},
		\label{eqn-GL2Components}
\end{equation}
where $m_{\text{c} i}$ and $q_i$ represent mass and charge of the respective Cooper pairs and $f_{\rm mix}$ contains mixed terms governing the coupling of individual components.
Employing Eqn.~\eqref{eqn-GL2Components} Babaev et al.~\cite{Babaev2005, Babaev2010} have shown that diverse flux distributions can be reproduced in systems with three or more
characteristic lengthscales if these satisfy a specific hierarchy. In the two-component case for example, this corresponds to the presence of two coherence lengths (shown
to correspond to two energy gaps in a microscopical model~\cite{Silaev2011, Silaev2012}) following an order $\xi_1 < \sqrt{2} \lambda < \xi_2$. This ultimately leads to the
coexistence of type-I and type-II behaviour which creates irregular flux patterns. Observational evidence for a two-component gap was found in MgB$_2$~\cite{Wang2001}, where
observed thermodynamical properties disagreed with a single isotropic energy gap. Experimental and numerical studies of MgB$_2$ have revealed a strong domain-like structure
that is similar to the intermediate states of conventional superconductors~\cite{Moshchalkov2009, Gutierrez2012}. Displaying fluxtube clusters and flux-free voids but being
physically distinct from the single-component type-I and type-II materials discussed above, MgB$_2$ has also been referred to as exhibiting a \textit{semi-Meissner} phase or
\textit{type-1.5 superconductivity}. Being able to study such systems in the laboratory would thus allow one to learn more about the superconductor in the high-density neutron
star interior, to investigate the formation of magnetic domains and to examine their macroscopic properties. Note that the idea of a two-gap superconducting neutron star core
was for example employed by Jones~\cite{Jones2006b}, who argues that the stars' interior should be described by a multi-component Ginzburg-Landau theory if several charged
baryonic components such as protons and hyperons were to be present.

Moreover, recent theoretical work predicts that the transition from type-I to type-II physics could be directly observable in systems with intermediate Ginzburg-Landau parameter,
because the interaction between individual fluxtubes undergoes a cross-over from attractive to repulsive as the temperature increases~\cite{Hove2002}. This thermally induced change
from a type-I state (fluxtubes clump together and build the normal conducting clusters) to a type-II state (fluxtubes are separated) illustrates the fact that the Ginzburg-Landau
parameter is temperature-dependent~\cite{Schelten1971, Auer1973, DiGrezia2008}. The behaviour of several hundred fluxtubes across a type-I/type-II interface has also been calculated
for superconducting bilayers exposed to an external field~\cite{Komendova2013}. As a result of forces competing on different lengthscales, simulations unearth a very rich palette
of mesoscopic patterns in the intermediate state (ranging from homogeneous flux distributions to fluxtube chains and flux-free zones within type-II domains). These arrangements
were present in both layers and could be controlled through the magnetic field, temperature and coupling strength between the two thin films.

While most laboratory studies are concerned with the interfaces of distinct superconducting regions, the transition between the type-II and type-I state in a neutron star is
governed by the density~\cite{Alford2008}, increasing continuously towards the stars' centre. Hence, in order to mimic the behaviour of such an interface with a terrestrial analogue,
it would be beneficial to control the type of superconductivity without creating a discontinuity. With this in mind, the work by Aegerter et al.~\cite{Aegerter2003}, studying the
influence of bismuth doping on the properties of a lead superconductor, is briefly mentioned. Whereas pure lead is characterised as a type-I medium, high doping of bismuth creates
type-II behaviour. For intermediate doping, on the other hand, both types of superconductivity can be observed depending on the temperature. An intuitive explanation for the
effects of doping is the following. An increase in the number of impurities causes a reduction in the electron mean free path and thus the coherence length. This in turn increases
the Ginzburg-Landau parameter and eventually leads to type-II characteristics. Therefore, if one could prepare a superconductor with gradually higher degrees of doping, this would
provide a promising way to test the physics of the type-II/type-I interface in a neutron star's inner core.

Finally, the normal-superconducting transition in laboratory media is addressed. In recent years, efforts have been concerned with analysing the heat and charge transport properties
of this interface~\cite{Cadden-Zimansky2006, Melin2009}. However, these studies are usually performed with mesoscopic \textit{hybrid structures}, as quantum mechanics play a crucial
role on small scales leading to new phenomena such as phase-coherent transport of electrons~\cite{Lambert1998}. While these results are less important for the macroscopic dynamics
of neutron stars, other aspects of research on the normal-superconducting interface could help to improve our understanding of the stars' magnetism. In particular, laboratory analyses
could provide information about the microscopic dynamics of the Meissner effect and growth of the superconducting phase. It is poorly understood how the superconducting state in
neutron stars is developed in the first place. Baym et al.'s standard argument~\cite{Baym1969} (based on the large conductivity of normal matter) assumes that flux cannot be expelled
from the stars' interior, which thus has to form a type-II state below the superconducting transition temperature. However, there have been various arguments from
observational~\cite{Link2003, Pons2007, Vigano2013} and theoretical~\cite{Buckley2004, Charbonneau2007, Alford2008} sides, casting doubt on the type-II scenario. Thus, experiments
and models linked to the corresponding phenomena in laboratory systems, which could improve our knowledge of neutron star superconductivity, are briefly reviewed here.

\begin{figure}[t]
\begin{center}
\includegraphics[width=5in]{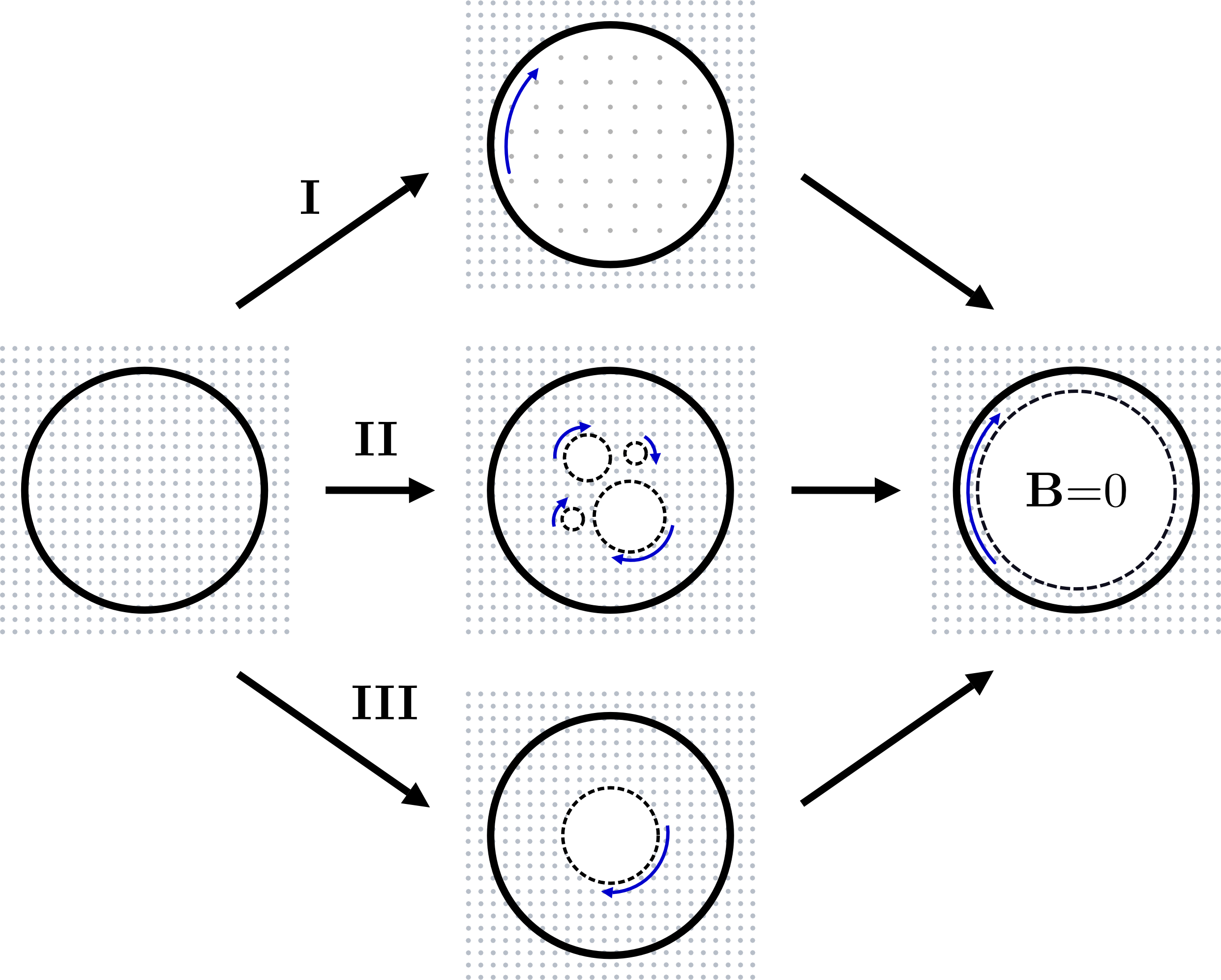}
\caption{Three evolutionary paths for flux expulsion from a cylinder, i.e.\ dynamics of the Meissner effect, as considered by Hirsch~\cite{Hirsch2015b}. The grey dots indicate
magnetic field lines coming out of the plane, whereas blue arrows mark the presence of surface currents, shielding the flux-free regions.}
\label{fig-SurfaceCurrents}
\end{center}
\end{figure}

Before proceeding, we point out that terrestrial experiments generally focus on the equilibrium state of superconductivity and study the system once the macroscopic phase has fully
developed. However, it is known that the superconducting transition occurs rather slowly, as it can take up to thirty minutes for the equilibrium to be established (see Liu et
al.~\cite{Liu1991} and references therein). This is usually explained by the fact that the superconducting-normal interface propagation is damped by the formation of eddy currents,
an interpretation based on work by Pippard from 1950~\cite{Pippard1950}. He was the first to study kinematic aspects of the superconducting transition and analysed the expansion
of the normal state into the superconducting phase, if a field larger than the critical field is applied. Pippard showed that the boundary's propagation is governed by
electromagnetic processes, as magnetic field changes in the normal region induce dissipative eddy currents slowing down the interface. For a simple plane-parallel
geometry with an external field in $z$-direction, an evolution equation for the induction $\mathbf{B} = B(x,y,t) \uz$ in the normal region can be derived by combining Maxwell's
equations and Ohm's law, i.e.
\begin{equation}
	\frac{\partial B}{\partial t} = \frac{c^2} {4 \pi \sigma\NF} \Delta B,
	\label{eqn-diffusion_pippard}
\end{equation}
where $\sigma\NF$ is the normal conductivity. Solving this diffusion equation with appropriate boundary conditions for the moving interface, the magnetic field dynamics in the
normal phase can be correctly modelled. Eqn.~\eqref{eqn-diffusion_pippard} has also been used to show that the interface propagation itself is stable~\cite{Pippard1950}.

Pippard's idea has recently been resumed by Hirsch~\cite{Hirsch2015b} to study the microphysics of the inverse problem, i.e.\ the formation of the superconducting state via Meissner
expulsion. According to Hirsch, there are three different ways for surface currents to expel flux from a cylindrical sample and establish a field-free superconducting region. These
paths are shown in Fig.~\ref{fig-SurfaceCurrents}: (I) a time-dependent current flows close to the cylinder's surface, gradually decreasing the flux density inside; (II) surface
currents shield several \textit{seed regions}, which expand until the interior is field free; (III) a single flux-free region in the centre of the cylinder expands until the
characteristic Meissner state is formed. As argued by Hirsch, path (I) is unphysical as it proceeds at non-zero magnetic field, thus not allowing the development of phase-coherence,
(i.e.\ the formation of a macroscopic quantum state). On the other hand, path (II) and path (III) are conceptually equivalent relying on the propagation of a superconducting-normal
interface. Observational evidence for the latter two was already obtained in the early stages of superconductivity research~\cite{Meissner1952, Faber1949} and Meissner himself noted
that the superconducting transition seemed to be initiated in areas of reduced magnetic field or temperature~\cite{Meissner1954}. Using Pippard's formalism, Hirsch discusses the
evolution of the superconducting phase into the normal one along paths (II) and (III). He concludes that dissipative currents can damp the interface motion but are not providing
a satisfactory explanation for the slow growth of the superconducting phase, since current theories of superconductivity seem to not explain the dissipation microscopically.

However, Pippard's linear diffusion model has been repeatedly applied to study the growth of the superconducting phase. Using appropriate boundary conditions, it was for example
shown that the planar type-I/normal interface becomes dynamically unstable to long-wavelength perturbations~\cite{Liu1991, Frahm1991}. This behaviour is similar to the instability
of the solid-liquid transition~\cite{Mullins1964, Dorsey1994} and has been confirmed by numerical analyses of the time-dependent Ginzburg-Landau model, which includes non-linear
modifications~\cite{Liu1991, Frahm1991}. These simulations have additionally shown that the expansion of the planar type-II/normal interface is stabilised by the absorption of
fluxtubes and, as expected, a triangular lattice is left behind. The time-dependent Ginzburg-Landau equations have further proven useful in analysing how a normal-metal coating of
tunable resistivity affects the magnetic properties of a two-dimensional, dirty type-II superconductor and how flux enters the sample if a magnetic field is
applied~\cite{Hernandez2002, Carty2005}. The results are again in accordance with Bean's critical state model.

As a concluding remark, we note that recent work by Martinello et al.~\cite{Martinello2015} gives observational evidence that the method of cooling the superconductor could also
significantly affect the equilibrium configuration. They found incomplete Meissner expulsion (i.e.\ the trapping of flux inside the sample), if the cooling was performed slowly,
whereas fast cooling resulted in the complete expulsion of flux. This could have implications for the formation of the superconducting state in neutron stars, since the transition
temperature is density-dependent and a spherical shell in the outer core should turn superconducting first~\cite{Ho2012b}. The gradual cooling of the neutron stars' interior below
$T_{\rm c}$ could indicate that several normal-conducting domains remain present in the outer core and the type-II state is not fully developed.


\section{Summary}

Following a brief description of the mathematical frameworks used to model laboratory condensates and neutron stars, we have outlined a number of ways in
which terrestrial condensates may be employed to improve our understanding of the dynamics of superfluid and superconducting components in neutron stars.
While not aimed at building realistic scale models (impossible due to the extreme conditions present in compact objects), terrestrial experiments could be
designed to capture characteristic features expected to affect the stars' behaviour. Prime candidates for such studies are superfluid helium, ultra-cold
gases and superconducting media. We will conclude this review by shortly summarising the advantages of the individual laboratory systems and highlighting
a number of experiments that could provide key information to improve the current state-of-the-art neutron star models. We do however point out that the
question of \textit{designing} laboratory neutron star analogues is addressed from the astrophysicists' point of view and hence our suggestions are
somewhat speculative and possibly oblivious to experimental difficulties that might arise when performing such experiments. Nonetheless, we see great
potential in using these to develop a better understanding of neutron star matter.

It was first discussed that helium is particularly useful in advancing two-fluid hydrodynamics, usually invoked to model the neutron star's rotational
evolution and in particular the glitch phenomenon. Specifically, an updated version of the spin-up and spin-down experiments originally performed by
Tsakadze and Tsakadze~\cite{Tsakadze1980} accompanied by modern vortex line simulations would be beneficial. New detailed studies of rotating superfluid
helium-4 or superfluid helium-3 would not only allow one to confirm the original laboratory glitch observations, but modern techniques should also
provide much better data which could be compared to analytical models like the ones developed by Reisenegger~\cite{Reisenegger1993} or van Eysden and
Melatos~\cite{VanEysden2011, VanEysden2014b}. In this experimental set-up, the container would mimic the solid neutron star crust, whereas the helium
superfluid would represent the neutron star's fluid core composed of the neutron condensate and the combined charged particle conglomerate. While these
analogues would ignore the presence of the superconducting component and thus not account for interactions between the neutron vortices and proton
fluxtubes, a comparison between theoretical models and laboratory observations could help to extract the key elements that influence the stars' dynamics.
This could for example give an indication on how important vortex nucleation, pinning, mutual friction and coupling to the container walls are for the
rotational evolution, providing useful to constrain the stars' superfluid properties, which are very difficult to determine with current astrophysical
observations. We would like to point out that spin-up/spin-down experiments and corresponding numerical simulations would be particularly valuable if they
were performed in spherical geometry. Despite the fact that neutron stars exhibit a high degree of sphericity, the majority of hydrodynamical models are
formulated in two dimensions (see e.g. Glampedakis et al.~\cite{Glampedakis2011a}), based on the assumption that vortices are straight and regularly
distributed. Being used for convenience and due to the lack of knowledge of more realistic vortex arrangements, these simplified models are likely to
miss important physics, which result due to the deviation from cylindrical geometry. Studying superfluid helium in rotating, spherical vessels would thus
provide more conclusive evidence on how the superfluid vortex array is distributed inside the star. This is one of the crucial factors that affects the
macroscopic, averaged models and specifically the coupling between the superfluid and the normal component. As discussed by Andersson et
al.~\cite{Andersson2007a}, the mathematical form of the vortex-mediated mutual friction force differs significantly if the superfluid is in a turbulent
regime, having an impact on various neutron star observables such as the glitch dynamics or the damping of free precession~\cite{Jones2001} and stellar
oscillations modes~\cite{Lindblom1995, Andersson2003a}. Although the spin-up and spin-down experiments described above would already represent a big step
forward, we attempt to take the idea of a spherical experiment one step further in order to account for the fact that neutron stars contain distinct
superfluid layers. By creating a two-phase sample in spherical geometry (i.e.\ using two helium-3 phases or a combination of helium-4 and helium-3 with
one component surrounding the other one to mimic the crust and the core), one could create an excellent set-up to analyse the rotational evolution of the
two neutron star superfluids. Specifically, the importance of the interface and associated phenomena such as the formation of turbulent vortex
tangles~\cite{Blaauwgeers2002, Finne2006, Walmsley2011} could be addressed in a spherical environment. Such studies should prove very valuable since
the crust-core interface is one of the elements of neutron star physics that is very poorly understood.

Chaotic superfluid flow can similarly be investigated with ultra-cold gases. BECs and Fermi gases could generally serve as excellent testing systems for
mesoscopic vortex dynamics (such as vortex-vortex interactions or pinning) due to the prospect of easily imaging the quantised structures. Further,
ultra-cold gases are particularly important to examine interacting condensates, the situation expected to be present in the neutron star interior. A
mixture of laboratory condensates could for example allow one to analyse how their interactions affect the superfluid phase transitions and vortex
formation. Coexisting ultra-cold gases could moreover provide insight into the onset of the superfluid two-stream instability~\cite{Andersson2013}
proposed to act between the neutron and proton condensate in the neutron stars' core as a possible driving mechanism for the glitch
phenomenon~\cite{Andersson2004}. Specifically extending the recent counterflow experiments by Delehaye et al.~\cite{Delehaye2015} from the linear relative
flow to the rotating regime would be beneficial, since such experiments would allow an analysis of the effects of rotation on the instability onset and
its damping. Furthermore, ultra-cold gases have the advantage that it is not only possible to analyse coexisting condensates but one may be able to
recreate non-dissipative entrainment behaviour~\cite{Kaurov2005, Hofer2012}, a key ingredient for neutron star dynamics, which cannot be easily
reproduced in superfluid helium or superconductors.

Superconductors primarily provide information about the dynamics of fluxtubes and vortices and could thus be particularly valuable for analysing pinning
and the different regimes associated with it. It could for example be achievable to perform detailed studies that give insight into the dynamics of
the inner neutron star crust, where lattice sites are expected to be non-spherical but complex structures, defects and impurities (manifest as the
\textit{pasta phase}) will be present. By designing a pinning geometry that mimics microscopic models of the inner crustal lattice~\cite{Ravenhall1983,
Horowitz2015} and studying the respective fluxtube dynamics, one could obtain the means to understand the pinning interaction of vortices at the bottom
of the crust. This would also improve our knowledge of the vortex transport properties in this region, which in turn impact on the glitch dynamics.
Secondly, we expect modern experimental techniques to open a new window to study the boundaries between the charged neutron star components, such as
the crust-core interface or the possible transition between a type-II and type-I state in the inner core. By creating samples where type-II and type-I
or type-II and normal media are spatially separated, one should be able to investigate the coupled electromagnetic characteristics in the respective layers.
This could for example provide information on how the interfaces influence the distribution of fluxtubes and help to derive a mathematical formalism for
the corresponding microphysics. Finally, charged quantum states might allow one to better comprehend the dynamical processes associated with the
transition into the superconducting phase and the resulting small-scale structure of the stars' interior magnetic field. In this context, it would be
desirable to carry out a systematic analysis of the formation of the superconducting phase in laboratory systems with $\kappa_{\rm GL} \sim \kappa_{\rm
NS}$ (see Fig.~\ref{fig-CrossSection}). This could specifically address how external parameters affect the flux distribution in the resulting
superconducting state and give more insight into the dynamics and microphysics of this process, which are very poorly constrained in the context of neutron
stars. We would expect such studies to help us better understand the nature of the metastable state at $B< H_{\rm c 1}$ thought to prevail in the interior
of most neutron stars, which is eventually related to the question if the outer core is indeed in a type-II state. This has been the general
consensus for over 40 years~\cite{Baym1969} but lacks a clear theoretical foundation and has been challenged in recent years~\cite{Link2003, Charbonneau2007,
Alford2008}. New terrestrial experiments with superconductors could thus provide inspiration for advancing our current understanding of
the neutron stars' magnetism in a similar manner as the observations of flux creep led to a generalisation of the same theory to the motion of vortices
in the neutron star crust~\cite{Alpar1984b, Alpar1984c}.

The range of promising examples demonstrates how well-known terrestrial condensates can serve as versatile analogues of neutron stars and mimic their
behaviour (albeit on much smaller lengthscales). By stimulating an exchange between the astrophysics and the low-temperature physics communities, we
hope it will be possible to identify those experiments that are within reach of current experimental limitations. A dialogue between the two research
fields should further help to identify the (most likely large number of) additional laboratory neutron star analogues that have not been considered in
this review or experiments that could already provide useful data but are not known to astrophysicists. By driving efforts in this direction, we are
optimistic that open questions in neutron star astrophysics can be answered.


\section*{Acknowledgements}

VG was partially supported by Ev. Studienwerk Villigst and NA gratefully acknowledges support from the STFC. We thank Jim Sauls for useful conversations
on a range of relevant topics.

\bibliographystyle{ws-ijmpd}

\bibliography{library_LabNS}

\end{document}